\documentstyle[12pt,psfig]{report}
\newcommand{\s}{\mathrm}
\newcommand{\bd}{\mathbf}
\newcommand{\ra}{\rightarrow}
\newcommand{\mn}{\mu \nu}

\newcommand{\be}{\begin{equation}}
\newcommand{\ee}{\end{equation}}
\newcommand{\ov}{\overline}
\newcommand{\bea}{\begin{eqnarray}}
\newcommand{\eea}{\end{eqnarray}}
\newcommand{\bef}{\begin{figure}}
\newcommand{\eef}{\end{figure}}

\newcommand{\lda}{\lambda}
\newcommand{\rpp}{\rho \pi \pi}
\newcommand{\opr}{\omega \pi \rho}
\newcommand{\D}{\Delta}
\newcommand{\eps}{\epsilon}
\newcommand{\ab}{\alpha\beta}
\newcommand{\sls}{\!\!\!/}
\newcommand{\lgl}{\langle}
\newcommand{\rgl}{\rangle}
\newcommand{\snn}{\sigma NN}
\newcommand{\wnn}{\omega NN}
\begin{document}
\addtolength{\baselineskip}{0.5\baselineskip}
\begin{titlepage}
\begin{center}
\Large
{\bf 
Photons from Ultra-Relativistic Heavy Ion Collisions
}
\vskip 2.5in
\large
{\bf
Thesis submitted in partial fulfilment of\\
the requirement for the Degree of\\
Doctor of Philosophy (Science)\\
of the \\
University of Calcutta\\
}

\vskip 3.0in
\large
{\bf
Saurabh Sarkar\\
Variable Energy Cyclotron Centre\\
Department of Atomic Energy\\
1/AF Bidhannagar, Calcutta-700 064\\
March 2000
}
\end{center}
\end{titlepage}

\setcounter{page}{0}
\pagenumbering{roman}
\centerline{\large {\bf Acknowledgements}}
\vskip .5 in

It gives me great pleasure to thank Prof. Bikash Sinha for being a constant source of
inspiration ever since I took up theoretical physics as my career. I have been very 
fortunate to have enjoyed his guidance as the Director of my Institute, VECC,
my thesis supervisor, and most importantly, as a close and warm-hearted colleague;
an experience which I will cherish throughout my life.

I express my sincere gratitude to Prof. Binayak Dutta-Roy for being a great 
teacher and an enthusiastic collaborator. Discussions with him has helped me 
to clarify numerous basic concepts of physics, particularly Quantum Field Theory.

I am very grateful to Dr. Dinesh Kumar Srivastava, with whom I have been
associated from the beginning, for always being very helpful to me.

It is a pleasure to thank 
Dr. Jan-e Alam and Dr.~Pradip Roy who have contributed immensely to
the work reported in this thesis.
I sincerely look forward to their association in future years. I am immensely
grateful to Prof. T. Hatsuda for many helpful discussions and guidance.

I also take this opportunity to thank Prof. Sibaji Raha, Dr. Abhijit
 Bhattacharyya,
Dr. Subhasis Chattopadhyay and Ms. Dipali Pal with whom I have enjoyed
working over the years. Special thanks are due for Dr. Tapan Nayak for
clarifying various aspects regarding experiments. I also express my
gratitude to Dr. Jadu Nath De, Dr.~Y. P. Viyogi and Dr. Santanu Pal
for their support and advice.

The encouragement, patience and tenderness from my wife Jui and little
daughter Sebanti  have been the essential ingredients to go through this 
task of writing my thesis.

\vskip 1.0 in
\begin{flushright}
Saurabh Sarkar\\
VECC, Calcutta\\
\end{flushright}

\newpage
\begin{center}
{\large{\bf List of Publications}}
\end{center}

\vskip 0.5in

\begin{enumerate}

\item Rapidity Distribution of Photons Emitted from a ~Hadronizing Quark
Gluon Plasma,\\ {\em S. Sarkar}, D. K. Srivastava, and B. Sinha.
 Phys. Rev. {\bf C51} (1995) 318.

\item Effects of  a Sharp Boundary on Thermal Photons from Quark Gluon 
Plasma,\\
{\em S. Sarkar}, P. K. Roy, D. K. Srivastava, and B. Sinha.\\
J. Phys. {\bf G22} (1996) 951.

\item A Scheme to Identify Transverse Collective Flow in Relativistic Heavy Ion Collisions at CERN SPS,\\
D. K. Srivastava, {\em S. Sarkar}, P. K. Roy, D. Pal, and B. Sinha,
Phys. Lett. {\bf B379} (1996) 54.

\item Soft Photons from Relativistic Heavy Ion collisions,\\
P. K. Roy, D. Pal, {\em S. Sarkar}, D. K. Srivastava, and B. Sinha,
Phys. Rev. {\bf C53} (1996) 2364.

\item Dissipative Effects in Photon Diagnostics of Quark Gluon Plasma,\\
{\em S. Sarkar}, P. K. Roy, J. Alam, S. Raha, and B. Sinha,
J. Phys. {\bf G23} (1997) 467.

\item Large Mass Diphotons from Relativistic Heavy Ion Collisions,\\
{\em S. Sarkar}, D. K. Srivastava, B. Sinha, P. K. Roy,
S. Chattopadhyay and D. Pal,
Phys. Lett. {\bf B402} (1997) 13.

\item Soft Electromagnetic Radiations from Relativistic Heavy Ion 
Collisions, \\
D. Pal, P. K. Roy, {\em S. Sarkar}, D. K. Srivastava, 
and B. Sinha,
Phys. Rev. {\bf C55} (1997) 1467.

\item  ~Thermal Masses and ~Equilibration Rates in ~Quark ~Gluon Plasma,\\
J. Alam, P. K. Roy, {\em S. Sarkar}, S. Raha, and B. Sinha,
Int. J. Mod. Phys.  {\bf A12} (1997) 5151.

\item  Quark Gluon Plasma Diagnostics in a Successive Equilibrium
Scenario,\\ P. K. Roy, J. Alam, {\em S. Sarkar}, B. Sinha, and S. Raha,
 Nucl. Phys. {\bf A624} (1997) 687. 

\item Photons from Hadronic Matter at Finite Temperature,\\
{\em S. Sarkar}, J. Alam, P. Roy, A. K. Dutt-Majumder,
B. Dutta-Roy, and B. Sinha,
Nucl. Phys. {\bf A634} (1998) 206.

\item Unstable Particles in Matter at a Finite Temperature: the
Rho and Omega Mesons,\\ J. Alam, {\em S. Sarkar}, P. Roy, B. Dutta-Roy, 
and B. Sinha, Phys. Rev. {\bf C59} (1999) 905.

\item Omega Meson as a Chronometer and Thermometer in Hot and Dense Hadronic 
Matter,\\ P. Roy, {\em S. Sarkar}, J. Alam, B. Dutta-Roy, and B. Sinha,
Phys. Rev. {\bf C59} (1999) 2778. 

\item Electromagnetic Radiations from Hot and Dense Hadronic Matter,\\
P. Roy, {\em S. Sarkar}, J. Alam, and B. Sinha,
Nucl. Phys. {\bf A653} (1999) 277. 

\item Photons from Pb+Pb and S+Au Collisions at Ultrarelativistic Energies,\\
{\em S. Sarkar}, P. Roy, J. Alam, and B. Sinha,  Phys. Rev.
{\bf C60} (1999) 054907.

\item Cosmological QCD Phase Transition and Dark Matter,\\ A. Bhattacharyya, 
J. Alam, {\em S. Sarkar}, P. Roy, B. Sinha, S. Raha, and P. Bhattacharjee,
Nucl. Phys. {\bf A661} (1999) 629c.

\end{enumerate}

\newpage
\centerline{\large{\bf Abstract}}
\vskip 0.5in

It is believed that a novel state of matter - Quark Gluon Plasma (QGP)
will be transiently produced if normal hadronic matter is subjected to sufficiently
high temperature and/or density. We have investigated the possibility
of QGP formation in the ultra-relativistic collisions of heavy ions
through the electromagnetic probes - photons and dileptons.
The formulation of the real and virtual photon production rate 
from strongly interacting matter is studied in the framework
of Thermal Field Theory. Since signals from the QGP will pick up
large backgrounds from hadronic matter we have performed a detailed study
of the changes in the hadronic
properties induced by temperature within the
ambit of the Quantum Hadrodynamic model, gauged linear and non-linear
sigma models, hidden local symmetry approach  and QCD sum rule approach.
The possibility of observing the direct thermal photons and lepton pairs
from quark gluon plasma has been contrasted with that from hot 
hadronic matter with and without medium effects for various
mass variation scenarios. 
The effects of medium induced modifications have also been incorporated in 
the evolution dynamics through the equation of state.
We find that the in-medium effects on the hadronic properties 
in the framework of the Quantum Hadrodynamic model, Brown-Rho scaling and Nambu scaling scenarios
are conspicuously visible through the low invariant mass 
distribution of dileptons and transverse momentum spectra of photons.
We have compared our evaluation of the photon and dilepton spectra
with experimental data obtained by the WA80, WA98 and CERES Collaborations
in the heavy ion experiments performed at the CERN SPS.
Predictions of electromagnetic spectra for RHIC energies have also
been made.

\newpage
\section*{List of Symbols}

\vskip 0.3 in
\renewcommand{\arraystretch}{1.2}
\begin{tabular}{ll}
$A_{\mn}$ & transverse projection tensor\\
$B_{\mn}$ & longitudinal projection tensor\\
${\bar D}_{\mn}^{0}$ & free vacuum propagator for spin 1 particles\\
${\bd D_{\mn}^0}$ & matrix propagator for spin 1 particles at finite temperature\\
$D_{\mn}^{R}$ & retarded propagator for spin 1 particles\\
${F_{\mn}}$ & field tensor for electromagnetic field\\
$f_{BE}$ & Bose distribution\\
$f_{FD}$ & Fermi distribution\\
$g_{\mn}$ & the metric tensor (1,-1,-1,-1)\\
$G^0$ & time-ordered free thermal propagator for nucleons\\
$G^H_F$ & Hartree nucleon propagator; vacuum part\\ 
$G^H_D$ & Hartree nucleon propagator; medium part\\ 
$J_\mu^h(J_\nu^l)$ & hadronic (leptonic) electromagnetic current\\
$L_{\mn}$ & leptonic tensor\\
${\cal M}$ & invariant amplitude\\
${M_B}$ & Borel mass\\
${M_N}$ & nucleon mass\\
${m_V}$ & mass of vector meson $V$\\
$M$ & the invariant mass\\
$p_T$ & transverse momentum\\
$P$ & thermodynamic pressure\\
$s$ & entropy density\\
$V$ & vector mesons\\
${\cal V}$ & three volume\\
$W_{\mn}$ & electromagnetic current correlation function\\
$Z(\beta)$ & partition function at temperature $T=1/\beta$\\
$\alpha$ & electromagnetic coupling constant, $e=\sqrt{4\pi\alpha}$\\
$\alpha_s$ & strong coupling constant, $g_s=\sqrt{4\pi\alpha_s}$\\
$\epsilon^{\alpha\beta\mn}$ & totally antisymmetric tensor,
with $\epsilon^{0123}=1$\\
$\epsilon$ & thermodynamic energy density\\
$\Gamma_V$ & width of the vector meson $V$\\
$\Omega$ & four-volume\\
$\Pi_{\mn}$ & proper self energy for spin 1 particles\\
$\rho_{\mn}$ & spectral function for spin 1 particles\\
$\varrho_{\mn}$ & non-abelian field tensor for the rho meson\\
\end{tabular}
\vskip .2in
\newpage
\tableofcontents
\newpage
\setcounter{page}{0}
\pagenumbering{arabic}
\pagestyle{myheadings}
\markright{}
\chapter{Introduction}

As per contemporary wisdom, hadrons {\it i.e.} baryons and mesons
are made up of quarks which interact by the exchange of gluons.
This interaction is governed by a non-abelian gauge theory
called Quantum Chromodynamics (QCD). Gluons, the gauge particles of QCD
unlike photons of Quantum Electrodynamics, carry colour charge and
are self interacting.
This endows the QCD interaction with remarkable properties as a 
function of the relative separation,
or equivalently, the exchanged momenta.
At short distances, or large momenta $(q)$, the effective coupling constant
$\alpha_s(q^2)$ decreases logarithmically. This means that the quarks and gluons appear 
to be weakly coupled at very short distances, a behaviour referred to as {\em asymptotic
freedom}~\cite{gross,pol}. At large separations, the effective coupling, 
it is believed, progressively turns stronger
resulting in the phenomena called quark {\em confinement}~\cite{tooft1,tooft2,wilson} which describe the 
observation that quarks do not occur isolated in nature, but only in colour
singlet hadronic bound states as mesons (quark-antiquark bound states) and
baryons (three quark bound states). In this case one also has {\em chiral symmetry 
breaking}~\cite{dashen} which expresses the fact that the quarks confined in hadrons do not appear
as nearly massless constituents but are endowed with a dynamically generated
mass of several hundred MeV. In nature there exist six quark flavours
of three colours in addition to eight bi-coloured gluons.

The QCD renormalization group calculation predicts that
as the temperature and/or density of strongly interacting hadronic matter is increased, the interactions among quanta occur effectively
at very short distances and is thus governed by weak coupling due to asymptotic freedom,
while the long range interactions become dynamically
screened due to Debye screening of colour charge~\cite{collins,km,evs}. 
The interaction thus decreases both at small distances as well as
for distances large in comparison to typical hadronic dimensions.
As a consequence, nuclear matter at very high temperature/density  exhibits neither
confinement nor chiral symmetry breaking. This new phase of QCD where the
bulk properties of strongly interacting matter are governed by 
the fundamental degrees of freedom -
the quarks and gluons, in
a finite volume is known as the state of {\em quark gluon plasma} (QGP).

There are sufficient reasons to expect that the transition between
the low and the high temperature manifestations of QCD is not smooth 
but exhibits a discontinuity, indicating the occurrence of a phase transition
at some intermediate value of temperature~\cite{hsatz,ldm}.
In this context one refers to the possibility of two different kinds of phase transitions, 
namely, the chiral symmetry restoring transition and the deconfinement
transition. 
Now, it is
necessary to realize that
though QCD is firmly established as the theory of strong interactions, the long
range behaviour of QCD is not well understood. This is because, at larger
separations the strong
coupling becomes large and invalidates perturbative approaches. As a result
the study of long range phenomena such as phase transitions and hadronic
properties and interactions using QCD is highly inhibited. One of the attempts to
study such phenomena non-perturbatively on a discrete space-time
lattice is the lattice gauge theory~\cite{kogut}.

At low energies, the QCD vacuum is characterized by non-vanishing
expectation values of certain composite operators, called vacuum condensates {\it e.g.} $\lgl\bar q q
\rgl$ which describe the non-perturbative physical properties of the QCD vacuum. They also
act as the order parameter for the chiral phase transition. 
QCD calculations on the lattice
suggest a first order (discontinuous) chiral transition for two massless quark
flavours and a second order (smooth) transition for three flavours. The chiral phase
transition is indicated by a strong drop in the value of the condensate 
$\lgl\bar q q\rgl$ at the transition temperature, $T_c$ believed to have values
between 130-160 MeV. Lattice calculations also indicate that there is a sharp change in the energy and entropy densities
at $T_c$~\cite{ukawa}. This can be interpreted as an increase in the number of degrees of freedom, 
corresponding to a deconfinement transition. Thus restoration of chiral symmetry and 
deconfinement seem to occur concomitantly.

\begin{figure}
\centerline{\psfig{figure=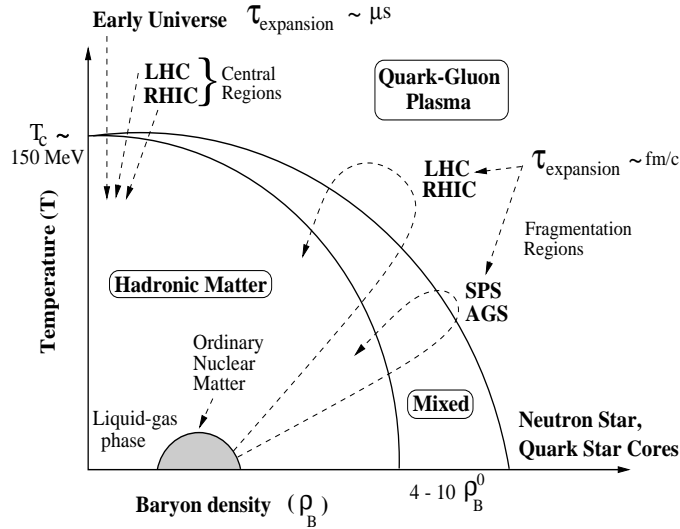,width=9cm,height=7cm}}
\caption{Phase diagram of strongly interacting matter.}
\label{phdiag}
\end{figure}

The quark-hadron phase transition is also expected to occur at high baryon
density even at zero temperature. There is however a large uncertainty in the 
critical baryon density of transition. Model calculations predict
critical densities to lie anywhere between 4 to 10 times normal nuclear
matter density. One expects a smooth connection between the high temperature 
and high baryon density phase transitions, giving rise to a continuous
phase boundary $T_c(\rho_B)$. Fig.~(\ref{phdiag}) shows a typical phase diagram
of strongly interacting matter. For $T<T_c(\rho_B)$, the effective description is in
terms of hadronic degrees of freedom, whereas for $T>T_c(\rho_B)$ the degrees of freedom
carry the quantum numbers of quarks and gluons. However, recently very
interesting theoretical developments regarding the QCD phase diagram has
taken place. At high densities, quarks may form Cooper pairs and a new colour
superconducting phase may exist. The $\rho_B-T$ phase diagram may have a critical 
or tricritical point somewhere along the phase transition line~\cite{bailin,iwasaki,raja}.

According to the
standard big bang model, the universe has gone through the QCD phase 
transition and matter is supposed to have existed as QGP
during the microsecond epoch after the big bang when the temperature was
about 100 MeV. This has important consequences in cosmology~\cite{bono}. 
The 
possible remnants that may have survived that primordial epoch till date
can provide valuable clues about the nature of the phase transition. 
A first order cosmological QCD phase transition
scenario~\cite{witten} could lead to the formation of quark nuggets made of $u$, $d$ and
$s$ quarks at a density somewhat larger than normal nuclear matter
density. Primordial quark nuggets with sufficiently large
baryon number could survive even today and could be a possible candidate for 
the baryonic component of cosmological dark matter~\cite{cosmo1}.
The QCD phase transition has its importance in astrophysics because
cold deconfined matter is also believed to
exist in the cores of neutron stars where the baryon density is
supposed to be high enough for the occurrence of the phase transition.
However, these cosmological
sources of quark matter can only provide indirect and limited information.
As a result, attempts to create 
conditions conducive for the production of QGP (energy density $\sim$ 1 GeV/fm$^3$)
in the laboratory through ultra-relativistic heavy ion collisions (URHICs)~\cite{muller,hwa,wong}
have been a major sphere of activity of physicists over the last several years. A
number of accelerators have been designed for this purpose. Experimental
data from Pb-Pb collisions at 158 GeV/nucleon at the Super Proton
Synchrotron (SPS) at CERN are in the last phases of analysis. 
The Relativistic Heavy Ion Collider (RHIC) at Brookhaven is about to
start functioning and experimental data will be
available in the very near future. 
Also, the Large
Hadron Collider (LHC) at CERN is expected to be ready
in about five years from now. Heavy ion ({\it e.g.} $^{208}$Pb and $^{197}$Au) collisions
at the RHIC and LHC with centre of mass energies 200 AGeV and 5500 AGeV
respectively are  expected to produce extremely high energy densities.
The question is whether one can produce a large enough and sufficiently long-lived
composite system so that collective and statistical phenomena can occur. An even
bigger challenge is to extract information on the dynamics and the properties of
the earliest and the hottest stage of such collisions, because even if 
QGP is produced it would only have a very transient existence. Due to 
colour confinement quarks and gluons cannot escape from the collision
and must combine to colour-neutral hadrons before travelling to the
detectors.
Hence, all signals emerging from the QCD plasma will receive substantial 
contribution from the hadronic phase. Therefore, a detailed study of the
hadronic phase is important in order to disseminate the emissions from
the QGP. 

It is necessary to understand that the main difference between
the QCD phase transition in the laboratory and in the 
early universe is that the effects of gravitation is very important
for the later. The dynamics of the 
QCD phase transition in the early universe is therefore governed by
the Einstein's equations in the Robertson-Walker space-time.
The solution of Einstein's equation with equation of states
for strongly interacting matter results in a characteristic 
time scale which is of the order of few micro seconds unlike
URHICs where time scales of the order of a few fm/c are involved.
Consequently, 
in the early universe the evolution takes place in a very
leisurely pace compared to
the interaction time scales for the quarks and gluons 
which is $\sim$ few fm/c. Therefore the QCD phase transition in the early
universe occurs in an 
environment of complete thermal equilibrium which might not be
satisfactorily realized in URHICs.

In the following Sections we will briefly
discuss the the general features of the evolution of relativistic 
heavy ion collisions and the 
proposed signals of the
quark-hadron phase transition. 

\section{Formation and Evolution of QGP in URHICs}

\begin{figure}
\centerline{\psfig{figure=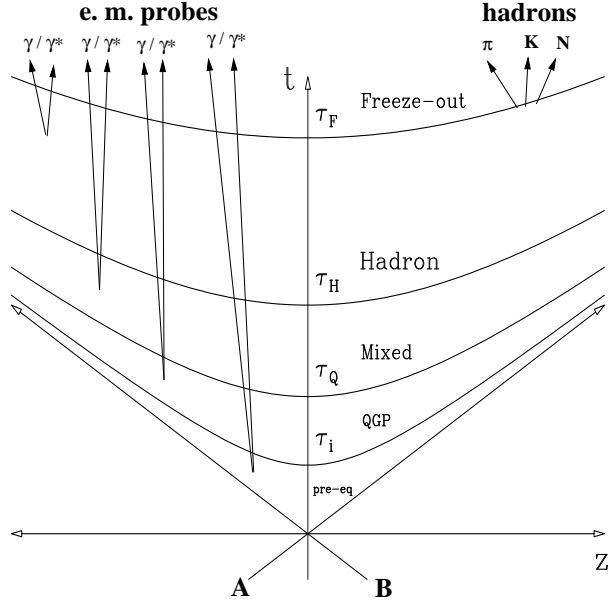,height=8cm,width=8cm}}
\caption{Space-time diagram of the collision in the $t-z$ plane.} 
\label{t-z}
\end{figure}

As the two heavy ions collide at very high energies, they deposit a substantial part of their kinetic energy
into a small region of space. 
Depending on the 
energy density achieved, the initial state of the system will be either 
in the form of a QGP or a hot/dense
hadronic gas. 
The evolution of an URHIC with an intermediate state of QGP can be 
described in terms of four regimes as discussed below. Since these vary widely in nature
the physics governing these stages are quite different. The evolution
appears as shown in Fig.~(\ref{t-z}) when projected in the plane of the
longitudinal coordinate $z$ and time $t$.
\vskip 0.15in
{\bf a) Formation of QGP:} Immediately after the collision 
of the two Lorentz contracted nuclei, the scattered partons decohere
and free stream in the longitudinal direction. Rescatterings of these
partons lead to locally thermalized QGP  after a time generally
taken to be $\sim$1 fm/c.
The value of the thermalization (realization of kinetic
equilibrium) time is still rather uncertain.
There are mainly two approaches to the microscopic description of the
process by which the fully coherent parton wavefunctions of the two nuclei
in their ground state just before the collision, evolve into a locally
thermal distribution of partons in a QGP - the QCD string breaking and the
partonic cascade. In the string breaking model colour strings formed
between mutually separating partons after the collision fragment producing
quark anti-quark pairs which eventually bring about thermalization.
Some well known realizations of this approach are VENUS~\cite{venus}, FRITIOF~\cite{fritiof}
 and RQMD~\cite{rqmd1}.
The parton cascade model~\cite{pcm} on the other hand is based on the parton picture and 
renormalization group improved perturbative QCD. Whereas the string picture
runs into conceptual difficulties at very high energy, the parton cascade
becomes invalid at lower energies where most partonic scatterings are too
soft to be described by perturbative QCD.

\vskip 0.15in
{\bf b) Expansion of the QGP:} 
Assuming that the interactions of quarks and gluons are 
sufficiently small at the temperatures achieved in URHICs, the energy density, pressure etc. can
be calculated in QCD using thermal perturbation theory.
Driven by the high internal pressure, 
the thermalized QGP expands according to the laws of relativistic
hydrodynamics~\cite{rhyd}. The hydrodynamic equations for an ultra-relativistic plasma
admit a boost invariant solution describing a longitudinally expanding
fireball with constant rapidity density~\cite{chiu,bjorken}. Relaxing the boost invariance
by assuming a Gaussian-like multiplicity density can lead to
interesting features~\cite{ssrap}. The most important question
that arises during this part of the evolution is that of chemical equilibration
of the partons. It is generally believed that gluons, because of their
larger colour degeneracy equilibrate chemically much faster than the quarks.
We have found that even light quark flavours fail to achieve chemical
equilibrium during the lifetime of the plasma~\cite{ijmpa,pkr}.

\vskip 0.15in
{\bf c) Hadronization and the mixed phase:} Expansion of the QGP proceeds till
the critical temperature $T_c$ is reached. At this instant, the phase
transition to hadronic matter starts. Through the process
of hadronization the coloured particles - quarks and gluons combine to
form colour-neutral hadrons. The order of the transition is still
somewhat controversial. Mostly, for the purpose of calculation of the
signals of QGP it is assumed that the phase transition is 
of first order. The released latent heat maintains the temperature of the system at
$T_c$ even though the system continues to expand. During this time a
coexistence of quark and hadronic matter follows by a Maxwell construction.
This mixed phase persists until all the matter has converted to the hadronic phase.
Though not generally considered in a simple picture, one realizes that
the large latent heat associated with a strong first order phase transition
may lead to deflagration waves during the expansion of the fireball along
with possible superheating and supercooling. A detailed microscopic
description of the hadronization process is yet to be achieved.

\vskip 0.15in
{\bf d) The hadronic phase and freeze-out:} The resulting hadronic matter expands
and cools as long as the system can sustain interactions. 
As we will see later in this thesis, the properties of hadrons at high temperature are modified
non-trivially due to interactions thereby changing the equation of state
and consequently the cooling process. 
Once the mean free path of the hadrons becomes comparable to 
the dimensions of the system they
decouple and free stream towards the detector. The temperature when this
occurs is the freeze-out temperature $T_f$. 
The value of the freeze-out temperature is still an unsettled
issue. The contribution to the signal output depends
substantially on the freeze-out temperature, even more so when transverse
expansion of the fireball is considered.

\section{Signals of QGP}

The size of the plasma volume is expected to be at most a few fermis in diameter.
It may live for a duration $\sim$ 1-10 fm/c out of an overall freeze-out time
of about 100 fm/c at the most. 
One realizes that once the system is
produced, its space-time evolution cannot be controlled.  In fact, the
only experimentally controllable initial parameters are the mass
numbers of the colliding nuclei and the collision energy. In addition,
a handle on the impact parameter in each collision can be obtained by
forming event classes of different multiplicities and transverse
energies with a correlation to the energies observed in the
zero-degree calorimeter. With these few controllable initial
parameters, information of the whole space-time evolution of the
system must be extracted from the various observables measured in the
final state. Regardless of whether or not  QGP is produced in the initial
stages, the system turns into a system of hadrons. Hence, the particles that 
are detected are mostly hadrons along with photons and leptons.
These hadrons, mostly light mesons like pions, kaons etc. make up the
large multiplicity in relativistic heavy ion collisions. In fact, out of
about 2500 particles created in central Pb-Pb collisions at the CERN SPS 
more than 99$\%$ have turned out to be pions.
The hadrons are emitted predominantly from the freeze-out
surface whereas photons and leptons are produced at all stages
of the evolution as indicated in Fig.~(\ref{t-z}). Being strongly interacting particles, hadrons
can provide only indirect evidence having undergone significant
reinteractions between the early collision stages and their final
observation. Leptons and photons in contrast are weakly interacting and are
considered to be direct probes.

The above discussions indicate the complexities involved in the identification
and investigation of the QGP. However, various signals have been proposed
which we will briefly discuss in the following.

\subsection{Probes of the Equation of State} 

Since hadronic matter and the QGP
is separated by a phase transition, most likely of first
order one looks for modifications in the dependence of energy density $\epsilon$,
pressure $P$, and entropy density $s$ of the evolving matter on the temperature
$T$ and/or the  baryon chemical potential $\mu_B$. One searches for a rapid rise in the
effective number of degrees of freedom, as expressed by the ratios $\epsilon/T^4$
or $s/T^4$ over a small temperature range.
However,the thermodynamic quantities $T$, $s$ and $\epsilon$ are 
not directly measured in experiments.
Usually, these are identified with the average transverse momentum $\lgl p_T\rgl$, the rapidity 
distribution of the hadron multiplicity $dN/dy$ and the transverse energy
$dE_T/dy$ respectively. When $\lgl p_T\rgl$ is plotted as a function of
$dN/dy$ or $dE_T/dy$, one expects first a rise corresponding to the increase
in the number of degrees of freedom in the hadronic phase, then a saturation
during the persistence of the mixed phase followed by a second rise when the 
change from colour-singlet hadrons to coloured partonic objects is completed~\cite{vhove}.
Another important feature which could indicate the collective nature of the 
evolution is the observation of {\it transverse flow} effects 
in the momentum spectra of final state particles.
This is due to the fact that if the lifetime of the 
produced collective system is long enough, a strong collective flow is 
generated~\cite{rhyd} and the heavier the particle is the more transverse momentum it
gains from the transverse flow. Again, non-central collisions, {\it i.e.} collisions with
non-zero impact parameter give rise to a different kind of flow - the {\it asymmetric
flow}. In this case transverse flow is generated by the pressure gradients
in the transverse plane leading to an azimuthally asymmetric flow~\cite{olit_qm97}. This
in turn causes the azimuthal angle distributions of final state hadrons
to be asymmetric. 
{\em Identical particle
interferometry} also provides
an independent source of information regarding the
space-time dynamics of the collision~\cite{heinz1}. 
By studying measured two-particle ({\it e.g.} $\pi\pi$, $KK$ or $NN$)
correlation functions in different directions of phase space, it is possible
to estimate the transverse and longitudinal size, the lifetime and flow patterns of
the hadronic fireball at the moment of freeze-out. The transverse sizes
measured in URHICs are found to be larger than the radii of the incident
particles clearly indicating that the produced hadrons have rescattered 
before emission~\cite{prat_qm97}.

\subsection{Signatures of Chiral Symmetry Restoration}

{\em Strangeness enhancement} and increase in antibaryon production 
relative to p-p or p-A collisions
are some of the proposed signatures of chiral symmetry restoration. The basic
argument in both cases is the lowering of the threshold for production
of strange hadrons and baryon-antibaryon pairs~\cite{rafel}. An optimal signal
is obtained by considering strange antibaryons which combine both effects.
The enhancement of strangeness in URHICs in simple terms is connected
to the fact that in an ideal baryon-dense quark matter the production of
$s\bar s$ is enhanced compared to that of light quark flavours $u$ and $d$
because the Fermi energy of the already present light quarks is higher that
the strange quark mass. In Pb-Pb collisions at the CERN SPS, an enhancement in strangeness
production by about a factor 2 has been observed from the $\lgl K+\bar K\rgl
/\lgl\pi\rgl$ ratio~\cite{odyn_qm97}. Also, a clear specific enhancement
in the yield of $\Omega^-+\Omega^+$ per negative hadrons by a factor $\sim$ 10
have been observed by WA97~\cite{WA97} relative to p-p and p-Be collisions. It
is believed that as multistrange hadrons are difficult to produce due to
high mass-thresholds, the strangeness increase could have an origin
at the partonic level before hadronization. It is believed that
domains of {\it disoriented chiral condensate} (DCC) could
provide a more direct signal for the restoration of chiral symmetry in URHICs~\cite{anselm}.
These correspond to isospin singlet, coherent excitations of the pion field
which would decay into neutral and charged pions in such a way
that there is a significant possibility of observing a large surplus of
charged pions over neutral pions in certain regions of phase space. 
Restoration of spontaneously broken chiral symmetry is also
reflected in the thermal modification of the hadronic spectral 
function~\cite{pisarski1,pisarski2,hk1,plb185,hatsuda1} particularly
through the mass shift of the vector mesons in hot/dense medium.
These modifications can be studied by analyzing photon, dilepton as well
as hadronic spectra. 

\subsection{Probes of Colour Deconfinement} 

The {\it suppression of} $J/\Psi$ production is
considered as a direct probe of the colour deconfinement phase transition. 
The $J/\Psi$ is a bound state of a $c\bar c$ pair dominantly produced by the fusion of 
gluons. In a deconfined environment like a QGP, the binding of a $c\bar c$
into a $J/\Psi$ is suppressed due to the fact that the screening length
is less than the bound state radius~\cite{matsui}. On the other hand, the $J/\Psi$ may also be
suppressed in a hadronic scenario due to nuclear absorption.
This description is a probabilistic one where one assumes that the probability
that a produced $J/\Psi$ escapes without making collisions is $\propto \exp(-L/\lambda)$
where $\lambda=1/\sigma_{\s abs}n_0$, $n_0$ is the nuclear density and $\sigma_{\s abs}$
is the cross section which controls the probability that the $J/\Psi$ gets destroyed 
in a nuclear collision. The quantity $L$ is the average length travelled by the $J/\Psi$ in
nuclear matter, is related to the transverse energy of the collision. This
picture explains both p-A data as well as A-A data upto the S-U system.
However, the Pb-Pb data from the CERN SPS has created a great deal of
excitement~\cite{blz_qm97}. This is due to the discontinuity observed in the ratio of $J/\Psi$
to the Drell-Yan cross section at $L\sim$ 8 fm (which corresponds to $E_T\sim$
50 MeV and impact parameter $b\sim$ 8 fm). Nuclear absorption apparently can not explain this discontinuity
and one needs to invoke partonic degrees of freedom and colour confinement.
It should be mentioned that there have been other attempts to explain this striking
feature in a hadronic description. It has been argued that even if the $J/\Psi$
 escapes the nuclei, it can be destroyed at a later stage
of the collision due to scattering on other produced particles, commonly
referred to as comovers. A trustable estimate of such contributions is still lacking.

Another possible way of probing the colour structure of the produced matter is by
studying the energy loss of a fast parton, also known as {\it jet quenching}~\cite{jquench1,jquench2}.
The parton loses its energy either by excitation of the penetrated medium
or by radiation.
The magnitude of the energy loss is proportional to the strong coupling
constant $\alpha_s^2$. It has been observed that the energy loss of a parton jet
is greater in A-A collisions than in p-p or p-A. Theoretical estimates
have inferred that the energy loss of a parton in hadronic matter ($-dE/dx\sim$1 GeV/fm)
is much more than in QGP ($-dE/dx\sim$0.1-0.2 GeV/fm). 
In this respect, since jets in hadronic matter is suppressed more in hadronic matter than
in QGP, jet ``unquenching'' is a signal of deconfinement.

\subsection{Electromagnetic Probes} 

These include real~($\gamma$) and virtual~($\gamma^*$) photons.
Virtual photons decay producing pairs of leptons. Photons and dileptons
 are considered
the cleanest signals of quark gluon plasma~\cite{mclerran}. Because of the very nature of
their interaction, they decouple immediately and leave the system without any
distortion of their energy-momentum carrying with them the information
from within the reaction zone. Hence, electromagnetic probes
are the only {\em direct} probes of QGP. They are emitted at all stages of
the evolution but do not get masked by the details of the evolution
process. Their production cross-section is strongly dependent on
temperature and hence are copiously produced from the hot phase of
the evolution. However, the photons and dileptons from the thermal phase
of the evolution, referred to as {\it thermal photons and dileptons} have
to compete with a large background~\cite{janepr,cassing,pvr,qm93} due to production from many other
mechanisms and this complicates the process of extraction of information 
about the basic thermal system we intend to study. The general features
of the photon and dilepton spectra will be discussed in the following Section. 

\section{Real and Virtual Photons: General Features}

Let us briefly
discuss the various sources of production of photons and dileptons
during various stages of an URHIC and the specific domains of phase
space where they are known to dominate.
Let us first consider {\em real photons}.
In the low transverse momentum region
the contribution from pseudoscalar meson decays involving $\pi^0$ and $\eta$
clearly dominate photon production. In fact they account for nearly 95$\%$ of
the total photon yield in a collision. Photons emitted due to primary 
interactions among the partons of the 
colliding nuclei form the principal background in the large transverse
momentum region. These are called prompt or QCD photons. The decay photons
can be isolated experimentally by invariant mass analysis and the prompt ones
can be accurately estimated by perturbative QCD calculations. 
These contributions are then subtracted out to get the thermal
photon spectra. Thermal photons from the QGP phase arise mainly due to
the QCD Compton $(qg\ra q\gamma)$ and annihilation
$(q\bar q\ra g\gamma)$ processes. The emission rate resulting
from these reactions turn out to be infra-red divergent and can be
evaluated~\cite{kapusta,baier,aurenche} in the framework of Hard Thermal Loop (HTL)~\cite{braaten,frenkel}
resummation in QCD. The hadronic matter is mainly composed of the 
pseudoscalar-isovector pion ($\pi$) and the spin-isospin vector rho ($\rho$) mesons. 
One also includes the vector-isoscalar omega ($\omega$), the pseudoscalar-isoscalar eta ($\eta$)
and the axialvector-isovector $a_1$ mesons. Photons from 
the hadronic matter are emitted from reactions of the type
$hh\ra h\gamma$ (where $h$ is one of the hadrons $\pi$, $\rho$ and $\eta$)
as well as from the decay of short-lived hadrons~\cite{kapusta}.
In a phase transition scenario these are emitted during the later stages of the collision
when the system has cooled to temperatures below $T_c$. Hence the photon spectra due to
the thermal hadronic phase is expected to have a steeper slope
compared to the QGP.
In Fig.~(\ref{gama_sche}) we have schematically shown the different contributions to
the overall photon yield in a relativistic collision of heavy ions as a function
of the transverse momentum of the emitted photons. One observes that the 
higher the transverse momentum of the photon the earlier they are produced
in the collision process.
\bef
\centerline{\psfig{figure=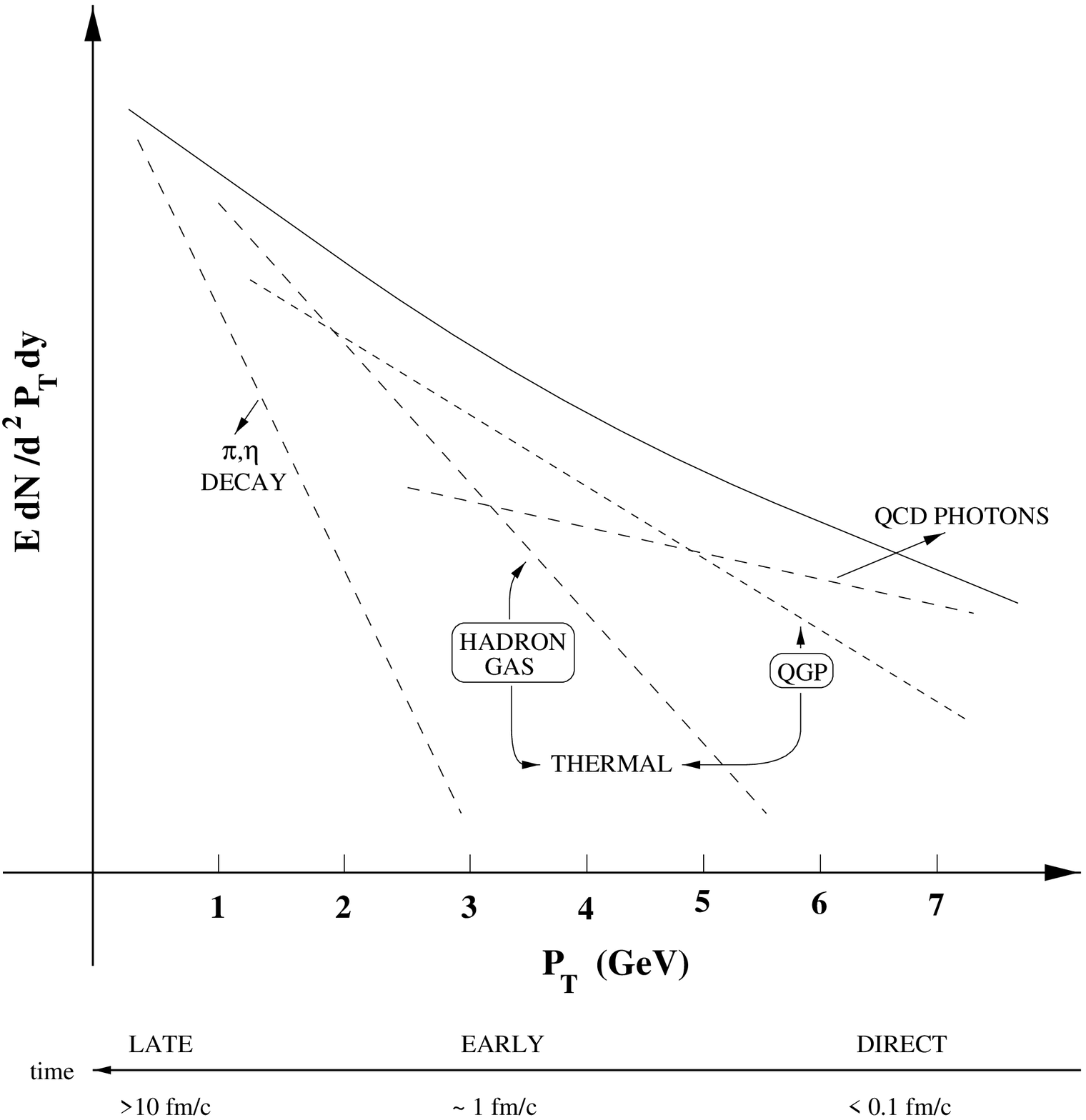,height=8cm,width=9cm}}
\caption{
Schematic diagram showing the different contributions to the total
photon yield in a ultra-relativistic heavy ion collision.
}
\label{gama_sche}
\eef

The different contributions to the production
of lepton pairs as a function of their invariant mass
are shown schematically in Fig.~(\ref{dil_sche}). 
The principal source of thermal {\em dileptons} from the QGP and hadronic phases 
are the quark-antiquark and pion annihilation processes respectively.
In the high mass region these compete with Drell-Yan pairs 
which are produced in primary interactions between incoming partons in the very
early stages of the collision.
The $J/\Psi$ peak marks the cut-off scale for thermal pairs from the plasma.
Around this region the decays of $D$ and $B$ mesons also become an important
source. The vector mesons which decay both during the expansion and after
freeze-out can be identified easily from their characteristic peaks in the
spectrum. Below the vector mesons, Dalitz decays of 
$\pi^0$, $\eta$, $\eta\prime$
and $\omega$ mesons provide the dominant source for dilepton production.
The fact that higher mass lepton pairs are produced earlier in the collision
process is evident from the time axis in Fig.~(\ref{dil_sche}).
We emphasize that whereas the dileptons from hadronic matter have distinct
features like the $\rho$, $\omega$ and $\phi$ peaks, 
the photon spectra is completely structureless. 
Hence it is very difficult to disentangle the thermal photons emitted from hadronic matter from 
those which have their origin in 
quark matter. In this case an accurate estimation of thermal photons
originating from hadronic matter is of utmost importance in order to 
comment on the formation of QGP in relativistic heavy ion collisions. Such
an estimation must incorporate medium modifications of hadronic properties in
the evaluation of emission rates as well as in the equation of state (EOS) of
the interacting hadronic matter. 
The change in the mass and decay width of a vector meson propagating in
a medium occurs due to its interaction with the real and virtual
excitations in the medium. 
Medium induced modifications of 
the properties of vector mesons, for example, the $\rho$ and $\omega$ are likely
to show up clearly in the invariant mass spectra of dileptons through the shifting and/or
broadening of the respective peaks. 
In the following Section we will give a brief
introduction to the study of hadronic properties in a thermal medium.
\bef
\centerline{\psfig{figure=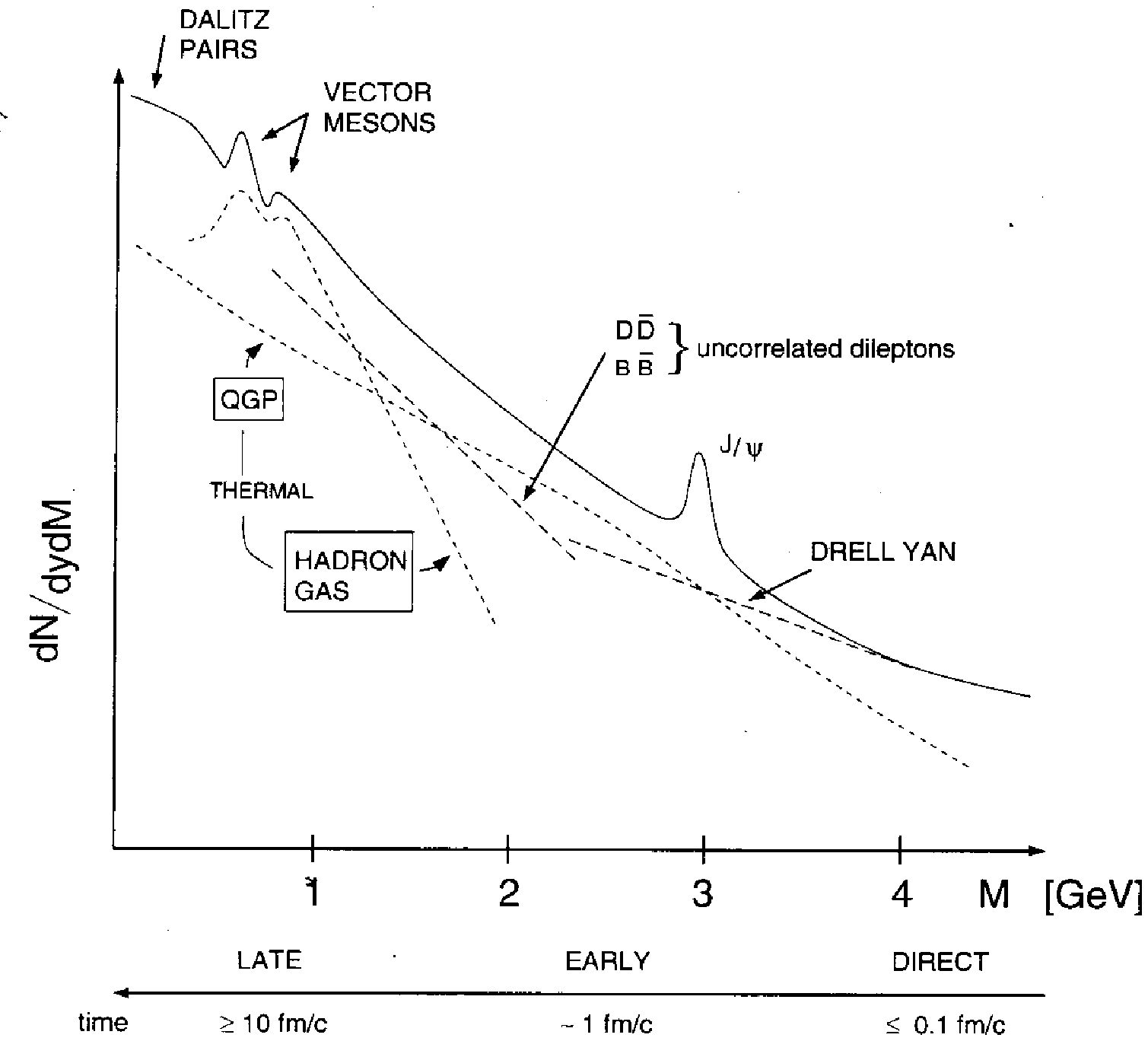,height=8cm,width=9cm}}
\caption{
Schematic diagram showing the different contributions to the total
dilepton yield in a ultra-relativistic heavy ion collision~\protect\cite{lect}.
}
\label{dil_sche}
\eef

\section{Medium Effects on Hadronic Properties}

In URHICs hadronic matter is expected to be formed
after a phase transition from QGP when the plasma has cooled beyond
the phase transition temperature $T_c$. Even if such a
phase transition does not occur, realization of hadronic matter at
high temperature ($\sim$ 150 -- 200 MeV) and/or baryon density (a few
times normal nuclear matter density) is inevitable. As a result
the study of hadronic interactions at high temperature and density
assumes great significance. 
However, progress in our understanding of hot and 
dense hadronic matter has
been retarded since the underlying theory of  strong interaction, QCD,
is nonperturbative in the low energy regime. This severe 
constraint has lead to considerable amount of work on model 
building~\cite{meissner,nowak,brhcnp,book}
in order to study the low energy hadronic states. 

The principal contention of this thesis is to study the medium 
effects on thermal photon and dilepton spectra. We will be
mainly concerned with the vector mesons $\rho$ and $\omega$.
Various investigations have addressed the issue of temperature and density
dependence of hadronic spectra  within different models over the
past several years.  
According to  Brown and Rho~\cite{rho} the requirement of chiral symmetry 
(in particular the QCD
trace  anomaly)  yields  an  approximate  scaling relation between
various effective hadronic masses,
which  implies that  all
hadronic   masses   decrease equally with temperature. 
The reduction in $\rho$ meson mass has also been observed
in the  gauged linear sigma model~\cite{rdp} at low temperature; however, 
near the chiral transition point it shows an upward trend.
The nonlinear sigma model claimed to be the closest
low energy description of QCD shows the opposite
trend, {\it {i.e.}} the effective $\rho$ mass increases with temperature~\cite{cs}. A 
similar qualitative behaviour of the $\rho$ mass has been observed in the  
hidden local symmetry approach~\cite{harada}. 
The relation between the self energy and the forward
scattering amplitude has also been utilized to study the 
change of hadronic properties in the medium~\cite{st,eik},
where the effects of non-zero temperature is rather small.

 In the Quantum Hadrodynamic Model (QHD) of Walecka~\cite{vol16,chin}
scalar and vector condensates generated by the nucleon sources are
themselves responsible for the modification of the nucleon mass. 
The vector meson mass gets shifted due to the 
decrease of the nucleon mass which appears through thermal loops in the
vector meson self energy \cite{sxk}, the imaginary part of which 
characterizes the response of the nuclear system to external (electromagnetic)
 probes.

In-medium  QCD sum rules 
are useful  to make constraints on the hadronic spectral functions 
at finite temperature and density~\cite{thatsuda}.
In the QCD sum rule approach the hadronic properties are related
to the scalar and tensor condensates of quark and gluon fields.
Due to lack of understanding of the behaviour of these condensates
near the critical point, the hadronic properties at finite temperature
in this approach is not firmly established.
The spectral function of vector mesons can be parametrized in vacuum from 
the experimental data obtained in $e^+e^-\rightarrow
hadrons$ (or from hadronic decays of 
the $\tau$ lepton) for various isovector and isoscalar
channels~\cite{shuryak,shifman}. The in-medium 
spectral function of the vector meson is then obtained  by modifying 
the pole and the continuum structure 
as a result of its interaction with the constituents of the thermal bath.
The vector
meson masses and the continuum thresholds 
are taken to vary with temperature according to Brown-Rho (BR)
and Nambu scaling~\cite{brpr} scenarios.

So we see that a wide range of variation of hadronic
properties with temperature are predicted by the models cited above. 
Other models {\it e.g.}, those 
proposed by Rapp {\it et al}~\cite{rcw} and by Klingl
{\it et al}~\cite{klingl}, where the effects of non-zero
baryon density (baryonic chemical potential) is dominant over 
non-zero temperature will not be discussed. 
This is because we intend to study the hot baryon free (central rapidity)
region of URHICs.
We shall consider various scenarios for the shift in the hadronic 
spectral function at finite temperature and evaluate its effects on the 
experimentally measurable quantities, the photon and dilepton
spectra originating from a thermalized system formed in URHICs.

\section{Organization of the Thesis}

The thesis is organized as follows. In Chapter~2 we shall discuss the
formalism of photon and dilepton production from a medium of interacting particles at
finite temperature and/or density starting from first principles using perturbation theory.
The rates of photon production from partonic interactions in the QGP 
as well as from hard primary interactions of partons will be considered.
Chapter~3 is devoted to the study of spectral modifications of
vector mesons in the medium and the evaluation of static (fixed temperature)
 rates of
photon and dilepton production from hadronic matter within various models.
The medium masses of the vector mesons are calculated from the pole positions of the
effective propagators which are
obtained in terms of 
the self energies calculated in the framework of thermal field
theory using a few well-known models. 
We have considered the QHD model, the gauged linear sigma model,
the gauged non-linear sigma model, and the hidden local symmetry and QCD sum rule 
approaches. 
The space-time evolution of these static rates using
relativistic hydrodynamics is discussed in Chapter~4.
The medium effects in the space-time evolution enters through the
equation of state which is manifested in the cooling rate as well as in the estimation of the initial temperature
of the produced matter. The final photon multiplicity with and without a QGP 
in the initial state is obtained. These are compared with data obtained 
by the WA80, WA98 and CERES experiments.  Chapter~5 contains the thesis  summary 
and related discussions. 
We have listed the invariant amplitudes of all the photon producing
hadronic reactions and decays that we have used 
in the Appendix.

\chapter{Formulation of Electromagnetic Emission Rates}

Electromagnetically interacting particles - photons and dileptons are excellent probes
of the thermodynamic state  of evolving strongly interacting matter
likely to be produced in ultra-relativistic nucleus-nucleus collisions.
This is because electromagnetic interactions are strong enough to lead to
a detectable signal and yet are weak enough to let the emitted photons
and leptons escape from the finite nuclear system without further
interactions. Hence the spectra of photons and dileptons can provide information
of the properties of the constituents from which they were emitted.

For most purposes  
the  emission rates of photons and dileptons
can be calculated in a classical framework.
It was shown by Feinberg~\cite{feinberg}
that  the  emission  rates  can   be   related   to   the
electromagnetic current  correlation  function  in  a
thermalized  system  in  a  quantum  picture   and,   more
importantly,  in  a nonperturbative manner. 
Generally, the production rate of a particle which interacts weakly 
with the constituents of the thermal bath (the constituents 
may interact strongly among themselves, 
the explicit form of their coupling strength is not important)
can always be expressed
in terms of the discontinuities or imaginary parts of the self
energies of that particle~\cite{ruuskanen,bellac}. 
In this Chapter we will make a detailed study of the
connection between the emission  rates of real
and virtual photons
and the spectral function of the photon which is 
connected with the discontinuities in self energies 
in a thermal system~\cite{weldon90}. 
This  in turn is connected to the electromagnetic current
correlation function~\cite{mclerran} through Maxwell equations. 
Alternatively, using the  kinetic theory approach the photon
emission rates can be written in terms of the equilibration rate of photons in a
thermal bath. The imaginary part of the photon self energy tensor in this case
is related to the exclusive kinetic rates of emission and absorption
in the system. 
We will discuss these facets regarding the formulation of
the photon and dilepton emission rates from a thermal medium  
extensively in Section~2.2 after a brief review of propagators in thermal
field theory in Section~2.1. The invariant rates 
of emission of both hard and soft thermal photons and dileptons
from QGP will be dealt with in Section~2.3 where we will compare
the static (fixed temperature) rates due to different processes contributing
to the thermal yield. 

\section{Review of Thermal Propagators}

As is well known, propagators play a central role in the description of
the dynamics of systems of particles using quantum field theory.
In this Section we will briefly discuss
the in-medium (thermal)
propagators in Thermal Field Theory~\cite{bellac,adas,kapustaft,landsman}
 which will be used extensively in
the thesis. We will begin by first defining the
propagators in vacuum.

The free propagator of a complex scalar field $\phi$ propagating with
a momentum $p$ in vacuum is defined as
\bea
i\bar\D^0(p)&\equiv&\int\,d^4x\,e^{ip\cdot x}\langle T\{\phi(x)\phi^*(0)\} \rangle_0
\nonumber\\
          &=&\frac{i}{p^2-m^2+i\eps}\,\,\,\,\,\,\,\,;\,\eps\ra 0^+
\label{sprop0}
\eea
The operator $T$ appearing within angular brackets ensures that the field
operators are time-ordered and the subscript `0' indicates that there are
no interactions.
The vacuum propagator for fermions is defined as
\bea
i\bar G^0_{\ab}(p)&\equiv&\int\,d^4x\,e^{ip\cdot x}\langle T\{\psi_\alpha(x)\bar\psi_\beta(0)\} \rangle_0
\nonumber\\
          &=&\frac{i(p\!\!\!/ + m)_{\ab}}{p^2-m^2+i\eps}
\label{fprop0}
\eea
where $\alpha$ and $\beta$ denote the spinor indices of the fermion field $\psi$.
In a similar way the free propagator of
a vector field $A^\mu$ is defined as
\be
i\bar D^0_{\mn}(p)\equiv\int\,d^4x\,e^{ip\cdot x}\langle T\{A_\mu(x)A_\nu(0)\} \rangle_0
\label{vprop0}
\ee
Depending on the nature of the quanta of the vector field $A^\mu$ the vector
propagator can take the following forms.
In the case of a vector particle of mass $m$ we get
\be
i\bar D^0_{\mn}(p)=\frac{i(-g^{\mn}+p^\mu p^\nu/m^2)}{p^2-m^2+i\eps}
\ee
which describes the free propagation of the $\omega$ meson, for example. 
For charged vector particles {\it e.g.} the $\rho$ meson,
the fields $A_\mu$ also carry isospin indices and we have
\be
i\bar D_{\mn}^{0\,ij}(p)=\frac{i\,\delta^{ij}\,(-g^{\mn}+p^\mu p^\nu/m^2)}{p^2-m^2+i\eps}
\ee
where $i$ and $j$ denote components in isospin space.
In the case of massless vector fields corresponding to photons or gluons the 
propagator will contain a gauge parameter as a reminder of the arbitrariness
of the gauge-fixing condition. The photon propagator in vacuum is obtained as
\be
i\bar D^0_{\mn}(p)=\frac{-i[g^{\mn}+(\xi-1)p^\mu p^\nu/p^2]}{p^2+i\eps}
\ee
Some well known choices are, $\xi=1$ which is the Feynman (or Lorentz)
gauge and $\xi=0$, the Landau gauge.
For gluons we need to add a Kronecker delta is colour space so that
the free gluon propagator is
\be
i\bar D_{\mn}^{0\,ab}(p)=\frac{-i\,\delta^{ab}\,[g^{\mn}+(\xi-1)p^\mu p^\nu/p^2]}{p^2+i\eps}
\ee
The gauge parameter $\xi$ must be absent from any physical quantity we calculate.

It must be noted that the fields $\phi$, $\psi$ and $A_\mu$ appearing above
are free fields and the expectation values are
calculated between noninteracting vacuum states. The propagators defined 
through Eqs.~(\ref{sprop0}), (\ref{fprop0}) and (\ref{vprop0}) are
referred to as Feynman propagators.

In the presence of interactions these propagators have to be redefined
with interacting Heisenberg fields in place of the free fields
and interacting vacua instead of the free vacua.
The interacting propagator can be expressed in terms of the bare (non-interacting) 
propagator using perturbation theory. 
In the scalar case, for example, the exact propagator in the presence
of interactions, $\bar \D$, is 
obtained as
\be
\bar\D=\bar\D_0 + \bar\D_0 \Pi \bar\D
\ee
Where, $\Pi$ is the self energy of the particle due to interactions.
This equation is known as the Dyson-Schwinger equation for propagators.

Let us now study the situation in a medium at finite temperature (and density).
We will be interested in a system in thermal equilibrium. Hence we will 
assume that the interaction slowly switches off as we go into the
remote past and the fields become noninteracting fields
satisfying the free equations of motion.
These fields appear in the definition of the free propagators in the
medium. The thermal propagator has more
structure than the vacuum case as a result of different combinations
of time-ordering on the real time contour~\cite{bellac,adas,kobes}.
In the real time formalism there are four non-trivial propagator structures
possible which are collected in a 2$\times$2 matrix~\cite{nieves,pvl}. 
For scalars the free thermal propagator is defined as
\renewcommand{\arraystretch}{1.7}
\bea
i\bd \Delta_0 &\equiv& \left[ \begin{array}{ll}
i\Delta^{11}_0(p)& 
i\Delta^{12}_0(p)\\ 
i\Delta^{21}_0(p)& 
i\Delta^{22}_0(p)\\
\end{array}\right] \nonumber\\
&=&\left[ \begin{array}{ll}
\int\,d^4x\,e^{ip\cdot x}\langle T\{\phi(x)\phi^*(0)\}\rangle_T^0 &
\int\,d^4x\,e^{ip\cdot x}\langle\{\phi^*(0)\phi(x)\}\rangle_T^0\\
\int\,d^4x\,e^{ip\cdot x}\langle\{\phi(x)\phi^*(0)\}\rangle_T^0 &
\int\,d^4x\,e^{ip\cdot x}\langle\bar T\{\phi(x)\phi^*(0)\}\rangle_T^0 \\
\end{array}\right]
\label{s0propft}
\eea
where the operator $\bar T$ denotes anti-time-ordered product. The subscript `$T$'
indicates that a thermal average is being performed.

In order to obtain the thermal propagators in momentum space one
follows the usual procedure of expanding the field operators in terms
of the creation and annihilation operators and making use of the 
commutation relations between them.
The four components are then obtained as
\bea
\Delta^{11}_0(p)&=&\frac{1}{p^2-m^2+i\eps}
-2\pi i\delta(p^2-m^2)\eta(p\cdot u)\nonumber\\ 
\Delta^{12}_0(p)&=&-2\pi i\delta(p^2-m^2)[\eta(p\cdot u)+\theta(-p\cdot u)]\nonumber\\
\Delta^{21}_0(p)&=&-2\pi i\delta(p^2-m^2)[\eta(p\cdot u)+\theta(p\cdot u)]\nonumber\\
\Delta^{22}_0(p)&=&\frac{-1}{p^2-m^2-i\eps}
-2\pi i\delta(p^2-m^2)\eta(p\cdot u)
\label{s0propft1}
\eea
where $\eta(p\cdot u)=\theta(p\cdot u)f_{BE}(z)+\theta(-p\cdot u)f_{BE}(-z)$.  
$f_{BE}=[e^z-1]^{-1}$ is the Bose distribution 
with $z=(p\cdot u-\mu)/T$, $u^{\mu}$ is
the four velocity of the thermal bath and $\mu$ is the chemical potential.
We observe that the elements of the matrix propagator $\bd \D_0$ are not independent. From
their definitions one can see that $\D_0^{11}$ and $\D_0^{22}$ can be expressed in terms of
$\D_0^{12}$ and $\D_0^{21}$. Also, the Kubo-Martin-Schwinger~\cite{kms} periodicity
condition yields $\D_0^{12}(p)=e^{-z}\D_0^{21}(p)$.

Similarly, for fermions the corresponding thermal propagators are,
\bea
i\bd G^0_{\ab} &\equiv& \left[ \begin{array}{ll}
iG_{\ab}^{0(11)}(p)& 
iG_{\ab}^{0(12)}(p)\\ 
iG_{\ab}^{0(21)}(p)& 
iG_{\ab}^{0(22)}(p)\\
\end{array}\right] \nonumber\\
&=&\left[ \begin{array}{ll}
\int\,d^4x\,e^{ip\cdot x}\langle T\{\psi_\alpha(x)\bar\psi_\beta(0)\}\rangle_T^0 &
-\int\,d^4x\,e^{ip\cdot x}\langle\{\bar\psi_\beta(0)\psi_\alpha(x)\}\rangle_T^0\\
\int\,d^4x\,e^{ip\cdot x}\langle\{\psi_\alpha(x)\bar\psi_\beta(0)\}\rangle_T^0 &
\int\,d^4x\,e^{ip\cdot x}\langle\bar T\{\psi_\alpha(x)\bar\psi_\beta(0)\}\rangle_T^0 \\
\end{array}\right] 
\label{f0propft}
\eea
Explicitly, the four components are
\bea
G_{\ab}^{0(11)}(p)&=&(p\sls+m)_{\ab}\left[\frac{1}{p^2-m^2+i\eps}
+2\pi i\delta(p^2-m^2)\eta(p\cdot u)\right]\nonumber\\ 
G_{\ab}^{0(12)}(p)&=&2\pi i(p\sls +m)_{\ab}\delta(p^2-m^2)[\eta(p\cdot u)-\theta(-p\cdot u)]\nonumber\\
G_{\ab}^{0(21)}(p)&=&2\pi i(p\sls +m)_{\ab}\delta(p^2-m^2)[\eta(p\cdot u)-\theta(p\cdot u)]\nonumber\\
G_{\ab}^{0(22)}(p)&=&(p\sls+m)_{\ab}\left[\frac{-1}{p^2-m^2-i\eps}
+2\pi i\delta(p^2-m^2)\eta(p\cdot u)\right] 
\label{f0propft1}
\eea
where,
$\eta(p\cdot u)=\theta(p\cdot u)f_{FD}(z)+\theta(-p\cdot u)f_{FD}(-z)$,
$f_{FD}=[e^z+1]^{-1}$, the Fermi-Dirac distribution.
For the fermions the KMS anti-periodicity condition leads to
$G^{12}_0=-e^{-z}G^{21}_0$. 

Lastly, we define the finite temperature propagators for vector particles:
\bea
i\bd D^0_{\mn} &\equiv& \left[ \begin{array}{ll}
iD_{\mn}^{0(11)}(p)& 
iD_{\mn}^{0(12)}(p)\\ 
iD_{\mn}^{0(21)}(p)& 
iD_{\mn}^{0(22)}(p)\\
\end{array}\right] \nonumber\\
&=&\left[ \begin{array}{ll}
\int\,d^4x\,e^{ip\cdot x}\langle T\{A_\mu(x)A_\nu(0)\}\rangle_T^0 &
\int\,d^4x\,e^{ip\cdot x}\langle\{A_\nu(0)A_\mu(x)\}\rangle_T^0\\
\int\,d^4x\,e^{ip\cdot x}\langle\{A_\mu(x)A_\nu(0)\}\rangle_T^0 &
\int\,d^4x\,e^{ip\cdot x}\langle\bar T\{A_\mu(x)A_\nu(0)\}\rangle_T^0 \\
\end{array}\right]\, .
\label{v0propft}
\eea
As before, one will have additional
indices corresponding to colour or electric charge; henceforth
we will not mention them explicitly.
Apart from the Lorentz indices, the
explicit forms of the thermal propagators will be similar to
the scalar case. For neutral particles, the chemical potential $\mu$
will be absent in the definition of the phase space factor $\eta(p\cdot u)$.

It is important to note that the real time propagators as given
by Eqs.~(\ref{s0propft1}) and (\ref{f0propft1}) consist of two
parts  - one corresponding to the vacuum, describing the exchange 
of virtual particles and the other, the temperature dependent part,
describing the participation of real (on-shell) particles present in the
thermal bath in the emission and absorption processes. The temperature
dependent part does not change the ultra-violet behaviour 
of the theory as it contains on-shell contributions
and has a natural cut-off due to the Boltzmann
factor. Therefore, the zero temperature counter term 
is adequate for the renormalization of the theory. 
However, the infra-red problem becomes more severe at finite
temperature~\cite{bellac,thoma}.

Thermal field theory in the real time approach can be reformulated by 
diagonalizing the 2$\times$2 matrix propagators described above.
A well-known possibility is to diagonalize to a matrix constructed from
the Feynman propagators~\cite{nieves}. The free thermal propagator defined by 
Eq.~(\ref{s0propft}) can be written as
\be
\bd \Delta_0 ={\bd U} \left[ \begin{array}{cc}
\bar\Delta_0 & 0 \\ 
0 & -\bar\Delta^*_0 \\
\end{array}\right]{\bd U} 
\label{diago}
\ee
where 
\[
{\bd U} = \left[ \begin{array}{cc}
\sqrt{1+\eta} & {\displaystyle \frac{\eta+\theta(-p\cdot u)}{\sqrt{1+\eta}}} \\ 
 & \\
{\displaystyle\frac{\eta+\theta(p\cdot u)}{\sqrt{1+\eta}}} & \sqrt{1+\eta} \\ 
\end{array}\right]. 
\]
The exact propagators in the medium
can be defined analogously as Eqs.~(\ref{s0propft}), (\ref{f0propft})
and (\ref{v0propft}) with interacting Heisenberg fields instead of the
free fields. In this case we write
\be
\bd \Delta ={\bd U} \left[ \begin{array}{cc}
\Delta & 0 \\ 
0 & -\Delta^* \\
\end{array}\right]{\bd U} 
\ee
where $\bd \D$ is the matrix of interacting thermal propagators.
Using thermal perturbation theory $\bd\D$ can be expressed in terms of $\bd\D_0$.
One obtains
\be
{\bd \D} = {\bd \D_0} + {\bd\D_0}{\bd \Pi}{\bd \D}
\label{d-sft}
\ee
where ${\bd \Pi}$ now is the self-energy matrix;
\be
\bd \Pi \equiv \left[ \begin{array}{cc}
\Pi_{11} & \Pi_{12} \\ 
\Pi_{21} & \Pi_{22} \\
\end{array}\right] 
={\bd U}^{-1} \left[ \begin{array}{cc}
\Pi & 0 \\ 
0 & -\Pi^* \\
\end{array}\right]{\bd U}^{-1} 
\label{matt}
\ee
Matching the elements appearing in the diagonal of Eq.~(\ref{d-sft}) we have
\be
\D =  \bar\D_0 + \bar\D_0 {\s \Pi} \D ~~~~~~~~~\Longrightarrow~~~~~~
\D=\frac{1}{p^2-m^2-\Pi+i\eps}
\ee
From Eq.~(\ref{matt}) it also follows that
\be
{\s Re}\Pi={\s Re}\Pi_{11}
\label{resc11}
\ee
and
\bea
{\s Im}\Pi&=&\eps(p_0)\tanh(p_0/2T)\,{\s Im}\Pi_{11}~~~~~~~~~~~~{\s {for~~bosons}}\nonumber\\
{\s Im}\Pi&=&\eps(p_0)\coth((p_0+\mu)/2T)\,{\s Im}\Pi_{11}~~~~~{\s {for~~fermions}}
\label{im11}
\eea

In the following we will discuss
the vector (spin 1) propagator in some detail.
It is very similar to the scalar case except now one has to take
into account the Lorentz structure of the propagator and the self energy.
The exact propagator (matrix) $\bd D_{\mn}$ can be diagonalized as above 
and the {\em diagonal} element satisfies Dyson equation
\be
D_{\mn}=\bar D^0_{\mn}+\bar D^0_{\mu\rho}{\s \Pi}^{\rho\sigma}D_{\sigma\nu}
\ee
which gives
\be
D_{\mu \nu}^{-1} = (\bar D_{\mu \nu}^{0})^{-1} - \Pi_{\mu \nu},
\label{dyson}
\ee
where 
$\bar D_{\mu \nu}^{0}$ is the vacuum propagator for vector
particles.
The quantity $-i\Pi_{\mu \nu}$ is the sum of all one particle irreducible (1PI)
self energy insertions. In has a vacuum and a medium part so that
\be
\Pi^{\mu \nu}=\Pi_{\s {vac}}^{\mu \nu}+\Pi_{\s {med}}^{\mu \nu},
\label{pitot}
\ee
where 
\be
\Pi_{\s {vac}}^{\mu \nu}=(g^{\mu \nu} - \frac{p^{\mu}p^{\nu}}{p^2})\,
\Pi_{\s {vac}}(p^2), 
\label{pivac}
\ee
is the vacuum contribution to the self energy. 

Naively, it would appear that finite temperature corrections to quantum 
field theory breaks Lorentz covariance since the rest frame of the heat
bath selects out a specific frame of reference. However, by defining
a fluid four-velocity $u_\mu$ with temperature defined in the fluid
rest frame where $u_\mu=(1,\vec 0)$, a manifestly covariant formulation
can be achieved~\cite{adas,h-weld}. Using the techniques of tensor decomposition it can
be shown that for a vector particle propagating with
four-momentum $p^{\mu} = (p_0, {\vec p})$,
\be
\Pi_{{\s med}}^{\mn}(p_0,{\vec p}) = A^{\mn}\Pi_{T,{{\s med}}} +
 B^{\mn}\Pi_{L,{\s {med}}}
\label{pimed}
\ee
where $\Pi_{T,{\s med}}$ and $\Pi_{L,{\s med}}$ are Lorentz invariant
self-energy functions which characterize the transverse and longitudinal
modes.
$A^{\mu \nu}$ and $ B^{\mu \nu}$ are the transverse and longitudinal
projection tensors given by
\be
A^{\mn}=\frac{1}{p^2-p_0^2}\left[(p^2-p_0^2)(g^{\mn}-u^{\mu}u^{\nu})
\frac{}{}-p^{\mu}p^{\nu}-p_0^2u^{\mu}u^{\nu}+p_0(u^{\mu}p^{\nu}+
p^{\mu}u^{\nu})\right],
\label{amunu}
\ee
and
\be
B^{\mn}=\frac{1}{p^2(p^2-p_0^2)}\left[\frac{}{}p_0^2p^{\mu}p^{\nu}+
p^4u^{\mu}u^{\nu}-p_0 p^2(u^{\mu}p^{\nu}+p^{\mu}u^{\nu})\right],
\label{bmunu}
\ee
which satisfy the following algebra:
\bea
A_{\mu \rho}A^{\rho \nu} &=& A^\nu_\mu\nonumber\\
B_{\mu \rho}B^{\rho \nu} &=& B^\nu_\mu\nonumber\\
A_{\mu \rho}B^{\rho \nu} &=& 0\nonumber\\
q^\mu A_{\mn}&=&0\nonumber\\
q^\mu B_{\mn}&=&0\nonumber\\
g^{\mn}A_{\mn}&=&2\nonumber\\
g^{\mn}B_{\mn}&=&1\nonumber\\
A^{\mu \nu} + B^{\mu \nu} &=& g^{\mu \nu} - \frac{p^{\mu}p^{\nu}}{p^2}.
\label{reduceAB}
\eea
The two functions $\Pi_{T,{\s med}}$ and $\Pi_{L,{\s med}}$ are 
obtained by contraction:
\bea
\Pi_{L,{\s med}}&=&-\frac{p^2}{|\vec p|^2}u_{\mu}u_{\nu}\Pi^{\mu \nu}_{\s med}\nonumber\\
\Pi_{T,{\s med}}&=&\frac{1}{2}(\Pi^{\mu\mu}_{\s med}-\Pi_{L,{\s med}}).
\label{contrac}
\eea
Now, for massive vector particles,
\be
(\bar D^0_{\mn})^{-1}=-(p^2-m^2)g_{\mn}+p_\mu p_\nu
\ee
Using Eqs.~(\ref{dyson}-\ref{reduceAB}) the effective propagator becomes
\be
D_{\mu \nu} = -\,\frac{A_{\mu \nu}}{p^2-m^2+\Pi_{T}}
-\,\frac{B_{\mu \nu}}{p^2-m^2+\Pi_{L}} + \frac{p_{\mu}p_{\nu}}{m^2p^2},
\label{deff}
\ee
where
\be
\Pi_{T(L)}=\Pi_{T(L),{\s {med}}}+\Pi_{\s {vac}}
\ee
and $m$, we recall, is the bare mass of the particle.
The real part of $\Pi_{T(L)}$
affects the dispersion relation of the particle in the medium.
The displaced pole position of the effective propagator
in the rest frame of the propagating particle ({\it i.e.} where the
three momentum of the particle is zero) gives the effective
mass of the particle in the medium. The imaginary part of ${\s \Pi}_{T(L)}$
is connected to the decay width.

A different scheme in the formulation of finite temperature field
theory known as the `R/A' formalism~\cite{bech}, is to diagonalize to a matrix composed of retarded and
advanced propagators which are known to have better analyticity properties 
than the Feynman ones. 
In this case the analogue of Eq.~(\ref{diago}) is
\be
\bd \Delta_0 ={\bd V} \left[ \begin{array}{cc}
\Delta^R_0 & 0 \\ 
0 & \Delta^A_0 \\
\end{array}\right]{\bd W}. 
\ee
The free retarded and advanced propagators are
defined as
\bea
i\D^R_0&\equiv&\int\,d^4x\,e^{ip\cdot x}\theta(x_0)\,\lgl[\phi(x),\phi(0)]\rangle_0 \nonumber\\
       &=&\frac{i}{p^2-m^2+i\epsilon p_0}\nonumber\\
i\D^A_0&\equiv&\int\,d^4x\,e^{ip\cdot x}\theta(-x_0)\,\lgl[\phi(x),\phi(0)]\rangle_0\nonumber\\
       &=&\frac{i}{p^2-m^2-i\epsilon p_0}.
\eea
The matrices $\bd V$ and $\bd W$ depend on the momentum as well as
the thermal factor containing the distribution functions. Their exact forms
are given in Ref.~\cite{bech}. 
For the case of massive vector particles one arrives at the following
equation for the effective retarded propagator at finite temperature:
\be
D^R_{\mu \nu} = -\,\frac{A_{\mu \nu}}{p^2-m^2+\Pi^R_{T}}
-\,\frac{B_{\mu \nu}}{p^2-m^2+\Pi^R_{L}} + \frac{p_{\mu}p_{\nu}}{m^2p^2},
\label{deffr1}
\ee
where ${\s \Pi}^R_T$ and ${\s \Pi}^R_L$ are respectively the retarded
transverse and longitudinal components of the self energy. For photons
we have~\cite{aurenche}
\be
D^R_{\mu \nu} = -\,\frac{A_{\mu \nu}}{p^2+\Pi^R_{T}}
-\,\frac{B_{\mu \nu}}{p^2+\Pi^R_{L}} -\xi \frac{p_{\mu}p_{\nu}}{p^4}.
\label{deffr2}
\ee

Before we end our discussion of finite temperature propagators let us briefly mention 
about the imaginary time formalism or Matsubara formalism
which has been used extensively in the literature.
In the imaginary
time formalism, the form of the propagator at finite temperature 
is the same as that in vacuum but the time component of the four-momentum
takes discrete values, {\it i.e.} $p_0=2n\pi\,iT (=(2n+1)\pi\,iT)$ for bosons
(fermions) with $n=-\infty$ to $+\infty$, the vertices
are the  same as the zero temperature theory and the loop integral 
$\int d^4p/(2\pi)^4$ is replaced by the sum $iT\sum_{n}\int d^3p/(2\pi)^3$.
There are standard methods to evaluate the sum over the 
frequencies~\cite{bellac}. 
The propagators in the imaginary time formalism
can also be obtained by proper analytic continuation of the
real time propagators~\cite{agd,fradkin}.
Another method, known as the SACLAY method 
has also been used extensively in the literature~\cite{braaten,npb309}. 
This method uses the mixed representation
of the propagator {\it i.e.} it depends on the three-momentum 
and Euclidean time.

\section{Thermal Emission Rates}

We note at the onset that the nature of emission of real and virtual
photons depends crucially on the size of the hot thermal system from
which they are emitted. If the system is large enough the photons will 
rescatter and thermalize and their momentum space distribution will be 
given by the Planck distribution. The corresponding emission rate will then
be that of black body radiation which depends only on the temperature and the
area of the emitting body but not on its microscopic properties. Since the
typical size of systems produced in heavy ion collisions is much less
than the mean free path of the photons, they are likely to escape
the hot zone without rescattering and the emission rate in this case depends on
the dynamics of the thermal constituents through the imaginary part of the
photon self energy. 
We will begin by demonstrating
this in a kinetic theory framework~\cite{lect,ruuskanen} before going on to a
 more rigorous scheme.
We will in general denote the four-momenta of the real and virtual photons
by $p^\mu$ and $q^\mu$ respectively.

\subsection{Emission Rate as Equilibration Rate}

We know that the probability of a photon of 4-momentum $p^\mu=(E,\vec p)$
to be absorbed in matter is given by
\be
\Gamma_a(E)=\frac{1}{2E}\sum_{\{i\},\{f\}}\int\,d\Omega_{\{i\},\{f\}}
\delta^4(p+\sum k_i-\sum k_f)|{\cal M}(p+\{i\}\ra\{f\})|^2
\prod_{\{i\},\{f\}}n_i(1\pm n_f),
\ee
and the probability of emission is given by 
\be
\Gamma_e(E)=\frac{1}{2E}\sum_{\{i\},\{f\}}\int\,d\Omega_{\{i\},\{f\}}
\delta^4(\sum k_f-p-\sum k_i)|{\cal M}(\{i\}\ra p+\{f\})|^2
\prod_{\{i\},\{f\}}n_f(1\pm n_i).
\ee
Here, $n_k$ is the equilibrium distribution function,
$\{i\}$ and $\{f\}$ denote the initial and final state particles,
 and $d\Omega_{\{i\},\{f\}}$
denotes the phase space integration including spin and polarization sums.
Now,
\be
\frac{1\pm n(E)}{n(E)}=\exp(\frac{E}{T})=\frac{\Gamma_a(E)}{\Gamma_e(E)}
\ee
where the squared matrix elements of the forward and backward processes
have been taken to be equal on account of time reversality.
If $f(E,t)$ is the momentum distribution function of photons, the
rate of decrease due to absorption is $f(E,t)\Gamma_a(E)$ and the rate of
increase due to emission is $(1+f(E,t))\Gamma_e(E)$. So the time evolution
equation for $f$ is 
\be
\frac{\partial f}{\partial t}=-f\Gamma_a+(1+f)\Gamma_e.
\label{dfdt}
\ee
The solution of this equation is 
\be
f(E,t)=f_{BE}(E)+c(E)e^{-\Gamma t}
\ee
where, $\Gamma=\Gamma_a-\Gamma_e$ is the equilibration rate,
$f_{BE}$ is the Bose distribution function 
and $c(E)$ is a function which depends on the initial conditions. 
Since the system under consideration is small enough for the photons to
escape immediately on production, we must have $f=0$. Consequently, Eq.~(\ref{dfdt})
reduces to
\be
\left.\frac{\partial f}{\partial t}\right|_{f=0}=\Gamma_e=f_{BE}\Gamma.
\label{dfdt1}
\ee
Now, the equilibration rate is related to the imaginary part of
the photon self energy through ${\s Im\Pi}=E\Gamma (E)$~\cite{weldon83}. 
For a real photon $\Pi_L=0$ and we have from Eq.~(\ref{contrac}),
 ${\s Im}\Pi={\s Im}\Pi_T={\s Im}\Pi_{\mu}^{\mu}/2$.
Using \[dN = \frac{2}{(2\pi)^3}\,f(x,p)\,d^3xd^3p\] for a real photon we get from Eq.~(\ref{dfdt1})
\be
E\frac{dN}{d^3x dt d^3p}=\frac{2\,f_{BE}(E)\,E\Gamma(E)}{(2\pi)^3}=\frac{f_{BE}(E)}
{(2\pi)^3}
{\s Im}\Pi^{\mu}_{\mu}
\label{impiphot}
\ee
The quantity ${\s Im\Pi}^\mu_\mu $ contains information about the
constituents of the thermal bath and thus is of great relevance. 
After this relatively simple but illuminating exercise, we will
proceed to arrive at this result from more general considerations in
the following.

\subsection{Emission Rate and Photon Spectral Function}

We begin our discussion with the {\em dilepton} production rate.
Following Weldon~\cite{weldon90} 
let us define $A^\mu$ as the exact Heisenberg photon field which 
is the source of the leptonic current $J^l_\mu$. 
To lowest order in the electromagnetic coupling, the scattering matrix
element, $S_{HI}$,  for the transition  
$\mid I\rangle\,\rightarrow\,\mid H;l^+l^-\rangle$ is given by
\be
S_{HI}=-ie\,\langle H;l^+l^-\mid\int\,d^4x\,J^l_\mu(x)\,A^\mu(x) \mid I\rangle\,,
\label{shi1}
\ee
where $\mid I\rangle$ is the initial state corresponding to the two
incoming nuclei, $\mid H;l^+l^-\rangle$ is the final state 
which corresponds to a lepton pair plus the rest of the interacting system.
The parameter $e$ is the
renormalized charge which couples the leptonic current with the
virtual photon field $A^\mu$. 
Since we assume that the lepton pair does not interact
with the emitting system, the matrix element can be factorized as 
\begin{equation}
\langle H;l^+l^-\mid J^l_\mu(x)\,A^\mu(x) \mid I\rangle=
\langle H\mid A^\mu(x)\mid I\rangle\langle\,l^+l^-\mid J^l_\mu(x)\mid 0\rangle,
\end{equation}
where $\mid 0\rangle$ is the vacuum state.
Putting $J^l_\mu=\bar\psi(x)\gamma_\mu\psi(x)$ where $\psi$ is the
field operator for the leptons, one obtains the expectation value
 in terms of the
Dirac spinors $\bar{u}(p_1)$ and $v(p_2)$ as
\be
\langle\,l^+l^-\mid J^l_\mu(x)\mid 0\rangle=
\frac{\bar{u}(p_1)\gamma_\mu\,v(p_2)}{{\cal V}\sqrt{2E_12E_2}}
\,e^{i(p_1+p_2)\cdot x}
\label{lepmat}
\ee
where $\gamma_\mu$ denote the Dirac matrices, $E_i=\sqrt{p_i^2+m^2}$, 
with $i=1,2$ are the energies of the leptons 
and ${\cal V}$ is the volume of the system.
Therefore,
\begin{equation}
S_{HI}=-ie\,\frac{\bar{u}(p_1)\gamma_\mu\,v(p_2)}{{\cal V}\sqrt{2E_12E_2}}\int
d^4x\,e^{iq\cdot x}\langle H\mid\,A^\mu(x)\mid I\rangle.
\end{equation}
The leptons of four-momenta $p_1$ and $p_2$ are produced from a single
virtual photon of four-momentum $q=(q_0,\vec{q})$ so that $q_0=E_1+E_2$
and $\vec q=\vec p_1+\vec p_2$.

Assuming that a thermalized system is produced in the collision,
the dilepton multiplicity $N$ is obtained 
by summing over the final states and averaging over the initial 
states with a weight factor $Z(\beta)^{-1}\,e^{-\beta\,E_I}$; 
\begin{equation}
N=\frac{1}{Z(\beta)}\sum_I\,\sum_H\,\mid S_{HI}\,\mid^2\,e^{-\beta\,E_I}
\frac{{\cal V}\,d^3p_1}{(2\pi)^3}\,\frac{{\cal V}\,d^3p_2}{(2\pi)^3},
\label{mult}
\end{equation}
where $E_I$ is the total energy in the initial state, $Z(\beta)$ is
the partition function and $\beta=1/T$ is the inverse temperature. 
After some algebra $N$ can be written in a compact form as follows:
\begin{equation}
N=e^2\,L^{\mu\nu}\,H_{\mu\nu} \frac{d^3p_1}{(2\pi)^3E_1}
\,\frac{d^3p_2}{(2\pi)^3E_2},
\end{equation}
where $L_{\mu\nu}$ is the leptonic tensor defined by
\begin{eqnarray}
L^{\mu\nu}&\equiv&\frac{1}{4}\,
\sum_{spins}\,\bar{u}(p_1)\gamma^\mu\,v(p_2)\bar{v}(p_2)
\gamma^\nu\,u(p_1)\nonumber\\
&=&\,p_1^\mu\,p_2^\nu+p_2^\mu\,p_1^\nu-\frac{q^2}{2}\,g^{\mn},
\end{eqnarray}
and $H_{\mu\nu}$ is the photon tensor
\be
H_{\mn}\equiv\frac{1}{Z(\beta)}\,
\sum_I\sum_H
e^{-\beta\,E_I}
\int d^4x\,d^4y\,
\,e^{iq\cdot x}
\,\langle H\mid A_\mu(x)\mid I\rangle\,
\,e^{-iq\cdot y}
\,\langle I\mid A_\nu(y)\mid H\rangle\, .
\ee
Using translational invariance we can write
\be
\langle I\mid A_\nu(y)\mid H\rangle\, =
\,\langle I\mid e^{i{\cal P}\cdot y}A_\nu(0)e^{-i{\cal P}\cdot y}\mid H\rangle\, =
\,e^{i(p_I-p_H)\cdot y}\langle I\mid A_\nu(0)\mid H\rangle\, ,
\ee
where ${\cal P}$ is the four-momentum operator and $p_I$ and $p_H$ are the
total four-momenta of the initial and final states respectively.
We now have
\be
H_{\mn}=\frac{1}{Z(\beta)}\,
\int d^4y\,e^{i(p_I-p_H-q)\cdot y}\,
\int d^4x\,e^{iq\cdot x}\sum_I\sum_H
e^{-\beta\,E_I}
\,\langle H\mid A_\mu(x)|\,I\rangle\langle\,I\,|A_\nu(0)\mid H\rangle\,.
\ee
Using the conservation of four-momentum, $p_I=p_H+q$ and 
the completeness relation $\displaystyle\sum_I \mid I\rangle\langle I\mid =1$
we get
\begin{equation}
H_{\mu\nu}=\Omega\,e^{-\beta\,q_0}D^{>}_{\mu\nu}(q),
\end{equation}
where $\Omega\,(={\cal V}.t)$ is the four-volume of the system and $D^>_{\mu\nu}$
is the component $iD^{21}_{\mn}$ of the exact
photon propagator $\bd D_{\mn}$ defined in the last Section.
The time ordered propagator is the $(1,1)$ component of $\bd D_{\mn}$. In 
coordinate space it is defined as
\bea
iD^{11}_{\mn}(x)&\equiv&\frac{1}{Z(\beta)}\sum_H
\,\langle H\mid T\{A_\mu(x)\,A_\nu(0)\}\mid H\rangle\,e^{-\beta\,E_H}\nonumber\\
&\equiv&\theta(x_0)D_{\mn}^>(x)+\theta(-x_0)D_{\mn}^<(x).
\eea
where $D_{\mn}^<$ is the component $iD_{\mn}^{12}$, $x_0$
is the time component of the four vector, 
$x_\mu= (x_0,\vec{x})$ and $\theta(x_0)$ is the step function.
Using the integral representation of the $\theta$-functions,
\[\theta(y)=i\int\frac{dz}{2\pi}\,\,\frac{e^{-iyz}}{z+i\eps}\]
and taking the Fourier transform we get~\cite{dolan}
\be 
D^{11}_{\mn}(q_0,\vec q)=\int_{-\infty}^{\infty}\,\frac{d\omega}{2\pi}\,
\left[\frac{D^>_{\mn}(\omega,\vec q)}{q^0-\omega+i\epsilon}
-\frac{D^<_{\mn}(\omega,\vec q)}{q^0-\omega-i\epsilon}\right].
\label{top}
\ee
Using the Kubo Martin Schwinger (KMS) relation in momentum space,
\be
D^>_{\mn}(q_0,\vec q)=e^{\beta\,q_0}D^<_{\mn}(q_0,\vec q),
\ee
we have
\be
D^>_{\mn}(q_0,\vec q)=-\frac{2}{1+e^{-\beta q_0}}{\s{Im}}D^{11}_{\mn}(q_0,\vec q).
\ee
The rate of dilepton production per unit volume ($N/\Omega$) is then obtained as 
\be
\frac{dN}{d^4x}=-\frac{2e^2}{e^{\beta q_0}+1}\,
L^{\mn}{\s Im}D^{11}_{\mn}(q_0,\vec q)
\frac{d^3p_1}{(2\pi)^3E_1}\,\frac{d^3p_2}{(2\pi)^3E_2}.
\label{dnd4x1}
\ee
Now, the spectral function of the (virtual) photon in the thermal bath is 
defined as
\be
\rho_{\mn}(q_0,\vec q)\equiv\frac{1}{2\pi Z(\beta)}\int d^4x\,e^{iq\cdot x}\sum_H
\,\langle H\mid[A_\mu(x),A_\nu(0)]\mid H\rangle e^{-\beta\,E_H},
\label{spectral1}
\ee
so that, we have~\cite{bellac,dolan}
\be
D^{11}_{\mn}(q_0,\vec q)=\int_{-\infty}^\infty 
d\omega\frac{\rho_{\mn}(\omega,\vec q)}
{q_0-\omega+i\eps}-2i\pi f_{BE}(q_0)\rho_{\mn}(q_0,q),
\ee
where $f_{BE}(q_0)=[e^{\beta q_0}-1]^{-1}$. This leads to
\be
{\s Im}D^{11}_{\mn}(q_0,\vec q)=-\pi[1+2f_{BE}(q_0)]\rho_{\mn}(q_0,\vec q).
\ee
In terms of the photon spectral function the dilepton emission rate
is obtained as
\be
\frac{dN}{d^4x}=2\pi e^2 L^{\mn}\rho_{\mn}(q_0,\vec q)
\frac{d^3p_1}{(2\pi)^3E_1}\,\frac{d^3p_2}{(2\pi)^3E_2}f_{BE}(q_0).
\label{dnd4x2}
\ee
This relation which expresses the dilepton emission rate 
in terms of the spectral function of the photon in the medium 
is an important result. 
Inserting, 
\[1=\int d^4q\, \delta^{(4)}(p_1+p_2-q)\]
and using the identity
\bea
\int\prod_{i=1,2}\,\frac{d^3p_i}{(2\pi)^3E_i}
\delta^4(p_1+p_2-q)\,L^{\mn}(p_1,p_2)&=&\frac{1}{(2\pi)^6}\frac{2\pi}{3}
(q^\mu\,q^\nu-q^2\,g^{\mn})\nonumber\\
&\times&(1+\frac{2m^2}{q^2})\sqrt{1-\frac{4m^2}{q^2}},
\label{red1}
\eea
the dilepton rate ($dR=dN/d^4x$) can be expressed as
\be
\frac{dR}{d^4q}=-\frac{\alpha}{12\pi^3}q^2(1+\frac{2m^2}{q^2})
\sqrt{1-\frac{4m^2}{q^2}}(g^{\mn}-q^\mu q^\nu/q^2)\rho_{\mn}f_{BE}(q_0),
\label{drd4q0}
\ee
where $m$ is the lepton mass and $\alpha$ denotes the fine structure constant. 
We will now proceed to evaluate the photon spectral function.

As is well known, it is not the time-ordered propagator that has the
required analytic properties in a heat bath, but rather the retarded one.
We thus introduce the retarded propagator which will enable us to express
the dilepton rate in terms of the retarded photon self energy.
The retarded photon propagator in momentum space is defined as
\be
iD^R_{\mn}(q_0,\vec q)\equiv\frac{1}{Z(\beta)}\int d^4x\,e^{iq\cdot x}\theta(x_0)\sum_H
\,\langle H\mid[A_\mu(x),A_\nu(0)]\mid H\rangle e^{-\beta\,E_H},
\label{retp}
\ee
which leads to the relations
\be
{\s Im}\,D^{11}_{\mn}=(1+2f_{BE}){\s Im}\,D^R_{\mn}
\label{topret}
\ee
and
\be
\rho_{\mn}=-\frac{1}{\pi}{\s Im}D^R_{\mn}.
\label{specdr}
\ee
The above equation implies that in order to evaluate the spectral function
at $T\neq 0$ we need to know the imaginary part of the retarded propagator.
We note that the above expression for spectral function
reduces to its vacuum value as $\beta\ra\,\infty$
since the only state which
enters in the spectral function is the vacuum~\cite{lsbrown}.

Now, as mentioned before, the exact retarded photon propagator can be expressed in terms of the proper self 
energy through the Dyson-Schwinger equation:
\be
D^R_{\mn}=-\frac{A_{\mn}}{q^2+\Pi^R_T}-
\frac{B_{\mn}}{q^2+\Pi^R_L}- \xi\frac{q_\mu\,q_\nu}{q^4},
\label{drself}
\ee
where, $-i\Pi_{\mn}^R$ is the sum of all 1PI (one particle irreducible)
retarded photon self energy insertions which can be decomposed as
\be
\Pi^R_{\mn}=A_{\mn}\Pi_T^R+B_{\mn}\Pi_L^R.
\label{trlong}
\ee
Here $A^{\mn}$ and $B^{\mn}$, as defined in the previous Section
 are the transverse and longitudinal 
projection tensors respectively  and $\Pi_T^R$ and
$\Pi_L^R$ are the transverse and longitudinal components of the retarded
photon self energy.
The presence of the parameter $\xi$ indicates the gauge 
dependence of the propagator. Although the gauge dependence cancels out
in the calculation of physical quantities,
one should, however, be careful when extracting
physical quantities from the propagator directly, especially in the
non-abelian gauge theory.

Inserting the imaginary part of the retarded photon propagator from 
Eq.~(\ref{drself}) in Eq.~(\ref{specdr}) we get
\be
\rho^{\mn}=A^{\mn}\rho_T+B^{\mn}\rho_L,
\ee
where
\be
\rho_{T,L}\equiv-\frac{1}{\pi}\frac{{\s Im}\,\Pi^R_{T,L}}
{(q^2+{\s Re}\,\Pi^R_{T,L})^2+({\s Im}\,\Pi^R_{T,L})^2}. 
\label{rhotl}
\ee 
Using,
\bea
g^{\mn}\rho_{\mn}&=&2\rho_T+\rho_L\nonumber\\
q^\mu q^\nu \rho_{\mn}&=&0\,,
\eea
the dilepton rate is finally obtained as
\be
\frac{dR}{d^4q}=-\frac{\alpha}{12\pi^3}q^2(1+\frac{2m^2}{q^2})
\sqrt{1-\frac{4m^2}{q^2}}(2\rho_T+\rho_L)f_{BE}(q_0).
\label{drd4q1}
\ee
This is the {\it exact} expression for the dilepton emission rate 
from a thermal medium of 
interacting particles.
It has been argued by Weldon~\cite{weldonprl} that the electromagnetic
plasma resonance occurring through the spectral function at $q^2=-{\s Re}\Pi_{T(L)}$
could be a signal of the deconfinement phase transition provided
the plasma life time is long enough for the establishment of the resonance. 

Since $\Pi^R_{L,T}$ and $\Pi^R_{L,T}$ are both proportional
to $\alpha$ (the fine structure constant) they are small
for all practical purposes.
Neglecting ${\s Re}\Pi^R_{T,L}$ and  ${\s Im}\Pi^R_{T,L}$
in the denominator of Eq.~(\ref{rhotl}) one obtains
\be
2\rho_T+\rho_L\simeq-\frac{1}{\pi}\frac{[2{\s Im}\Pi^R_T+{\s Im}\Pi^R_L]}{q^4}
=-\frac{1}{\pi}\frac{{\s Im}\Pi^{R\mu}_\mu}{q^4}.
\label{approx}
\ee
This corresponds to the free 
propagation of the virtual photon in the thermal bath. 
Using Eqs.~(\ref{drd4q1}) and ~(\ref{approx})
we get
\be
\frac{dR}{d^4q}=\frac{\alpha}{12\pi^4\,q^2}(1+\frac{2m^2}{q^2})
\sqrt{1-\frac{4m^2}{q^2}}{\s Im}\Pi^{R\mu}_\mu\,f_{BE}(q_0).
\label{drd4q2}
\ee
This is the familiar result most widely used for the dilepton emission
rate~\cite{bellac}. 
It must be emphasized that this relation is valid only
to $O(e^2)$ since 
it does not account for the possible reinteractions of the virtual
photon on its way out of the bath.
The possibility of emission of more than one
photon has also been neglected here. However, the expression is true to all 
orders in strong interaction.

\subsection{Emission Rate and Current Correlation Function}

The emission rate of dileptons can also be obtained 
in terms of the electromagnetic
current correlation function~\cite{mclerran}.
We denote the electromagnetic current of the strongly interacting
particles (quarks or hadrons) by the operator
$J^h_\mu$ and the leptonic current by $J^l_\nu$.
As before, $e$ is the coupling
between $J^l_\nu$ and the virtual photon. For the present,
the current $J^h_\mu$ is taken to contain the coupling constant.
The matrix element  for dilepton production
is
\be
S_{HI}=-ie\langle\,H;l^+\,l^-\,\mid\int d^4x d^4y
J^l_{\mu}(x){\bar D}_0^{\mn}(x-y)\,J^h_{\nu}(y)\mid I\rangle
\label{shicor}
\ee
where ${\bar D}_0^{\mn}$ is the free photon propagator. 
As in the earlier case the leptonic part of the current 
can be easily factored out and we get Eq.~(\ref{lepmat}). 
The photon propagator is written in momentum space as
\be
\bar D^{\mn}_0(x-y)=\int\frac{d^4q}{(2\pi)^4}\,e^{-iq\cdot(x-y)}\bar D^{\mn}_0(q)
\ee
to  obtain
\begin{equation}
S_{HI}=-ie\,\frac{\bar{u}(p_1)\gamma_\mu\,v(p_2)}{{\cal V}\sqrt{2E_12E_2}}
{\bar D}_0^{\mn}(q)\int
d^4x\,e^{iq\cdot x}\langle H\mid\,J^h_\nu(x)\mid I\rangle.
\end{equation}
Squaring
the matrix elements and using Eq.~(\ref{mult})
one obtains the rate of dilepton
production
\be
dR=e^2\,L^{\mn}\,W^>_{\mn}\,\frac{e^{-\beta q_0}}{q^4}\,
\frac{d^3\,p_1}{(2\pi)^3\,E_1}
\frac{d^3\,p_2}{(2\pi)^3\,E_2},
\label{drcor}
\ee
where $W^>_{\mu\nu}(q)$ is the Fourier transform of
the electromagnetic current correlation function defined as
\be
W^>_{\mu\nu}(q)\equiv\int d^4x e^{iq\cdot x}\sum_H\,
\langle H\,\mid J^h_{\mu}(x)J^h_{\nu}(0)\mid\,H\,\rangle\,
\frac{e^{-\beta\,E_H}}{Z(\beta)}.
\label{correl}
\ee
Note that this definition is different from that of Mclerran and Toimela~\cite{mclerran}
where
\[
\tilde{W}_{\mu\nu}(q)\equiv\int d^4x e^{-iq\cdot x}\sum_H\,
\langle H\,\mid J^h_{\mu}(x)J^h_{\nu}(0)\mid\,H\,\rangle\,
\frac{e^{-\beta\,E_H}}{Z(\beta)}.
\]
The correlation function is symmetric in $\mu$ and $\nu$. One can use
translational invariance and the KMS relation to show that
\[
\tilde{W}_{\mu\nu}(q)= W^>_{\mu\nu}(-q)= e^{-\beta q_0} W^>_{\mu\nu}(q).
\]
In this case the dilepton rate is
\[
dR=e^2\,L^{\mn}\,\frac{\tilde{W}_{\mn}(q)}{q^4}\,
\frac{d^3\,p_1}{(2\pi)^3\,E_1}
\frac{d^3\,p_2}{(2\pi)^3\,E_2},
\]
as in Ref.~\cite{mclerran}.

It is seen from Eq.~(\ref{drcor}) that from the measured dilepton and photon 
distributions the full tensor structure
of $W^{\mu\nu}$ can in principle be determined.
This will yield considerable information about the thermal state 
of the strongly interacting system.

Now, $W^>_{\mn}$ is related to the retarded correlator by
\be
W^>_{\mn}=2e^{\beta q_0}f_{BE}(q_0){\s Im}W^R_{\mn}
\label{wmng}
\ee
where
\be
W^R_{\mu\nu}(q)\equiv i\int d^4x e^{iq\cdot x}\theta(x_0)\sum_H\,
\langle H\,\mid [J^h_{\mu}(x),J^h_{\nu}(0)]\mid\,H\,\rangle\,
\frac{e^{-\beta\,E_H}}{Z(\beta)}.
\label{retcor}
\ee
Using the identity Eq.~(\ref{red1}) and the transversality of the correlation
function, the dilepton emission rate is obtained as
\be
\frac{dR}{d^4q}=-\frac{\alpha}{12\pi^4\,q^2}(1+\frac{2m^2}{q^2})
\sqrt{1-\frac{4m^2}{q^2}}g^{\mn}{\s Im}W^R_{\mn}\,f_{BE}(q_0).
\label{drretcor}
\ee
We now define the improper photon self energy through the relation
\be
D^{R,\,\alpha\beta}=
D_0^{R,\,\alpha\beta}+D_0^{R,\alpha\mu}\,P_{\mn}^R\,D_0^{R,\nu\beta}
\ee
where $-iP^R_{\mn}$ is the sum of {\it all} self energy diagrams. 
The advantage is that $P^R_{\mn}$ can be defined in coordinate space 
as~\cite{beres}
\be
iP^R_{\mu\nu}(x)\equiv\theta(x_0)\sum_H\,
\langle H\,\mid [J^h_{\mu}(x),J^h_{\nu}(0)]\mid\,H\,\rangle\,
\frac{e^{-\beta\,E_H}}{Z(\beta)}.
\label{impself}
\ee
Taking the Fourier transform and 
comparing with Eq.~(\ref{retcor}) we see that 
\be
P^R_{\mn}(q)=-W^R_{\mn}(q)\,.
\label{impw}
\ee
Therefore, the dilepton rate can also be expressed as~\cite{gale}
\be
\frac{dR}{d^4q}=\frac{\alpha}{12\pi^4\,q^2}(1+\frac{2m^2}{q^2})
\sqrt{1-\frac{4m^2}{q^2}}{\s Im}P^{R\mu}_\mu\,f_{BE}(q_0).
\label{drd4qimp}
\ee
It is important to realize 
that the analysis is essentially nonperturbative
up to this point.
To $O(e^2)$ we note that $P$ reduces to the proper self energy $\Pi$
($=P\,D_0\,D^{-1}$)
and consequently Eq.~(\ref{drd4qimp}) reduces to Eq.~(\ref{drd4q2}).
This approximation is the same as implied in Eq.~(\ref{approx}).

Let us now try to relate the approach discussed in this Section with the previous one.
We know that the equation of motion of the photon field is given by the Maxwell equation
\[\partial_\alpha\partial^\alpha\,A_\mu
-(1-\xi^{-1})\partial_\mu\,(\partial_\alpha\,A^\alpha)=J^h_\mu
\]
which has the formal solution 
\[
A^\mu(x)=\int d^4y {\bar D}_0^{\mn}(x-y)J^h_\nu(y),
\]
where ${\bar D}_0$ is the free photon propagator.
This can be used in Eq.~(\ref{shi1}) to get Eq.~(\ref{shicor}).  
Again, the connection between 
the electromagnetic current correlation function and the spectral function 
can be expressed in a straight forward way
by substituting $J^h_\mu$ and $J^h_\nu$ from the Maxwell equation
in Eq.~(\ref{correl}) to obtain
\bea
W^>_{\mn}&=&\left(q^2g_{\mu\alpha}-(1-\xi^{-1})q_\mu q_\alpha\right)
D_>^{\alpha\beta}
\left( q^2g_{\beta\nu}-(1-\xi^{-1})q_\beta q_\nu\right)\nonumber\\
&=&2\pi\left(q^2g_{\mu\alpha}-(1-\xi^{-1})q_\mu q_\alpha\right)
\rho^{\alpha\beta}\left(q^2g_{\beta\nu}-(1-\xi^{-1})q_\beta q_\nu\right)
(1+f_{BE}).
\eea
The gauge dependent terms will not contribute due to current
conservation and we have
\be
W^>_{\mn}(q)=2\pi q^4\rho_{\mn}(1+f_{BE}).
\ee
Substituting this in 
Eq.~(\ref{drcor}) 
we can recover Eq.~(\ref{dnd4x2}).
This establishes the connection between the approaches of 
Refs.~\cite{weldon90} and \cite{mclerran}.

Let us now consider {\em real photon} emission
from a system in thermal equilibrium. 
The matrix element is given by
\be
S_{HI}=-i\,\lgl H;\gamma|\int\,d^4x\,J^h_\mu(x)A_0^\mu(x)|I\rgl
\ee
where, $J^h_\mu(x)$ is the electromagnetic current 
of the strongly interacting particles 
which produces the photon and $A_0^\mu(x)$ is the free photon field.
Since the photon escapes without re-interacting,
the matrix element can be taken to
factorize as,
\be
\lgl H;\gamma|J^h_\mu(x)A_0^\mu(x)|I\rgl=\lgl H|J^h_\mu(x)|I\rgl\
\,\lgl \gamma|A_0^\mu(x)|0\rgl.
\ee
In the case of a single photon of four-momentum $p=(E,\vec p)$
the free field can be expanded as
\be
A_0^\mu(x)=\frac{\eps^{\mu}(p)}{\sqrt{2E{\cal V}}}\,\,e^{ip\cdot x},
\label{freea}
\ee
so as to obtain
\be
S_{HI}=-i\frac{\eps^{\mu}(p)}{\sqrt{2E{\cal V}}}\int\,d^4x\,e^{ip\cdot x}\lgl H|J^h_\mu(x)|I\rgl.
\ee
The thermally averaged photon multiplicity is given by
\be
N_\gamma=\frac{1}{Z(\beta)}\,\sum_I\sum_H|S_{HI}|^2\,e^{-\beta E_I}\,
\frac{{\cal V}d^3p}{(2\pi)^3}\,.
\ee
The sum over the photon polarization  is performed
using ${\displaystyle\sum_{pol}}\eps^{\mu}\eps^{\nu}=-g^{\mn}$. Proceeding
as before we obtain the photon emission rate as 
\be
E\frac{dR_\gamma}{d^3p}=-\frac{g^{\mn}}{2(2\pi)^3}
e^{-\beta E}\, W^>(p).
\ee
Note that this expression also follows from the dilepton emission rate given by
Eq.~(\ref{drcor}) with a few modifications.
The factor 
$e^2\,L_{\mn}/q^4$ which arises from the lepton spin sum
of the square modulus of the product of the electromagnetic vertex 
$\gamma^\ast\,\ra\,l^+\,l^-$, the leptonic current involving Dirac spinors
and the square of the photon propagator is to be replaced by the factor 
$-g_{\mn}/2$ from the polarization sum. The factor $1/2$ follows from the
normalization of the photon field (Eq.~(\ref{freea})).
Finally the phase space factor for the lepton pair,
$d^3p_1/[(2\pi)^3\,E_1]\,d^3p_2/[(2\pi)^3E_2]$
is replaced by $d^3p/[(2\pi)^3E]$. 

\begin{figure}
\centerline{\psfig{figure=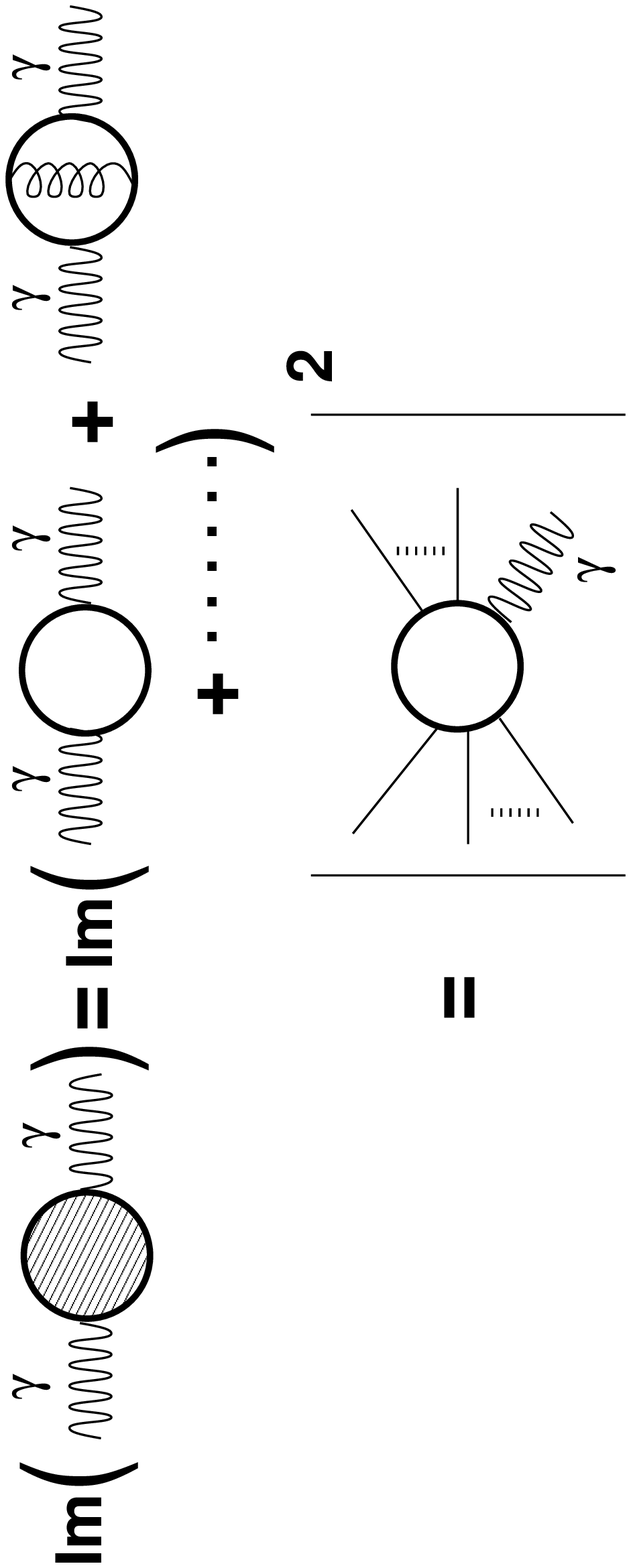,width=13cm,height=5cm,angle=-90}}
\caption{Optical Theorem in Quantum Field Theory}
\label{opt}
\end{figure}

Using Eqs.~(\ref{wmng}) and (\ref{impw}) and the fact that to lowest
order in $e$ the improper and proper photon self energies are equal, we get
\be
E\frac{dR_\gamma}{d^3p}=\frac{g^{\mn}}{(2\pi)^3}\,{\s Im}\Pi^R_{\mn}f_{BE}(E). 
\ee
In this form, the real photon emission rate is correct up to order $e^2$ in 
electromagnetic interaction but exact, in principle, to all orders
in strong interaction. However, for all practical purposes 
one is able to evaluate up to a finite order of loop expansion.
It is clear from the above 
that in order to deduce the photon and dilepton emission rate from a 
thermal system we need to evaluate the 
imaginary part of the photon self energy.
The Cutkosky rules at finite temperature
or the thermal cutting rules~\cite{adas,kobes,gelis}
give a systematic procedure to calculate the imaginary part of a 
Feynman diagram. The Cutkosky rule expresses the 
imaginary part of the $n$-loop amplitude in terms of physical 
amplitude of lower order ($n-1$ loop or lower). This is shown schematically
in Fig.~(\ref{opt}).  When the imaginary part of the self energy is calculated
up to and including $L$ order loops where $L$ satisfies $x\,+\,y\,<\,L\,+\,1$,
then one obtains the photon emission rate for the reaction $x$ particles
$\ra$ $y$ particles $+\,\gamma$ and the above formalism becomes 
equivalent to the relativistic kinetic theory formalism~\cite{gk}. 

For a reaction $1\,+\,2\,\ra\,3\,+\gamma$ the photon 
emission rate is given by~\cite{sourav}
\begin{eqnarray}
E\frac{dR}{d^3p}&=&\frac{{\cal{N}}}{16(2\pi)^7E}\,\int_{(m_1+m_2)^2}^{\infty}
\,ds\,\int_{t_{{\s min}}}^{t_{{\s max}}}\,dt\,|{\cal M}|^2\,
\int\,dE_1\nonumber\\
&&\times\int\,dE_2\frac{f(E_1)\,f(E_2)\left[1+f(E_3)\right]}{\sqrt{aE_2^2+
2bE_2+c}},
\label{rktp}
\end{eqnarray}
where
\begin{eqnarray}
a&=&-(s+t-m_2^2-m_3^2)^2\nonumber\\
b&=&E_1(s+t-m_2^2-m_3^2)(m_2^2-t)+E[(s+t-m_2^2-m_3^2)(s-m_1^2-m_2^2)\nonumber\\
&&-2m_1^2(m_2^2-t)]\nonumber\\
c&=&-E_1^2(m_2^2-t)^2-2E_1E[2m_2^2(s+t-m_2^2-m_3^2)-(m_2^2-t)(s-m_1^2-m_2^2)]
\nonumber\\
&&-E^2[(s-m_1^2-m_2^2)^2-4m_1^2m_2^2]-(s+t-m_2^2-m_3^2)(m_2^2-t)\nonumber\\
&&\times(s-m_1^2-m_2^2)
+m_2^2(s+t-m_2^2-m_3^2)^2+m_1^2(m_2^2-t)^2\nonumber\\
E_{1{{\s min}}}&=&\frac{(s+t-m_2^2-m_3^2)}{4E}+
\frac{Em_1^2}{s+t-m_2^2-m_3^2}
\nonumber\\
E_{2{{\s min}}}&=&\frac{Em_2^2}{m_2^2-t}+\frac{m_2^2-t}{4E}\nonumber\\
E_{2{{\s max}}}&=&-\frac{b}{a}+\frac{\sqrt{b^2-ac}}{a}.\nonumber
\end{eqnarray}
${\cal N}$ is the overall degeneracy of the particles 1 and 2,
${\cal M}$ is the invariant amplitude of the reaction (summed over
final states and averaged over initial states), $f$ denotes
the distribution functions and $s$, $t$, $u$ are the usual Mandelstam variables.
Now, the rapidity of a particle is defined as
\[y=\frac{1}{2}\ln\frac{E+p_z}{E-p_z}\]
where $E$ and $p_z$ are the energy and longitudinal momentum of the particle
respectively.
For massless particles the transverse momentum is
\[p_T=(E^2-p_z^2)^{1/2}.\]
We then have ~$E=p_T\cosh y$, ~$p_z=p_T\sinh y$ and $d^3p/E=d^2p_Tdy$
in the case of real photons.

The dilepton emission rate derived above in terms of the photon self-energy
can be connected by the
optical theorem to the kinetic theory rate for a reaction
$a\,\bar a\,\ra\,l^+\,l^-$ which is given by
\bea
\frac{dR}{d^4q}&=&\int {d^3p_a\over 2E_a(2\pi)^3}f(p_a)
\int {d^3p_{\bar a}\over 2E_{\bar a}(2\pi)^3}f(p_{\bar a})
\int {d^3p_1\over 2E_1(2\pi)^3}
\int {d^3p_2\over 2E_2(2\pi)^3}\nonumber\\
&&\mid {\cal M}\mid_{a\bar a\rightarrow l^+l^-}^2
(2\pi)^4\delta^{(4)}(p_a+p_{\bar a}-p_1-p_2)
\delta^{(4)}(q-p_a-p_{\bar a}).
\label{dilrate1}
\eea
where  $f(p_a)$ is the appropriate 
occupation probability for bosons or fermions.
The Pauli blocking of the lepton pair in the final state has been
neglected in the above equation. 
The transverse mass of the lepton pair is defined as
~$M_T^2=q_T^2+M^2=q_0^2-q_z^2$ where ~$q^2=M^2$; ~$M$ being the invariant mass
of the lepton pair.
Using the above definition of rapidity, we have
~$q_0=M_T\cosh y$, ~$q_z=M_T\sinh y$ and ~$d^4q=MdMd^2M_Tdy$. 
In terms of the cross-section for the production of a lepton pair
the dilepton production rate can be expressed as
\be
\frac{dR}{d^4q}=\int {d^3p_a\over (2\pi)^3}f(p_a)
\int {d^3p_{\bar a}\over (2\pi)^3}f(p_{\bar a})\,\,
v_{\s rel}^{a\bar a}(p_a,p_{\bar a})\,\sigma^{a\bar a}_{l^+l^-}(p_a,p_{\bar a})
\,\,\delta^{(4)}(q-p_a-p_{\bar a})
\label{dilrate2}
\ee
where
\[
\sigma^{a\bar a}_{l^+l^-}(p_a,p_{\bar a})=
\frac{1}{v_{\s rel}^{a\bar a}}
\int {d^3p_1\over 2E_1(2\pi)^3}
\int {d^3p_2\over 2E_2(2\pi)^3}
\mid {\cal M}\mid_{a\bar a\rightarrow l^+l^-}^2
(2\pi)^4\delta^{(4)}(p_a+p_{\bar a}-p_1-p_2)
\]
and 
\[v_{\s rel}^{a\bar a}=\frac{(p_a\cdot p_{\bar a}-m_a^4)^{1/2}}{E_a E_{\bar a}}\] 
is the relative velocity of the
colliding particles $a$ and $\bar a$.

\section{Emission Rates from Quark Matter}

In this Section we will discuss the rates of photon and dilepton emission
from a thermal system composed of quarks, antiquarks and gluons at a 
temperature $T$ due to various processes which are known to contribute
substantially to the yield from QGP. 
Since the constituents of the system are massless for all practical
purposes, we will encounter infrared singularities in the evaluation
of the rates. We will discuss how these can be screened by summing 
the Hard Thermal Loops (HTLs)~\cite{braaten,frenkel} in the theory.
The yield due to the primary scattering of partons embedded in
the colliding hadrons will also be discussed since they are expected to
constitute the principal background to the thermal photons (dileptons) in
the region of large transverse momentum (invariant mass).
  
\subsection{Thermal Photon Emission Rates from QGP}

Naively, one expects that the properties of QGP at high temperature
($T>>T_c$) can be studied by applying perturbation theory due to the small
value of the strong coupling constant, $\alpha_s(T)$ which
is given by the parametrized form~\cite{karsch}
\be
\alpha_s(T)=\frac{6\pi}{(33-2N_f)\ln(8T/T_c)}.
\ee
However, QCD 
perturbation theory at high temperature is plagued by infra-red problems 
and gauge dependence of physical quantities, {\it e.g.} the
gluon damping rate~\cite{thoma,rkobes,npa525}. 
The gauge dependence of the gluon damping rate was cured by
Braaten and Pisarski~\cite{braaten} by an effective expansion in terms of hard
thermal loops - {\it i.e.} including all the relevant loop effects in a given order 
of the coupling constant in a systematic way.
The idea of  HTL is based on the observation that at non-zero
temperature there are two energy scales -  one associated with the
temperature $T$, referred to as the hard scale and the other connected 
with the fermionic mass $\sim g_sT$ ($g_s<<1$), induced by the temperature,
known as the soft scale.  A momentum $p^\mu$ appearing in the self 
energy diagram of
photon would be called soft (hard) if both the temporal and the 
spatial components are $\sim g_sT$ (any component is $\sim T$).
If any physical quantity is sensitive to the soft scale then
HTL resummation becomes essential, {\it i.e.} in such cases
the correlation function has to be expanded in terms of the
effective vertices and propagators, where the effective quantities are
the corresponding bare quantities plus the high temperature
limit of one loop corrections.

However, the problem of infra-red 
divergences in QCD is not solved completely by the HTL framework.
The quantities which are quadratically divergent in naive perturbation
theory such as the damping rate of fast moving fermions in QGP becomes 
logarithmically divergent in effective perturbation theory. On the 
other hand  quantities which are logarithmically divergent in the
naive perturbation theory turns out to be finite if one applies
HTL resummation method. The hard photon ($E>T$) emission rate which falls
in the second category, is the relevant quantity for the present discussions.

The thermal photon emission rate from QGP is governed 
by the following Lagrangian density:
\be
{\cal L}_{QGP}={\cal L}_{QCD}+{\cal L}_{\gamma q},
\ee
where
\bea
{\cal L}_{QCD}&=&-\frac{1}{4}\sum_{a=1}^{8}G_{\mn}^aG^{a\mn}+
\sum_{f=1}^{N_f}\bar \psi_f(i\partial\sls-
g_s\gamma^\mu G^a_\mu\frac{\lda^a}{2})\psi_f,\nonumber\\
{\cal L}_{\gamma q}&=& 
-\frac{1}{4}F_{\mn}F^{\mn}-
\sum_{f=1}^{N_f}e_f\bar \psi_f\gamma^\mu A_\mu\psi_f.
\eea
\bef
\centerline{\psfig{figure=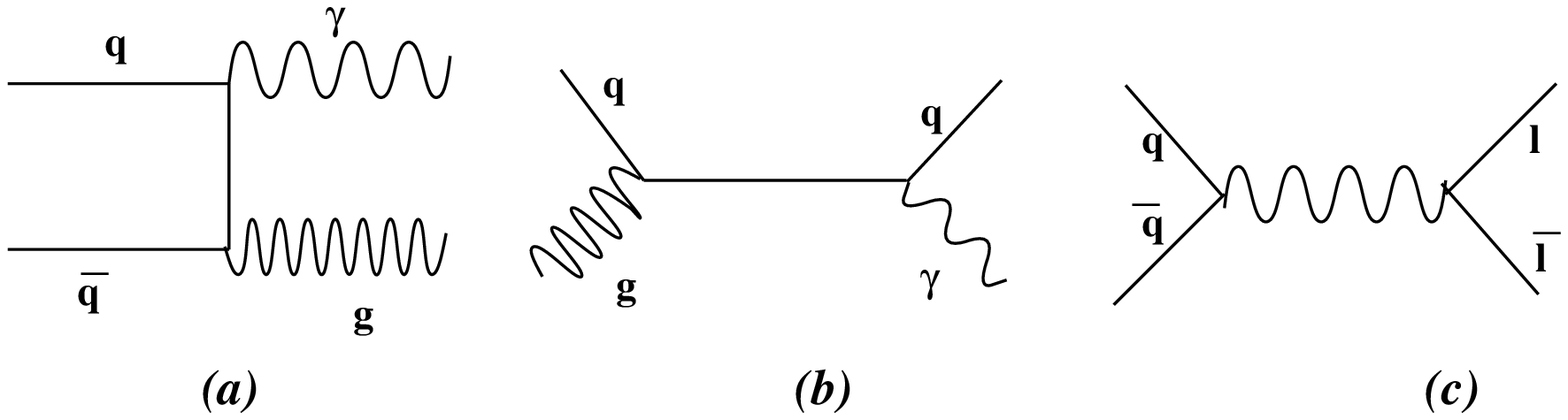,height=3cm,width=10cm}}
\caption{
Lowest order diagrams for photon and dilepton production from QGP.
}
\label{loqgp}
\eef
In the above, $G_{\mn}^a$ is the non-abelian field tensor for
the gluon field $G_{\mu}^a$ of color $a$, $\psi_f$ is the
Dirac field for the quark of flavour $f$, $g_s$ is the color
charge, $e_f$ is the (fractional) electric charge of quark flavor $f$, 
$\lambda^a$'s are the Gell-Mann matrices, $F_{\mn}$ is the electromagnetic 
field tensor and $A^\mu$ is the photon field.
As mentioned in the introduction, the dominant processes for 
photon production from QGP are the annihilation 
($q\bar{q}\rightarrow g\gamma$)
and the Compton processes ($q(\bar{q})\rightarrow q(\bar{q})\gamma$)
as shown in Figs.~(\ref{loqgp}a) and (\ref{loqgp}b) respectively.
However, the production rate from these processes diverges due to
the exchange of massless particles. 
This is a well-known problem in thermal perturbative expansion of non-abelian
gauge theory which suffers from infra-red divergences. 
One type of the divergences could be cured by taking into
account the `electric type' screening through 
the HTL approximation~\cite{braaten}. 
The non-abelian gauge theory also contains `magnetic type' divergences,
which can be eliminated if there is a screening
of the magnetic field~\cite{rmp,nair,blaizot}.  This is in sharp contrast to
Quantum Electrodynamics, which is free from screening of static magnetic
field. However, the study of magnetic screening is beyond 
the scope of HTL approximation 
as the transverse component of the gluon self energy  
vanishes in the static limit in this framework.  
Magnetic screening is relevant if any physical quantity is sensitive
to the scale $g_s^2T$, where all the loop contributions are of the same order
\cite{asmilga} and hence the perturbation theory breaks down~\cite{alinde}.
The production of soft photons ($E \leq g_sT$) 
from QGP is non-perturbative because it 
is sensitive to the magnetic screening mass of the gluons~\cite{agz} and
consequently the soft photon emission rate is poorly known.
The production of hard photons ($E\ge T$) is insensitive
to the scale $g_s^2T$ and hence
infra-red divergences can be eliminated within the framework of HTL  
as discussed below.

Let us try to understand the notion of HTL using
massless $\phi^4$ theory described by the Lagrangian density
\be
{\cal L}=\frac{1}{2}(\partial\phi)^2-g^2\phi^4.
\ee
The  thermal mass (self energy)
resulting from the one loop tadpole diagram in this model is
$m_{\s th}^2\sim g^2T^2$. At soft momentum scale ($p^\mu\sim gT$)
the inverse of the bare propagator goes as $\sim g^2T^2$.
Thus, the one loop (tadpole) correction is as large as the
tree amplitude. Therefore, this tadpole is a HTL by definition. 
Braaten and Pisarski \cite{braaten} have argued that these 
HTL contributions should be taken into account consistently  
by re-ordering the perturbation series in terms of effective 
vertices and propagators.
Therefore, according to their prescription we have
\be
{\cal L}=\frac{1}{2}(\partial\phi)^2-g^2\phi^4 - \frac{1}{2}m_{\s th}^2\phi^2
+ \frac{1}{2}m_{\s th}^2\phi^2={\cal L}_{\s eff} +{\cal L}_{\s ct},
\ee
where ${\cal L}_{\s ct}=m_{\s th}^2\phi^2/2$ is the counter term which should
be treated in the same footing as the $\phi^4$ term.
${\cal L}_{\s ct}$ has been introduced in order 
to avoid thermal corrections at higher order which has already been
included in the tree level. With the counter term the Lagrangian 
remains unchanged, so the effective theory is a mere re-ordering
of the perturbative expansion. A similar exercise has to be carried 
out in gauge theory keeping in mind that an addition and subtraction of local mass
terms will violate gauge invariance. 
The effective action for hot gauge
theories have been derived in Refs.~\cite{taylor,bpea,en,vpn,flechsig},
whereas the authors of Refs.~\cite{bi,kelly} follow the classical kinetic 
theory approach for the derivation of the HTL contributions. It has 
been shown in Ref.~\cite{nairvp} that the contribution of HTL to the
energy of the QGP is positive. The counter term required to avoid double
counting in evaluating the virtual photon production from QGP  in the
two-loop approximation has been derived in~\cite{lapth} recently.

\begin{figure} 
\centerline{\psfig{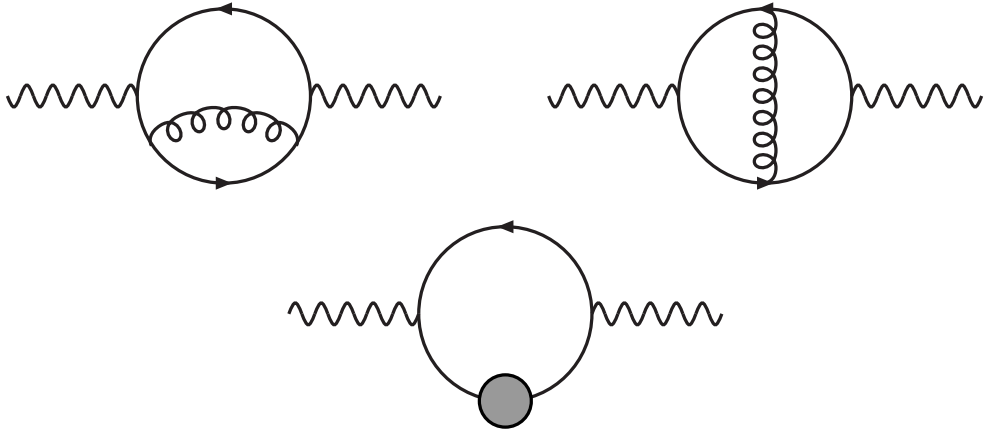}}
\caption{Two loop contribution to the photon self energy. A diagram
interchanging the blob in the
internal line of the third diagram should also be considered.}
\label{loop1}
\end{figure}

\begin{figure} 
\centerline{\psfig{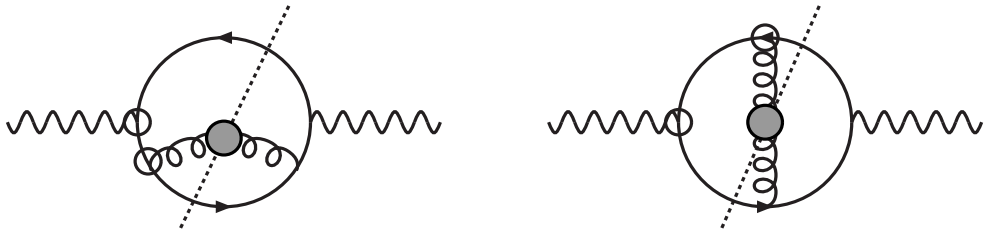}}
\caption{Two loop photon diagram relevant for bremsstrahlung processes.
The blob on the gluon (spiral line) indicates effective gluon propagator.
The circle on the vertices represent those required 
to evaluate the imaginary part of the photon self energy in the 
framework of thermal cutting rules (see Refs.~(\protect\cite{aurenche}) 
and also~(\protect\cite{gelis})).
}
\label{loop2}
\end{figure}

The photon emission from Compton and annihilation processes can be 
calculated from the imaginary parts of the first two diagrams in Fig.~(\ref{loop1}).
Since these processes involve exchange of massless quarks in the $t/u$
channels the rate becomes infrared divergent. One then 
obtains the hard contribution by introducing a lower
cut-off to render the integrals finite.
In doing so, some part of the
phase space is left out and the rate becomes cut-off dependent.
The photon rate from this (soft) part of the phase space is then 
handled using HTL resummation technique.
The application of HTL to hard photon emission rate was first 
performed in Refs.~\cite{kapusta,baier}. For hard photon
emission, one of the (soft) quark propagators in the photon
self energy diagram should be replaced by effective 
quark propagators (third diagram in Fig.~(\ref{loop1})), 
which consists of the bare propagator and
the high temperature limit of one loop corrections~\cite{klimov,hweldon}.
When the hard and the soft contributions are added,
the emission rate becomes finite because of the Landau damping
of the exchanged quark in the thermal bath and the cut-off scale
is cancelled. 
The rate of hard photon emission is then obtained as~\cite{kapusta}
\be
E\frac{dR_\gamma^{QGP}}{d^3p}=\frac{5}{9}\frac{\alpha\alpha_s}{2\pi^2}
T^2\,e^{-E/T}\ln(2.912E/g_s^2T).
\ee
where $\alpha_s$ is the strong coupling constant.
Recently, the bremsstrahlung contribution to photon emission 
rate has been computed~\cite{aurenche} 
by evaluating the photon self energy in two loop HTL approximation.
The physical processes arising from two loop 
contribution (Fig.~(\ref{loop2})) 
are the bremsstrahlung of quarks, antiquarks and
quark anti-quark annihilation with scattering in the thermal bath. 
The rate of 
photon production due to bremsstrahlung process for a two-flavour
thermal system with $E>T$ is given by~\cite{aurenche}
\be
E\frac{dR_\gamma^{QGP}}{d^3p}=\frac{40}{9\pi^5}\,\alpha\alpha_s
T^2\,e^{-E/T}\left(J_T-J_L\right)\ln 2,
\label{brm2}
\ee
and the rate due to $q-\bar q$ annihilation with scattering 
in the thermal bath is given by,
\be
E\frac{dR_\gamma^{QGP}}{d^3p}=\frac{40}{27\pi^5}\,\alpha\alpha_s
ET\,e^{-E/T}\left(J_T-J_L\right),
\label{ann2}
\ee
where $J_T\approx 4.45$ and $J_L\approx -4.26$.
The most important implication of this work is that the magnitude of the two loop
contribution comes out to be  of the same order as those evaluated
at one loop~\cite{kapusta,baier} due to the larger size of the
available phase space. 
In case of soft thermal photon ($E\sim g_sT$) emission,
all the vertices and the propagators have to be replaced
by the corresponding effective quantities. It has been
shown~\cite{baier2,aurenche1} that the result is divergent
due to the exchange of massless quarks introduced through the HTL 
effective vertices itself. 
However, such collinear singularities
for light-like external momentum could be removed with
an improved action~\cite{flechsig}.
It is also shown that such infrared singularities could be
removed through KLN (Kinoshita - Lee - Nauenberg) theorem~\cite{kln1,kln2} 
by including appropriate diagrams and summing
over all degenerate initial and final states~\cite{niegawa1,niegawa2}, but
the rate is non-perturbative because it is sensitive 
to the scale $g_s^2T$~\cite{agz}. 

The emission rate of hard photons 
is well under control within the framework of HTL resummation. However, there
are important issues in hot gauge theories which cannot be addressed
within the HTL resummation method~\cite{rkobes,banff}.
For example, (i) HTL resummation is based on the
weak coupling limit ($g_s<<1$) to distinguish between hard ($T$) and soft
momentum scale ($g_sT$) but such a limit may not be achievable in URHIC even for
the highest energy to be available at the CERN LHC in the near future. Extrapolation
of results obtained in HTL approximation to higher values of coupling constant
will be demonstrated in Section 3.2 of Chapter~3 through the photon spectra. (ii) It 
cannot cure the infra-red divergence problem that arises in the damping rate of fast 
fermions, (iii) it cannot remove the mass shell singularities in the  soft
photon (real) emission rate, (iv) the next to leading order correction to the Debye
mass diverges unless one includes magnetic screening, which is beyond the
scope of HTL approximation and finally (v) HTL works for a system in equilibrium 
; extension of the formalism to non-equilibrium processes is still in the
early stages of development. Results from other methods  
such as ladder approximation~\cite{cornwall}, 
renormalization group equation~\cite{km} etc. will be  very important in these
cases. 

\bef
\centerline{\psfig{figure=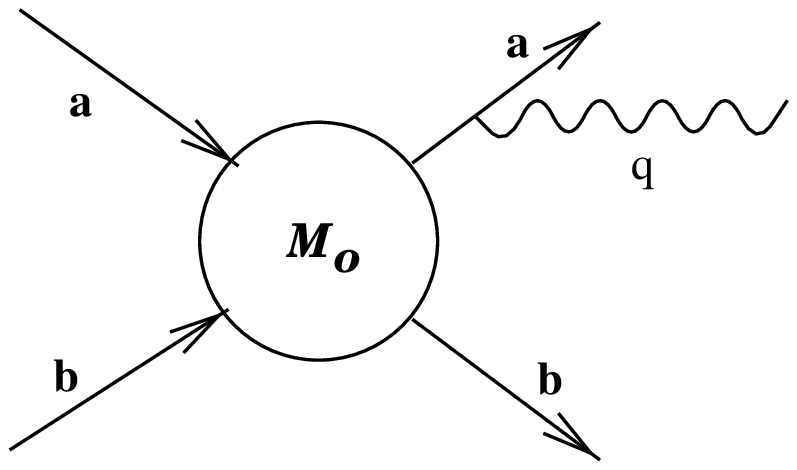,height=3cm,width=6cm}}
\caption{
Feynman diagram for soft photon production in the SPA.
}
\label{soft_fd}
\eef
We have studied soft (low transverse momentum) photon production 
due to parton bremsstrahlung 
in a thermal system  
within the soft photon approximation (SPA)~\cite{soft1}.
These photons are produced from one of the external legs in a parton scattering diagram
(Fig.~(\ref{soft_fd}))
{\it i.e.} due to processes like $ab\ra ab\gamma$ where $a$ is a quark or
antiquark and $b$ can be
a quark, antiquark or a gluon. These are hence $O(\alpha\alpha_s^2)$
processes compared to the $O(\alpha\alpha_s)$ processes described
earlier. Naively, the soft photon approximation provides
that the cross section for the process factorizes into a scattering part and
a photon production part. The photon can be emitted from any of the 
external lines in Fig.~(\ref{soft_fd}). The emission of photons from
the interior of the scattering vertex (the central blob in the figure) is neglected
because in the limit of very low energy of the emitted photon its contribution 
is very small. The dependence of the photon momentum $p$ 
is neglected both in
the strong part of the matrix element ${\cal M}_0$ as well as in
the phase space delta function. The latter is however taken care of through
a correction factor.  The matrix element is then  written as
\be
{\cal{M}}=e{\cal{M}}_0J^\mu\epsilon_\mu
\ee
and the rate of production of soft thermal photons at temperature $T$
is given by 
\begin{eqnarray}
E\frac{dR}{d^3p}&=&\frac{T^6g_{ab}}{16\pi^4} 
\int_{z_{\s {min}}}^{\infty} \,dz
\frac{\lambda(z^2T^2,m_a^2,m_b^2)}{T^4}\nonumber\\
&&\times 
\Phi(s,s_2,m_a^2,m_b^2)
K_1(z)\, E\frac{d\sigma^{\gamma}}{d^3p},
\end{eqnarray}
where
$z_{\s {min}}=(m_a+m_b)/T$, $z=\sqrt s/T$ and $g_{ab}$ is the colour and
spin degeneracy.
The cross-section for the process $ab \rightarrow cd \gamma$
is given by
\begin{equation}
E\frac{d\sigma^{\gamma}}{d^3p} =
 \frac{\alpha}{4 \pi^2} \,\frac{\widehat{\sigma}(s)}{E^2},
\end{equation}
with
\begin{equation}
\widehat{\sigma}(s)=\int_{- \lambda(s,m_a^2,m_b^2)/s}^0 \,dt\,
\frac{d\sigma_{ab \rightarrow cd}}{dt}\,
\left(E^2\left|\epsilon\cdot J\right|^2_{ab \rightarrow cd}\right).
\label{sftsig}
\end{equation}
Here $J$ is the real photon current and
\begin{equation}
\Phi(s,s_2,m_a^2,m_b^2)=\frac{\lambda^{1/2}(s_2,m_a^2,m_b^2)}
                         {\lambda^{1/2}(s,m_a^2,m_b^2)}\,\frac{s}{s_2}
\end{equation}
where $\lambda(x,y,z)=x^2-2(y+z)x+(y-z)^2$.
The strong interaction differential cross-section
$d\sigma_{qq}/dt$ and $d\sigma_{qg}/dt$ for scattering of
quarks and gluons are obtained from semi-phenomenological expressions
used earlier by several authors for this purpose.
\bef
\centerline{\psfig{figure=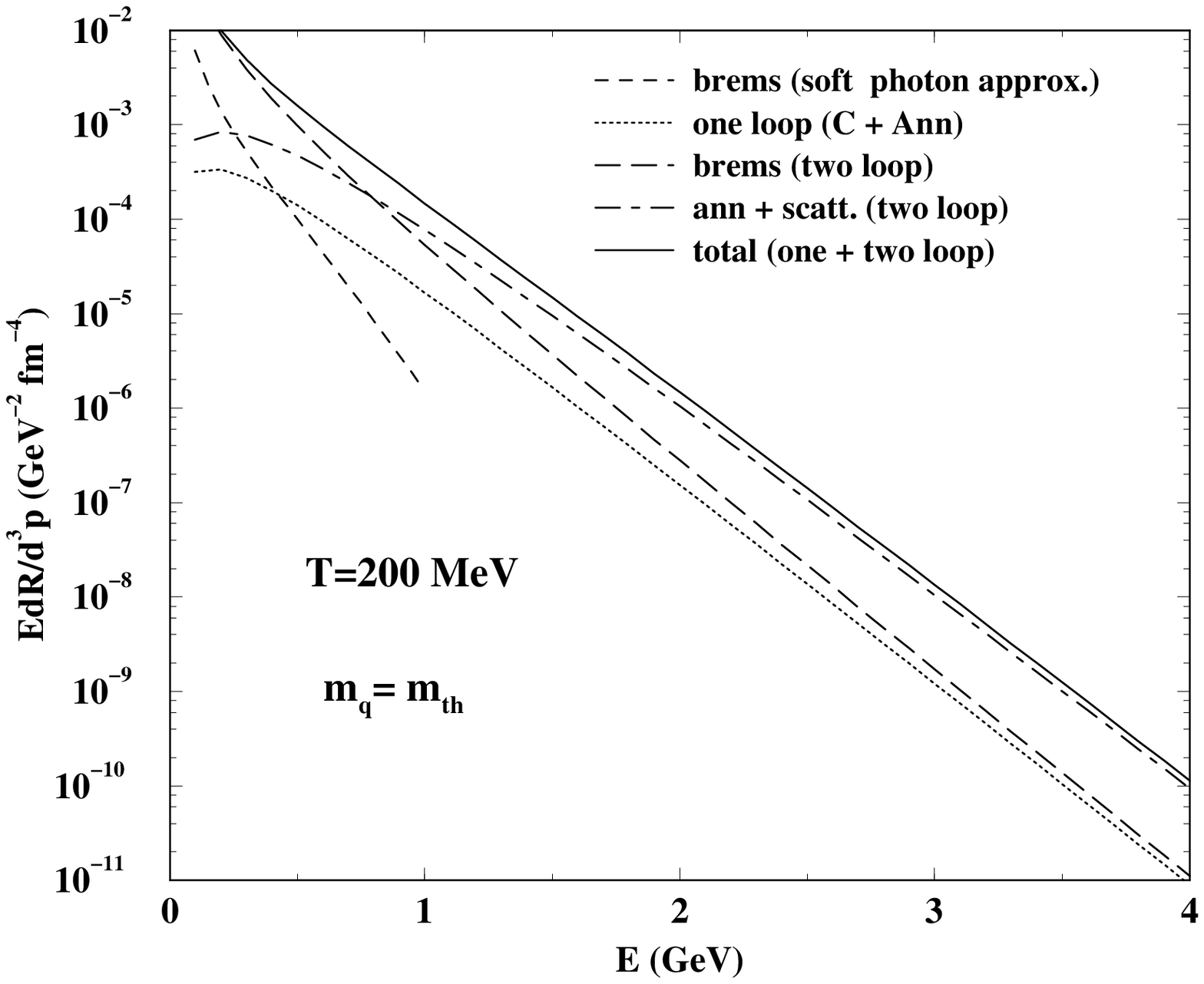,height=7cm,width=9cm}}
\caption{
Thermal photon emission rates from QGP at $T$= 200 MeV.
}
\label{qgpgamrate}
\eef
 The emission
rates of hard and soft thermal photons are shown in Fig.~(\ref{qgpgamrate}).
One observes that the 2-loop corrections to the photon rate 
namely, the bremsstrahlung contribution given by Eq.~(\ref{brm2}) (the long-dashed curve)
and the contribution from $\bar q q$ annihilation with scattering as given by
Eq.~(\ref{ann2}) (dot-dashed line) are of the
same order as the 1-loop result (dotted line). Soft photons calculated in the SPA is
plotted only up to $E=$1 GeV for consistency~\cite{soft1} regarding the Landau-Pomeranchuk
effect~\cite{lpm}.

\subsection{Hard QCD Photons}

As remarked earlier, the large transverse momentum window for the photons
will have a
contribution due to  partonic interactions which occur when nuclei start
to interpenetrate at very high collision energy
This contribution can
be understood in terms of perturbative QCD, and can provide reliable
information about the partonic distribution in the nuclei, as well as a
means of providing near-absolute normalization of the photon measurements
at larger transverse momenta. Since these photons are emitted at the very early
stages of the collision with large momenta they are usually referred to
as hard or prompt QCD photons. The cross section to leading order of
perturbative QCD for the production of a photon in a hadronic collision, is
obtained  by
convoluting the cross-section for the elementary processes {\it e.g.}
Compton and annihilation with the gluon or quark contents of the participating
hadrons~\cite{owen}. 
Neglecting the correction due to neutron-proton asymmetry the nucleus-nucleus 
($A-B$) collision
is then built up as an incoherent sum of independent nucleon-nucleon ($N-N$) collisions. 
The $N-N$ cross section for the QCD Compton process at $y=0$ is given by
\bea
\frac{d\sigma_N^{Comp}}{d^2p_Tdy}&=&\frac{\alpha\alpha_s}{3s^2(x_T/2)}
\int_{x_{min}}^1\frac{dx_a\,x_a\,x_b}{x_a-(x_T/2)}\nonumber\\
&&\times \sum_q e_q^2\left[[q(x_a)+\bar q(x_a)]g(x_b)\frac{x_b^2+(x_T/2)^2}{x_a^2x_b^3}
+(x_a\leftrightarrow x_b)\right]
\eea
where, $q(x)$ and $g(x)$ are the quark and gluon structure functions of
the nucleon respectively and
\begin{eqnarray*}
x_b&=&\frac{x_ax_T}{2x_a-x_T}\nonumber\\
x_T&=&2p_T/\sqrt{s}\nonumber\\
x_{\s min}&=&\frac{x_T}{2-x_T}.
\end{eqnarray*}
For the annihilation process we have
\bea
\frac{d\sigma_N^{ann}}{d^2p_Tdy}&=&\frac{8\alpha\alpha_s}{9s^2}
\int_{x_{min}}^1\frac{dx_a\,x_a\,x_b}{x_a-(x_T/2)}\nonumber\\
&&\times \sum_q e_q^2\left[q(x_a)\bar q(x_b)\frac{x_a^2+x_b^2}{x_a^3x_b^3}
+(x_a\leftrightarrow x_b)\right].
\eea

A large contribution to the prompt photons also comes from the fragmentation of
partons in the final state in a parton-parton
scattering, $ab\,\rightarrow\,cd$. These are bremsstrahlung processes
and  though of higher order ($O(\alpha\alpha_s^2)$) compared to 
Compton and annihilation ($O(\alpha\alpha_s$)), is  found to contribute in
the same order of magnitude at all values of the transverse momenta.
Using an effective structure function the cross section for $N-N$ collisions is 
obtained as~\cite{owen}
\bea
\frac{d\sigma_N^{brem}}{d^2p_T\,dy}&=&K\frac{\alpha\alpha_s^2}{2\pi s^2}
\ln\frac{p_T^2}{\Lambda^2}\frac{1}{x_T}\int_{x_T}^1\frac{dy_T}{(y_T/2)^2}
[1+(1-x_T/y_T)^2]\int_{y_T/(2-y_T)}^1\frac{dx_a}{x_a-y_T/2}\nonumber\\
\nonumber\\
&&\times\left[F_2(x_a)[G(x_b)+\frac{4}{9}Q(x_b)]\frac{x_a^2+(y_T/2)^2}{x_a^4}
+(x_a\leftrightarrow x_b)\right],
\eea
where $x_b=x_ay_T/(2x_a-y_T)$ and
\[
F_2(x)=x\sum_qe_q^2[q(x)+\bar q(x)],~~~~~~
Q(x)=x\sum_q[q(x)+\bar q(x)]~~~~{\s and}~~~
G(x)=xg(x).
\]
The strong coupling is given by 
\be
\alpha_s(Q^2)=\frac{12\pi}{(33-2N_f)\ln(Q^2/\Lambda^2_{QCD})},
\ee
with $Q^2=p_T^2$, $N_f=4$ and $\Lambda_{QCD}=0.23$ GeV.

Neglecting the effects of shadowing, the photon distribution in a 
collision of nuclei $A$ and $B$ at an impact
parameter $b$ due to hard scattering of partons is given by
\be
E\frac{dN^{AB}}{d^3p}(b)=T_{AB}(b)\,E\frac{d\sigma^{NN}}{d^3p}
\ee
where $T_{AB}(b)$ is the nuclear overlap integral. It is defined as
\be
T_{AB}(b)=\int\,d^2s\,T_A(s)T_B(|\vec b-\vec s|).
\ee
$T_{A}=\int\,dz\,\rho_{A}(z,\vec s)$ is the nuclear thickness function
and $\rho_A$ is the nuclear number 
density normalized to the mass number $A$. Also, $T_{AB}$ is normalized such that
\be
\int\,d^2b\, T_{AB}(b)=AB.
\ee
Let us now consider central ($b=0$) collisions of two identical nuclei of mass
number $A$.
We will assume a constant nuclear number density $\rho_0$, so that
\[\frac{4}{3}\pi R_A^3\rho_0=A\]
and
\[T_A(s)=2\rho_0(R_A^2-s^2)^{1/2}.\]
Therefore,
\begin{eqnarray*}
T_{AA}(0)&=&\int_0^{R_A}\,d^2s\,T_A(s)T_A(s)\\
        &=&\frac{9}{8}\,\frac{A^2}{\pi R_A^2}.
\end{eqnarray*}
Writing $T_{AA}(0)\sim {A^2}/{\pi R_A^2}$
we obtain
\be
\frac{dN^{AA}}{d^2p_Tdy}=\frac{A^2}{\pi R_A^2}\frac{d\sigma^{NN}}{d^2p_Tdy}.
\ee
In Fig.~(\ref{hardgam}) we have plotted the photon yield due to 
the three processes described above using the MRSD-$\prime$~\cite{mrsd}
 set of structure
functions. One observes that the yield from bremsstrahlung is of the same 
order as the sum of the Compton and annihilation processes. We have used
$K=2$ in order to take into account the higher order QCD corrections.
\bef
\centerline{\psfig{figure=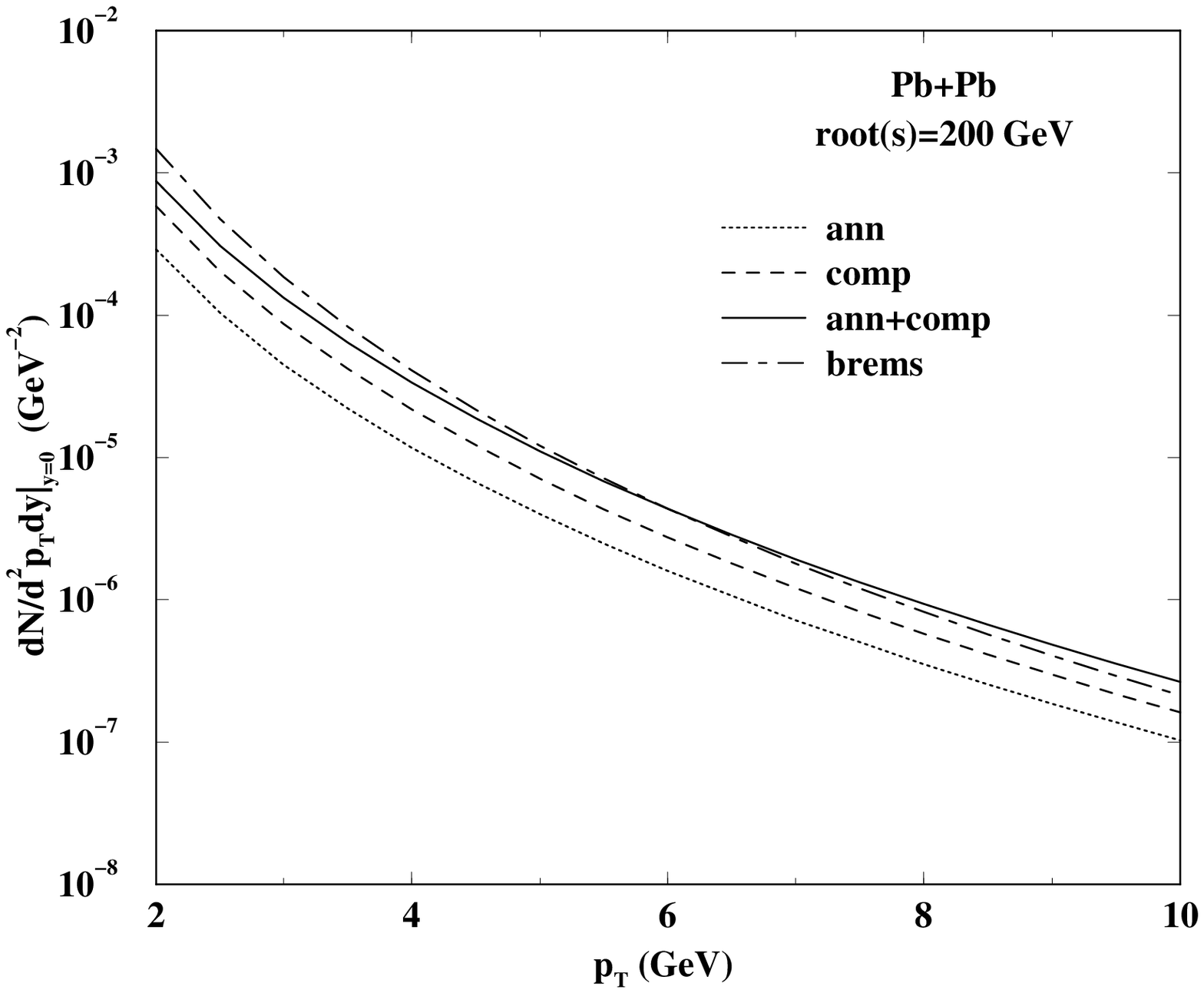,height=7cm,width=9cm}}
\caption{QCD (prompt) photons from Pb-Pb collisions at
200 A GeV.
}
\label{hardgam}
\eef

\subsection{Thermal Dilepton Emission Rates from QGP}

The most dominant contribution to  thermal dilepton production from QGP
comes from quark anti-quark annihilation as shown if Fig.~(\ref{loqgp}c). 
To obtain the differential number of lepton pairs emitted per
unit four-volume  we start from Eq.~(\ref{dilrate2}).
Since quantum effects in the QGP is known to have a negligible effect we will
use Boltzmann distribution for the quarks. 
The phase-space integrals are performed using the delta function
to obtain
\be
\frac{dR}{dM^2d^2M_Tdy}=\frac{1}{4(2\pi)^5}M^2(1-\frac{4m_q^2}{M^2})\,\sigma_{q\bar q}(M)
\exp(-M_T\cosh y/T).
\ee
Integrating over $M_T$ we arrive at
\be
\frac{dR}{dM^2dy}=\frac{1}{4(2\pi)^4}(1-\frac{4m_q^2}{M^2})\,\sigma_{q\bar q}(M)
\frac{M^2T^2}{\cosh^2 y}
\left(1+\frac{M\cosh y}{T}\right)\exp(-M\cosh y/T)
\label{kkmm1}
\ee
and on integration over $y$ we get
\be
\frac{dR}{dM} = \frac{1}{(2\pi)^4}\,M^4\,T\,
(1-\frac{4m_q^2}{M^2})\,\sigma_{q\bar q}(M)
K_1(M/T).
\label{kkmm2}
\ee
The  cross section for two light quark flavours can be evaluated using the 
diagram Fig.~(\ref{loqgp}c) to obtain~\cite{wong} 
\be
\sigma_{q\bar{q}\rightarrow e^+e^-}=\frac{80\pi}{9}\frac{\alpha^2}{M^2}
\left(1-\frac{4m^2}{M^2}\right)^{1/2}
\left(1+\frac{2m^2}{M^2}\right)
\left(1+\frac{2m_q^2}{M^2}\right)
\left(1-\frac{4m_q^2}{M^2}\right)^{-1/2}
\ee
where $m$ is the lepton mass and $m_q$ is the quark mass. The
thermal mass of the quarks will be used in the calculations.
It is worth mentioning that the $\alpha_s$ corrections to this rate comes from the Compton and
annihilation diagrams {\it i.e.} the diagrams labelled
$a$ and $b$ in Fig.~(\ref{loqgp})
with a virtual photon in place of the real one~\cite{alth}. One encounters 
mass (collinear) singularities in the evaluation of these processes which
can be cured through the use of KLN theorem.

A significant contribution to the dilepton yield in the lower invariant mass region comes
from soft (bremsstrahlung) processes.
The rate of production of soft thermal dileptons 
can be evaluated analogously as the soft photons using the SPA. The photon emitted 
from the external lines in Fig.~(\ref{soft_fd}) in this case is a virtual one which eventually
produces a lepton pair. The rate is given by~\cite{soft3} 
\begin{eqnarray}
\frac{dR}{dM^2dy}&=&\frac{T^6g_{ab}}{16\pi^4} 
\int_{(m_a+m_b+M)/T}^{\infty} \,dz\int_M^{\sqrt s -m_a-m_b}2\pi M_T dM_T
\nonumber\\ 
&&\frac{\lambda(z^2T^2,m_a^2,m_b^2)}{T^4}
\Phi(s,s_2,m_a^2,m_b^2)
K_1(z)\frac{d\sigma^{e^+e^-}}{dM^2d^2M_Tdy},
\end{eqnarray}
where
\begin{equation}
\frac{d\sigma^{e^+e^-}}{dM^2d^2M_Tdy}
=\frac{\alpha^2}{12 \pi^3 M^2} \,\frac{\widehat{\sigma}(s)}{M_T^2\cosh^2 y}.
\end{equation}
$\widehat{\sigma}(s)$ is as given in Eq.~(\ref{sftsig}) with $J$ appropriately
modified to account for virtual photon emission. The rates for thermal dilepton
production from $q\bar q$ annihilation ($O(\alpha^2)$) is plotted 
 in Fig.~(\ref{dildig_q}) 
along with the soft $O(\alpha^2\alpha_s^2)$ contributions discussed
above.
\bef
\centerline{\psfig{figure=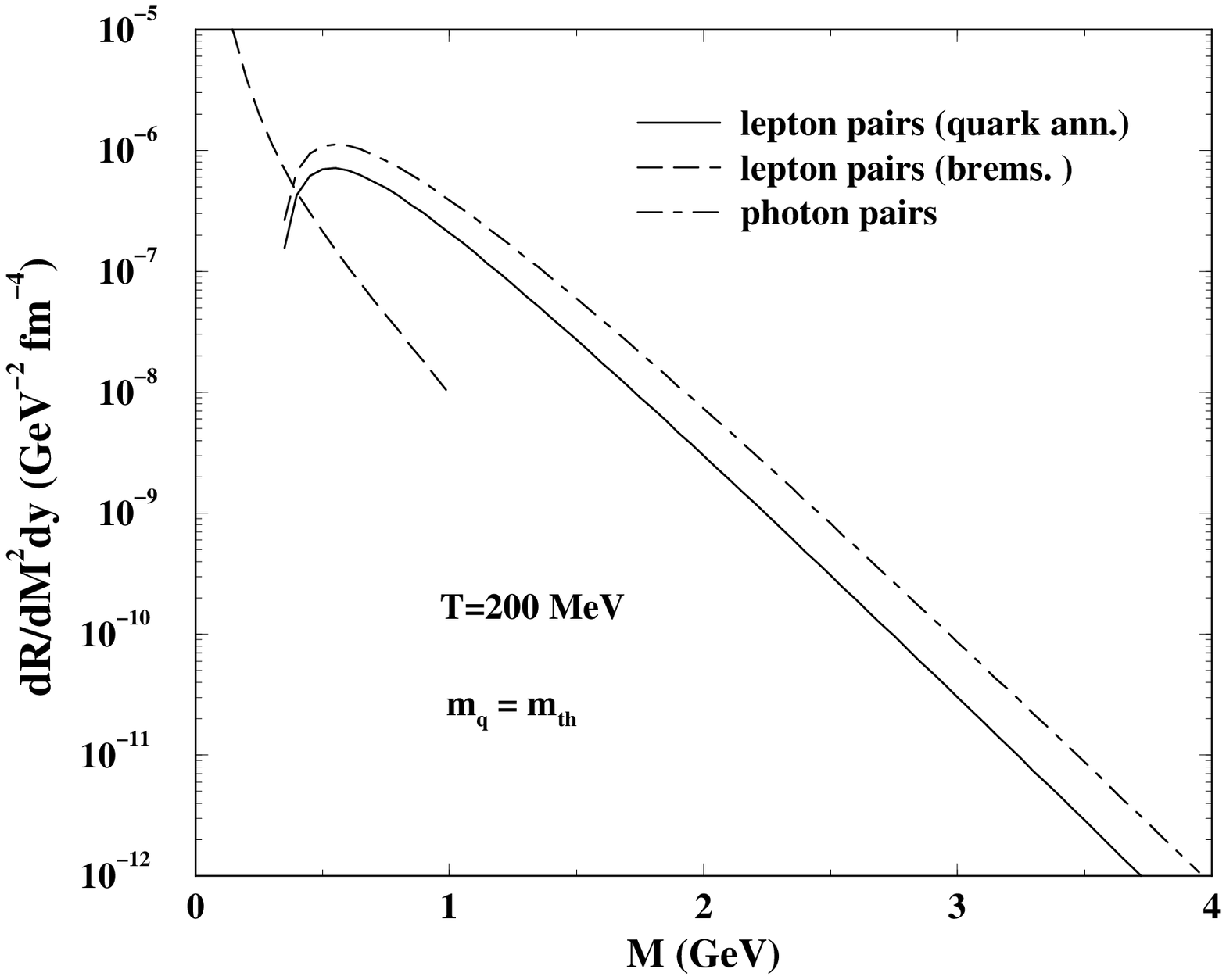,height=7cm,width=9cm}}
\caption{
Thermal $e^+e^-$ and $\gamma\gamma$ emission rates from QGP at $T$=200 MeV.
}
\label{dildig_q}
\eef

\subsection{Drell-Yan Dileptons}

In the region of large invariant mass the dilepton yield is dominated
by the Drell-Yan process $A\, +\, A \ra l^+l^- + X$ and thus forms the
principal background to dileptons emitted from the QGP. Here, a quark from one
of the nucleons in nucleus $A$ annihilates an antiquark from one of
the nucleons in the other nucleus to produce a virtual photon which subsequently
decays into a lepton pair. The differential
yield of such lepton pairs produced is $ A-A$ collisions is
obtained by an incoherent sum of the contributions from
independent nucleon-nucleon collisions.
For central collisions, the Drell-Yan yield is given by
\be
\frac{dN}{dM^2dy}=\frac{A^2}{\pi R_A^2}K\frac{4\pi\alpha^2}{9M^2s}
\sum_qe_q^2[q(x_a)\bar q(x_b) + (x_a\leftrightarrow x_b) ]
\ee
where as in the case of prompt photons, $q(x)$ and $\bar q(x)$ are
the quark distribution functions of a nucleon.
At $y=0$, $x_a=x_b=M/\sqrt s$
where $s$ is the square of the center of mass energy of the
colliding nucleons. 

\subsection{Diphotons}

We will briefly discuss the production of diphotons  produced
due to the annihilation of quarks and antiquarks.
The importance of thermal diphoton emission
lies in the fact that an experimental detection of large mass
diphotons can possibly provide a valuable confirmation of the results 
obtained from the measurement of dileptons~\cite{ssdipho}. 
The diphoton cross section is given by:
\begin{eqnarray}
\sigma_{q\bar{q}}^{\gamma\gamma}(M)&=&2\pi\alpha^2\,N_c\,(2S+1)^2\,
\sum_q\frac{e_q^4}{M^2-4m_q^2}\nonumber\\
& &\left[\left[1+\frac{4m_q^2}{M^2}-\frac{8m_q^4}{M^4} \right]
\,\ln\left\{\frac{M^2}{2m_q^2}
 \left[1+\left[1-\frac{4m_q^2}{M^2}\right]^{1/2}\right]-1\right\}
\right.\nonumber\\
& &\left.-\left[1+\frac{4m_q^2}{M^2}\right]\left[1-\frac{4m_q^2}{M^2}\right]
^{1/2}\right],
\end{eqnarray}
In the above $N_c=3$, $S=1/2$ and $e_q$ 
is the charge of the quark and we have used 
$m_q=m_{\s th}=\sqrt{(2\pi\alpha_s/3)}\,T$. 
The emission rate is obtained by using this cross-section in 
Eq.~(\ref{kkmm1}) or Eq.~(\ref{kkmm2}). The rate of emission of thermal
$\gamma\gamma$ pairs from QGP at $T$=200 MeV is shown in Fig.~(\ref{dildig_q}).

\bef
\centerline{\psfig{figure=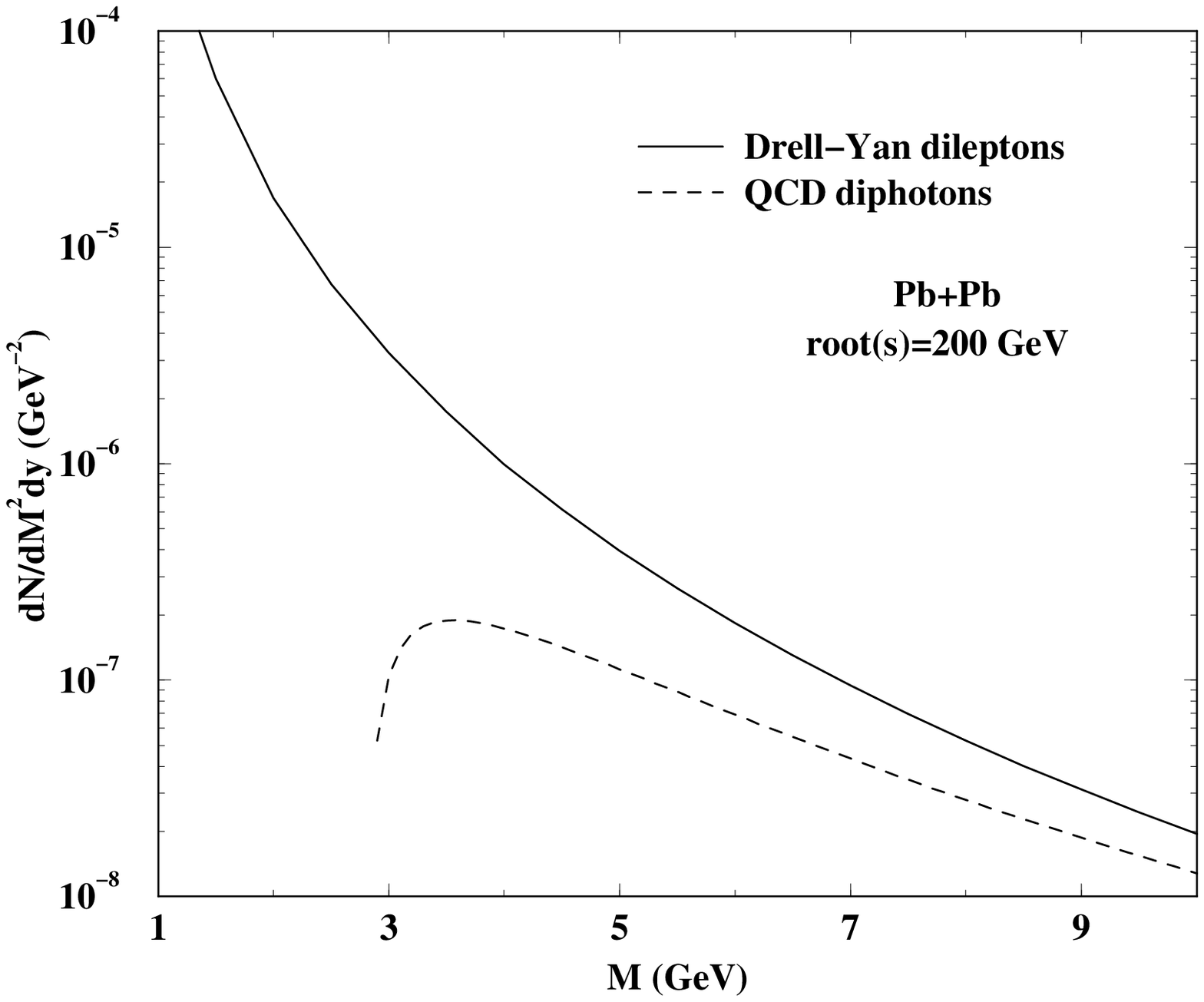,height=7cm,width=9cm}}
\caption{
QCD dielectrons and diphotons from $q\bar q$ annihilation.
}
\label{qcddildig}
\eef
Diphotons with large invariant mass are also produced from the hard QCD 
annihilation of quarks and antiquarks in the colliding nuclei. The yield can be
calculated analogously as the Drell-Yan pairs to get~\cite{ssdipho} 
\begin{eqnarray}
\frac{dN}{dM^2dy}&=&\frac{A^2}{\pi R_A^2}\frac{2\pi \alpha^2}{3sM^2}\left[
\ln\left(\frac{M^2-p_c^2}{p_c^2}\right)-\left(1-2p_c^2/M^2\right)
\right]\nonumber\\
&&\times \sum_q e_q^4
\left[q(x_a)\bar q(x_b)+(x_a\leftrightarrow x_b)\right].
\end{eqnarray}
At $y=0$, $x_a=x_b=M/\sqrt s$. An arbitrary cut-off on the momentum transfer
$p_c (=2$ GeV) has been introduced so that perturbative QCD remains valid in
this case. In Fig.~(\ref{qcddildig}) we have shown the dilepton yield due
to the Drell-Yan process for Pb-Pb collisions at 200 A GeV. Also shown is
an estimate of the diphoton yield in such a collision. As before we have used 
the MRSD-$\prime$ set of nucleon structure functions.

\chapter{Medium Effects and Emission Rates from Hot Hadronic Matter} 

In this Chapter we will consider photon and dilepton emission from
a thermal system of interacting hadrons. 
We have seen that the photon and dilepton emission rates
are related to the imaginary part of the photon self energy in the medium.
As a result the rates will depend on
the in-medium modifications of the
hadrons appearing in the internal loop of the photon
self energy diagram.
Here the hadronic medium consists of
mesons and baryons at a finite temperature. Due to
the interactions with real and virtual excitations, the properties of
these hadrons are expected to get modified. As a result the 
propagators appearing in the photon self energy undergo modifications.
These are studied in the framework of Thermal Field 
Theory.

Many of the hadrons in a hot hadronic gas are electrically charged and
hence couple to the electromagnetic field. Pions and $\rho$ mesons form the most
important constituents of such a system. This is because pions are light and
the $\rho$ mesons have large spin-isospin degeneracy. 
Lagrangian densities constructed with the $\pi$, $\rho$, $\omega$, $\eta$ and
$a_1$ fields have been used to calculate the amplitudes for photon production.
Of these, the medium modifications of the $\rho$ and $\omega$ mesons are 
known to affect the photon and dilepton spectra significantly.
Though we have included the presence of nucleons and antinucleons for the
evaluation of the medium effects of vector mesons, we have neglected their
contribution to the production of photons and dileptons. Also, we will
assume the net baryon number to be zero.
In Section~3.1 we will study
the in-medium modifications of hadronic masses and decay widths 
using well known models.
Thereafter, in Secs.~3.2 and 3.3 we shall discuss the emission rates of 
photons and dileptons where these medium modifications are taken into account.

\section{Hadronic Properties at Finite Temperature}

In this Section we will consider the in-medium modifications of the mass
and decay widths of the $\rho$ and $\omega$ mesons. 
We will discuss the Quantum Hadrodynamic (QHD) model, the gauged linear and non-linear sigma models,
and the hidden local symmetry approach. We will also show how the QCD
sum rules can be used to constrain the spectral functions of the vector
mesons in the medium.

The change in the hadronic mass in the medium 
can be understood from 
the following phenomenological
arguments~\cite{eletsky}. 
Let us consider the propagation of a vector meson in a nuclear medium. 
The attenuation of the amplitude at a distance $z$, in a Fermi gas 
approximation, is given by $e^{-n\sigma z}$,  where $n$ is the density 
of nucleons and $\sigma$ is the meson-nucleon interaction cross section. 
The optical theorem relates $\sigma$ to the imaginary part of the
forward scattering amplitude; $\sigma=4\pi{\s Im}{\cal F}(E)/k$. It then follows that
the meson wave function $\psi \sim \exp[2\pi inz{\cal F}(E)/k]$,
where the imaginary part of ${\cal F}$ accounts for the attenuation
and the real part modifies the dispersion relation of the propagating particle. In terms of an
effective mass $(m_{\s eff}=m+\D m)$, the propagation can also be described by
$\psi\sim \exp[i\sqrt{E^2-m_{\s eff}^2}z]$. Comparing the arguments of
the exponential we get
\be
\D m=-\frac{2\pi n k}{m}{\s Re}{\cal F}(E).
\ee
This relation clearly shows that the enhancement or reduction
of hadronic masses depends on the sign of ${\s Re}{\cal F}(E)$.

\subsection{Quantum Hadrodynamics}

In the Quantum Hadrodynamic model~\cite{vol16,chin} of nuclear matter
the vector meson properties are modified due to coupling with
nucleonic excitations. The discussion has two parts. 
We will first study how the properties of nucleons are modified in matter 
at finite temperature. 
The nucleons interact through the exchange of scalar $\sigma$ and the vector $\omega$ 
mesons and their mass is modified due to the scalar condensate. 
Thereafter, we will consider the changes in the $\rho$ and $\omega$ meson masses 
due to coupling with these modified nucleonic excitations.

\subsubsection{a) The Nucleon Mass}

The interaction in QHD is described by the Lagrangian
\be
{\cal L}^{\s int}_{QHD} = -g_{\wnn}\,{\bar N}\gamma_{\mu}\,N\,\omega^{\mu}+
g_{\snn}\,{\bar N}\,\sigma\,N,
\label{lqhd}
\ee
where $N(x)$, $\sigma(x)$, and $\omega(x)$ are the nucleon, $\sigma$, and
$\omega$ meson fields respectively. The $\sigma (\omega)$ field couples to the
scalar (vector) current of the nucleon with the coupling constant $g_{\snn} 
(g_{\wnn})$ which will be specified later. 

As discussed in the previous Chapter (Section~2.1), the free nucleon propagator at finite temperature and density in
general has four components. The time-ordered {\it i.e.}
the (11)-component is physically relevant for our purpose and we will
denote this as $G^0(p)$ where $p$ denotes the four-momentum of the nucleon.
So we have 
\bea
G^0(p)&\equiv&G^{0(11)}(p)\nonumber\\
&=&(p\sls+M_N)\left[\frac{1}{p^2-M_N^2+i\eps}
+2\pi i\delta(p^2-M_N^2)\eta(p.u)\right]\nonumber\\
&\equiv&G_F^0(p) + G_D^0(p),
\eea 
where the first term ($G_F^0$) describes the free propagation of 
nucleon-antinucleon pairs and the second term ($G_D^0$) allows for the on-shell
propagation of particle-hole pairs. $M_N$ in the above equation is the free
nucleon mass.

 The effective mass of the nucleon in matter at finite temperature in 
presence of the interaction described by Eq.~(\ref{lqhd}) will appear as a pole of
the effective nucleon propagator. 
In the Relativistic Hartree Approximation (RHA)~\cite{vol16,chin} one obtains 
the 
effective propagator by summing up scalar and vector tadpole diagrams 
self-consistently {\it i.e.} by using the interacting propagators to 
determine the self energy. The effective propagator referred to as the  
Hartree propagator is given by 
\begin{figure}
\centerline{\psfig{figure=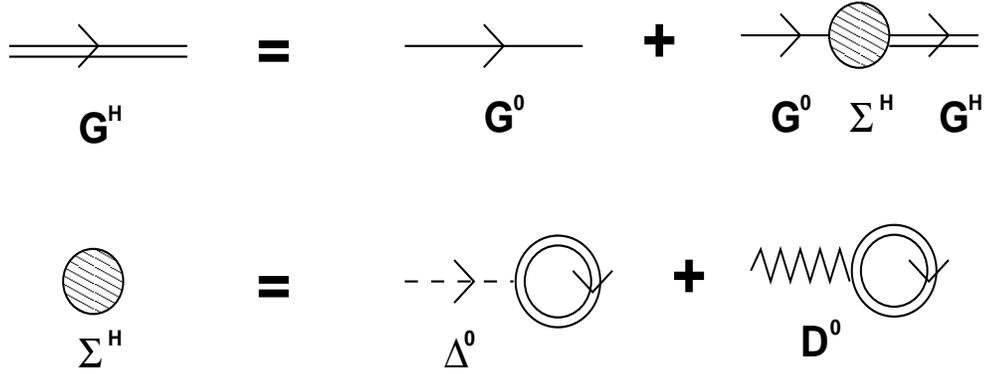,width=13cm,height=5cm,angle=-90}
}
\caption{Diagrammatic representation of Dyson-Schwinger equation for nucleons
in RHA }
\label{rhafig}
\end{figure}
\be
G^H(p) = G^0(p) + G^0(p)\Sigma^H(p)G^H(p).
\label{rhadyson}
\ee 
This is pictorially shown in Fig.~(\ref{rhafig}).
Here $\Sigma^H(p)$ is the nucleon self energy which contains contributions
from both scalar ($\Sigma_s$) and vector ($\Sigma_v^{\mu}$) tadpole
diagrams~\cite{vol16,chin} and is given by
\be
\Sigma^H = \Sigma_s^H - \gamma^{\mu}\Sigma_{\mu v}^H,
\label{sigma}
\ee
where
\be
\Sigma_s^H = i\frac{g_{\snn}^2}{m_{\sigma}^2}\,\int\,\frac{d^4p}{(2\pi)^4}\,
{\s {Tr}}[G^H(p)] 
\label{sigmas1}
\ee 
and
\be
\Sigma_{\mu v}^H = i\frac{g_{\wnn}^2}{m_\omega^2}\,\int\,\frac{d^4p}{(2\pi)^4}\,
{\s {Tr}}[\gamma_{\mu}\,G^H(p)].
\label{sigmav1}
\ee
Here, $m_\sigma\,\,(m_\omega)$ is the mass of the neutral scalar (vector) meson.
The solution of Eq.~(\ref{rhadyson}) now reads,
\bea
G^H(p)&=&(\bar p\sls+M_N^{\ast})\left[\frac{1}{\bar p^2-M_N^{\ast 2}+i\eps}
+2\pi i\delta(\bar p^2-M_N^{\ast 2})\eta(\bar p.u)\right]\nonumber\\ 
&\equiv&G_F^H(p) + G_D^H(p)
\label{gH}
\eea
One observes that the pole structure of the
effective nucleon propagator in RHA resembles that of the non-interacting
propagator with shifted mass and four-momentum {\it i.e.} $\bar p=p+\Sigma_v^H$
and $M_N^\ast=M_N+\Sigma_s^H$, is the effective mass. 
Using $G_D^H$ in place of the full Hartree 
propagator in Eqs.~(\ref{sigmas1}) and (\ref{sigmav1}) defines the Mean
Field Theory (MFT) values of the self energies. This is equivalent to 
solving the meson field equations with the replacement of the meson
field operators by their expectation values which become classical
fields {\it i.e.} $\sigma \ra \langle\sigma\rangle$ and 
$\omega \ra \langle\omega\rangle$. This yields 
\bea
\langle\sigma\rangle
&=&g_{\snn}\,\rho_s/m_\sigma^2\nonumber\\
\langle\omega^\mu\rangle &=& 
g_{\wnn}\,\delta^{\mu 0}\rho_B/m_\omega^2
\eea
which indicate that the nuclear ground state contains scalar and vector
meson condensates generated by baryon sources. $\rho_B$ is the
baryon density of the medium and $\rho_s$ is the (Lorentz) scalar density. The spatial part of
the $\omega$ condensate vanishes due to rotational symmetry in infinite
nuclear medium. These condensates 
are related to the scalar and vector self energies generated by summing
tadpole diagrams in QHD as
\bea
\Sigma_s&=&-g_{\snn}\langle\sigma\rangle\nonumber\\
\Sigma_v^0&=&-g_{\wnn}\langle\omega^0\rangle. 
\eea
The mean field approximation is thus to neglect the fluctuations in the
meson fields which themselves are generated by the nucleons. 

RHA is obtained when one includes the vacuum fluctuation corrections 
to the MFT results. This amounts to the inclusion of the Dirac
part of the propagator $G_F^H$ in the calculation of the self energies.
Summing over the vacuum tadpoles  results in a sum over all occupied states
in the negative energy sea of nucleons. Vacuum (or quantum) fluctuations,
as these are called, form an essential ingredient in a relativistic
theory of many particle systems. Since there are infinite number of
negative energy states in the vacuum one expects that the vacuum contribution
to the self energy is infinite.

Let us now find the Hartree self energy of the nucleon with the full
nucleon propagator consisting of a medium as well as a vacuum part.
The vector part of the self energy is obtained from Eq.~(\ref{sigmav1}) as
\be
\Sigma_v^{H\mu}=8i\frac{g_{\wnn}^2}{m_\omega^2}\int \frac{d^4p}{(2\pi)^4}\,
\frac{\bar p^\mu}{\bar p^2-M_N^{\ast 2}+i\eps}-
\frac{g_{\wnn}^2}{m_\omega^2}\delta^{\mu 0}
\rho_B.
\label{sigmav2}
\ee
The first term of this equation appears to be divergent. The usual
procedure is to regularize the integral in $n$ dimensions by dimensional
regularization to render the integral finite. One can then shift the
integration variable from $p$ to $\bar p$. The resulting integral vanishes
by symmetric integration. 
The vector self energy then reduces to
\be
\Sigma_v^{H\mu}=-{g_{\wnn}^2}\delta^{\mu 0}\rho_B/{m_\omega^2}
\ee
and gives
rise to a an effective chemical potential, 
\be
\mu^\ast=\mu-{g_{\wnn}^2\,\rho_B}/
{m_\omega^2}.
\ee
The scalar part of the self energy follows from Eq.~(\ref{sigmas1}):
\bea
\Sigma_s^H&=&8i\frac{g_{\snn}^2}{m_\sigma^2}\int\frac{d^4p}{(2\pi)^4}
\,\frac{M_N^{\ast 2}}{\bar p^2-M_N^{\ast 2} +
i\eps}- \frac{4g_{\snn}^2}{m_\sigma^2}
\int\frac{d^3p}{(2\pi)^3}\frac{M_N^{\ast}}{E^\ast}\,\nonumber\\
&&\times\,\left[\frac{}{}f_{FD}(\mu^{\ast},T)+{\bar f}_{FD}(\mu^{\ast},T)\right]
\label{sigmas2}
\eea
where
\bea
f_{FD}(\mu^{\ast},T)& = &\frac{1}{{\s exp}[(E^{\ast}-\mu^{\ast})/T]+1}\nonumber\\
{\bar f}_{FD}(\mu^{\ast},T)& = &\frac{1}{{\s exp}[(E^{\ast}+\mu^{\ast})/T]+1}
\nonumber\\
E^{\ast} & = & \sqrt{({\vec p}^2+M_N^{\ast 2})}\nonumber\\
\eea
The baryon density 
of the medium is given by
\be
\rho_B = \frac{4}{(2\pi)^3}\int\,d^3p\,
[f_{FD}(\mu^{\ast},T)-{\bar f}_ {FD}(\mu^{\ast},T)].
\label{bden}
\ee
The first term in Eq.~(\ref{sigmas2}), to be denoted by $\Sigma_s^{(1)}$,
 represents the contribution to the
scalar self energy from the filled Dirac sea and is ultraviolet divergent.
We will now proceed to renormalize this divergent contribution. The first step
is to isolate the divergent part through dimensional regularization. 
This gives
\bea
\Sigma_s^{(1)}&=&-\frac{g_{\snn}^2}{m_\sigma^2}\frac{\Gamma(2-n/2)}{2\pi^2}
M_N^{\ast 3}
\nonumber\\
&=&-\frac{g_{\snn}^2}{m_\sigma^2}\frac{\Gamma(2-n/2)}{2\pi^2}
(M_N^3+3M_N^2\Sigma_s^H+ 3M_N{\Sigma_s^H}^2+{\Sigma_s^H}^3)
\eea
since $M_N^{\ast}=M_N+\Sigma_s^H$.
The divergence in $\Sigma_s^{(1)}$ now appears as the pole of the 
$\Gamma$-function for physical dimension $n=4$.
The counter terms needed to remove the divergent contributions
from the loop corrections to the measurable amplitudes are
\be
{\cal L}_{CT}=\sum_{n=1}^4\alpha_n\,\sigma^{n}/n!
\label{lCT}
\ee
Including the contributions from the counter terms the renormalized
self energy becomes
\be
\Sigma_s^{(1) ren}=\Sigma_s^{(1)}+\Sigma_s^{CTC}, 
\ee
where
\be
\Sigma_s^{CTC}=\sum_{n=0}^3\frac{1}{n!}\left(\frac{-g_{\snn}}{m_\sigma^2}\right)
\left(\frac{-\Sigma_s^H}{g_{\snn}}\right)^n\alpha_{n+1}.
\ee
The coefficients ($\alpha_i$) are fixed by defining a set of
renormalization conditions. Since the scalar density $\rho_s$
(=$\langle\bar\psi\psi\rangle$) is not a conserved quantity the
tadpole diagrams appear in the self energy. The tadpole contribution
must vanish in normal vacuum (free space) {\it i.e.} $\langle\sigma\rangle_0=0$. 
This is ensured by the term $\alpha_1\sigma$ in ${\cal L}_{CT}$.
$\alpha_2\sigma^2$ is the meson mass counter term which ensures that
$m_\sigma$ is the physical (measured) mass. Since the original Lagrangian
of QHD~\cite{vol16} does not contain $\sigma^3$ and $\sigma^4$ terms,
three and four point meson amplitudes must vanish
at the tree level. The
last two counter terms in Eq.~(\ref{lCT}) are chosen to maintain this condition
at zero external momenta for the $\sigma$ meson when nucleon loop corrections are
included. We thus have
\be
\alpha_n=-i(-g_{\snn})^n(n-1)!\int\frac{d^4p}{(2\pi)^4}{\s Tr}[G_F^0(p)^n].
\ee

Consequently the effective nucleon mass reads
\bea
\Sigma_s^H & = &M_N^{\ast} - M_N\nonumber\\
&=& -\frac{4g_{\snn}^2}{m_\sigma^2}\int\,\frac{d^3p}{(2\pi)^3}\,
\frac{M_N^{\ast}}{E^{\ast}}
\left[\frac{}{}f_{FD}(\mu^{\ast},T)+{\bar f}_ {FD}(\mu^{\ast},T)
\right]\nonumber\\
           & + & \frac{g_{\snn}^2}{m_\sigma^2}\,\frac{1}{\pi^2}\left[M_N^{\ast 3}
{\s ln}\left(\frac{M_N^{\ast}}{M_N}\right)-
M_N^2(M_N^{\ast}-M_N)\right.\nonumber\\
           & - & \left.\frac{5}{2}M_N(M_N^{\ast}-M_N)^2-
\frac{11}{6}(M_N^{\ast}-M_N)^3
\right].
\label{nmass}
\eea
The solution of this equation (with $g_{\snn}^2=54.3$ and $m_\sigma=458$ MeV)
gives the effective nucleon mass $M_N^\ast$ as
a function of temperature and baryon density. At zero baryon density it can
be parametrized as~\cite{sspbpb}
\be
M_N^\ast=M_N\left[1-0.0264\left(\frac{T}{0.16}\right)^{8.94}\right].
\ee
where $T$ is in GeV. 
We thus observe that in nuclear matter, scalar($\sigma$) and vector($\omega$) mean fields
induced by nucleon sources give back-reactions to the nucleon propagation itself
and modify its self energy. This is the origin of $M_N^\ast < M_N$ in QHD.
\bef
\centerline{\psfig{figure=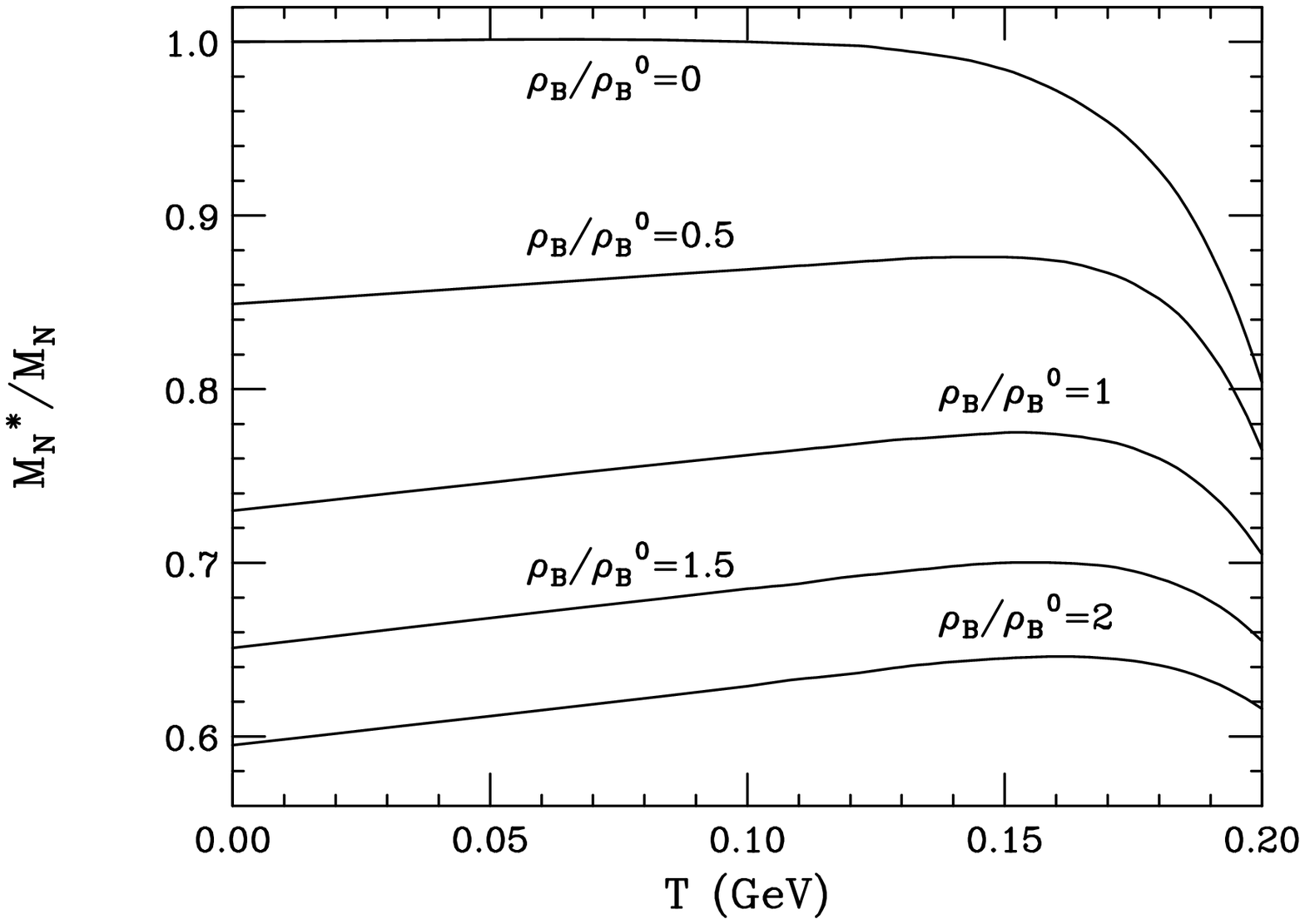,height=7cm,width=10cm}}
\caption{
Variation of nucleon mass with temperature 
for different values of baryon densities.
The normal nuclear matter density $\rho_B^0$=0.1484 fm$^{-3}$. 
}
\label{nmass1}
\eef
\bef
\centerline{\psfig{figure=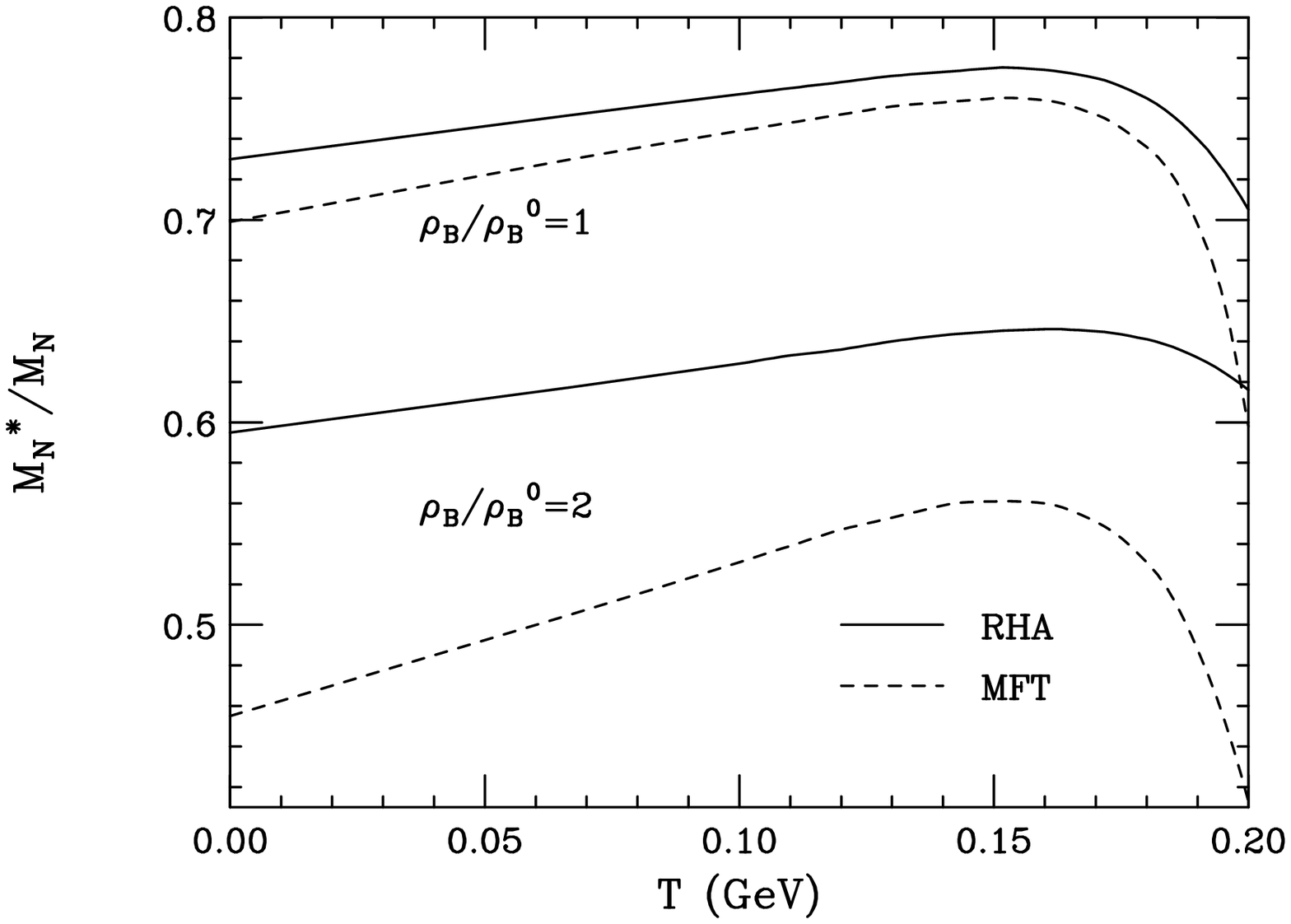,height=7cm,width=10cm}}
\caption{
Same as Fig.~(\protect\ref{nmass1}) with (solid) and without( dashed)  
vacuum fluctuation corrections to MFT.
}
\label{nmass2}
\eef

In Fig.~(\ref{nmass1}) we depict the
variation of nucleon mass with temperature for a set of baryon
densities.  We observe that the nucleon mass falls steadily 
with density for a fixed temperature. However, the variation
with temperature for given values of baryon densities shows
interesting features. At zero baryon density the nucleon mass  
decreases monotonically as a function of temperature, but 
for finite densities it increases slightly before falling.
This trend is similar to that obtained by Li {\it et al}~\cite{lkb}, and
may be attributed to the modification of the Fermi-sea
at finite temperature and density. Our calculation shows a 35\%
reduction of the effective nucleon mass at $T$ = 160 MeV and
two times normal nuclear matter density 
compared to its free mass. In order to highlight the effect of vacuum 
fluctuation corrections we compare the MFT results with those
obtained using RHA. This is plotted in Fig.~(\ref{nmass2}). We observe that
the effect of vacuum fluctuation is substantial for higher values of the
baryon density. The contribution of the antinucleons from the
Dirac sea is responsible for such an effect.     

\subsubsection{b) The Vector Meson Mass}

In a medium, meson properties get modified due to its coupling to nuclear
excitations as shown in Fig.~(\ref{vec_self_nn}). This modification is contained in the meson self energy 
which appears in the Dyson-Schwinger equation for the effective propagator
in the medium. The interaction vertices are provided by the Lagrangian
\be
{\cal L}^{\s int}_{VNN} = g_{VNN}\,\left({\bar N}\gamma_{\mu}
\tau^a N{V}_{a}^{\mu} - \frac{\kappa_V}{2M_N}{\bar N}
\sigma_{\mu \nu}\tau^a N\partial^{\nu}V_{a}^{\mu}\right),
\label{lagVNN}
\ee
where $V_a^{\mu} = \{\omega^{\mu},{\vec {\rho}}^{\mu}\}$,
$M_N$ is the free nucleon mass, $N$ is the nucleon field
and $\tau_a=\{1,{\vec {\tau}}\}$, $\vec\tau$ being the Pauli matrices.
\bef
\centerline{\psfig{figure=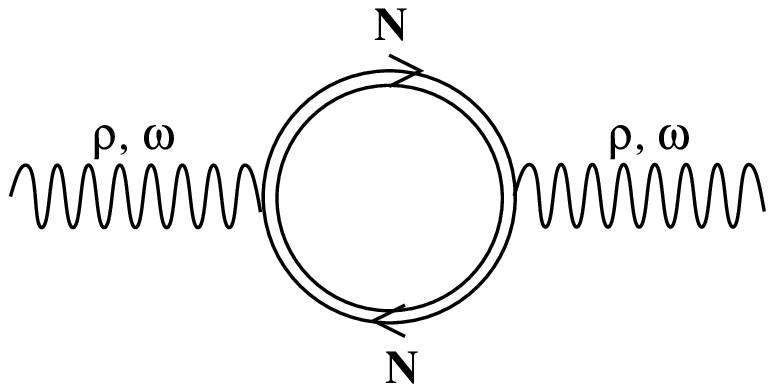,height=3cm,width=6cm}}
\caption{
The vector meson self energy due to $N\bar N$ polarization. The double
lines in the loop indicate nucleon propagators in RHA.
}
\label{vec_self_nn}
\eef

The lowest order contribution to the vector meson self energy is expressed
in terms of the self-consistent nucleon propagator described 
in Eq.~(\ref {gH}). This is given by
\be
\Pi^{\mn}(k)=-2ig_{VNN}^2\int\frac{d^4p}{(2\pi)^4}{\s Tr}\left[\Gamma^\mu(k)\,
G^H(p)\Gamma^\nu(-k)\,G^H(p+k)\right],
\ee
where $\Gamma^\mu$ represents the meson-nucleon vertex function obtained from
Eq.~(\ref{lagVNN}) and is given by
\bea
\Gamma^\mu(k)&=&\gamma^\mu;~~~~~~~~~~~~~~~~~~~~~~~~~~{\s for}~~ \omega\nonumber\\
\Gamma^\mu(k)&=&\gamma^\mu+i\frac{\kappa_{\rho}}{2M_N}\sigma^{\mu\alpha}k_\alpha;
~~~~~~~~ {\s for}~~\rho
\label{vtxfn}
\eea
where $\sigma^{\mu\alpha}=\frac{i}{2}[\gamma^\mu,\gamma^\alpha]$.
The vector meson self energy can be written as a sum of two parts
\be
\Pi^{\mn}(k)=\Pi_F^{\mn}(k)+\Pi_D^{\mn}(k),
\ee
where
\bea
\Pi_F^{\mn}(k)&=&-2ig_{VNN}^2\int\frac{d^4p}{(2\pi)^4}{\s Tr}\left[\Gamma^\mu(k)\,
G_F^H(p)\Gamma^\nu(-k)\,G_F^H(p+k)\right]\nonumber\\
\Pi_D^{\mn}(k)&=&-2ig_{VNN}^2\int\frac{d^4p}{(2\pi)^4}{\s Tr}\left[
\Gamma^\mu(k)\,G_F^H(p)\Gamma^\nu(-k)\,G_D^H(p+k)\right.\nonumber\\
&&+\Gamma^\mu(k)\,G_D^H(p)\Gamma^\nu(-k)\,G_F^H(p+k)\nonumber\\
&&+\left.\Gamma^\mu(k)\,G_D^H(p)\Gamma^\nu(-k)\,G_D^H(p+k)\right].
\label{fullpi}
\eea
$\Pi_F^{\mn}$ is the vacuum polarization. This is a bilinear function of 
$G_F^H$ and hence describes the correction to the meson propagators 
due to coupling to $N\bar N$ excitations. The $N\bar N$ pairs can
be excited only if the four-momentum carried by the mesons
is in the time-like region $(k^2>0)$. Hence the shift in the mass of the 
vector mesons due to vacuum polarization is caused by processes like
$V\ra N\bar N \ra V$ where $N$ represents nucleons in the modified Dirac
sea having an effective mass $M_N^\ast$, smaller than what it would be in
free space. 
From Eqs.~(\ref{fullpi}) and (\ref{gH}) we have
\be
\Pi_F^{\mn}(k)=-2ig_{VNN}^2\int\frac{d^4p}{(2\pi)^4}\frac{
{\s Tr}[\Gamma^\mu(p\sls +M_N^\ast)\Gamma^\nu(p\sls+k\sls+M_N^\ast)]}
{(p^2-M_N^{\ast 2})[(p+k)^2-M_N^{\ast 2}]}.
\ee
From naive power 
counting it can be seen that this part of the self energy is ultraviolet
divergent and has to be renormalized. A few comments about renormalizability
of the interaction given by Eq.~(\ref{lagVNN}) is in order here.
At very large momenta the propagator for massless
bosons $\sim O(k^{-2})$, whereas for massive vector bosons it goes as $\sim O(1)$.
This poses severe problems to the renormalizability of the theory with
massive vector bosons.
However, in a gauge theory with spontaneous 
symmetry breaking the vector gauge bosons acquire mass in such a way
that the renormalizability of the theory is always preserved. The theory
which involves neutral massive vector bosons coupled to a
conserved current is also renormalizable. This is because in a physical
process the propagator
$\bar D_0^{\mn}=(-g^{\mn}+k^\mu k^\nu/m^2)/(k^2-m^2+i\eps)$ appears 
between two conserved currents $J_\mu$ and $J_\nu$ and the offending
term $k^\mu k^\nu/m^2$ does not contribute because of current conservation
$(k_\mu J^\mu = 0)$, making the theory renormalizable. This is the case
for the $\omega$ meson~\cite{jean,caillon}
which we shall consider first.
The counter term required in this case is
\be 
{\cal L}_{VNN}^{CT}=-\frac{1}{4}\zeta V^{\mn}\,V_{\mn}.
\ee
We use dimensional regularization to separate the divergent and the
finite parts. The divergences now appear as a pole in the $\Gamma$-function
at the physical dimension $n=4$. The renormalized vacuum polarization
tensor for the $\omega$ is then given by
\be
\Pi_F^{\mn}(k)=(g^{\mn}-k^\mu k^\nu/k^2)\Pi_F^{ren}(k^2),
\ee
where
\bea
\Pi_F^{ren}(k^2)&=&\frac{g_{\omega NN}^2}{\pi^2}\left\{\Gamma(2-n/2)\int_0^1\,
dz\,z(1-z)\right.\nonumber\\
&&-\left.\int_0^1\,dz\,z(1-z)\ln[M_N^{\ast 2}-k^2z(1-z)]\right\}-\zeta
\eea
in which the counter term contribution 
\be
\Pi_F^{\mn CTC}=-\zeta(g^{\mn}-k^\mu k^\nu/k^2)
\ee
has been included. $\zeta$ is now determined by the renormalization condition
\be
\Pi_F^{ren}(k^2)|_{M_N^\ast\ra M_N}=0. 
\ee
Finally, we arrive at
\bea
\Pi_F^\omega (k^2)&=&\frac{1}{3}{\s Re}(\Pi_F^{ren})^\mu_\mu\nonumber\\
&=&-\frac{g_{\omega NN}^2}{\pi^2}k^2\int_0^1\,dz\,z(1-z)\ln\left[
\frac{M_N^{\ast 2}-k^2z(1-z)}{M_N^2-k^2z(1-z)}\right].
\eea

Because of the tensor interaction in Eq.~(\ref{lagVNN}) 
the vacuum self energy for the $\rho$ meson is not renormalizable.
We employ a
phenomenological subtraction procedure~\cite{shiomi,hatsuda2} to
extract the finite part using the condition;
\be
\left.\frac{\partial^n\Pi_F^\rho(k^2)}{\partial(k^2)^n}\right|_{M_N^\ast\ra M_N}=0
\ee
with ($n=$ 0,1,2,....$\infty$).
Using dimensional regularization and the above subtraction procedure we
arrive at the following expressions:
\be
\Pi_F^\rho (k^2)=-\frac{g_{\rho NN}^2}{\pi^2}
k^2\,\left[I_1+M_N^{\ast}\frac{\kappa_\rho}{2M_N}I_2+\frac{1}{2}\,(\frac{\kappa_\rho}
{2M_N})^2\,(k^2I_1+M_N^{\ast 2}I_2)\right],
\label{pik2}
\ee
where
\be
I_1=\int_{0}^{1}\,dz\,z(1-z)\,\ln\left[\frac{M_N^{\ast 2}-k^2\,z(1-z)}
{M_N^2-k^2\,z(1-z)}\right],
\ee
\be
I_2=\int_{0}^{1}\,dz\,\ln\left[\frac{M_N^{\ast 2}-k^2\,z(1-z)}
{M_N^2-k^2\,z(1-z)}\right].
\ee

The medium dependent part of the polarization, $\Pi_D^{\mn}$, describes
the coupling of the vector mesons to particle-hole excitations. It
contains at least one on-shell nucleon propagator which provides
a natural ultraviolet cut-off in the loop momenta. This part of the
self energy leads to an increased effective mass of the vector mesons
in the medium. 

We recall that in a hot and dense medium because of Lorentz invariance and current
conservation the general structure of the polarization tensor takes the form
\[
\Pi^{\mu \nu} = \Pi_T(k_0,\vec k)A^{\mu \nu}+\Pi_L(k_0,\vec k)B^{\mu \nu}
\]
where the two Lorentz invariant functions $\Pi_T$ and $\Pi_L$ are 
obtained by contraction as
\begin{eqnarray*}
\Pi_L&=&-\frac{k^2}{|\vec k|^2}u^{\mu}u^{\nu}\Pi_{\mu \nu}\\
\Pi_T&=&\frac{1}{2}(\Pi_{\mu}^{\mu}-\Pi_L),
\end{eqnarray*}
$u_{\mu}$ being the four velocity if the thermal bath.

In the case of a vector meson of four-momentum $k$ interacting with real particle-hole
excitations in the nuclear medium these are obtained as
\bea
\Pi_{\mu \nu}^D &=& -2ig_{VNN}^2\,\int\,\frac{d^4p}{(2\pi)^4}\,{\s{Tr}}\left[
\frac{}{}\Gamma_{\mu}(k)G_F(p)\Gamma_{\nu}(-k)G_D(p+k)+(F\leftrightarrow D)
\right]\nonumber\\
&=&(\Pi^{D,v}+\Pi^{D,vt}+\Pi^{D,t})_{\mu \nu}
\eea
with
\bea
(\Pi^{D,v})_{\mu}^{\mu} &=& \frac{g_{VNN}^2}
{2\pi^2}\,\frac{1}
{|\vec k|}\,\int\,\frac{pdp}{\omega_p}\,\left[(k^2+2M_N^{\ast 2})
\ln\left\{
\frac{(k^2+2|\vec p||\vec k|)^2-4k_0^2\omega_p^2}{(k^2-2|\vec p||\vec k|)^2-
4k_0^2\omega_p^2}\right\}\right.\nonumber\\
&&-\left.\frac{}{}8|\vec p||\vec k|\right]
\left[\frac{}{}f_{FD}(\mu^{\ast},T)+{\bar f}_{FD}(\mu^{\ast},T)\right]\,,
\eea
\bea
(\Pi^{D,vt})_{\mu}^{\mu} &=& \frac{3g_{VNN}^2}{\pi^2}\,
M_N^{\ast}\left(\frac{\kappa_V}{2M_N}\right)\,\frac{k^2}{|\vec k|}\,\int\,
\frac{pdp}{\omega_p}\,
\ln\left\{
\frac{(k^2+2|\vec p||\vec k|)^2-4k_0^2\omega_p^2}{(k^2-2|\vec p||\vec k|)^2-
4k_0^2\omega_p^2}\right\}\nonumber\\
&&\times
\left[\frac{}{}f_{FD}(\mu^{\ast},T)+{\bar f}_{FD}(\mu^{\ast},T)\right]\,,
\eea
and
\bea
(\Pi^{D,t})_{\mu}^{\mu}&=& \frac{g_{VNN}^2}{4\pi^2}\,
\left(\frac{\kappa_V}{2M_N}\right)^2\frac{k^2}
{|\vec k|}\,\int\,\frac{pdp}{\omega_p}\,\left[\frac{}{}(k^2+8M_N^{\ast 2})
\right.\nonumber\\
&&\times\left.
\ln\left\{
\frac{(k^2+2|\vec p||\vec k|)^2-4k_0^2\omega_p^2}{(k^2-2|\vec p||\vec k|)^2-
4k_0^2\omega_p^2}\right\}
-4|\vec p||\vec k|\right]\nonumber\\
&&\times\left[\frac{}{}f_{FD}(\mu^{\ast},T)+{\bar f}_{FD}(\mu^{\ast},T)\right].
\eea
The longitudinal component of the polarization tensor is given by
\be
\Pi_L^D=\Pi_L^{D,v}+\Pi_L^{D,vt}+\Pi_L^{D,t}
\ee
with
\bea
\Pi_L^{D,v}&=&-\frac{g_{VNN}^2}{4\pi^2}\,
\frac{k^2}{|\vec k|^3}
\int\,\frac{pdp}{\omega_p}\,\left[\frac{}{}\{(k_0-2\omega_p)^2-|\vec k|^2\}
\ln{\frac{k^2-2k_0\omega_p+2|\vec p||\vec k|}{k^2-2k_0\omega_p-2|\vec p||\vec k|}}
\right.\nonumber\\
&&+\left.\{(k_0+2\omega_p)^2-|\vec k|^2\}
\ln{\frac{k^2+2k_0\omega_p+2|\vec p||\vec k|}{k^2+2k_0\omega_p-2|\vec p||\vec k|}}
-8|\vec p||\vec k|\right]
\nonumber\\
&&\times\left[\frac{}{}f_{FD}(\mu^{\ast},T)+
{\bar f}_{FD}(\mu^{\ast},T)\right]\,,
\eea
\bea
\Pi_L^{D,vt} &=&\frac{g_{VNN}^2}{\pi^2}\,
M_N^{\ast}\left(\frac{\kappa_V}{2M_N}\right)
\frac{k^2}{|\vec k|}
\int\,\frac{pdp}{\omega_p}\,
\ln\left\{
\frac{(k^2+2|\vec p||\vec k|)^2-4k_0^2\omega_p^2}{(k^2-2|\vec p||\vec k|)^2-
4k_0^2\omega_p^2}\right\}\nonumber\\
&&\times\left[\frac{}{}f_{FD}(\mu^{\ast},T)+{\bar f}_{FD}(\mu^{\ast},T)\right],
\eea
and finally
\bea
\Pi_L^{D,t}&=&-\frac{g_{VNN}^2}{2\pi^2}\,
\left(\frac{\kappa_V}{2M_N}\right)^2
\frac{k^2}{|\vec k|}
\int\,\frac{pdp}{\omega_p}\,
\left[\left\{2|\vec p|^2-\frac{k^2}{2}-\frac{(k^2-2k_0\omega_p)^2}{2|\vec k|^2}
\right\}
\right.\nonumber\\
&&\times\left.\ln{\frac{k^2-2k_0\omega_p+2|\vec p||\vec k|}
{k^2-2k_0\omega_p-2|\vec p||\vec k|}}
+\left\{2|\vec p|^2-k^2-\frac{(k^2+2k_0\omega_p)^2}{|\vec k|^2}\right\}
\right.\nonumber\\
&&\times\left.\ln{\frac{k^2+2k_0\omega_p+2|\vec p||\vec k|}
{k^2+2k_0\omega_p-2|\vec p||\vec k|}}
-\frac{4|\vec p|k_0^2}{|\vec k|}\right]\nonumber\\
&&\times\left[\frac{}{}f_{FD}(\mu^{\ast},T)+
{\bar f}_{FD}(\mu^{\ast},T)\right]\,
\eea
where $\omega_p^2={\vec p}^2+M_N^{\ast 2}$ and $f_{FD}$ stands for Fermi-Dirac
distribution functions for the nucleons.
In the above the superscripts `$v$', `$vt$' and `$t$' represent
the vector-vector, vector-tensor and tensor-tensor components respectively
arising from the product of vector and tensor terms in Eq.~(\ref{vtxfn}).
The dispersion relation for the longitudinal (transverse) mode now reads 
\be
k_0^2-|\vec k|^2-m_V^2+{{\s Re}}\Pi_{L(T)}^D(k_0,{\vec k})+{\s {Re}}
\Pi^F(k^2) = 0.
\label{disp}
\ee
\bef
\centerline{\psfig{figure=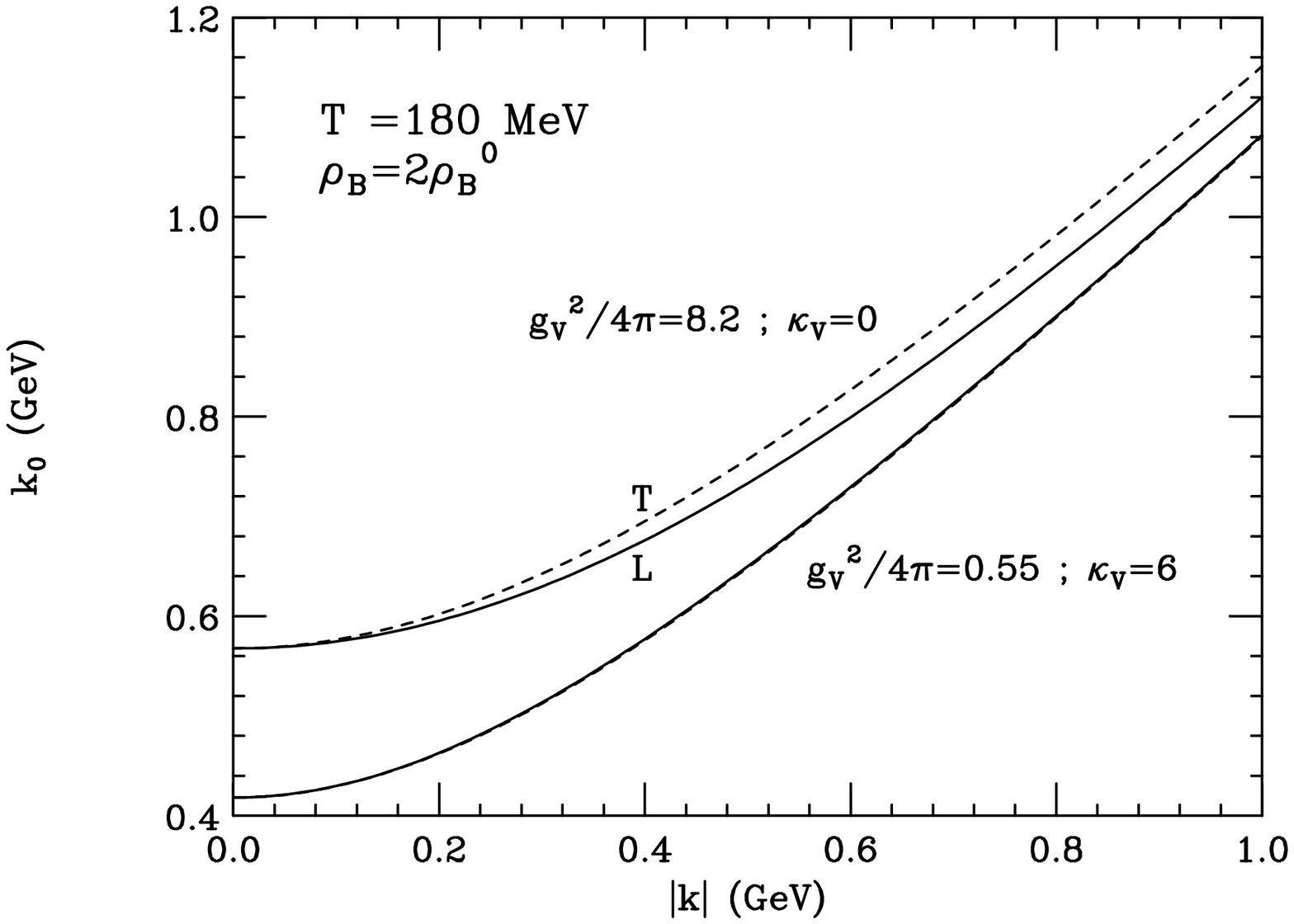,height=7cm,width=10cm}}
\caption{
Transverse and longitudinal dispersion relations of the $\rho$ and
$\omega$ mesons. The solid and dashed curves pertains to the transverse and
longitudinal modes, respectively.
}
\label{dispfig}
\eef
It is important to point out that the self-energy functions of the nucleon as
well as that of the vector mesons calculated in this section are actually 
the 11-component of the 2$\times$2 self-energy matrix in the real time 
formulation of thermal field theory. Since here we are concerned about the 
real(dispersive) part of the self-energy in our discussion on effective masses
we do not make any distiction between these and the scalar self energy 
function following Eq.~(\ref{resc11}) of Chapter 2. 

In Fig.~(\ref{dispfig}) we plot the dispersion relations for $\rho$ and $\omega$
mesons at $T=180$ MeV and twice normal nuclear matter density. This is obtained by
solving Eq.~(\ref{disp}) for the $\rho$ and $\omega$ mesons.
One observes a small difference between the longitudinal(L)
and transverse(T) modes in case of the $\omega$ meson
but in case of the $\rho$ this splitting is negligible
(attributable to the smaller vector coupling constant). 
We have observed that 
the quantity $k_0^2 - |\vec k|^2$ along the dispersion curve 
remains almost constant $\sim m_{V}^{\ast 2}$ which is defined as the value
of $k_0^2$ at ${\vec k = 0}$ on the mass hyperbola. 
This means that a simple pole
approximation of the $\rho$ and $\omega$ propagator at $k^2=m_{V}^{\ast 2}$
is good enough for our calculations.  
The splitting between the transverse and longitudinal components of
the self energy of vector mesons with both vector and tensor 
interactions can be shown to be~\cite{ja},
\bea
\Pi_T-\Pi_L &=& \frac{2g_{VNN}^2}{\pi^2}\left(1-k^2(\frac{\kappa_V}{2M_N})^2
\right)\,\int\frac{p^2\,dp\,d(\cos\theta)}{\sqrt{|\vec p|^2
+M_N^{\ast 2}}}\left[\frac{}{}f_{FD}+\bar f_{FD}\right]\nonumber\\
&& \times\,\left[\frac{u\cos^2\theta-v\cos\theta+w}
{C+8p_0k_0|\vec p|\,|\vec k|\cos\theta -
4|\vec p|^2\,|\vec k|^2\cos^2\theta}\right]
\eea
where
\begin{eqnarray*}
u&=&3k_0^2|\vec p|^2-|\vec k|^2|\vec p|^2\,\nonumber\\
v&=&4k_0p_0|\vec p|\,|\vec k|\,\nonumber\\
w&=&2p_0^2|\vec k|^2+|\vec k|^2|\vec p|^2-k_0^2|\vec p|^2\nonumber\\
C&=&|\vec k|^4-k_0^2p_0^2.
\end{eqnarray*}
The following values of the coupling constants
and masses~\cite{jean,shiomi} have been used in our calculations so as to
reproduce the nuclear saturation density: 
$\kappa_{\rho} = 6.1$, $g_{\rho NN}^2 = 6.91$, $m_{\rho} = 
770$ MeV, $M_N = 939$ MeV,  $\kappa_{\omega} = 0$,
and $g_{\omega NN}^2 = 102$. 

Thus the physical mass $(m_V^{\ast})$ is defined as the lowest
zero of Eq.~(\ref{disp}) in the limit ${\vec k}\ra 0$. 
In this limit
$\Pi_T^D = \Pi_L^D = \Pi^D$, and we have,
\be
\frac{1}{3}\Pi_{\mu}^{\mu}=\Pi= \Pi^D+\Pi^F
\ee
where
\be
\Pi^D(k_0,{\vec k}\ra 0) =  
-\frac{4g_{VNN}^2}{\pi^2}\,\int\,p^2dp\,F(|\vec p|,M_N^{\ast})\,
[\frac{}{}f_{FD}(\mu^{\ast},T)+{\bar f}_{FD}(\mu^{\ast},T)]
\ee
with
\bea
F(|\vec p|,M_N^{\ast})&=&\frac{1}{\omega_p(4\omega_p^2-k_0^2)}
\,\left[\frac{2}{3}(2|\vec p|^2+3M_N^{\ast 2})+k_0^2\left\{2M_N^{\ast}
(\frac{\kappa_V}{2M_N})\right.\right.\nonumber\\
&&+\left.\left.\,\frac{2}{3}(\frac{\kappa_V}{2M_N})^2(|\vec p|^2
+3M_N^{\ast 2})\right\}\right]
\eea
and $\omega_p^2={\vec p}^2+M_N^{\ast 2}$.
The effective mass of the vector meson is then obtained by solving the
equation:
\be
k_0^2 - m_V^2 + {\s {Re}}\Pi = 0.
\label{mass}
\ee
The effective masses take the following parametrized forms~\cite{sspbpb}:
\bea
m_\rho^\ast&=&m_\rho\left[1-
0.127\left(\frac{T({\s GeV})}{0.16}\right)^{5.24}\right]\nonumber\\
m_\omega^\ast&=&m_\omega\left[1-
0.0438\left(\frac{T({\s GeV})}{0.16}\right)^{7.09}\right].
\label{pmasswal}
\eea
\bef
\centerline{\psfig{figure=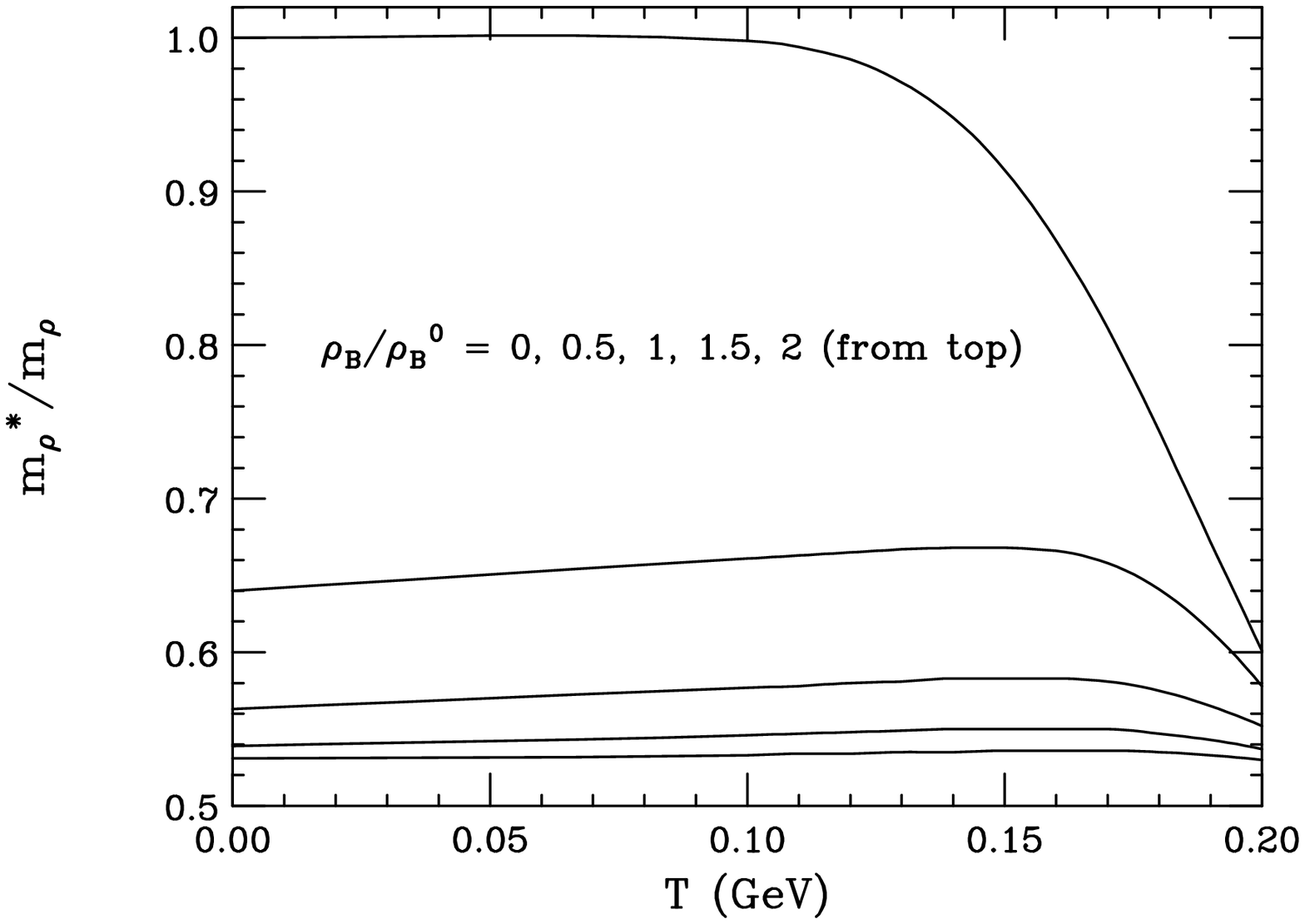,height=7cm,width=10cm}}
\caption{
Variation of $\rho$ meson mass with temperature for various baryon densities.
}
\label{rmass}
\eef

The effective mass of the $a_1$ meson has been estimated from that of the $\rho$
mass using Weinberg's sum rule~\cite{weinberg}.
One finds reference to two other kinds of masses in the literature. 
The invariant mass is defined
as the lowest order zero of Eq.~(\ref{disp}) with $\Pi^D$ neglected. 
Again, the screening mass of a vector
meson is obtained from the pure imaginary zero of the quantity
on the left hand side of the same equation with $k_0=0$.
These two masses are different because of the non-analyticity
of the polarization tensor at the origin {\it i.e.} at
($p_0,\vec{p})=(0,\vec{0}$).
\bef
\centerline{\psfig{figure=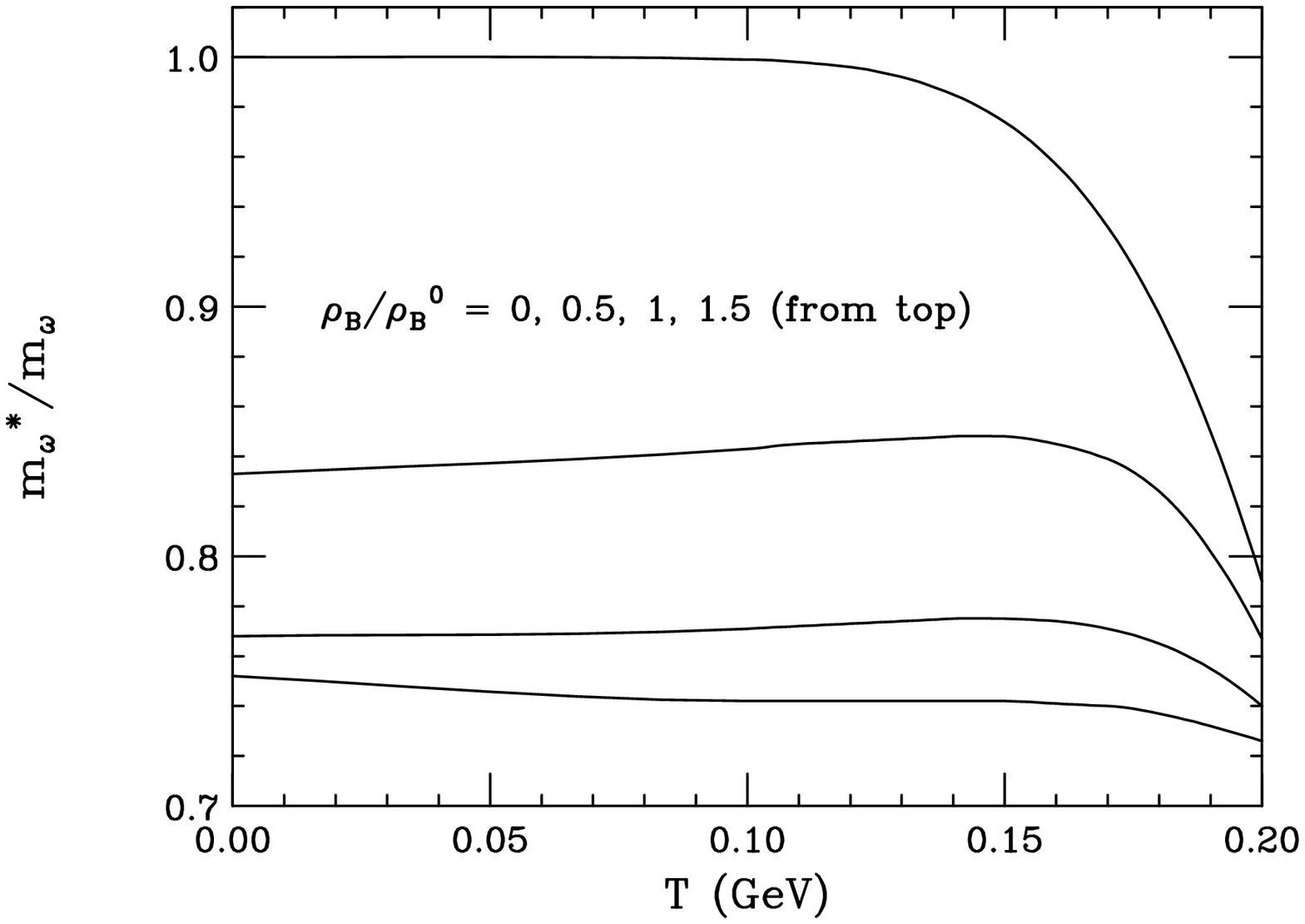,height=7cm,width=10cm}}
\caption{
Variation of $\omega$ meson mass with temperature for various baryon densities.
}
\label{omass}
\eef

In Fig.~(\ref{rmass}) the effective mass of the $\rho$ meson is plotted against
temperature for various values of baryon density. We observe that the
variation of the $\rho$ mass follows qualitatively the same trend as that of the 
nucleon. In this case, the $\rho$ mass decreases by 45\% at $T$= 160 MeV
and two times normal nuclear matter density compared to its free space value.
This is due to the fact that the large  decrease of the modified
Dirac sea contribution to the $\rho$ self energy dominates over the in-medium
contribution which is seen to increase with temperature.
We then evaluate the effective $\omega$ mass with the
values of the  coupling constants mentioned above. The results are plotted
in  Fig.~(\ref{omass}).
The quantitative difference in the $\rho$ and $\omega$ meson masses 
is due to the different numerical values of the coupling constants {\it e.g.} 
the tensor interaction is absent in case of the $\omega$ meson and quite
significant for the $\rho$ meson. 

Before we proceed further a few comments on the QHD model calculations
are in order.
In this model the major contribution to the change in the masses of the $\rho$ 
 and $\omega$ mesons
arises from the nucleon-loop diagram.
For the dressing of internal lines in matter we restrict ourselves to 
the Mean Field Theory (MFT) to avoid a plethora of diagrams and to maintain
internal consistency. 
It has been shown~\cite{gale,sourav} that the change in the $\rho$ mass due
to $\rho\pi\pi$ interaction is negligibly small at non-zero temperature
and zero baryon density. 
Therefore the change in the $\rho$ meson mass
due to $\rho\pi\pi$ interaction is neglected here.
 At finite baryon density, the dynamics is more involved
 due to the   medium effects on the $\rho\pi\pi$ vertex,
 the pion propagator coupled with  delta-hole excitation,
 and the coupling of the $\rho$ meson  with $N^*$-hole excitations
 \cite{klingl,asakawa,friman,chanfray,fp,rubw,ppllm,blrrw}.
The effect of such medium modifications is to broaden the
 $\rho$-peak as well as to produce complicated structure
 around the peak.
Here we have 
restricted our calculations within the realm of MFT, {\it i.e} 
the internal nucleon loop in the $\rho$ and $\omega$ self energy 
is modified due to tadpole diagram only.

\subsubsection{c) The Vector Meson Width}

\begin{figure}
\centerline{\psfig{figure=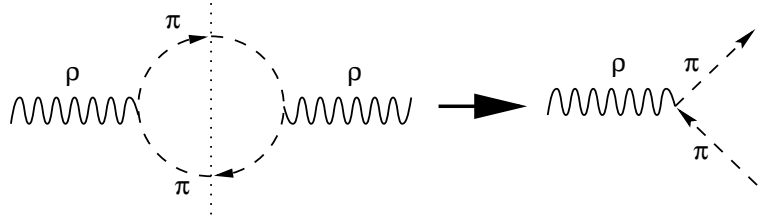,height=3cm,width=10cm}}
\caption{Decay of $\rho$ meson.}
\label{cut}
\end{figure}
The physical decay width of an unstable particle is related to the
imaginary part of its self energy. For a particle at rest the width is 
obtained from the relation
\be
{\s {Im}}\Pi(k_0) = k_0\Gamma(k_0)
\label{impi}
\ee
where $k_0$ is the energy of the decaying particle. 
There are various ways to calculate the imaginary part.
The Cutkosky
rules at finite temperature~\cite{adas,kobes,gelis} provides
a simple and systematic way to calculate the imaginary part of 
the self-energy. Again, use can be made of the fact that
the self-energy function develops cuts along the real axis when the 
particles in the internal loop  
become on-mass shell. The discontinuity across these cuts is pure
imaginary for real $k_0$ so that we have
\be
{\s {Disc}}\Pi(k_0) = [\Pi(k_0+i\epsilon)-\Pi(k_0-i\epsilon)]
=2i{\s {Im}}\Pi(k_0).
\label{discpi}
\ee
Let us consider the $\rho$ meson width.
The imaginary part
of the $\rho$ self energy is totally dominated by the pion loop which is borne 
out by the fact that the two pion decay mode of the $\rho$ has a branching
ratio of $\sim$ 100 \%. In contrast, the real part of the $\rho$ self energy 
which is responsible for the mass modification has a negligible role
to play as far as the pion loop is concerned. It is found to cause a small 
positive shift of the $\rho$ pole.
The $\rho-\pi$ interaction is described by the Lagrangian
\be
{\cal L}^{\s int}_{\rpp} = -g_{\rho \pi \pi}{\vec {\rho}}^{\mu}\cdot
({\vec \pi}\times\partial_{\mu}{\vec \pi}). 
\label{lagrhopipi}
\ee
The 11-component of the self energy of the $\rho$ meson due to pion loop 
(Fig.~(\ref{cut}))is given by
\be
-i\Pi^{\mn}_{11}(k)=-g_{\rho \pi \pi}^2\,\int\frac{d^4p}{(2\pi)^4}\,
(2p-k)^{\mu}\,i\Delta_{11}^\beta(p)\,(2p-k)^{\nu}\,i\Delta_{11}^\beta(p-k),
\label{pipimn}
\ee
where the pion propagator is
\begin{eqnarray*}
\Delta_{11}^\beta(q)&=&\frac{1}{q^2-m_\pi^2+i\eps}-2\pi i \delta(q^2-m_\pi^2)
f_{BE}(|q_0|)\\
&=&\frac{1+f_{BE}(|q_0|)}{q^2-m_\pi^2+i\eps}-\frac{f_{BE}(|q_0|)}{q^2-m_\pi^2-i\eps}
\end{eqnarray*}
After integration over $p_0$, the imaginary part of the self-energy
function is obtained as
\bea
{\s Im}\Pi^{\mn}(k_0,\vec k)&=&-\pi g_{\rpp}^2\,
\int\frac{d^3p}{(2\pi)^3}\left[\frac{(2p-k)^\mu (2p-k)^\nu}{2\omega_p\,
2\omega_{p-k}}\right.\times\nonumber\\
&&\!\!\!\!\!\!\!\!\left\{\frac{}{}(1+f_{BE}(\omega_p)+f_{BE}(\omega_{p-k}))
\delta(k_0-\omega_p-\omega_{p-k})+
\right.\nonumber\\
&&\!\!\!\!\!\!\!\!\left.\left.(f_{BE}(\omega_{p-k})-f_{BE}(\omega_p))\delta(k_0-\omega_p+\omega_{p-k})
\frac{}{}\right\}
+\left \{k_\mu \rightarrow -k_\mu\frac{}{}\right\}\right]
\label{fullpirho}
\eea
where use has been made of Eq.~(\ref{im11}). Using the relations
given in Eq.~(\ref{contrac}) the longitudinal and transverse
components can now be worked out. 
The terms involving the thermal distribution functions in the above 
equation can be interpreted in terms of pion absorption from and 
emission into the medium. The first and second $\delta$-functions
correspond to time-like and space-like regions of $k$ respectively.
Restricting to the time-like region, we define the spin-averaged 
quantity,
\bea
\frac{g^{\mn}}{3}[A_{\mn}{\s Im}\Pi_T+B_{\mn}{\s Im}\Pi_L]&=&\frac{1}{3}
[2{\s Im}\Pi_T+{\s Im}\Pi_L]
=\frac{1}{3}{\s Im}\Pi^\mu_\mu\nonumber\\
&=&
{g_{\rho \pi \pi}^2 \over 48\pi}\,k^2\,W^3(k^2)\,
\left[1+\frac{2T}{|\vec k|\,W(k^2)}\right.\nonumber\\
&&\times\left.\ln\left\{\frac{1-\exp[-\frac{\beta}{2}(k_0+|\vec k|W(k^2))]}
{1-\exp[-\frac{\beta}{2}(k_0-|\vec k|W(k^2))]}\right\}
\right]
\eea
where $W(k^2) = \sqrt{1-4m_{\pi}^2/k^2}$.
In the rest frame of the $\rho$ ($\vec k =0$) this reduces to
\be 
{\s {Im}}\Pi_{\rpp}(k_0)=\frac{{\s {Im}}\Pi_\mu^\mu(k_0)}{3} = 
\frac{g_{\rho \pi \pi}^2}{48\pi k_0}
(k_0^2-4m_{\pi}^2)^{3/2}
\left[2f_{BE}(\frac{k_0}{2})+1\right].
\label{rpp-pi}
\ee
Using Eq.~(\ref{impi}) the $\rho$ decay width is obtained as
\be
\Gamma_{\rho}(k_0) = 
\frac{g_{\rho\,\pi\,\pi}^2}{48\pi}\,\frac{(k_0^2-4m_\pi^2)^{3/2}}{k_0^2}
\left[\left(1+f_{BE}(\frac{k_0}{2})\right)\,\left(1+f_{BE}(\frac{
k_0}{2})\right)-f_{BE}(\frac{k_0}{2})f_{BE}(\frac{k_0}{2})
\right]
\label{width}
\ee
with $f_{BE}(x) = [e^x -1]^{-1}$. 
It is interesting to note that the phase space factor $(2f_{BE}+1)$
when written in this form clearly shows that the in-medium width is
actually the difference between the rates of decay and formation of the
resonance.

In Fig.~(\ref{rwidth1})
we demonstrate the in-medium effect on the decay width of
$\rho$ meson. 
The observed enhancement of the decay width with temperature at
non-zero values of the baryon density is solely due to the stimulated
emission of pions in the medium; a consequence of the $(1+f_{BE})$ terms
in the decay width. This is just a manifestation of
the well known Bose enhancement (BE) effect which
is more clearly observed in Fig.~(\ref{rwidth2}). 
\bef
\centerline{\psfig{figure=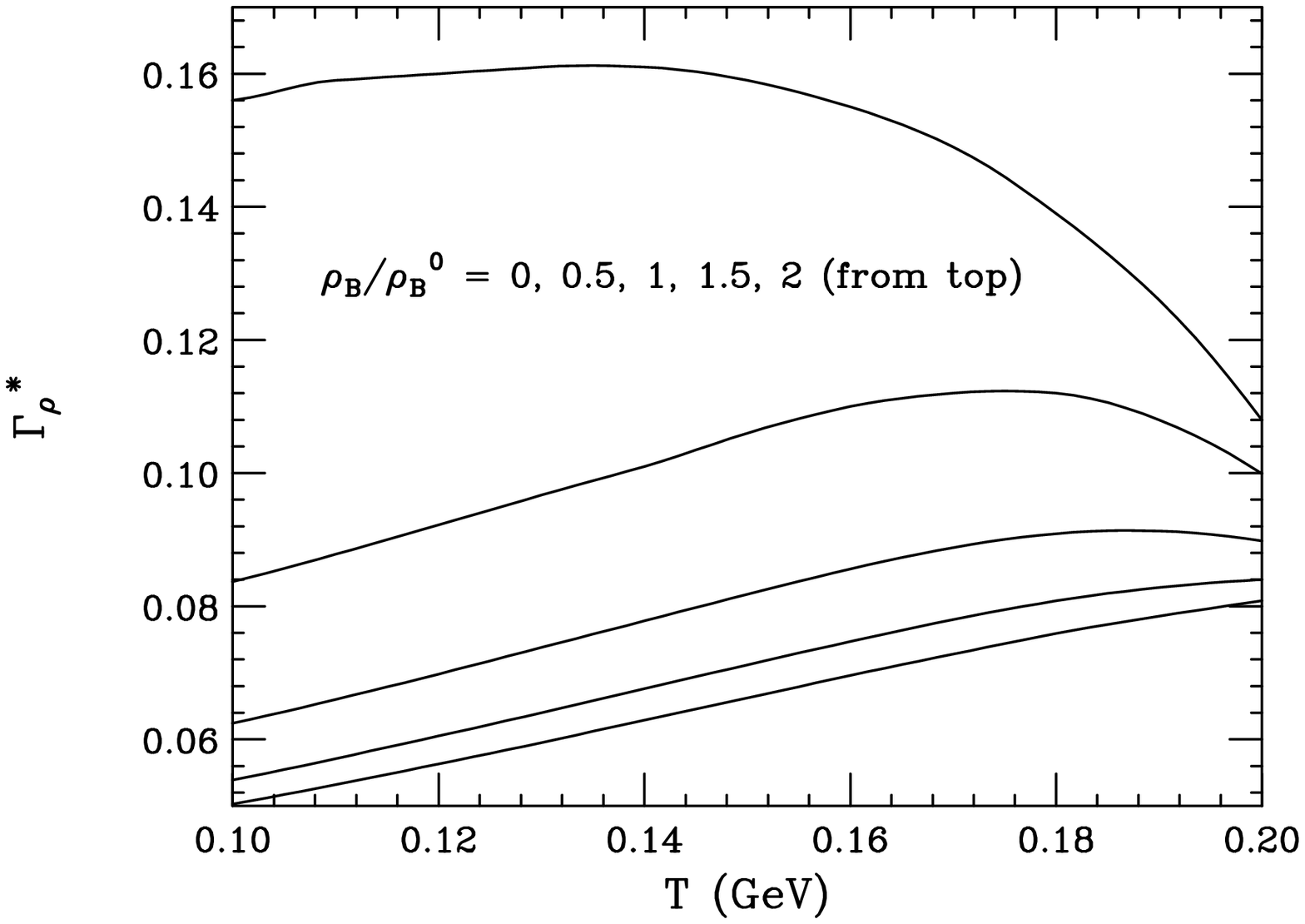,height=7cm,width=10cm}}
\caption{
The $\rho\,\ra\,\pi\,\pi$ decay width as a function of temperature
for different values of baryon densities.
}
\label{rwidth1}
\eef
\bef
\centerline{\psfig{figure=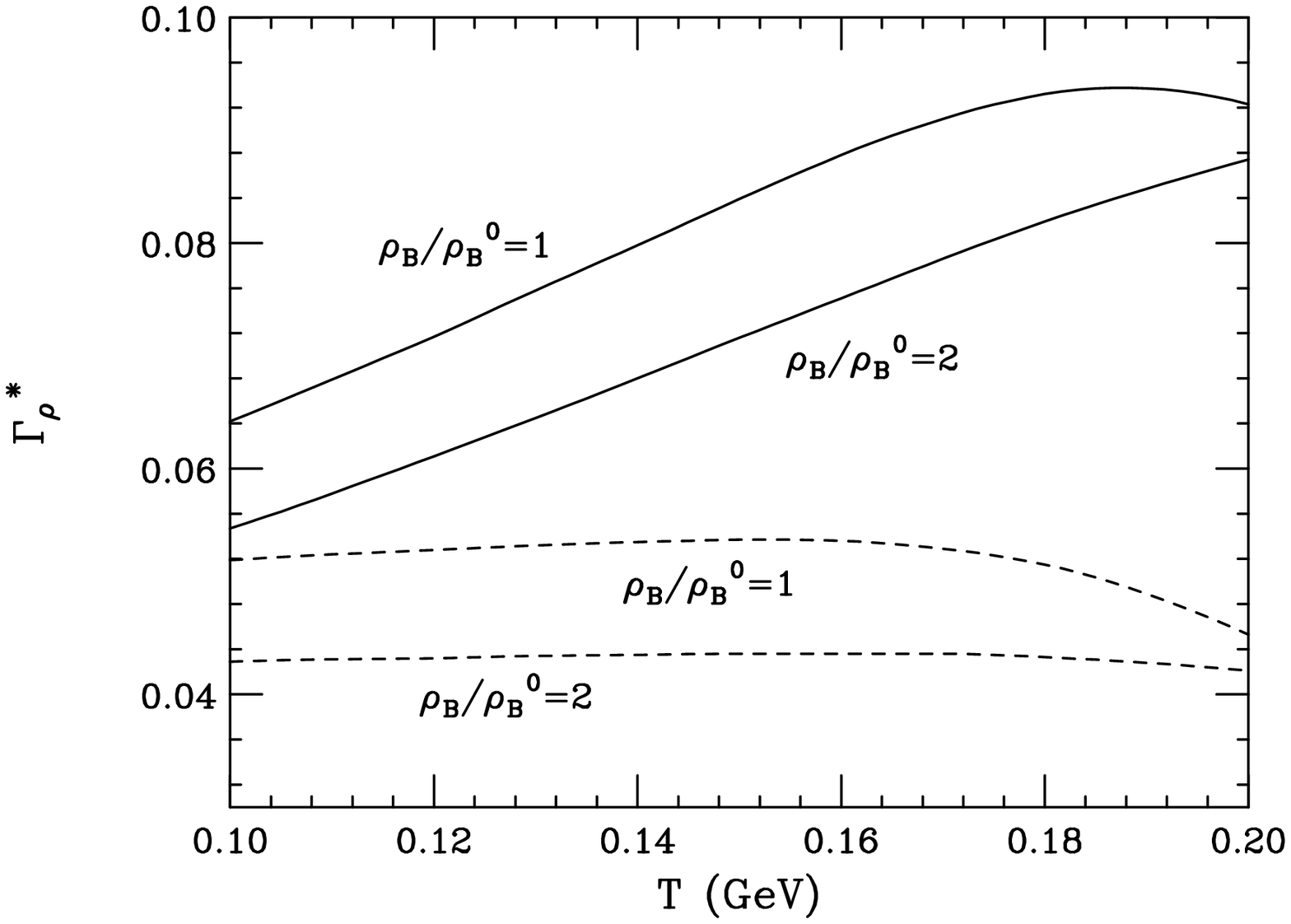,height=7cm,width=10cm}}
\caption{
Same as Fig.~(\protect\ref{rwidth1}) with (solid) and without (dashed)  
BE effect.
}
\label{rwidth2}
\eef
\bef
\centerline{\psfig{figure=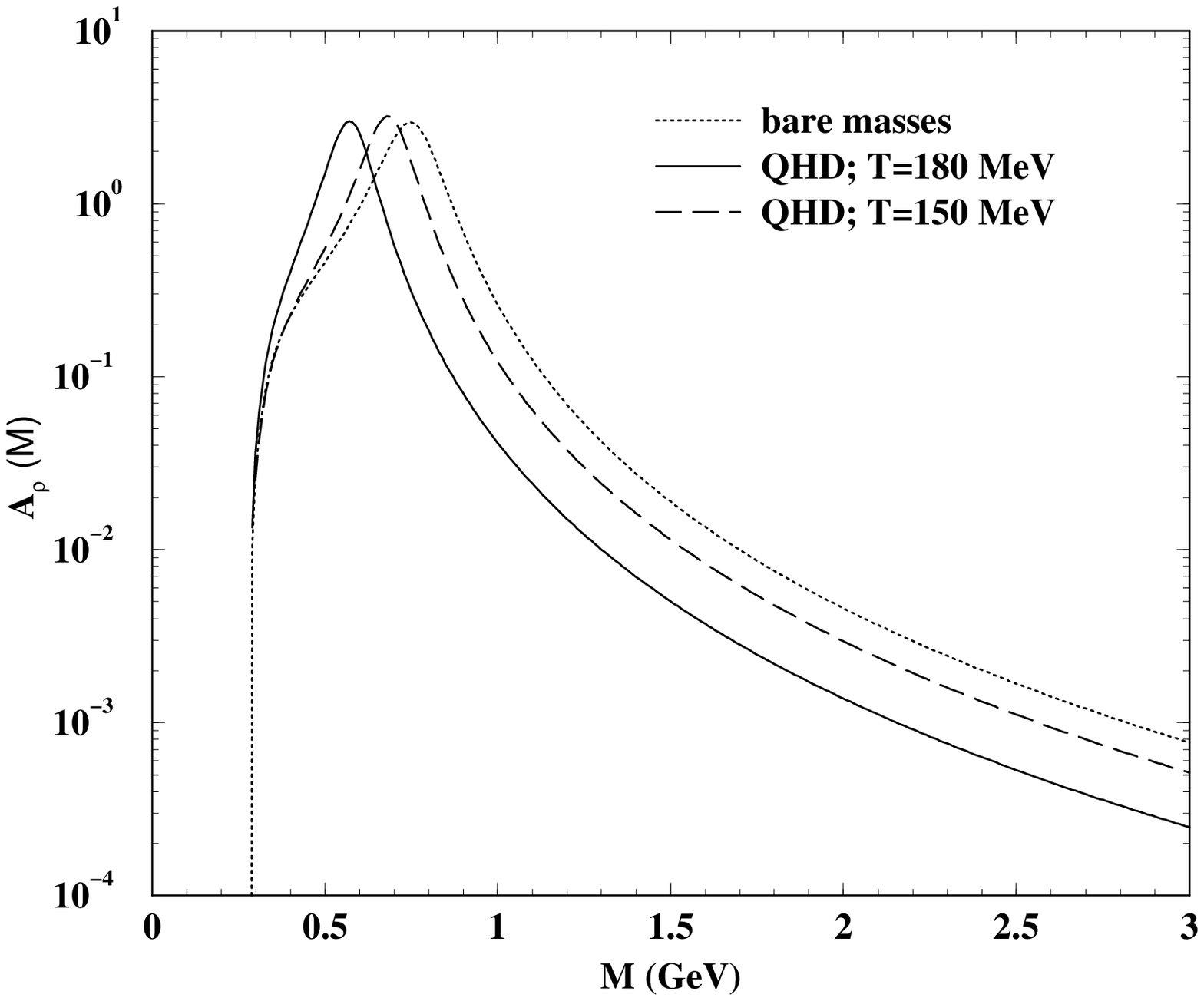,height=7cm,width=9cm}}
\caption{ Spectral function of $\rho$ meson in QHD.
Solid (long dashed) line corresponds to $T=$180 MeV ($T=$150 MeV).
The spectral function in vacuum is shown by the dotted line.
}
\label{spwro}
\eef

The effect of the modifications in the mass and decay width of an
unstable particle is conveniently illustrated
through the dimensionless quantity $A_V$ which is basically the spectral function
apart from some factors. It is defined as
\be
A_V=\frac{8\pi m_V^4}{g_V^2M^2}\,\frac{M\Gamma_V}{[(M^2-m_V^{\ast\,2})^2+M^2\Gamma_V^2]}
\label{av}
\ee
where $M$ is the invariant mass of a lepton pair and $m^{\ast\,2}_V=m^2_V-{\s Re}\Pi$.
The $\rho$ spectral function $A_\rho$, in units of $e$ is plotted in 
Fig.~(\ref{spwro}).
The shifts in both the spectral functions
towards the lower invariant mass region correspond to the reduction
of their masses due to thermal interactions. 

\begin{figure}
\centerline{\psfig{figure=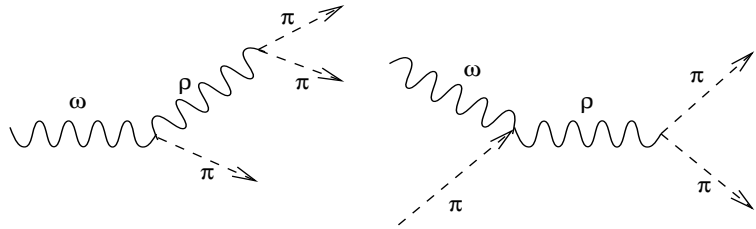,height=3cm,width=10cm}}
\caption{Diagrams contributing to the width of the $\omega$ meson.}
\label{omdec}
\end{figure}

Let us now consider the $\omega$ meson. The vacuum width ($8.5$ MeV) of the $\omega$
is known to be dominated by the $\omega\ra 3\pi$ mode. 
A substantial contribution to the $\omega$ width also comes from the process
$\omega\pi\ra\pi\pi$ in a thermal bath~\cite{ja,pr}. 
The Feynman diagrams are shown in Fig.~(\ref{omdec}).
We shall use the Gell-Mann Sharp Wagner (GSW) interaction~\cite{GSW} given by
\be
{\cal L}_{GSW} =
\frac{g_{\omega \rho \pi}}{m_{\pi}}\,
\epsilon_{\mu \nu \alpha \beta}\partial^{\mu}{\omega}^{\nu}\partial^{\alpha}
\rho^{\beta}\pi
\ee
for the $\omega\rho\pi$ vertex and the Lagrangian given by Eq.~(\ref{lagrhopipi})
for the $\rho\pi\pi$ vertex.
The $\omega\ra 3\pi$ width is obtained as
\be
\Gamma_{\omega\ra 3\pi}(k_0)=C\,\int_{z_{min}}^{z_{max}}\,dz\,
\int_{x_{min}}^{x_{max}}\,dx
\mid\,F\,\mid^2\,S
\ee
where $S$ is the phase space factor, given by
\[
S=\left[(1+f_{BE}(E_1))(1+f_{BE}(E_2))(1+f_{BE}(E_3))-
f_{BE}(E_1)f_{BE}(E_2)f_{BE}(E_3)\right]
\]
and
\[
C=\frac{g_{\omega\rho\pi}^2\,
g_{\rho\pi\pi}^2\,k_0}{48\pi^3\,m_{\pi}^2}.
\]
The limits of integration are 
\begin{eqnarray*}
z_{\s min}&=&m_\pi,\nonumber\\
z_{\s max}&=&{(k_0^2-3m_\pi^2)}/{2\,k_0},\nonumber\\ 
x_{\s max}&=&\sqrt{{0.5k_0\,(z-z_{\s max})(z^2-m_\pi^2)}
/{(2k_0 \,z-k_0^2-m_\pi^2)}},\nonumber\\
x_{\s min}&=&-x_{\s max}, \nonumber\\
\end{eqnarray*}
and
\begin{eqnarray*}
E_1&=&z,\nonumber\\
E_2&=&x+(k_0-z)/2,\nonumber\\
E_3&=&-x+(k_0-z)/2,\nonumber\\
\mid\,\vec p_i\,\mid&=&\sqrt{E_i^2-m_\pi^2},
\end{eqnarray*}
where $\vec p_i$ is the pion 3-momentum.
\bef
\centerline{\psfig{figure=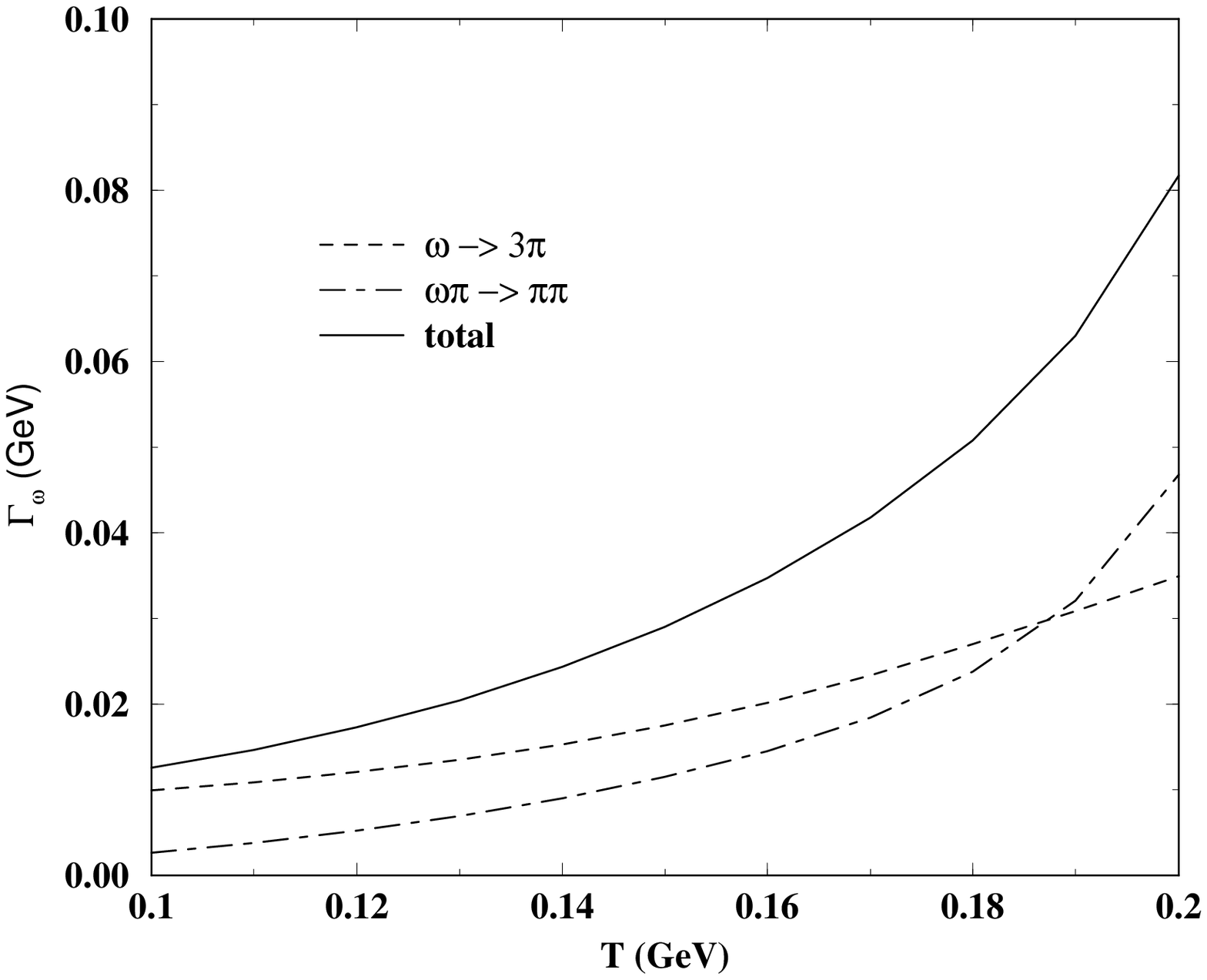,height=7cm,width=9cm}}
\caption{The $\omega$ width as a function of temperature.}
\label{omwid}
\eef
The amplitude for the process is
\[
\mid F \mid^2=\mid \vec p_1 \mid^2\mid \vec p_2 \mid^2
(1-Z_0^2)H
\]
where
\[
Z_0=\frac{k_0^2+m_\pi^2-2k_0(E_1+E_2)+2E_1E_2}
{2\mid \vec p_1||\vec p_2 \mid}
\]
and
\[
H=\sum_{i=1}^6\,h_i
\]
with
\begin{eqnarray*}
h_1&=&\frac{1}{d_{12}^2+m_\rho^2\,\Gamma_\rho^2}\nonumber\\
h_2&=&\frac{1}{d_{13}^2+m_\rho^2\,\Gamma_\rho^2}\nonumber\\
h_3&=&\frac{1}{d_{23}^2+m_\rho^2\,\Gamma_\rho^2}\nonumber\\
h_4&=&2(d_{12}d_{13}+m_\rho^2\Gamma_\rho^2)h_1h_2\nonumber\\
h_5&=&2(d_{13}d_{23}+m_\rho^2\Gamma_\rho^2)h_2h_3\nonumber\\
h_6&=&2(d_{12}d_{23}+m_\rho^2\Gamma_\rho^2)h_1h_3\nonumber\\
d_{12}&=&(E_1+E_2)^2-\vec p_3^2-m_\rho^2\nonumber\\ 
d_{13}&=&(E_1+E_3)^2-\vec p_2^2-m_\rho^2\nonumber\\ 
d_{23}&=&(E_2+E_3)^2-\vec p_1^2-m_\rho^2. 
\end{eqnarray*}
The contribution from the reaction $\omega\pi\ra\pi\pi$ to the decay
 width of the $\omega$ is calculated analogously.

In Fig.~(\ref{omwid}) we have shown how the decay width of the $\omega$ meson 
increases with temperature. 
The narrow peak of the $\omega$ in vacuum is broadened
substantially due to interactions with the thermal pions.
Both the modes discussed above are seen to contribute
almost equally to a ten-fold broadening of the $\omega$ in the medium.
This is also observed in the spectral function of the $\omega$ shown
in Fig.~(\ref{spwom}) where $\Gamma_\omega=\Gamma_{\omega\ra 3\pi}
+\Gamma_{\omega\pi\ra\pi\pi}$ and $g_\omega=3g_{\rho\pi\pi}$.
\bef
\centerline{\psfig{figure=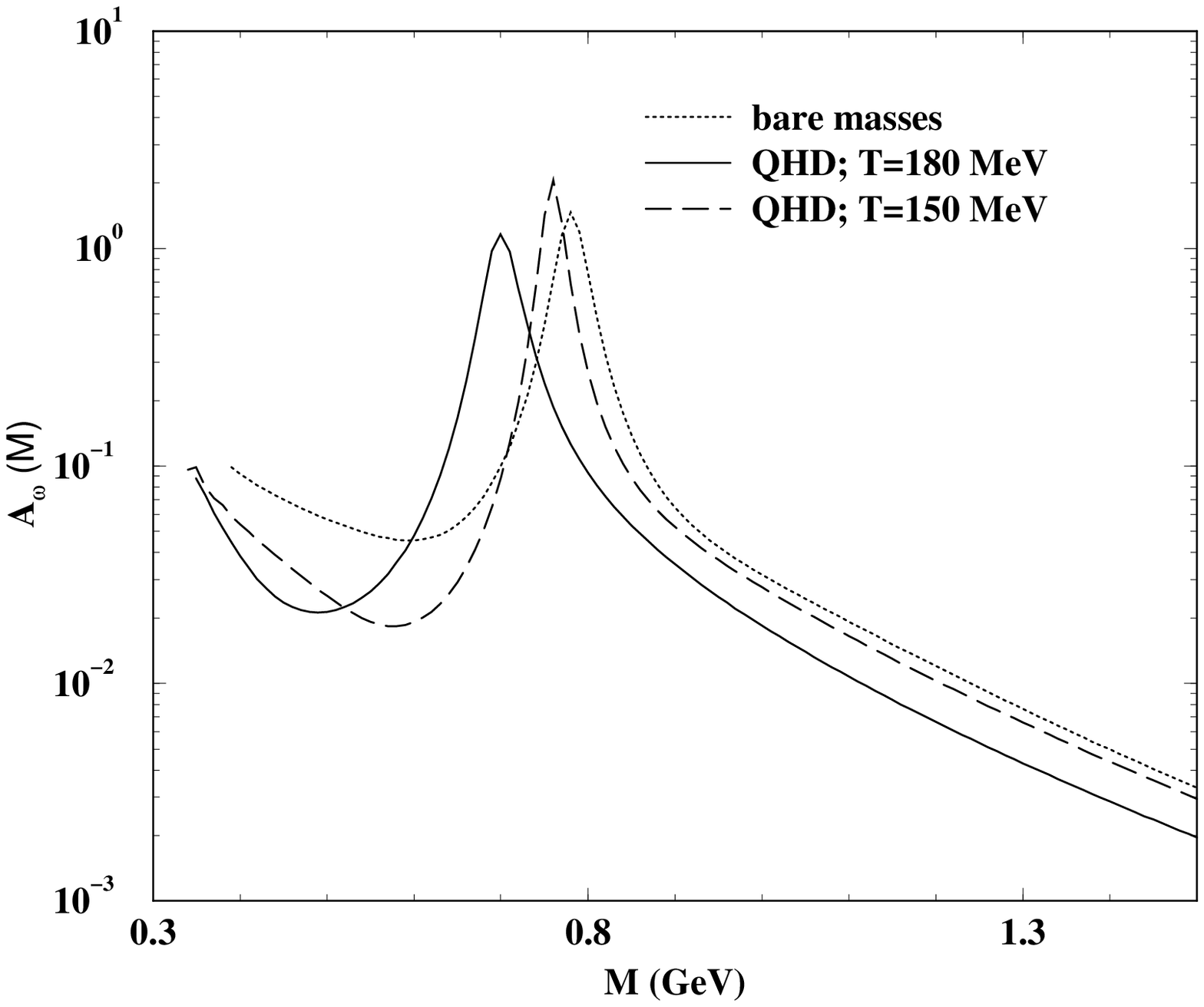,height=7cm,width=9cm}}
\caption{ Spectral function of $\omega$ meson in QHD.
Solid (long dashed) line corresponds to $T=$180 MeV ($T=$150 MeV).
The spectral function in vacuum is shown by the dotted line.
}
\label{spwom}
\eef

\subsection{Models with Chiral Symmetry}

In this Section we will discuss finite temperature effects
on the vector meson properties considering a few models which respect
chiral symmetry. The effects of in-medium properties of vector mesons
on the electromagnetic probes will be presented in Section~3.2.

\subsubsection{a) The Gauged Linear Sigma Model}

The linear sigma model (LSM) is a convenient tool to describe
the low energy dynamics of pions, because it shows 
explicitly how the spontaneous symmetry breaking (SSB) 
of global chiral symmetry (SU(2)$\bigotimes$SU(2)) by the isosinglet 
sigma field generates pions, the Nambu-Goldstone (NG) bosons.
However, there are reservations about the description of the $\sigma$ meson as
a well-defined degree of freedom because of its large decay width which is 
comparable to its mass.    
But, it has been argued~\cite{plb185,raja1} 
that in the limit of
chiral symmetry restoration, the decay of $\sigma$ into
two pions should be disallowed as $\sigma$ and $\pi$
become degenerate in mass in this limit. It has been explicitly shown that 
the width of $\sigma$ due to $\sigma\ra 2\pi$ decay vanishes as 
$T\ra T_\chi$, where $T_\chi$ is the critical temperature for
chiral transition.
 Though it is still not known whether or not the critical
temperature for deconfinement and chiral transition are the same; 
presently we will make no distinction between them.

The simplest version of LSM contains isosinglet $\sigma$ field and 
isotriplet pion field and respects the charge conjugation,
parity and time reversal symmetry (CPT)~\cite{pisarski2}. The Lagrangian obeying
these constraints is
\be
{\cal L}_{LSM}=tr\mid\partial_{\mu}\Phi\mid^2 +\mu^2\,tr \mid\Phi\mid^2
+\frac{1}{2}\lambda\,tr(\mid\Phi\mid^2)^2-h\,tr(\Phi)
\label{lsm}
\ee
where $\Phi$ is defined as
\be
\Phi=\frac{1}{2}(\sigma + i\vec{\pi}\cdot\,\vec{\tau})
\ee
with $\vec{\tau}$ being the Pauli matrices. The non-zero value of $h$ ensures
that the pions are massive and consequently 
PCAC (partially conserved axial current) relation is satisfied. 
Note that $\mid\Phi\mid^2=\sigma^2+\vec\pi^2$ is chirally invariant. 
Elimination of the $\sigma$ field by imposing the condition 
$\sigma^2+\vec\pi^2=f_\pi^2$  results in the Non-Linear Sigma Model,
which will be discussed in the next Section.

In the gauged LSM~\cite{pisarski2,rdp,gg}
one introduces the vectors and 
their chiral partners (axial
vector) through left and right handed fields as 
\be
V^{\mu}_l=(\vec\rho^{\mu}+\vec a^{\mu})\cdot\vec t +(\omega^\mu+f_1^\mu)
\ee
\be
V^{\mu}_r=(\vec\rho^{\mu}-\vec a^{\mu})\cdot\vec t +(\omega^\mu-f_1^\mu)
\ee
where $a_1$ and $f_1$ are the chiral partners of the $\rho$ and $\omega$
mesons respectively and $\vec{t}=\vec{\tau}/2$.

The inclusion of axial vector mesons will increase the number
of possible couplings and hence the number of arbitrary parameters
become large. In order to include vector mesons in LSM with minimal 
coupling to the matter fields $\pi$ and $\sigma$,
one requires~\cite{kroll},
that the Lagrangian and its chiral transformation
properties are such that the current generated by the chiral transformation
is proportional to the vector field itself. This leads to 
the field-current identities and eventually 
the idea of VMD~\cite{sakurai}. The field-current
identity is achieved by promoting the SU(2) global chiral symmetry
of vector fields to a local gauge symmetry as was done 
by Yang and Mills~\cite{yang} for the isospin symmetry.
The Lagrangian for the vector field reads,
\be
{\cal L}_{lr}=\frac{1}{2}tr\mid\,F^{\mn}_l\,\mid^2 
+\frac{1}{2}tr\mid\,F^{\mn}_r\,\mid^2 
+m_0^2\,tr[(V^{\mu}_l)^2+(V^{\mu}_r)^2]
\label{lr}
\ee
where $F^{\mn}_{l,r}=\partial^\mu\,V^{\nu}_{l,r}
-\partial^\nu\,V^{\mu}_{l,r}-ig\left[V^{\mu}_{l,r},V^{\nu}_{l,r}\right]$.
In the above Lagrangian the kinetic term for the gauge fields
remains invariant under the transformation and the field-current
identity is obtained through Gell-Mann Levy theorem 
from the mass term of the gauge fields as
\be
J^{\mu}_{l,r}=-\frac{m_0^2}{g}V^{\mu}_{l,r}
\ee
We note that chiral symmetry is a global one in QCD. Therefore, the
local symmetry has to be broken and this is achieved by the
mass term of the vector fields in the Lagrangian.
Next, one has to introduce the interaction of matter fields ($\pi$ and $\sigma$)
and the gauge fields preserving the field-current identity.
Noting that the ordinary derivatives occurring in Eq.~(\ref{lsm})
spoils the field-current identity, we introduce the required interactions
consistent with the gauge principle {\it i.e.} by replacing the partial
derivatives by covariant derivatives:
\be
D^\mu\Phi=\partial^\mu\Phi-ig(V^{\mu}_l\Phi-\Phi\,V^{\mu}_r).
\label{covar}
\ee
Finally the Lagrangian density for the gauged LSM is 
obtained from Eqs.~(\ref{covar}), ~(\ref{lsm}) and ~(\ref{lr}) as
\be
{\cal L}_{glsm}=tr\mid\,D_{\mu}\Phi\mid^2 +\mu^2\,tr \mid\Phi\mid^2
+\frac{1}{2}\lambda\,tr(\mid\Phi\mid^2)^2-h\,tr(\Phi)
+{\cal L}_{lr}.
\label{glsm}
\ee
Expanding the kinetic term for the matter field one finds
\be
tr\mid\,D_{\mu}\Phi\mid^2=\frac{1}{2}
\left[(\partial_{\mu}\sigma+g\vec a\cdot\vec\pi)^2+
(\partial_{\mu}\pi+g\vec \rho_\mu\times\vec\pi-ga^{\mu}\sigma)^2
+g^2(\sigma^2+\pi^2)(f_1^\mu)^2\right].
\label{keeq}
\ee
The above equation indicates that (i) a shift in the $\sigma$ 
field ($\sigma\ra\sigma_0+\sigma$)
gives rise to mixing between $\pi$ and $a_1$ fields 
(a term $\sim\,g\sigma_0\partial_\mu\pi\cdot\,a_1^\mu$ arises from the
second term of the r.h.s. of Eq.~(\ref{keeq})), which has  
to be eliminated by an appropriate shift in the $a_1$ field, (ii)there
is no interaction term involving $\omega$, which can only be introduced
through anomaly and (iii) the kinetic term for pion gets modified
because of the shift in the $a_1$ field. Thus to get back the
canonical form of this term one has to renormalize the
pion field $\pi\ra\pi/\sqrt{Z_\pi}$, where $Z_\pi=m_\rho^2/m_{a_1}^2$. 
$Z_\pi{^2}=1/2$ gives the Kawarabayashi - Suzuki 
- Riazuddin- Fayyazuddin (KSRF) relation~\cite{ks,rf}.
After some algebra one gets~\cite{rdp},
\bea
m_{\pi}^2&=&h/(Z_{\pi}\sigma_0),\nonumber\\
m_{\sigma}^2&=&h/\sigma_0+2\lambda\sigma_{0}^2,\nonumber\\
f_\pi&=&\sqrt{Z_\pi}\sigma_0.
\eea
Taking $m_\pi=137$ MeV, $m_\sigma=600$ MeV,
$m_\rho=770$ MeV and $m_{a_1}=1260$ MeV,  
we get $\sigma_0=152$ MeV, $h=(102$ MeV)$^3$,
$\mu=412$ MeV and $\lambda=7.6$.

With these inputs
the thermal masses of $\rho$  and $a_1$ mesons to
lowest order in $g$ at low temperatures are obtained as~\cite{rdp} 
\be
m_{\rho}^2(T)\approx m_\rho^2-\frac{g^2\pi^2T^4}{45m_\rho^2}
\left[\frac{4m_{a_1}^2(3m_\rho^2+4k^2)}{(m_{a_1}^2-m_\rho^2)^2}-3\right]
\ee
\be
m_{a_1}^2(T)\approx m_{a_1}^2+\frac{g^2\pi^2T^4}{45m_\rho^2}
\left[\frac{4m_{a_1}^2(3m_{a_1}^2+4k^2)}{(m_{a_1}^2-m_{\rho}^2)^2}
+\frac{2m_{\rho}^4}{m_{a_1}^2(m_{a_1}^2-m_{\sigma}^2)}-
\frac{m_{a_1}^2}{m_{\rho}^2}\right].
\ee

In the chiral limit $\sigma_0$ goes to zero and many of the
couplings vanish. Assuming the validity of VMD in the medium
Pisarski has showed that~\cite{rdp9505} $\rho$ and $a_1$ become
degenerate with a mass value $\sim 962$ MeV ({\it i.e.} $\rho$ mass increases).
On the other hand, if one adopts a scenario where vector meson dominance
(VMD) is not 
valid in the medium then $m_\rho(T_\chi)=m_{a_1}(T_\chi)=630$
MeV ($\rho$ mass decreases). 
However, it is important to mention at this point that chiral
symmetry can also be realized via the Georgi limit~\cite{georgi}
where the $\rho$ meson becomes massless. Pisarski~\cite{pisarski2}
has argued that the results obtained by Georgi in the non-linear
sigma model can be translated in terms of the gauged linear sigma 
model without the validity of VMD, for which there is no
unique prediction for the behaviour of the $\rho$ mass at non-zero
temperature. Thus the behaviour of in-medium 
$\rho$ depends on the validity of VMD in the medium. 

\subsubsection{b) The Gauged Non-Linear Sigma Model}

It is well-known that the global $SU(2)_l\bigotimes\,SU(2)_r$
symmetry of two-flavour QCD is expected to be spontaneously 
broken to the subgroup $SU(2)_V$ and the pions
appear as the N-G bosons. The non-linear sigma model
with $SU(2)_l\bigotimes\,SU(2)_r/SU(2)_V$ is an effective
theory of QCD for the description of pion dynamics.
The in-medium properties of vector mesons have been studied
by Song~\cite{cs,cs93} in the framework of gauged non-linear
sigma model (NLSM)~\cite{meissner}. We 
will discuss this model briefly
because it is very similar to the gauged LSM; the
main difference is that the $\sigma$ degree of freedom
is eliminated in NLSM by the non-linear realization of
chiral symmetry as mentioned in the previous Section.
We start with the observation that a perfectly valid 
parametrization of the matter field $\Phi$ could be
\be
U=\exp\left[\frac{2i}{F_\pi}\sum_a\frac{\phi_a\tau^a}{\sqrt{2}}\right]\equiv
\exp\left[\frac{2i}{F_\pi}\phi\right]
\label{uu}
\ee 
where $\phi=\phi^a\tau^a/\sqrt{2}$ is the pseudoscalar 
field and $F_\pi=\sqrt{2}f_\pi$.
The Lagrangian for the NLSM
based on the manifold 
$SU(2)_l\bigotimes\,SU(2)_r/SU(2)_V$ is given by
\be
{\cal L}_0=\frac{f_{\pi}^2}{4}
tr\left[\partial_{\mu}U\partial^{\mu}U^{\dagger}\right].
\label{nlspi}
\ee

The vector and the axial vector fields can be introduced 
as the Yang-Mills gauge fields as before to minimize the 
number of arbitrary parameters in 
the model. The resulting Lagrangian
is given by
\bea
{\cal L}_{NLSM}&=&\frac{f_{\pi}^2}{4}
tr\left[\,D_{\mu}UD^{\mu}U^{\dagger}\right]
-\frac{1}{2}tr\mid\,F^{\mn}_l\,\mid^2 
-\frac{1}{2}tr\mid\,F^{\mn}_r\,\mid^2 \nonumber\\
&&+m_0^2\,tr[(V^{\mu}_l)^2+(V^{\mu}_r)^2],
\label{nlsm}
\eea
where $V_{\mu}^{l,r}=(v_\mu\pm\,a_\mu)/2$, $v_\mu$ and $a_\mu$
denote vector and axial vector fields.
To improve the phenomenology of the model, 
the following higher dimensional terms  can be added to 
the Lagrangian~\cite{cs,brh} without spoiling the 
symmetry under consideration
\be
{\cal L}_{6dim}=-i\xi\,tr\left[\,D_{\mu}UD_{\nu}U^{\dagger}\,F^{l,\mn}
+\,D_{\mu}U^{\dagger}D_{\nu}U\,F^{r,\mn}\right],
\label{non}
\ee
where $\xi$ is a constant determined form the
decay of vector mesons~\cite{cs}. The thermal shift of the
$\rho$-mass evaluated with pion loop, pion
tadpole and pion-$a_1$ loop resulting from the
interaction given by Eqs.~(\ref{nlsm}) and (\ref{non})
shows negligible change in the $\rho$-mass from its vacuum value~\cite{cs}.

The effective masses of $\rho$, $a_1$ and $\omega$ at non-zero 
temperature have also been evaluated~\cite{cs93} with
a $SU(3)_l\,\bigotimes SU(3)_r$ symmetric Lagrangian:
\bea
{\cal L}_{NLSM}&=&\frac{f_\pi^2}{4}tr\left[\,D_{\mu}UD^{\mu}U^{\dagger}\right]
-\frac{1}{2}tr\mid\,F^{\mn}_l\,\mid^2 
-\frac{1}{2}tr\mid\,F^{\mn}_r\,\mid^2 \nonumber\\
&&+m_0^2\,tr[(V^{\mu}_l)^2+(V^{\mu}_r)^2]
+\frac{1}{4}f_\pi^2\,tr\left[M(U+U^{\dagger}-2)\right]\nonumber\\
&&-i\xi\,tr\left[\,D_{\mu}UD_{\nu}U^{\dagger}\,F^{l,\mn}
+\,D_{\mu}U^{\dagger}D_{\nu}U\,F^{r,\mn}\right]\nonumber\\
&&+\kappa\,tr\left[F_{\mn}^lU\,F^{r,\mn}U^\dagger\right],
\label{su3}
\eea
where $U$ is defined as
in Eq.~(\ref{uu}) with the Pauli matrices $\tau^a$ replaced by the Gell-Mann
matrices $\lambda^a$.
The two higher dimensional
terms with co-efficients $\xi$ and $\kappa$ are added to improve
the phenomenology. It may be noted that although these terms retain
the gauge invariance of the model the renormalizability
of the model is spoiled.

The dynamics of the $\omega$ meson is governed by the 
anomalous interaction,  also known as Wess-Zumino interaction
given by
\be
{\cal L}_{anomaly}=
\frac{3g^2}{8\pi^2\,F_\pi}\,
\epsilon_{\mu \nu \alpha \beta}\partial^{\mu}{\omega}^{\nu}
\,tr[\partial^{\alpha}
\rho^{\beta}\pi].
\label{anomaly}
\ee
This is very similar to the Gell-Mann Sharp Wagner~\cite{GSW}
interaction already described before.

The following values 
of the parameters consistent
with the vacuum properties of the vector and axial 
vector mesons have been considered~\cite{cs93}: 
$(g,\kappa,\xi)=(10.30,0.34,0.45)$ and
$(6.45,-0.29,0.06)$, referred to as set I and II 
respectively.
The calculation of the thermal mass shift of the vector and axial
vector mesons with these inputs reveal that: 
(i)for parameter set I $\rho$ and $\omega$ masses increase
with different rate and $a_1$ mass decreases,
(ii) for parameter set
II the thermal mass shift of $\rho$ and $\omega$ is negligibly small 
but $a_1$ mass decreases slightly. 

\subsubsection{c) The Hidden Local Symmetry Approach}

In case of the two chiral models described above the vector 
mesons are introduced as Yang-Mills field and the mass term for the
gauge bosons are put in by hand which is not entirely satisfactory.
In the hidden local symmetry
(HLS) approach the $\rho$ meson is generated as a dynamical
gauge boson of a hidden symmetry in the NLSM~\cite{bandoprl,bandopr}.
It has been explicitly shown that, in general,
any NLSM corresponding to the manifold $G/H$ is gauge equivalent 
to a ``linear'' model having $G_{global}\bigotimes\,H_{local}$
symmetry. Accordingly, the Lagrangian of Eq.~(\ref{nlspi})
can be written in a form that exhibits, besides 
$SU(2)_l\bigotimes\,SU(2)_r$ global, a local
$SU(2)_V$ symmetry - the hidden symmetry and the $\rho$ 
meson appears as a gauge boson corresponding to this symmetry.
(The axial vector $a_1$ is not included in the minimal
version of HLS Lagrangian.) To make it more explicit, one introduces
two $SU(2)$ matrix-valued variables $\xi_l(x)$ and $\xi_r(x)$
with the transformation properties~\cite{bandoprl},
\be
\xi_{l,r}(x) \ra h(x)\xi_{l,r}(x)g^\dagger_{l,r}
\ee
with 
\be
U=\xi_l^\dagger\xi_r
\label{capu}
\ee
where $h(x)\in\,[SU(2)_V]_{local}$ and 
$g_{l,r}\in\,[SU(2)_{l,r}]_{global}$.
$\xi_{l,r}$ is parametrized as
\be
\xi_{l,r}=\exp[i\Sigma(x)/f_\Sigma\mp\,i\pi/f_\pi\,]
\label{unph}
\ee
where $\pi=\pi^a\,t^a$ and  
$\Sigma=\Sigma^a\,t^a$.
The unwanted degrees of freedom
$\Sigma$, which have entered through
Eqs.~(\ref{capu}) and (\ref{unph}) are known as the ``compensators''- the would be
N-G bosons  which have to be ``eaten up'' by the hidden gauge boson,
$\rho$. These extra degrees of freedom then reappear as the longitudinal
polarization of the (massive) $\rho$.  

Now the covariant derivatives are defined as
\bea
D_\mu\xi_l&=&\partial_\mu\xi_l-igV_\mu\xi_l+i\xi_l\,l_\mu\nonumber\\
D_\mu\xi_r&=&\partial_\mu\xi_r-igV_\mu\xi_r+i\xi_r\,r_\mu\,,
\eea
where $l_\mu (r_\mu)$ is the external field corresponding to
the gauging of $SU(2)_l\bigotimes SU(2)_r$ and $V_\mu$ is the gauge field
corresponding to the symmetry $[SU(2)_V]_{local}$. 
With these fields two types of 
$[SU(2)_l\bigotimes\,SU(2)_r]_{global}$$\bigotimes[SU(2)_V]_{local}$
invariants can be constructed~\cite{bandoprl,bandopr,fuji} which are
\bea
{\cal L}_{V}&=&-\frac{f_{\pi}^2}{4}\,
tr\,\left[D_\mu\xi_l\cdot\xi_l^{\dagger}
+D_\mu\xi_r\cdot\xi_r^{\dagger}\right]^2\nonumber\\
{\cal L}_{A}&=&-\frac{f_{\pi}^2}{4}\,
tr\,\left[D_\mu\xi_l\cdot\xi_l^{\dagger}
-D_\mu\xi_r\cdot\xi_r^{\dagger}\right]^2.
\eea
A linear combination ${\cal L}={\cal L}_{A}+a{\cal L}_{V}$
is equivalent to the original Lagrangian given in Eq.~(\ref{nlspi}).
By fixing the gauge $\xi_l^\dagger=\xi_r=\exp(i\pi/f_\pi)$
and hence eliminating the unphysical degrees of freedom,
$\Sigma$) one can show that
${\cal L}_{A}={\cal L}_{0}$ 
while ${\cal L}_{V}$ vanishes when the equation of motion for
$V_\mu$ is used. So far, $V_\mu$ has been treated as an auxiliary field.
It is assumed that the kinetic term for this field is
generated by quantum effects or by QCD dynamics. The full
Lagrangian with the kinetic term is
\be
{\cal L}_{HLS}={\cal L}_{A}+a{\cal L}_{V}-\frac{1}{4}\vec\varrho_{\mn}\vec\varrho^{\mn}
\label{hlslag}
\ee
where $\vec\varrho_{\mn}$ is the non-abelian field tensor for the $\rho$ meson.
The Lagrangian of Eq.~(\ref{hlslag}) is then written as~\cite{bandoprl},
\be
{\cal L}_{HLS}=\frac{1}{2}(\partial_\mu\vec\pi)^2+
\frac{1}{2}ag\vec\rho^\mu\cdot\vec\pi\times\partial_\mu\vec\pi+
\frac{1}{2}g^2af_\pi^2\vec\rho_\mu^2
-\frac{1}{4}\vec\varrho_{\mn}\vec\varrho^{\mn}+......
\ee
The above equation implies that the mass of the $\rho$ meson  
($m_\rho^2=ag^2f_\pi^2$) is generated due to SSB via Higgs mechanism
and the unphysical N-G modes $\Sigma$ (and not $\pi$) are ``eaten-up''
by the gauge boson
{\it i.e.} the three extra degrees of freedom
get converted to the three degrees of polarization
appropriate for the massive gauge boson. For $a=2$ one recovers the KSRF relation. This 
value of $a$ also results in universal coupling of $\rho$.

Harada {\it et al}~\cite{harada} have evaluated the finite temperature
effects on the $\rho$-mass upto one loop order in the HLS approach
due to the thermal pion and $\rho$ meson interactions.
Their results reveal that at high temperature the 
reduction in $\rho$ mass due to pion loop is overwhelmed by
the increase due to thermal $\rho$ loop contribution, although
the net shift is rather small. The contribution of thermal
pions to the $\rho$ self energy in this model is different from 
other calculations because in HLS approach there is no pion
tadpole contribution. 

\bef
\centerline{\psfig{figure=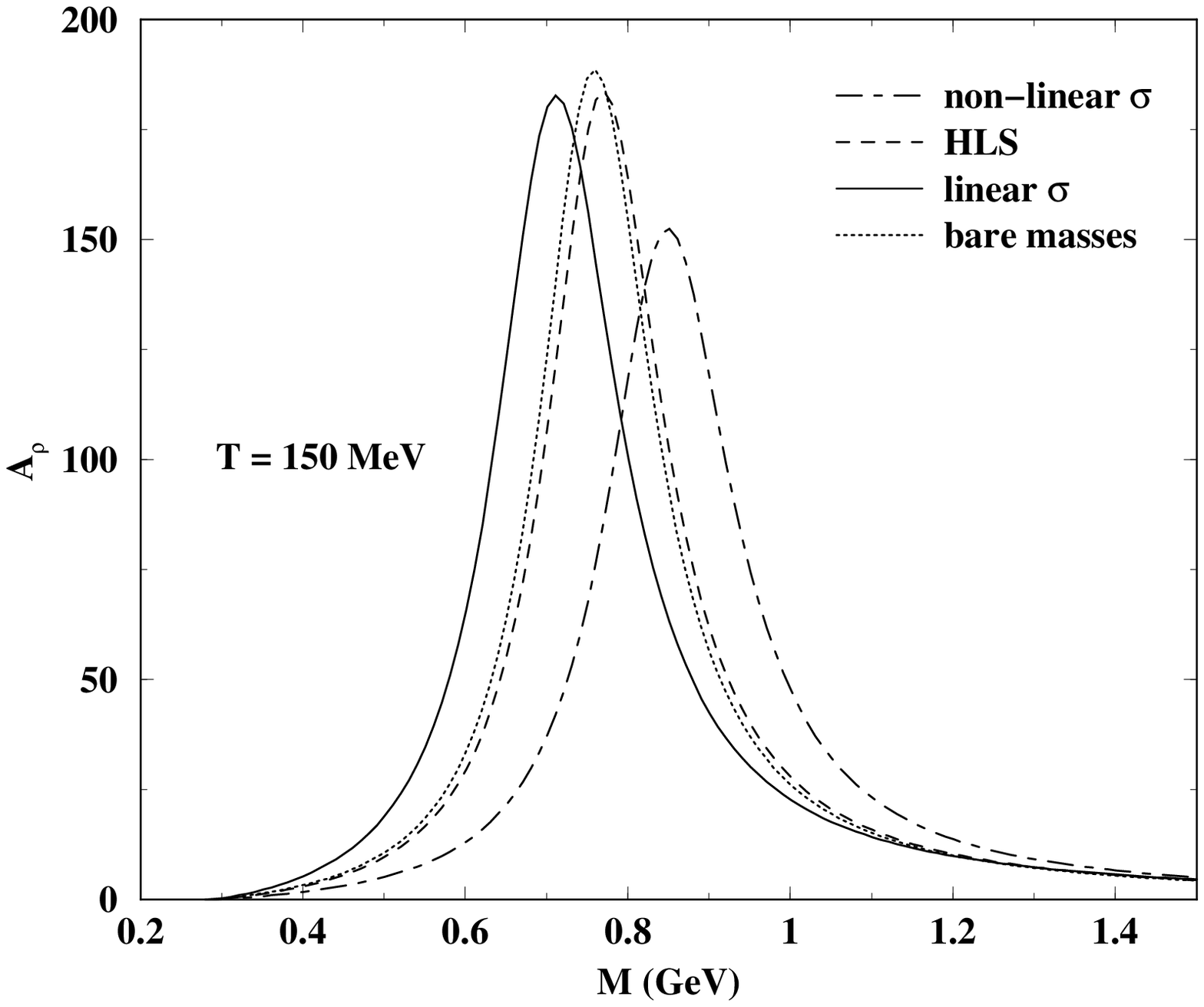,height=7cm,width=9cm}}
\caption{ Shift in the pole position of the $\rho$
spectral function for gauged Linear and Non-Linear Sigma Models and
Hidden Local Symmetry Lagrangian at $T=150$ MeV.
}
\label{spro_oth}
\eef
In Fig.~(\ref{spro_oth}) the shift in the pole position of 
the $\rho$-spectral function is depicted for the 
Linear Sigma Model (LSM), Non-Linear Sigma Model (NLSM), and 
Hiddden Local Symmetry (HLS) approach. 
For both the NLSM and HLS
interactions the $\rho$ mass increases by an amount 90 MeV and 10 MeV
respectively. Due to the enhancement of $\rho$ mass in the NLSM, a larger
phase space is available for the decay process $\rho\ra\pi\pi$ 
and consequently the $\rho$ appears to be broader in this
case compared to HLS interaction.
On the other hand, in the gauged LSM the $\rho$ mass
reduces by about 45 MeV at $T=150$ MeV.
It may be noted that $\rho$ mass decreases in gauged
LSM for low temperatures and increases for temperatures in the vicinity
of the chiral phase transition. 

\subsection{Spectral Constraints at Finite $T$}

We will now discuss the medium modifications of hadronic properties
in the QCD sum rule (QSR) approach.
We will briefly discuss the basic principles of 
QSR in vacuum~\cite{shifman,reinders,ioffe,ms,narison,svz}  
and at finite temperature~\cite{BS,hl,HKL}
and see how these ideas
can be utilized to constrain the spectral function of the vector mesons
$\rho$ and $\omega$.

\subsubsection{a) QCD Sum Rules at $T=0$}

Because of the complex infrared properties of QCD it is very difficult
to extract information on hadronic spectra from the QCD Lagrangian.
The idea of QSR formalism is to approach the bound state problem in QCD
from the asymptotic freedom side {\it i.e.} to start at short distances 
and move to larger distances where confinement effects become important,
asymptotic freedom starts to break down and resonances emerge as a
reflection of the fact that quarks and gluons are permanently
confined within hadrons. The breakdown of asymptotic freedom is signalled by
the emergence of power corrections due to non-perturbative effects of the QCD
vacuum and are known to be more important than higher order $\alpha_s$ 
corrections~\cite{reinders}. 
These are introduced via non-vanishing expectation values of 
quark and gluon condensate operators such as, $\langle 0| {\bar q}q |0\rangle$, 
$\langle 0| G_{\mu \nu}^a G^{\mu \nu a}| 0\rangle$ 
etc.  where $q(x)$ is the
quark field and $G_{\mu \nu}^a(x)$ is the gluon field tensor. 
In standard perturbation theory these matrix elements vanish after normal ordering.
In the following we will discuss
how the QCD sum rule approach connects the
perturbative and non-perturbative domains and 
leads to the determination of hadronic resonance parameters like mass, coupling
constant etc. in terms of the condensates.

The QCD sum rule approach starts with the Wilson operator 
product expansion (OPE) of a
product of suitable currents. 
The gluon and quark condensates
appear as higher dimensional operators in the expansion. The coefficients
of this expansion contain the short distance part and the long range part
is contained in the vacuum expectation values. The coefficients can be 
evaluated perturbatively in terms of the parameters 
($\alpha_s$ and the quark masses) of the Lagrangian used. 
In our discussions on QSR in vacuum, we shall consider the 
time-ordered or causal current correlator
\be
W^F_{\mu \nu}(q)\equiv i\,\int\,d^4x\,e^{iq\cdot x}\,\langle 0|T\{
J_{\mu}(x)J_{\nu}(0)\}|0
\label{eq1}
\rangle
\ee
which has a tensor structure of the form
\be
W^F_{\mn}(q)=-q^2\,(g_{\mn}-q_{\mu}q_{\nu}/q^2) W(q^2).
\label{altdef}
\ee
The source (electromagnetic) currents $J_\mu$ can be defined 
in terms of the quark fields (in units of $e$);
\be
J_{\mu}  =   {2 \over 3} \bar{u} \gamma_{\mu} u
 - {1 \over 3} \bar{d} \gamma_{\mu} d
 - {1 \over 3} \bar{s} \gamma_{\mu} s. 
\ee
Defining the current in the $\rho$, $\omega$ and $\phi$ channels as
\bea
J_{\mu}^{\rho}& = &(1/2)(\bar{u} \gamma_{\mu} u - 
\bar{d} \gamma_{\mu} d),\nonumber\\ 
J_{\mu}^{\omega}& = &(1/2)(\bar{u} \gamma_{\mu} u + 
\bar{d} \gamma_{\mu} d),\nonumber\\ 
J_{\mu}^{\phi}& = &\bar{s} \gamma_{\mu} s ,
\eea
one can express the electromagnetic current in terms of 
$\rho$, $\omega$ and $\phi$ fields as,
\be
 J_{\mu} = J_{\mu}^{\rho} + {1 \over 3} J_{\mu}^{\omega} -
      { 1 \over 3}  J_{\mu}^{\phi}.
\label{emhad}
\ee
Presently, we will confine our discussions to the $\rho$ meson 
($J^{PC}=1^{--}, I=1$) only.
The analytic structure of the correlator ($W$), for spacelike $Q^2 = -q^2$,
can be expressed through a dispersion relation:
\begin{equation}
W(Q^2) = \frac{1}{\pi}\,\int\,\frac{{\s Im}W(s)\,ds}{s+Q^2}
+({\s subtraction}). 
\label{eq3}
\end{equation} 
The imaginary part of $W$ is proportional to the spectral density which can be 
modelled as consisting of a conspicuous resonance and a continuum with a 
sharp threshold $\omega_0$,
\begin{equation}
{\s Im}W(s) = \pi\,\sum_{\s Res}\,{\cal G}_R\,m_R^2\,
\delta(s-m_R^2)+\frac{1}{8\pi}\left(1+
\frac{\alpha_s}{\pi}\right)\,\theta(s-\omega_0)
\label{eq4}
\end{equation} 
 ${\cal G}_R$ is the resonance strength and the pole position is at $m_R^2$.

The theoretical side of the sum rule is derived from an operator product
expansion for large $Q^2 = -q^2$ (deep Euclidean region) where asymptotic
freedom is realized.
Thus we write
\begin{equation}
i\,\int\,d^4x\,e^{iq\cdot x}\,\lgl\,T\{J_{\mu}(x)J_{\nu}(x)\}\rgl
= C_I(q) + \sum_{n}\,C_n(q){\cal O}_n
\label{eq5}
\end{equation} 
where $I$ is the identity operator, $C$'s are the Wilson coefficients,
and ${\cal O}_n$'s are the local gauge invariant operators constructed from the 
quark and gluon fields. The operators are ordered by their increasing
dimensions and therefore, the coefficients fall off by corresponding 
powers of $q^2$. On dimensional grounds one observes that the operators of dimension
$d > 0$ leads to $1/q^d$ power corrections. However, for large $Q^2 = -q^2$
a fewer number of power corrections ($d = 6$) is sufficient to converge the series.
Taking the vacuum expectation value of Eq.~(\ref{eq5}) we 
obtain~\cite{reinders}
\begin{eqnarray}
W(Q^2)&=&-\frac{1}{8\pi^2}\left(1+\frac{\alpha_s}{\pi}\right)\,
\ln{\frac{Q^2}{\mu^2}}+\frac{1}{Q^4}\,\langle 0|m_u{\bar u}u+m_d{\bar d}d
|0\rangle \nonumber\\
&&+\frac{1}{24Q^4}\langle 0|\frac{\alpha_s}{\pi}G_{\mu \nu}^aG^{\mu \nu a}|0
\rangle - \frac{\pi \alpha_s}{2Q^6}\langle 0|({\bar u}\gamma_{\mu}\gamma_5
\lambda^a u-{\bar d}\gamma_{\mu}\gamma_5\lambda^a d)^2|0\rangle \nonumber\\
&&-\frac{\pi \alpha_s}{9Q^4}\,\langle 0|({\bar u}\gamma_{\mu}\lambda^a u+
{\bar d}\gamma_{\mu}\lambda^a d)\,\sum_{q=u,d,s}\,{\bar q}\gamma_{\mu}\
\lambda^a q|0 \rangle.
\label{eq7}
\end{eqnarray}

The left hand side (l.h.s.) is the well known two point Greens function which can be 
expressed in terms of the phenomenological
parameters characterizing the strong interaction processes via the 
dispersion relations, consistent
with the current under consideration.
The right hand side (r.h.s.) has been evaluated by using OPE in the short distance (asymptotic
freedom) region. The vacuum expectation value of the higher dimensional
operator appears as a power correction to the asymptotic contribution
(the first logarithmic term in the r.h.s. of the above equation).

The sum rule therefore, becomes (modulo subtractions)
\begin{equation}
\frac{1}{\pi}\,\int\,\frac{{\s Im}W(s)\,ds}{s+Q^2} = W(Q^2).
\label{eq8}
\end{equation} 

In Eq.~(\ref{eq8}) the r.h.s. corresponds to large $Q^2$ or small distance
scale with fewer power corrections and l.h.s should be saturated by the lowest
resonance, which is a long distance phenomenon. Therefore, in order to get a
balance between the two sides we would like to have a weight function which
enhances the low $Q^2$ contribution relative to the high $Q^2$ one. This can
be done by taking additional derivative with respect to $Q^2$, and then taking
$Q^2$ and the number of derivatives $n$ to infinity. This yields the Borel 
transformed sum rule. Borel transformation is equivalent to the following
mathematical operation:
\begin{equation}
{\hat L}_M\,\frac{1}{s+Q^2} = \frac{1}{M_B^2}\,e^{-s/M^2}
\label{eq9}
\end{equation} 
where
\begin{equation}
{\hat L}_M = lim_{_{_{_{\!\!\!\!\!\!\!\!\!\!\!\!\!\!\!\!\!\!\!\!\!\!\!\!\!\!\!
\stackrel{Q^2,n\rightarrow \infty} {Q^2/n=M_B^2=
{\s const.}}}}}}
 \frac{1}{(n-1)!}\,Q^{2n}\left(-\frac{\partial}{\partial Q^2}\right)^n
\label{eq10}
\end{equation} 
and $M_B$ is the Borel mass.
Applying Eq.~(\ref{eq9}) on the l.h.s. of Eq.~(\ref{eq8}) and Eq.~(\ref{eq10})
on the r.h.s. and expressing vacuum expectation value of four fermion  
operators in terms of two fermions, we  obtain~\cite{reinders}
\bea
\int\,e^{-s/M_B^2}\,{\s Im}W(s)\,ds&=&\frac{1}{8\pi}\,M_B^2\left[
1+\frac{\alpha_s}{\pi}+\frac{8\pi^2}{M_B^4}\langle 0| m_q{\bar q}q|0\rangle\right.
\nonumber\\
&&+\left.\frac{\pi^2}{3M_B^4}\langle0|\frac{\alpha_s}{\pi}G_{\mu \nu}^a
G^{\mu \nu a}|0\rangle\right.\nonumber\\
&&-\left.\frac{448}{81}\,\frac{\pi^3\alpha_s}{M_B^6}\,
{\langle 0|{\bar q}q|0\rangle}^2\right] .
\label{eq11}
\eea
Substituting the various values of the matrix elements as given in 
Ref.~\cite{reinders} we obtain
\begin{equation}
\int\,e^{-s/M_B^2}\,{\s Im}W(s)\,ds = \frac{1}{8\pi}\,M_B^2\left[
1+\frac{\alpha_s}{\pi}+\frac{0.04}{M_B^4}-\frac{0.03}{M_B^6}\right].
\label{eq12}
\end{equation} 
Differentiating with respect to $1/M_B^2$ we obtain another sum
rule:
\begin{equation}
\int\,e^{-s/M_B^2}\,{\s Im}W(s)\,sds = \frac{1}{8\pi}\,M_B^4\left[
1+\frac{\alpha_s}{\pi}-\frac{0.04}{M_B^4}+\frac{0.06}{M_B^6}\right].
\label{eq13}
\end{equation} 
In Eqs.~(\ref{eq12}) and (\ref{eq13}) the terms $M_B^{-4}$ and $M_B^{-6}$
arise due to gluon and quark condensates respectively.
Assuming that r.h.s. of Eq.~(\ref{eq4}) is saturated by the $\rho$ resonance  
we get from Eqs.~(\ref{eq12}) and 
(\ref{eq13}) 
\begin{equation}
m_{\rho}^2 = M_B^2\,\frac{\left(1+\alpha_s/\pi\right)\,\left[1-\left(1+
\omega_0/M_B^2\right)e^{-\omega_0/M_B^2}\right]
-0.04/M_B^4+0.06/M_B^6}
{\left(1+\alpha_s/\pi\right)\,[1-
e^{-\omega_0/M_B^2}]+0.04/M_B^4-0.03/M_B^6}.
\label{eq14}
\end{equation} 
The above expression still depends on the Borel mass $M_B$ and the continuum
threshold $\omega_0$. 
The value of $\omega_0$ can be inferred from the data of $e^+e^-$  
annihilation. The absolute value of $\rho$ mass is then
obtained by looking for the stability plateau {\it i.e.} choosing $M_B^2$ such
that ${\partial m_{\rho}(M_B^2)}/{\partial M_B^2} = 0$. To determine the 
resonance strength for the $\rho$ meson we keep only the 
$\rho$ resonance in the sum of Eq.~(\ref{eq4}) and substitute it in Eq.~(\ref{eq12})
to obtain, after an elementary integration,
\be
4\pi\,{\cal G}_\rho=\frac{M_B^2e^{m_\rho^2/M_B^2}}{2\pi\,m_\rho^2}
\left[1+\frac{\alpha_s}{\pi}+\frac{0.04}{M_B^4}-\frac{0.03}{M_B^6}
-(1+\alpha_s/\pi)e^{-\omega_0/M_B^2}\right]
\label{eq15}
\ee
where ${\cal G}_\rho=1/g_\rho^2$.

Eqs.~(\ref{eq14}) and (\ref{eq15}) indicate how the resonance 
parameters of vector mesons can be extracted 
by using QCD sum rules in vacuum. In the next Section we will
briefly discuss the QCD sum rules at non-zero temperature.

\subsubsection{b) QCD Sum Rules at Finite $T$}

As mentioned earlier, it is not the causal (time-ordered) but the retarded current
correlator has the required analytic properties in a thermal system.
QCD sum rules for vector mesons in medium~\cite{hl,HKL} start with  the
retarded current correlation function,
\begin{eqnarray}
W_{\mu \nu}^R (q_0 , {\vec q})
=i \int d^4x e^{iq\cdot x}  \theta(x_0)\langle\,[J_{\mu}(x), J_{\nu}(0)]\,\rangle\ ,
\label{correlator}
\end{eqnarray}
where  $q^\mu \equiv (q_0 , {\vec q})$ is the four momentum and the currents
$J_\mu$ are defined in Eq.~(\ref{emhad}). 
As discussed earlier
there are two independent
invariants in medium, the transverse ($W^R_T$) and 
the longitudinal ($W^R_L$) 
components of the polarization tensor  
both of which satisfy fixed ${\vec q}$ dispersion relations. 
These are defined  through
\be
W_{\mn}^R=-q^2(A_{\mn}W^R_T+B_{\mn}W^R_L)
\label{newdef}
\ee
where $A_{\mn}$ and $B_{\mn}$ are defined by Eqs.~(\ref{amunu}) and (\ref{bmunu})
respectively.
In the limit  ${\vec q} \rightarrow 0$, as there is no spatial direction, we have
\be
W^R_T=W^R_L\equiv W^R
=W_{\mu\mu}^R/(-3 q_0^2)
\label{newdef1}
\ee
 where the last relation follows from the trace of Eq.~(\ref{newdef}).
In this limit,
\begin{eqnarray}
\label{dispersion2}
{\s Re} W^{R} (q_0) =
 {1 \over \pi} {\s P} \int_0^{\infty} du^2
{ {\s Im} W^R(u) \over u^2-q_0^2} + ({\s subtraction}). 
\label{disp0}
\end{eqnarray}
 ${\s Re} W^R$ can be calculated using perturbation
 theory with power corrections 
 in the deep Euclidean region $q_0^2 \rightarrow - \infty $ using OPE.
  For example, OPE for ${\s Re} W^R(q_0)$, which is the same as
 the OPE for the causal (Feynman) correlator  $W^F(q_0)$,
 has a general form at $q_0^2 \equiv - Q^2 \rightarrow - \infty$;
\begin{eqnarray}
 {\s Re} W^R(q_0^2 \rightarrow - \infty)
 =  - C_0 \ln Q^2 + \sum_{n=1}^{\infty}
 {C_n (\alpha_s(\mu^2), \ln (\mu^2/Q^2)) \over Q^{2n}} \langle {\cal O}_n
(\mu^2)
\rangle_{_T}
\
\
\ ,
\label{ope}
\end{eqnarray}
where $\mu$ is the renormalization point of the local  operators which 
separates the  hard scale $|Q|$ and soft scales such 
as  $\Lambda_{QCD}$ and $T$.
 $C_n$ are the c-number Wilson coefficients which are
 $T$ independent. All the medium effects are contained in the
 thermal average of the local operators ${\cal O}_n$.
 Since $\langle {\cal O}_n \rangle_{_T} \sim T^{2l}\cdot \Lambda_{QCD}^{2m}$
 with $l+m=n$ due to dimensional reasons, Eq.~(\ref{ope})
 is a valid asymptotic expansion as long as
 $Q^2 \gg T^2$ and $\Lambda_{QCD}^2$. 
 The local operators ${\cal O}_n(\mu^2)$
 in the vector meson sum rule are essentially 
 the same with those 
 in the lepton-nucleon deep inelastic scattering (DIS) and
 can be characterized by their canonical dimension ($d$) and the 
twist ($\tau$=dimension-spin).   They are given in Ref.~\cite{HKL}
  up to dimension 6 operators and we will not recapitulate them here.
 For ${\vec q} \rightarrow 0$,
 Eq.~(\ref{ope}) is  an asymptotic series in $1/Q^2$
 or equivalently  an expansion with respect to  $d$.
  The  medium condensates $\langle {\cal O}_n (\mu^2) \rangle_{_T}$  may be
evaluated by
 low energy theorems, the parton distribution of hadrons
 and lattice QCD simulations. 

 Matching the left and right hand sides of Eq.~(\ref{dispersion2}) 
 in the asymptotic region $q_0^2 \rightarrow - \infty $
 is the essential part of QSR. This procedure 
 gives constraints on the spectral integral and hence the 
 hadronic properties in the medium as well as in the vacuum.
 There are two well-known procedures for this matching, namely 
 the Borel sum rules (BSR) \cite{svz} and the
  finite energy sum rules (FESR) \cite{kpt},
 which can be summarized as
\begin{eqnarray}
\label{sumrules}
\int_0^{\infty}
 & dq_0^2\ \ V(q_0^2)& \ [{\s Im} W^R(q_0)
 - {\s Im} W^R_{_{OPE}}(q_0) ] =0 ,\\ 
 & V(s) & = \left\{ \begin{array}{ll}
                    q_0^{2n} \ 
 \theta(\omega_0 -q_0^2) & \ \ \ \ \ \ \ \ \  ({\s FESR}), \\ 
\nonumber
                    e^{-q_0^2/M_B^2} & \ \ \ \ \ \ \ \ \ ({\s BSR}).
                   \end{array}     \right.
\end{eqnarray}
Here ${\s Im} W^R_{_{OPE}}(q_0)$ is a hypothetical imaginary part of
 $W^R$ obtained from OPE and $M_B$ is the Borel mass.
 
We have seen in the previous Section that
in QSR in the vacuum, the spectral function
 ({\it i.e.} ${\s Im} W^R$ in  Eq.~(\ref{disp0}))
 is usually modelled with a  resonance pole and the 
 continuum to extract the mass and decay constant of hadrons.
 In the medium, such a simple parametrization is not always
 justified because of the thermal broadening of the spectrum and
 also because of the new spectral structure due to Landau damping and the 
 thermal mixing among mesons. 
  Therefore, the model independent constraints obtained from
 QSR are only for the weighted spectral integral.
 
 For example, the first three finite energy sum rules at finite $T$
 read \cite{HKL}
\begin{eqnarray}
I_1 & = & \int_0^{\infty}
 [{\s Im} W^R(q_0) - {\s Im} W^R_{_{OPE}}(q_0)] \ dq_0^2 = 0,
  \\
 I_2 & = & \int_0^{\infty}
 [{\s Im} W^R(q_0) - {\s Im} W^R_{_{OPE}}(q_0)]\ q_0^2 
  \ dq_0^2 = - C_2 \langle {\cal O}_2 \rangle_{_T}, \\
I_3 & = & \int_0^{\infty}
 [{\s Im} W^R(q_0) - {\s Im} W^R_{_{OPE}}(q_0)] \ q_0^4 
 \ dq_0^2 = C_3 \langle {\cal O}_3 \rangle_{_T} .
\end{eqnarray}
 Similar sum rules hold  for the axial vector
 channel (in the chiral limit)
 except that one has a different operator for ${\cal O}_3$.
 One can also generalize the above  sum rules to finite
 ${\vec q}$ \cite{KS,shlee}.
  
 Explicit forms of $C_n \langle {\cal O}_n \rangle_{_T} $ have been
 calculated as \cite{HKL}
\begin{eqnarray}
C_0 & = & - {1 \over 8 \pi} (1+ {\alpha_s \over \pi}) , \ \ \ 
 C_1  =  0 ,  \\
\label{dim4}
C_2 \langle {\cal O}_2 \rangle_{_T} & = & {1 \over 24} 
\langle {\alpha_s \over \pi} G^2 \rangle_{_T} +
 {4 \over 3} \langle {\cal S} \bar{q} i \gamma_0 D_0 q \rangle_{_T} , \\
C_3 \langle {\cal O}_3 \rangle_{_T} & = & -
 \langle {\s scalar\ 4-quark}) \rangle_{_T}
 +
 {16 \over 3} \langle {\cal S} \bar{q} i \gamma_0 D_0 D_0 D_0 q \rangle_{_T} . 
\end{eqnarray}
Here we have neglected the terms proportional to the
 light quark masses and the quark-gluon mixed operators.
${\cal S}$ is used to make the operators symmetric and traceless.
At low $T$, one may use the soft pion theorems
 and the parton distribution of the pion to estimate the r.h.s.
 of the above equations.
 When $T$ is close to $T_c$, one has to look for a totally different
 way of estimation; the simplest approach is to assume a resonance gas
 to evaluate the r.h.s., while the direct lattice simulations
  will be the most reliable way in the future.

 The sum rules $I_i$ can be used to check the 
 validity of the calculations of the spectral functions
 using effective theories of QCD.  This is in fact quite 
 useful for the spectral function at finite baryon density.
 At finite $T$, especially near the critical point,
 the behavior of the condensates with $d\ge 4$ is
 not known precisely.  Therefore, it is rather difficult to
 make a strong argument on the spectral constraints near $T_c$
 at present. The future lattice simulations of these
 condensates are highly called for.

\subsubsection{c) Parametrization of the Spectral Function}

We will now introduce a 
 parametrization of the correlator at finite $T$.
 The parametrization
 should be consistent with the experimental
  data from $e^+e^-\rightarrow
hadrons$. It should
 also be consistent with the  
 high energy behaviour known from perturbative QCD at $q_0 \gg T$.

As the vector mesons appear as resonances in the electromagnetic 
correlator, using Eqs.~(\ref{correlator}) and (\ref{emhad}) we can write  
\be
{\s Im} W_{\mu \nu}^{R} = 
 {\s Im} W_{\mu \nu}^{R,\rho} +
 {1 \over 9} {\s Im} W_{\mu \nu}^{R,\omega} +
 {1 \over 9} {\s Im} W_{\mu \nu}^{R,\phi},
\label{rcorr}
\ee
which shows that the contributions of $\omega$ 
and $\phi$ mesons to the electromagnetic probes are less by
almost an order of magnitude compared to $\rho$ mesons. 
In the limit $\vec q=0$, we have
from Eq.~(\ref{newdef1}) 
\be
{\s Im} \ W^R_{\mu \mu}(q_0) 
= -3 q_0^2 \ {\s Im}W^R (q_0).
\label{cormm}
\ee
Our next task is to parametrize ${\s Im} W^R(q_0)$
which is now a positive dimensionless quantity.
 We take a
Breit-Wigner form with an energy-dependent width for the resonance
along with a continuum:
\be
 {\s Im} W^R_{\rho} (q_0, {\vec q}=0)
 = f_{\rho}^{\ast 2} {{\s Im}\Pi^R_{\rho} \over (q_0^2 - m_{\rho}^{\ast 2})^2
 + ({\s Im}\Pi^R_{\rho})^2 } + {1 \over 8 \pi}(1+ {\alpha_s \over \pi })
 {1 \over 1 + e^{(\omega_0^\ast - q_0)/\delta}}.
\label{parametrho}
\ee
where `$\ast$' indicates the in-medium values of the parameters. 
 At $T=0$, the above form 
reduces to a relativistic generalization of the parametrization
used by Shuryak~\cite{shuryak} to fit the experimental data of
$e^+e^-\rightarrow  hadrons$. 
 Here ${\s Im}\Pi^R_{\rho}$ is 
the imaginary part of the self-energy which should in principle
contain all the channels which can destroy or create
a $\rho$ in the thermal bath. 
Hence ${\s Im}\Pi^R_\rho$ is the difference of the decay-width and the formation 
width and is given by 
${\s Im}\Pi^R_\rho=q_0 \Gamma(q_0)$.  
However, we have seen
that for baryon free matter the most dominant
contribution to ${\s Im}\Pi^R_\rho$ comes from the pion-loop~\cite{ja}.
The quantity $\omega_0$ in Eq.~(\ref{parametrho}) 
is the continuum threshold above which the asymptotic 
freedom is restored and
$f_\rho$ is the coupling between the electromagnetic current and
the $\rho$ field in vacuum; 
\be
\langle 0\mid J_\mu^\rho \mid \rho \rangle=f_\rho m_\rho\epsilon_\mu. 
\ee
Assuming vector dominance in the medium  we obtain
\be
g_{\rho} = m_{\rho}/ f_{\rho}.
\ee
In vacuum, the standard parameters for the $\rho$ spectral
function are given by,
$m_{\rho}=0.77$ GeV, $m_{\pi}=0.14$ GeV ,
$f_{\rho}=0.141$ GeV, 
$g_{\rho}=5.46$,
$\omega_0=1.3$ GeV,
$\delta=0.2$ GeV and
$\alpha_s = 0.3$. 

Let us now concentrate on the spectral function in the $\omega$ channel.
We again take a Breit-Wigner form along with a continuum:
\be
 {\s Im} \ W^R_\omega (q_0, \vec q=0)
 = f_{\omega}^{\ast 2} {{\s Im}\Pi^R_{\omega} \over (q_0^2 - m_{\omega}^{\ast 2})^2
 + ({\s Im}\Pi^R_{\omega})^2 } + {1 \over 8 \pi}(1+ {\alpha_s \over \pi })
 {1 \over 1 + e^{(\omega_0^\ast - q_0)/\delta}}.
\label{parametomg}
\ee
In  vacuum
$f_{\omega}$ is the coupling of the current with the $\omega$ meson
 defined as 
\be 
\langle 0 \mid J_{\mu}^{\omega} \mid \omega \rangle =
 f_{\omega} m_{\omega} \epsilon_{\mu}.
\ee
Note that $f_{\omega}$ here is defined as 
 factor 3 larger than Shuryak's definition~\cite{shuryak} .
${\s Im}\Pi^R_{\omega}$, which is 
 the imaginary part of the $\omega$ self-energy is given by,
\be
{\s Im}\Pi^R_{\omega}(q_0) = 
q_0 (\Gamma_{\omega\ra 3\pi}+\Gamma_{\omega\pi\ra\pi\pi}).
\ee
In vacuum the standard parameters for $\omega$ are as follows.
$m_{\omega} =  0.782$ GeV,  
$m_{\pi}  =  0.14$  GeV, 
$f_{\omega}  = 0.138$ GeV, 
$\omega_0  = 1.1$ GeV, 
$\delta  =  0.2$ GeV and
$\alpha_s  =  0.3$. 

Since not much  is known about the critical behavior of the 
 scalar and tensor condensates at finite $T$ in QCD sum rules we take
 a simple ansatz for in-medium quantities for their $T$-dependence.
A possible parametrization of $*$-quantities at finite $T$ is
\be
{m_{V}^* \over m_{V}}  = 
{f_{V}^* \over f_{V}} = 
{\omega_{0}^* \over \omega_{0}}  =
 \left( 1 - {T^2 \over T_c^2} \right) ^{\lambda},
\label{anst}
\ee
where $\lambda$ is a sort of {\em dynamical} critical exponent and
$V$ stands for vector mesons ($\rho$ and $\omega$).
It may be noted that there is no definite reason to believe that all the in-medium
 dynamical quantities are dictated by a single exponent $\lambda$.
 Since the numerical value of $\lambda$ is not known precisely, 
 we take two typical cases:
 $\lambda=1/6$ (BR scaling) and $1/2$ 
 (Nambu scaling)~\cite{brpr}
with the following remarks:

\noindent
(i) Eq.~(\ref{anst}) for $m_{\rho}^*$ is not entirely consistent 
 with the low temperature theorem~\cite{mdey},
 which says there should be no $O(T^2)$ correction to the mass.
 Therefore, one cannot take the ansatz too seriously at low $T$.
 For our purposes, however,
  $T < 100$ MeV is not relevant in any way since it is below
 the freeze-out temperature (130 MeV) we have considered. 

\noindent
(ii) Local duality constraint $I_1$ in QCD sum rules implies that
 $(f_{\rho}^*)^2 = 8 \pi^2 (1+\alpha_s/\pi) 
 (\omega_0^*)^2$ + (scattering (Landau damping) term)~\cite{HKL}
which is slightly violated for $f_{\rho}^*$ as defined in Eq.~(\ref{anst}).
 
\noindent
(iii) The vector dominance assumption in the medium together with 
 Eq.~(\ref{anst}) simply leads to  $g_{\rho}^* = g_{\rho}$.

Under these reservations, we will use the parametrized spectral
 functions with the BR scaling and Nambu scaling ansatz in the calculation of 
 the  lepton and photon
 production in the following Section.
 The principal qualitative difference between the spectral function
 in QHD and that described in this Section is
 the existence of the continuum and its medium modification 
 at finite $T$.

\bef
\centerline{\psfig{figure=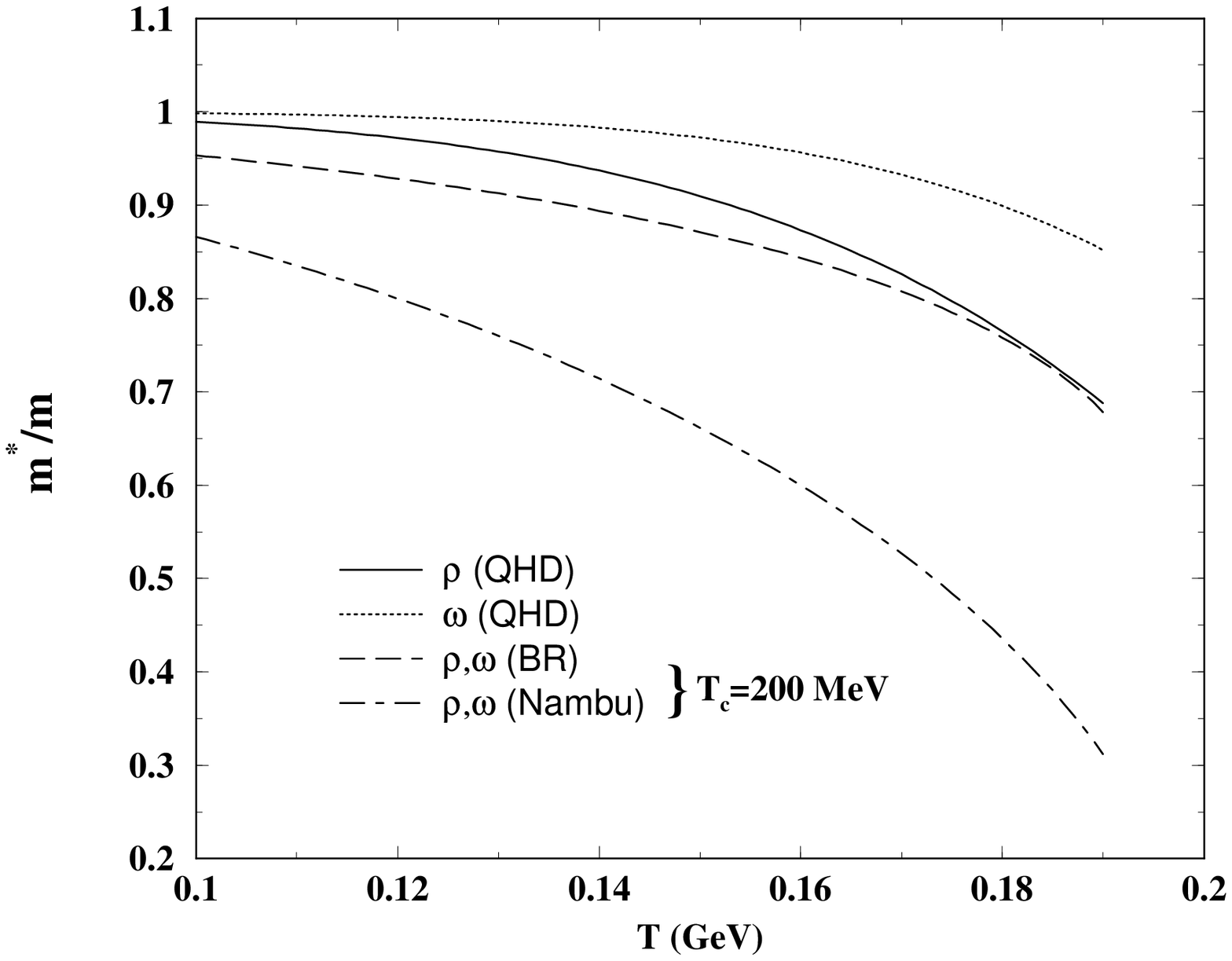,height=7cm,width=9cm}}
\caption{
Variation of vector meson mass with temperature 
for BR (long-dashed line), 
Nambu (dot-dashed line) scaling with $T_c$=200 MeV and
in the QHD model for $\rho$ (solid line) and $\omega$
(dotted line).
}
\label{massall}
\eef
In Fig.~(\ref{massall}) we depict the
variation of vector meson masses as a function of temperature
in the BR and Nambu scaling scenarios.
Results in the QHD model is also shown for the sake of comparison. 
The mass variation in the QHD model and BR scaling
is slower than the Nambu scaling scenario. At higher temperature (near $T_c$)
the QHD and the BR scaling results tend to converge. 
However, such a small difference in the mass
variation in the above two scenarios may not be 
visible through photon spectra.
We also note at this point that in QHD unlike the scaling scenarios
the $\rho$ and $\omega$
masses show different rate of reduction~\cite{npa99} due to
different values of their coupling constants with the nucleons.
In Fig.~(\ref{spr15qsr}) the spectral function 
($8\pi$ times Eq.~(\ref{parametrho})) for
the isovector ($\rho$) channel is plotted as a function 
of invariant mass at $T=150$ MeV and $T_c=160$ MeV. 
\bef
\centerline{\psfig{figure=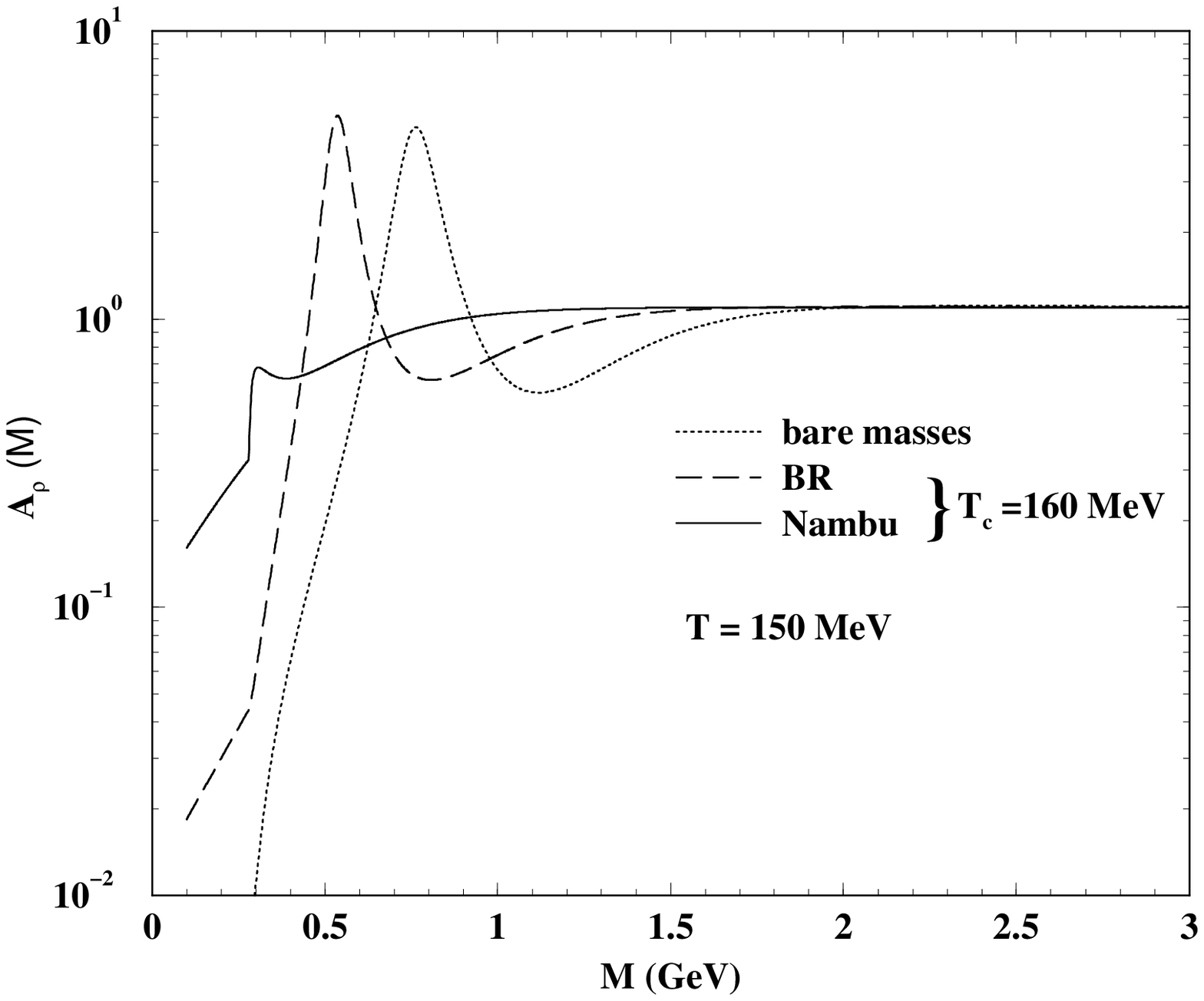,height=7cm,width=9cm}}
\caption{
Spectral function for the isovector ($\rho$)
channel extracted from $e^+e^-$ collisions (dotted line)
as a function of invariant mass. The dashed (solid) line indicates
the spectral function when $m_\rho$ and $\omega_0$ 
vary according to BR (Nambu) scaling.
}
\label{spr15qsr}
\eef
\bef
\centerline{\psfig{figure=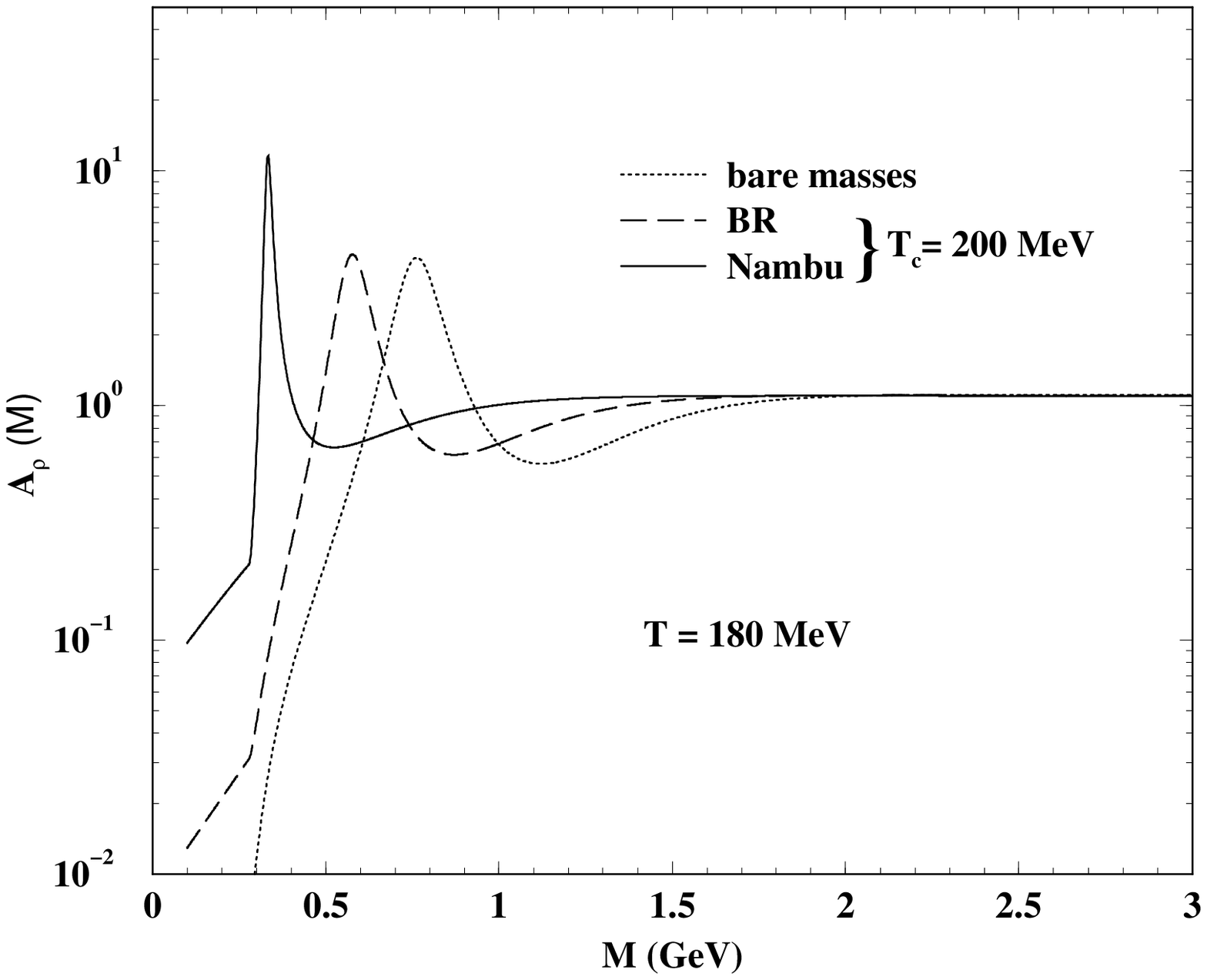,height=7cm,width=9cm}}
\caption{
Same as Fig.~(\protect\ref{spr15qsr}) at $T=180$ MeV and $T_c=200$ MeV.
}
\label{spr18qsr}
\eef
We find that both the peak and
the continuum threshold of the spectral function 
move towards lower invariant mass. In the case 
of Nambu scaling scenario the shift is more compared to BR scaling.  In the 
Nambu scaling scenario the peak of the spectral function and the 
continuum are not well separated; a merging of the two 
would take place at $T=T_c$. This could possibly indicate the 
onset of a deconfinement phase transition.
Fig.~(\ref{spr18qsr}) shows the spectral function at $T=180$ MeV
and $T_c=200$ MeV. Due to a larger separation between 
$T_c$ and $T$ compared to the 
previous case the peaks in the spectral function in all the
cases are well separated from the continuum.       
In Figs.~(\ref{spo15qsr}) and (\ref{spo18qsr}) the spectral functions for 
the isoscalar ($\omega$) channel 
obtained by multiplying Eq.~(\ref{parametomg}) by 8$\pi$ are shown
for $T$=150 and 180 MeV respectively. In both cases
the peaks in the spectral function corresponding to the BR and Nambu scalings
is distinctly visible.
The larger width in the isoscalar channel
is due to the combined processes $\omega\rightarrow\,3\pi$ and
$\omega\,\pi\rightarrow\,\pi\,\pi$ as discussed before.
\bef
\centerline{\psfig{figure=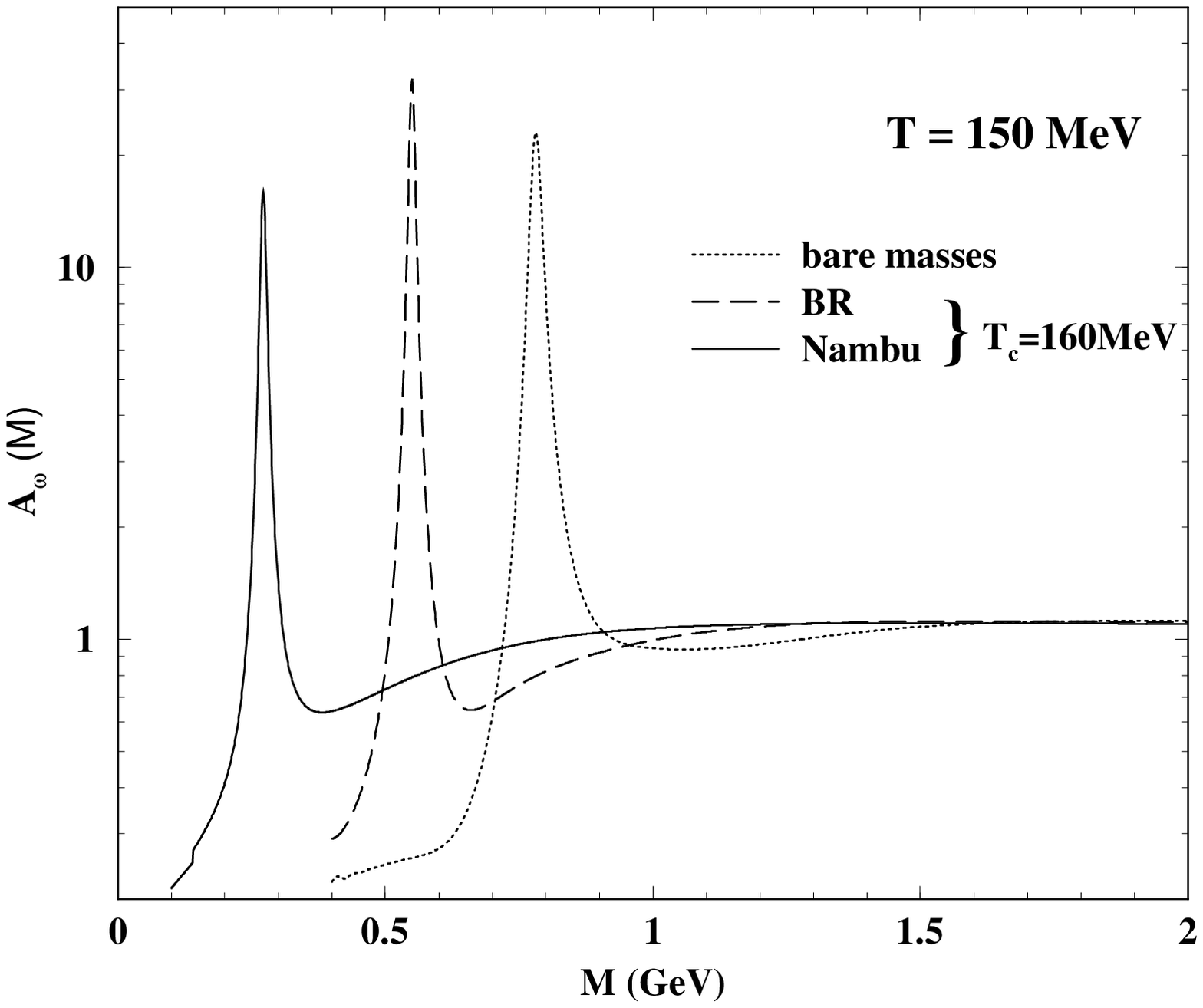,height=7cm,width=9cm}}
\caption{
Spectral function for the isoscalar ($\omega$)
channel extracted from $e^+e^-$ collisions (dotted line)
as a function of invariant mass. The dashed (solid) line indicates
the spectral function when $m_\rho$ and $\omega_0$ 
vary according to BR (Nambu) scaling.
}
\label{spo15qsr}
\eef
\bef
\centerline{\psfig{figure=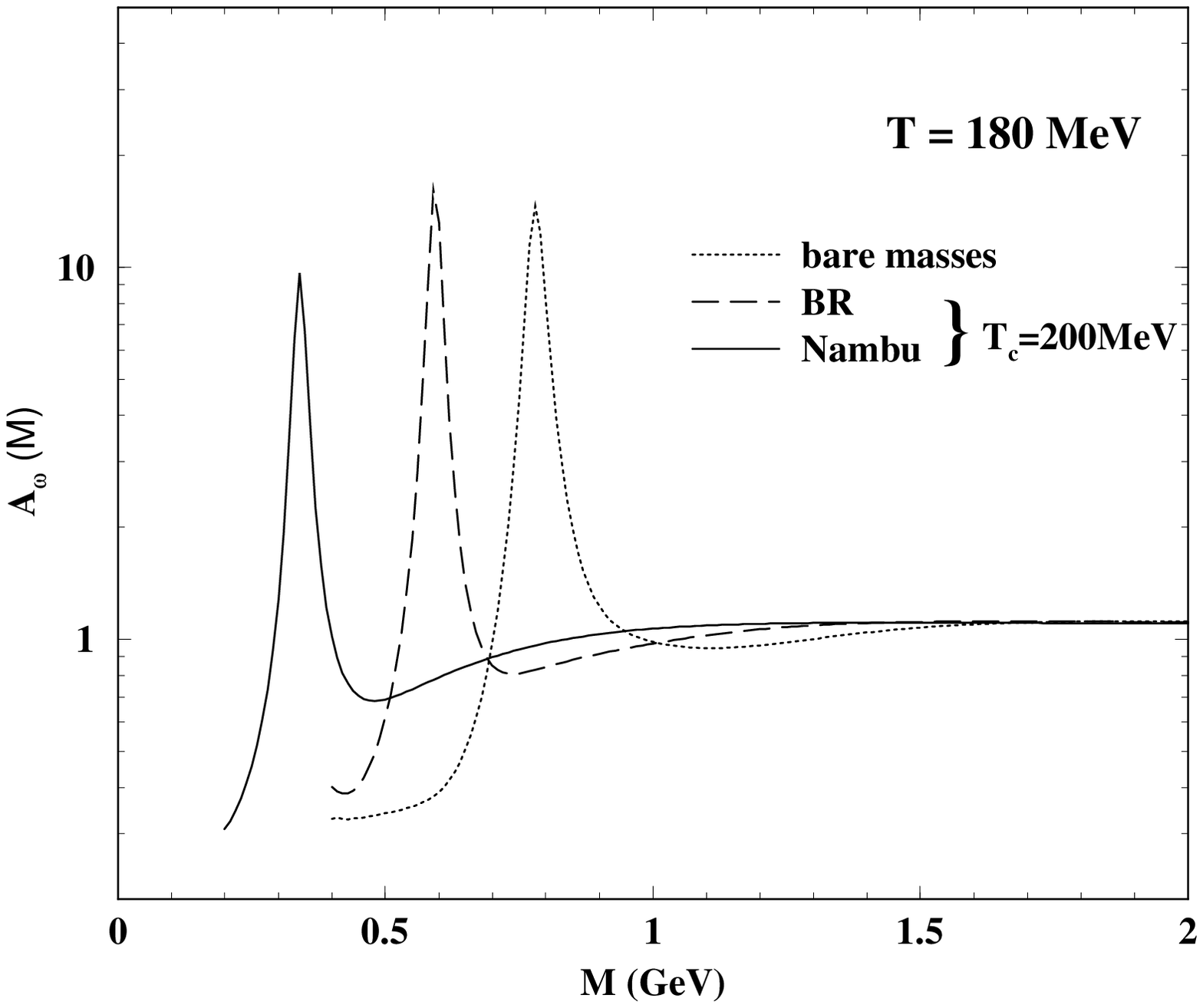,height=7cm,width=9cm}}
\caption{
Same as Fig.~(\protect\ref{spo15qsr}) at $T$=180 MeV and $T_c$= 200 MeV. 
}
\label{spo18qsr}
\eef

The spectral functions for the vector mesons both in the isoscalar 
and isovector channels are plotted in Fig.~(\ref{spomro}) 
at a temperature $T\sim T_c$. As expected from the scaling law, the 
Breit-Wigner peak has 
vanished due to its 
overlap with the continuum (see Eqs.~(\ref{parametrho}) and (\ref{parametomg})). All the hadrons in the thermal bath
have melted to their fundamental constituents - the 
quarks and gluons. Such a spectral function would indicate
	a transition from hot hadronic matter to QGP. 
This behaviour should,
	in principle, be reflected in the dilepton spectrum
	originating from these channels. 
	\bef
	\centerline{\psfig{figure=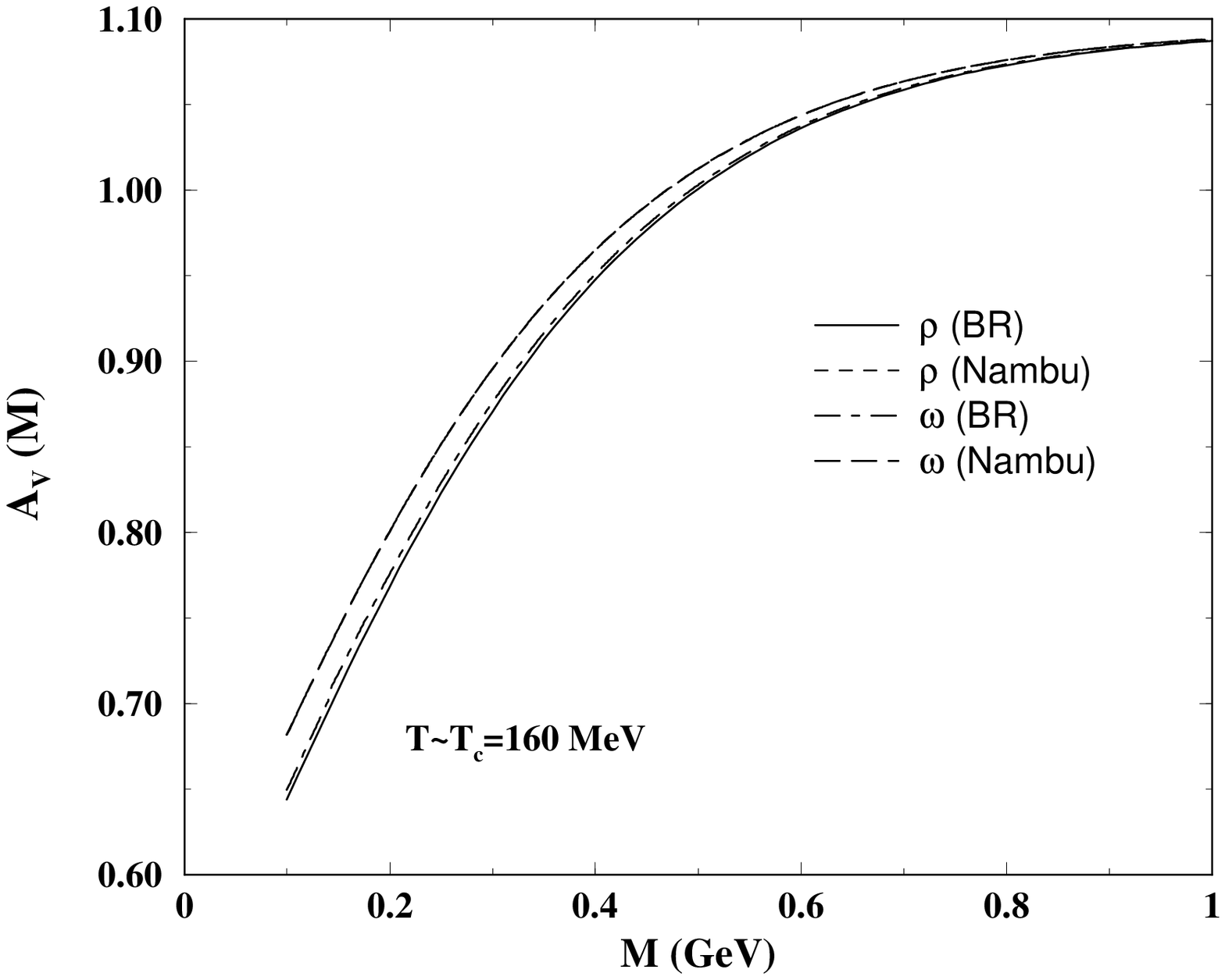,height=7cm,width=9cm}}
	\caption{
	Spectral functions for isovector ($\rho$) and
	isoscalar ($\omega$) channels  at $T_c$ .
	}
	\label{spomro}
	\eef

\section{Photon Emission Rates ~from ~Hot Hadronic ~Matter}

	In the energy regime of our interest
the most important  sources of
photon production from hadronic matter are
	the reactions 
	$\pi\,\rho\,\ra\, \pi\,\gamma$, 
	$\pi\,\pi\,\ra\, \rho\,\gamma$, $\pi\,\pi\,\ra\, \eta\,\gamma$, 
	$\pi\,\eta\,\ra\, \pi\,\gamma$, and the decays $\rho\,\ra\,\pi\,\pi\,\gamma$
	and $\omega\,\ra\,\pi\,\gamma$~\cite{kapusta,sourav,npa99}.
Apart from these we have also included those reactions
	which produce photons via the intermediary axial vector $a_1$.
A non-zero width of vector and axial vector mesons in the
	intermediate state has been taken into account.

 The relevant vertices for the reactions 
$\pi\,\pi\,\ra\,\rho\,\gamma$ and $\pi\,\rho\,\ra\,\pi\,\gamma$
and the decay $\rho\,\ra\,\pi\,\pi\,\gamma$
are obtained from the following Lagrangian:
\be
{\cal L} = -g_{\rho \pi \pi}{\vec {\rho}}^{\mu}\cdot
({\vec \pi}\times\partial_{\mu}{\vec \pi}) - eJ^{\mu}A_{\mu} + \frac{e}{2}
F^{\mu \nu}\,({\vec \rho}_{\mu}\,\times\,{\vec \rho}_{\nu})_3,
\label{photlag}
\ee
where $F_{\mu \nu} = \partial_{\mu}A_{\nu}-\partial_{\nu}A_{\mu}$ is the
field tensor for electromagnetic field
and $J^{\mu}$ is the hadronic part of the electromagnetic
current given by
\be
J^{\mu} = ({\vec \rho}_{\nu}\times{\vec \varrho^{\nu \mu}})_3 + (
{\vec \pi}\times(\partial^{\mu}\vec \pi+g_{\rho \pi \pi}{\vec \pi}\times{\vec
\rho}^{\mu}))_3,
\label{jmu}
\ee
with ${\vec \varrho_{\mu \nu}} = \partial_{\mu}{\vec \rho}_{\nu}-\partial_{\nu}
{\vec \rho}_{\mu}-g_{\rho \pi \pi}(\vec \rho_{\mu}\times\vec \rho_{\nu})$.
The coupling strength of the $\rho\pi\pi$ vertex, 
$g_{\rho\pi\pi}$, is
fixed from the observed decay $\rho\ra\pi\pi$.

Photon 
emission rates due to the reactions $\pi\,\eta\,\rightarrow\,\pi\,\gamma$, 
$\pi\,\pi\,\rightarrow\,\eta\,\gamma$ and the decay 
$\omega\,\ra\,\pi\,\gamma$ have been evaluated 
using the following interaction~\cite{GSW}:
\be
{\cal L} =
\frac{g_{\omega \rho \pi}}{m_{\pi}}\,
\epsilon_{\mu \nu \alpha \beta}\partial^{\mu}{\omega}^{\nu}\partial^{\alpha}
\rho^{\beta}\pi
+\frac{g_{\rho \rho \eta}}{m_{\eta}}\,
\epsilon_{\mu \nu \alpha \beta}\partial^{\mu}{\rho}^{\nu}\partial^{\alpha}
\rho^{\beta}\eta
+\frac{em_{\rho}^2}{g_{\rho}}A_{\mu}\rho^{\mu}
\label{etaro}
\ee
where $\epsilon_{\mu \nu \alpha \beta}$ is the 
totally antisymmetric Levi-Civita tensor.
The second term is constructed analogously as 
the first term which is the familiar GSW Lagrangian. 
The last term is written down on the basis
of Vector Meson Dominance (VMD)~\cite{sakurai}.
The values of $g_{\rho\rho\eta}$ and
$g_{\omega\rho\pi}$ are fixed from the observed decays, $\rho\,\ra\,\eta\,\gamma$
and $\omega\,\ra\,\pi\,\gamma$ respectively~\cite{npa99}.
The constant $g_{\rho}$ is determined from the 
decay, $\rho^0\ra\,e^+e^-$. 

The importance of the role of $a_1$ as an intermediary meson in the
process $\pi\,\rho\,\rightarrow\,\pi\,\gamma$ was emphasized 
in Refs.~\cite{xiong,song}.
Recently it has been shown~\cite{halasz} that this contribution is not so large.
We use the
following interaction Lagrangian for the $\pi\rho a_1$ and 
$\pi a_1\gamma$ vertices~\cite{kim,rudaz}:
\bea
{\cal L}&=& 
\frac{g_\rho^2f_\pi}{Z_\pi}
\left[(2c+Z_\pi)\vec \pi\cdot\vec \rho_\mu\times\,\vec a^\mu+
\frac{1}{2m_{a_1}^2}
\vec \pi\cdot
(\partial_{\mu}{\vec \rho}_{\nu}-\partial_{\nu}{\vec \rho}_{\mu})
\times\,
(\partial^{\mu}{\vec a}^{\nu}-\partial^{\nu}{\vec a}^{\mu})
\right.\nonumber\\
&&\left. +\frac{\kappa_6\,Z_\pi}{m_\rho^2}
\partial^\mu\vec \pi\cdot
(\partial_{\mu}{\vec \rho}_{\nu}-\partial_{\nu}{\vec \rho}_{\mu})
\times\,\vec a^\nu\right]\nonumber\\
&&+\frac{eg_\rho\kappa_6f_\pi}{m_\rho^2}F^{\mu\nu}
( \partial_{\mu}{\vec a}_{\nu}-\partial_{\nu}{\vec a}_{\mu}\times\vec \pi)_3
\label{aro}
\eea
where $a_\mu$ corresponds to the $a_1$ field,
$Z_\pi$ is the renormalization constant for pion fields and $f_\pi(=93$ MeV) 
is the pion decay constant. The interaction terms with coefficients
$c$ and $\kappa_6$ are introduced to improve the phenomenology of the model.
The following values of the parameters,  $m_{a_1}=1260$ MeV, $g_\rho=5.04$,
$c=-0.12$, $Z_{\pi}=0.17$ and
$\kappa_6=1.25$~\cite{rudaz,kim} are considered which reproduce 
the width of the $a_1$ meson in vacuum. 
	\bef
	\centerline{\psfig{figure=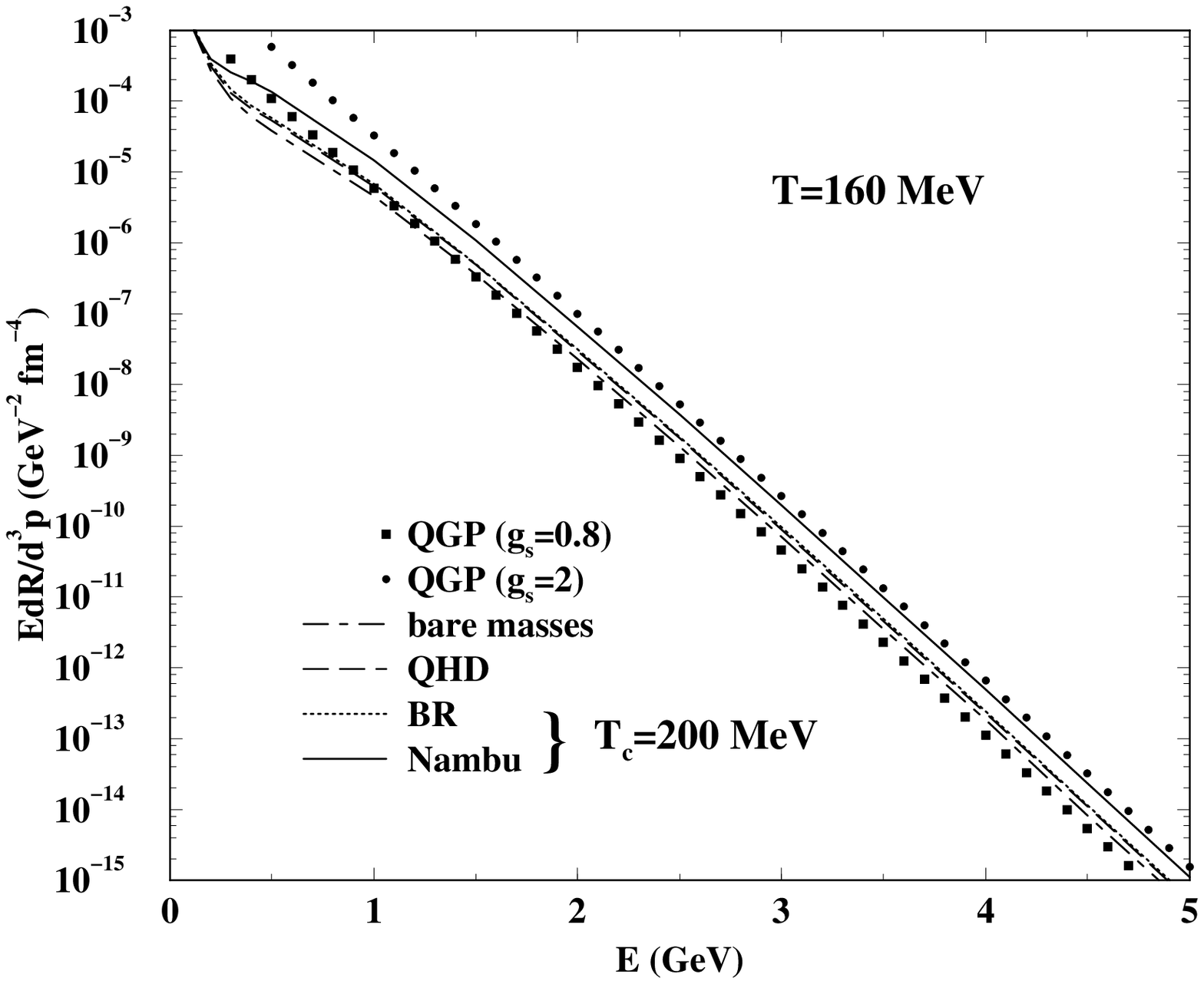,height=7cm,width=9cm}}
	\caption{
	Thermal photon spectra at $T=160$ MeV. Thick dots (squares) indicate 
	photon emission rate from QGP including one and two loop 
	contributions for $g_s=2 (0.8)$. Dot-dash line represents photon spectrum
	from hot hadronic gas without medium effects. The result with the in-medium 
	effects in the QHD model 
	is shown by the long-dashed line. Dotted (solid) line indicates
	photon spectrum with BR (Nambu) scaling mass variation scenario.
	}
	\label{prate160}
	\eef

The invariant amplitude for all the
reactions mentioned above are listed in the Appendix along
with the corresponding Feynman diagrams. These are used in Eq.~(\ref{rktp})
to obtain the rates of photon production incorporating
the effective masses and
decay widths of the participating hadrons.
	While evaluating the photons from QGP we have considered both one loop and
	two loop contributions 
	(as shown in Figs.~(\ref{loop1}) and (\ref{loop2}))
to the photon self energy.
	 
	The total photon emission rate from QGP and hadronic matter 
	at $T=160$ MeV is plotted in Fig.~(\ref{prate160})
	as a function of the energy of the emitted photon for
	different values of strong charge $g_s$ in the QGP phase and for various 
	mass variation scenarios in the hadronic sector. 
	\bef
	\centerline{\psfig{figure=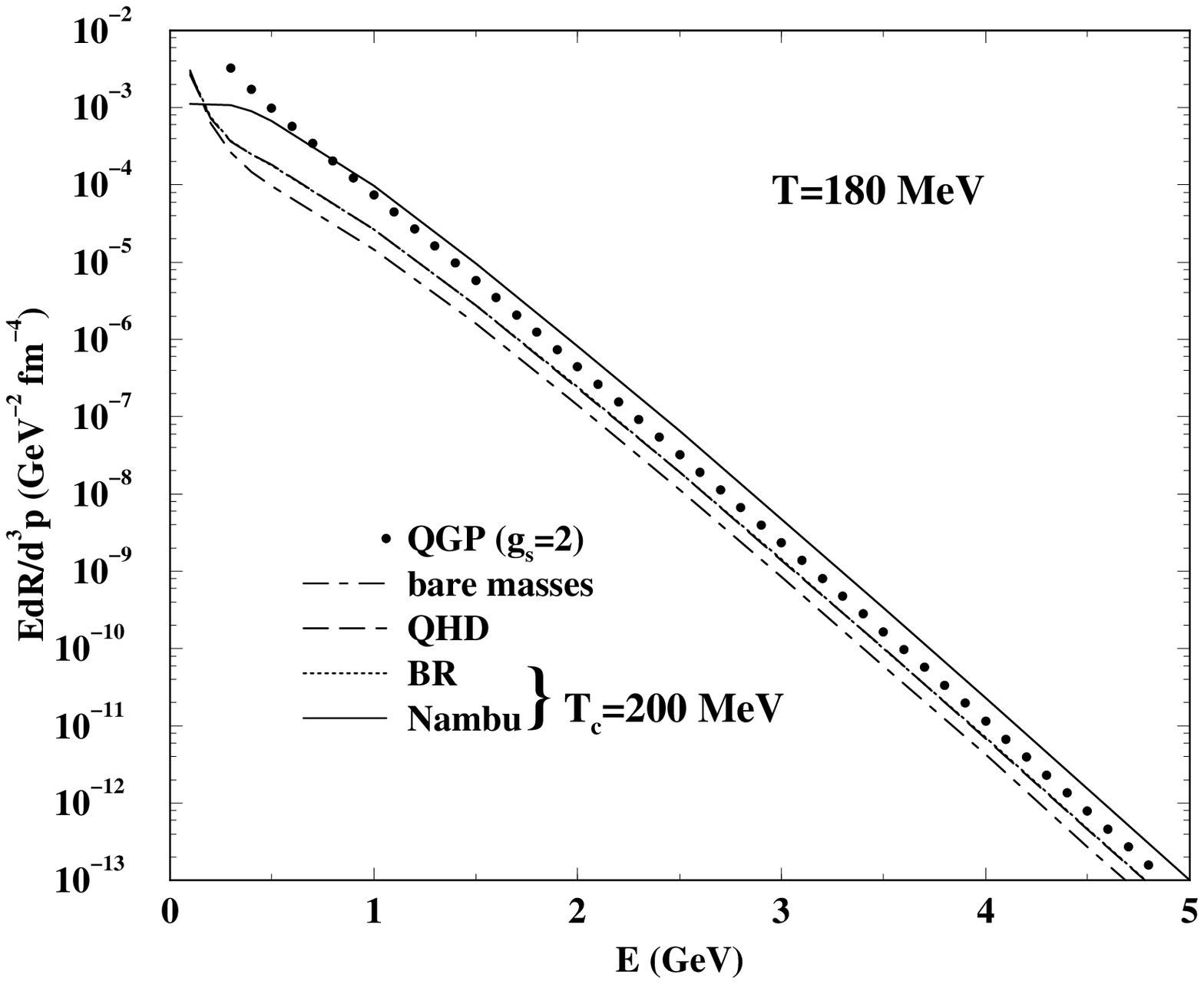,height=7cm,width=9cm}}
	\caption{
	Thermal photon spectra at $T=180$ MeV and $g_s=2$.
	}
	\label{prate180}
	\eef
	The photon production rate from QGP has been
	evaluated in the HTL approximation, which is valid if
	the hard and soft scales are well separated, {\it i.e.}
	for $g_s<<1$ (which corresponds to $\alpha_s<<0.08$). 
	However, lattice QCD calculations~\cite{karsch}
	suggest that $\alpha_s\sim 0.2-0.3$
	at the temperatures achievable in URHICs.  We emphasize
	that the extrapolation of  results obtained under HTL approximation 
	to higher values of $g_s$ (or $\alpha_s$) may be dubious. 
	We have evaluated the photon spectra for two values of the strong coupling 
	constants $g_s=0.8$ (thick squares) and $2$ (thick dots)
	to demonstrate the sensitivity
	of the photon spectra to the value of the strong charge
	and to show the uncertainties involved in the problem.
	In the hadronic sector the photon
	yield is seen to be  enhanced 
	in comparison to the case when the effects of the thermal interaction
	on the hadronic properties are neglected. 
	This is true for almost the
	entire energy range of the emitted photon under consideration. 
	As a result of the similar mass shift in QHD and BR scaling
	the photon spectra in these two scenarios (long-dashed and dotted lines 
	respectively) have a negligible difference,
	whereas the enhancement in the spectrum due to hadronic mass shift
	according to Nambu scaling is clearly visible (solid line).
	In Fig.~(\ref{prate180}) we show the photon emission rate at $T=180$ MeV.
	Photon spectra from hadronic matter with mass variation
	according to the Nambu scaling scenario overshine the 
	photons from QGP even for a larger value of $g_s$ ($\sim 2$).

The photon spectra  at $T=$ 150 MeV with
in-medium masses calculated in the framework of 
gauged LSM, NLSM and HLS approaches is shown in Fig.~(\ref{phorate_oth}). 
An increase of $\rho$ mass in the NLSM reduces its number due to 
Boltzmann suppression which leads to a suppression in the photon emission rate
(dotted line). The production rate is enhanced due to a reduction in the
$\rho$ mass (solid line) in LSM. 
(We recollect that the $\rho$ mass decreases in gauged
LSM for low temperature and increases for temperatures close to the chiral
transition temperature $T_\chi$. Therefore, for $T\sim T_\chi$ we will 
observe a reduction in
photon emission rate and the net yield would be a superposition of
all temperatures, from initial to freeze-out.) The change in the 
mass of $\rho$ is so small in the  HLS approach (short-dashed line) 
that the production rate is 
almost indistinguishable
from the spectra with vacuum masses of the hadrons.
We have also demonstrated 
in Fig.~(\ref{phorate_oth})
how the photon spectra
is modified for a drastic change in the width of the $\rho$
meson ($\Gamma_\rho\sim 400$ MeV)
without any appreciable change 
in the pole mass ($m_\rho\sim 770$ MeV). 
Such a large value of the width
have been proposed in Refs.~\cite{rcw,ek}. 
We observe
that the effects of such modifications in the properties
of $\rho$ on the photon spectra is rather negligible (long-dashed line). 
\bef
\centerline{\psfig{figure=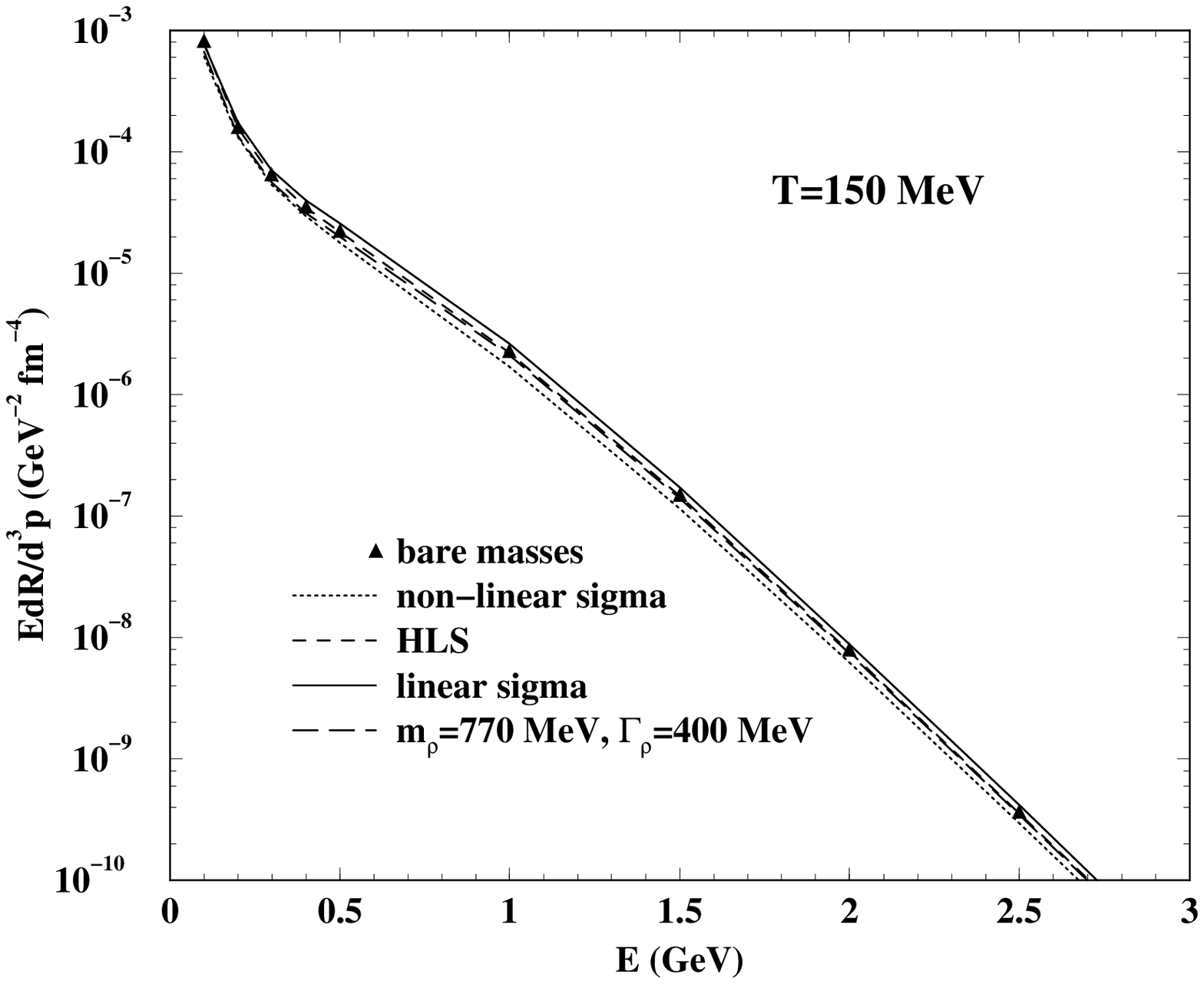,height=7cm,width=9cm}}
\caption{ The change in the photon spectra due to the finite temperature
effects on the hadronic masses in Linear, Non-Linear Sigma Model and
Hidden Local Symmetry approach at T= 150 MeV.
}
\label{phorate_oth}
\eef

So, from the above discussions we can infer that the relative photon yields from
QGP and hot hadron gas depends on:
(i) the value of the strong 
	coupling constant, (ii) the degree of hotness of the medium and (iii) how 
	adversely the hadrons are affected in the medium.  

We will now focus on a few other aspects that might affect the thermal photon emission
from hot hadronic matter.  

The first is concerned with the fact that most of the photon producing
reactions which we have considered contain unstable particles ($\rho$
and $\omega$) in the external lines. 
Now, we know that the density of a stable hadron of mass $m$ in a thermal
bath is completely
determined by the temperature, chemical potential and the statistics
obeyed by the species through the distribution function
\be
\frac{dN}{d^3x\,d^3k\,ds}=\frac{\cal N}{(2\pi)^3}
\frac{1}{\exp(k_0-\mu)/T\pm\,1}\,\delta(s-m^2)
\label{stable}
\ee
where ${\cal N}$ is the degeneracy and $k_0=\sqrt{\vec k^2+s}$ is the
energy of the particle in the rest frame of the thermal bath.
To account for ``particles'' like the $\rho$ and $\omega$ mesons
whose life-times are such that they can decay within the thermal
system, the above expression is replaced by~\cite{npa99}
\be
\frac{dN}{d^3x\,d^3k\,ds}=\frac{\cal N}{(2\pi)^3}
\frac{1}{\exp(k_0-\mu)/T\pm\,1}\,P(s)
\label{unstable}
\ee
where $P(s)$ is the spectral function of the species under consideration and
is given by
\be
P(s)=\frac{1}{\pi}\frac{{\s Im}\Pi}{(s-m^2+{\s Re}\Pi)^2+({\s Im}\Pi)^2}.
\ee
[Note that $P(s)$ reduces to the Dirac delta function in the limit ${\s Im}\Pi\ra 0$.]
The above exercise has been carried out wherever the $\rho$ or $\omega$ 
appears in the photon producing reactions in the external lines. 
In the case these particles appear in the internal lines, the propagators
have been suitably modified, as discussed in the previous Section.
The limits of the $s$ integration in the case of different reaction 
channels have been determined from kinematical considerations. 
The results for the reactions $\pi\,\pi\,\ra\,\rho\,\gamma$, $\pi\,\pi\,\ra\,
\pi\,\gamma$, $\rho\,\ra\,\pi\,\pi\,\gamma$ and $\omega\,\ra\,\pi\,\gamma$ have
been shown in Fig.~(\ref{fold1}). The difference caused by the inclusion of the
spectral function is observed to be $\sim$ a few percent.
This is reflected in the total photon spectra as shown in Fig.~(\ref{fold2}).
The effective mass and decay widths have been considered in QHD for
the sake of illustration.
\bef
\centerline{\psfig{figure=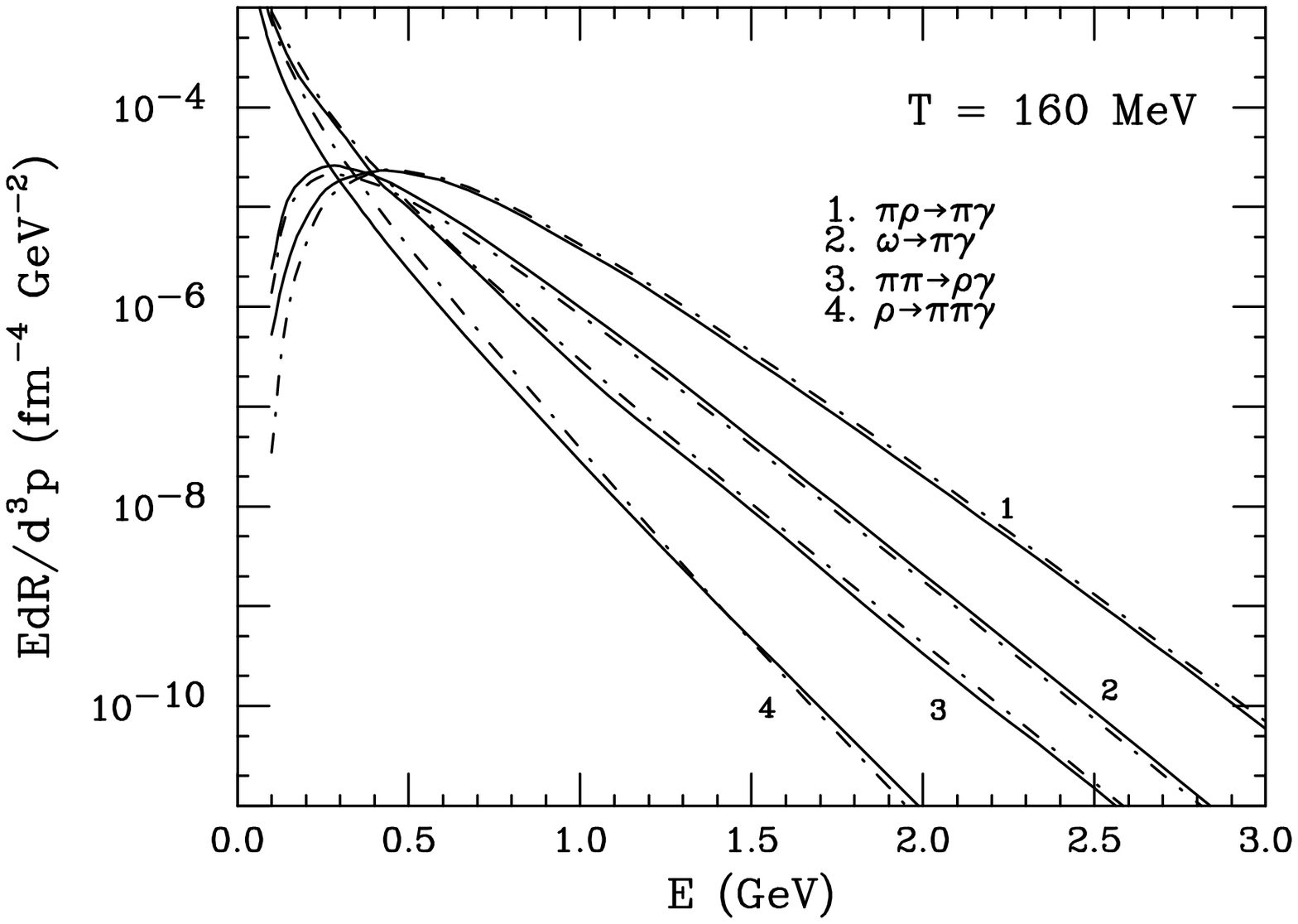,height=7cm,width=10cm}}
\caption{Effect of spectral function of vector mesons on photon emission
rates at $T$ = 160 MeV. Solid (dot-dashed) line shows
results with (without) the inclusion of spectral function.
}
\label{fold1}
\eef
\bef
\centerline{\psfig{figure=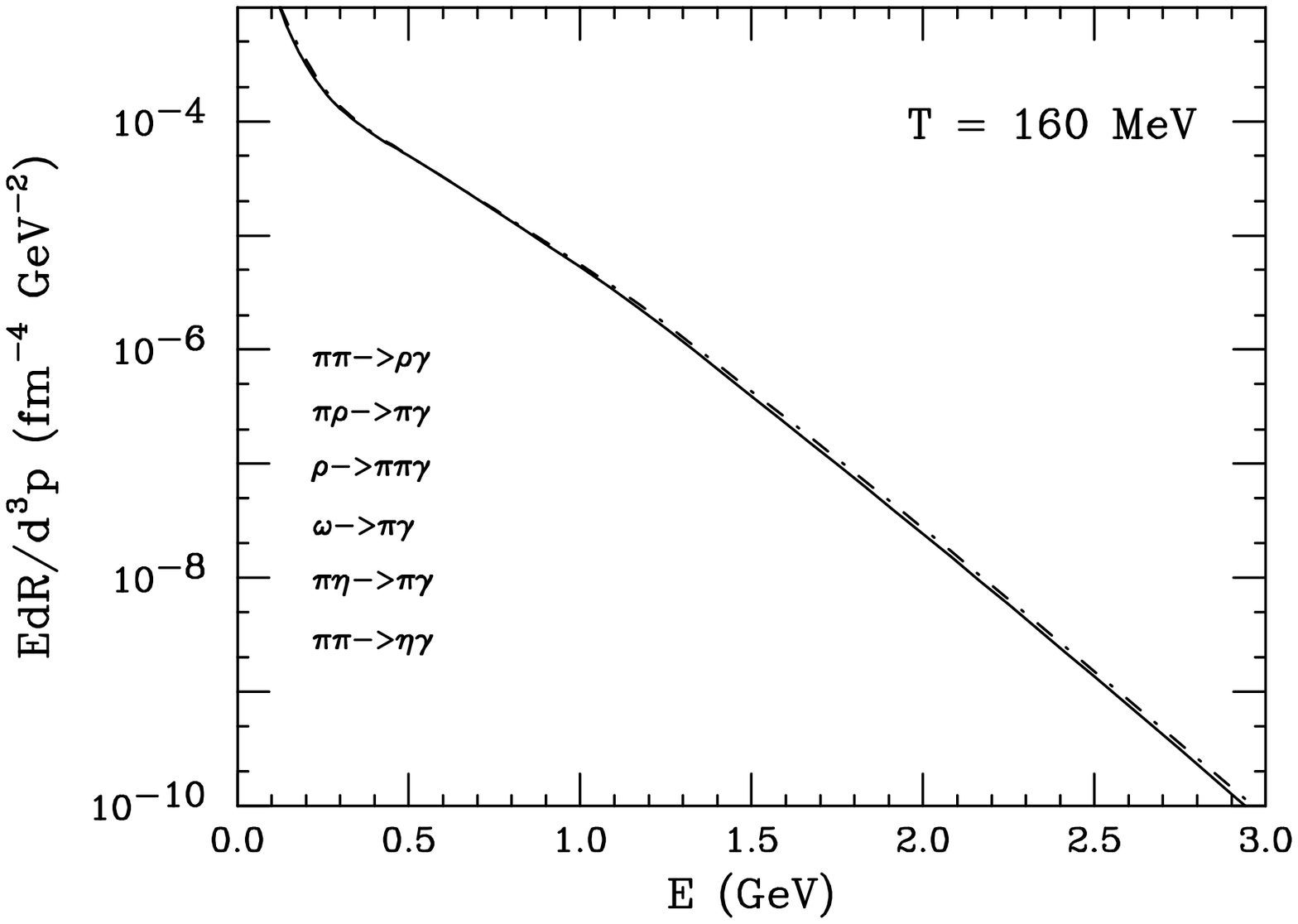,height=6.5cm,width=10cm}}
\caption{Total photon emission
rate at $T$ = 160 MeV.
}
\label{fold2}
\eef

The second point to consider is that though we have been treating
the hadrons ($\rho$, $\omega$ etc.) as point particles, in reality
they are composite objects and may need form factors at high momentum 
transfer. We have investigated how the inclusion of form factors influence
the results. Using the reaction $\pi\pi\rightarrow \rho\gamma$ this has been
demonstrated in Fig.~(\ref{prateform}).
	We have taken the same monopole form factor for both $\pi\pi\rho$
	and $\pi\pi\gamma$ vertices~\cite{kapusta} to suppress the contribution from
	very high momentum region where the quark structure of the hadrons
	could be relevant. The Ward-Takahashi identity has been used to obtain   
	the dressed propagator. The in-medium mass of the $\rho$ meson
	has been taken from QHD. The form factor effects 
	for the above reaction reduces the photon production rate by about
	10-15\% at $T=$180 MeV. Once the space-time evolution is carried out
        such an effect will turn out to be negligible. We have therefore neglected
        it in our discussions. 
	\bef[h]
	\centerline{\psfig{figure=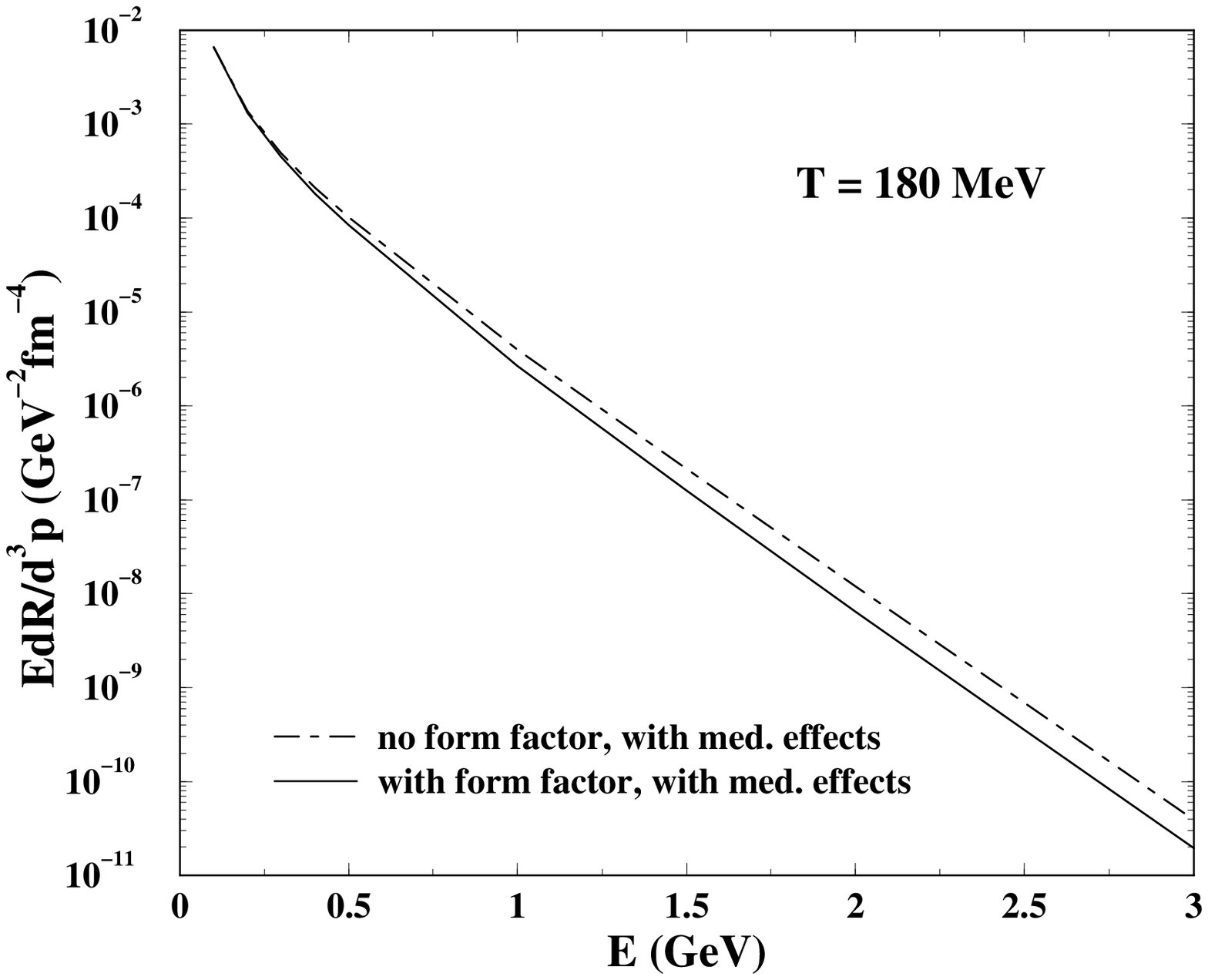,height=6.5cm,width=9.5cm}}
	\caption{
	The effect of the monopole form factor on the photon emission rate
	from the reaction $\pi\pi\rightarrow \rho\gamma$.
	}
	\label{prateform}
	\eef

\vskip 0.2in
\noindent{\bf Decay Photons}
\vskip 0.2in

Photons from $\pi^0$ and
$\eta$ decays constitute almost the entire background to the spectra of
thermal photons in the low transverse momentum region. We will briefly
discuss how the transverse momentum spectra of decay photons can be estimated.
One has to first obtain the
spectra of the $\pi^0$ and $\eta$ mesons at the time of freeze-out. For the
case of  no transverse flow the hadronic spectra is given by
\begin{equation}
\frac{dN_h}{d^2p_T\,dy} =\pi\,R_T^2\,\tau_f\,\frac{\cal N}{4\pi^3}\,m_T\,
\sum_{n=1}^{\infty}(\pm1)^{n+1}\,K_1\left(n\frac{m_T}{T_f}\right)~,
\end{equation}
where, ${\cal N}$ is the degeneracy of the particles, $\tau_f$ is the freeze-out
time of the hadron and
the plus and minus signs are for bosons and fermions respectively.
Additionally, $m_T=\sqrt{p_T^2+m^2}$, where $m$ is mass of the hadron
and $p_T$ is its transverse momentum.
It is worthwhile to mention that the
transverse expansion of the system may considerably enhance the transverse
momentum of the the particles at the time of freeze-out. 
In any case, the
decay photons have to be subtracted out by an invariant mass analysis, 
for the thermal photons to  be identified.
The decay photons from the process $\pi^0\,\rightarrow\,\gamma\gamma$,
for example are then obtained as~\cite{cahn},
\begin{equation}
\frac{dN}{d^2k_T\,dy}=\int\frac{d^3p}{E}\left(E\frac{dN_\pi}{d^3p}\right)
\frac{1}{\pi}\delta\left(p \cdot k -\frac{1}{2}m^2\right)~,
\end{equation}
where $p$ is the four-momentum of the $\pi^0$, and $k$ is the momentum of the
detected photon.
After some algebra, this reduces to,
\begin{equation}
\frac{dN}{d^2k_T\,dy}=\frac{1}{\pi\,k}\int^\infty_{E_0}\,dE
\int_0^{2\pi}\,d\phi\, E\,\frac{dN_\pi}{d^3p},
\end{equation}
where,
\begin{eqnarray}
p_T^2&=&\left[\left(\frac{m^2}{k}(E-E_0)\right)^{1/2}\frac{k_\parallel}{k}\cos
\phi +\left(E-\frac{m^2}{2k}\right)\frac{k_T}{k}\right]^2
+\frac{m^2}{k}(E-E_0)\sin^2\phi~~,\nonumber\\
k_\parallel&=&k_T\, \sinh y~~,\nonumber\\
E&=&m_T\,\cosh y_\pi~~, \nonumber\\
E_0&=&k+\frac{m^2}{k}~~.
\end{eqnarray}

\section{Dilepton Emission Rates from Hot Hadronic Matter}

In order to discuss dilepton emission from hadronic matter we
will use 
Eq.~(\ref{drretcor}) which expresses the emission rate
in terms of the retarded current correlation function. 
\[
\frac{dR}{d^4q}=-\frac{\alpha}{12\pi^4\,q^2}(1+\frac{2m^2}{q^2})
\sqrt{1-\frac{4m^2}{q^2}}{\s Im}W^R_{\mu\mu}\,f_{BE}(q_0).
\label{drcor1}
\]
Putting $d^4q=2\pi\,MdM\,M_TdM_T\,dy$ this is written as
\be
\frac{dR}{dM}=-\frac{\alpha}{6\pi^3}\frac{1}{M}
\left(1+\frac{2m^2}{M^2}\right)\sqrt{1-\frac{4m^2}{M^2}}
\,\int \frac{M_TdM_Tdy}{ e^{M_T\cosh y/T}-1}
\,\,{\s Im}W^{R}_{\mu\mu}.
\label{drcor_bol}
\ee
The parametrized form of the electromagnetic current correlation function 
in the $\rho$ and $\omega$ channels as discussed in Section~3.1.3c
can be used in the above equation to obtain the dilepton emission rate from
a thermal hadronic medium.
However, instead of using the current correlation function directly in the above
equation one can use vector meson dominance (VMD) to obtain the dilepton
yield from $\pi^+\pi^-\,\ra\,l^+l^-$ which is known to be the most dominant
source of dilepton production. 
In order to make a comparative study 
we state briefly how the dilepton emission rate from pion annihilation can be
derived using Eq.~(\ref{drcor_bol}). Recall that VMD relates the hadronic electromagnetic
current to the vector meson field through the field current identity as
\be
J^h_\mu=-\sum_{V}\, \frac{e}{g_V}m_V^2 V_\mu
\ee
where, $V=\rho,\omega,\phi$ and $m_V$ stands for the bare masses. We shall keep only the $\rho$ meson in the following.
The electromagnetic current correlator can then be expressed in terms of
the propagator of the vector particle in the following way:
\be
{\s Im}W_{\mn}^R=-\frac{e^2m_{\rho}^4}{g_\rho^2}{\s Im}D_{\mn}^{\rho R}
\label{cor-prop}
\ee
where
\bea
{\s Im}D_{\mn}^{\rho R}& = &A_{\mn}\left[\frac{{\s Im}\Pi_T^{\rho R}}{(q^2-m_\rho^2+{\s Re}
\Pi_T^{\rho R})^2+[{\s Im}\Pi_T^{\rho R}]^2}\right]\nonumber\\
&&+B_{\mn}\left[\frac{{\s Im}\Pi_L^{\rho R}}{(q^2-m_\rho^2+{\s Re}
\Pi_L^{\rho R})^2+[{\s Im}\Pi_L^{\rho R}]^2}\right].
\eea
The complete expression for the dilepton emission rate due to pion 
annihilation is then obtained from Eq.~(\ref{drcor_bol}) as
\bea
\frac{dR}{dM}&=&\frac{2\alpha^2}{\pi^2}\frac{m_\rho^4}{g_\rho^2}\frac{1}{M}
\left(1+\frac{2m^2}{M^2}\right)\sqrt{1-\frac{4m^2}{M^2}}\,
\int \frac{M_TdM_Tdy}{ e^{M_T\cosh y/T}-1}\times\nonumber\\
&&\left[\frac{2{\s Im}\Pi^{\rho R}_T}{(M^2-m_\rho^2+{\s Re}\Pi^{\rho R}_T)^2+
[{\s Im}\Pi^{\rho R}_T]^2}+
\frac{{\s Im}\Pi^{\rho R}_L}{(M^2-m_\rho^2+{\s Re}\Pi^{\rho R}_L)^2+
[{\s Im}\Pi^{\rho R}_L]^2}\right].\nonumber\\
\eea
The quantities ${\s Im}\Pi^{\rho R}_{T,L}$ can be obtained from Eq.~(\ref{fullpirho}).
It is however found that the difference
between the longitudinal and transverse
polarizations is very small upto
reasonably high temperatures~\cite{gale} for the interaction considered here.
We hence make the approximation $\Pi_T^{\rho R}=\Pi_L^{\rho R}=\Pi^{\rho R}$, 
so that 
\be
{\s Im}W_{\mu\mu}^R=-\frac{3e^2 m_\rho^4}{g_\rho^2}
\left[\frac{{\s Im}\Pi^{\rho R}}{(q^2-m_\rho^2+{\s Re}
\Pi^{\rho R})^2+[{\s Im}\Pi^{\rho R}]^2}\right].
\label{walcor}
\ee
Writing ${\s Im}\Pi^{\rho R}=M\Gamma^\rho(M)$ and neglecting the 
thermal factor in the $\rho$ width one obtains
\be
\frac{dR}{dM} = \frac{\sigma(M)}{(2\pi)^4}\,M^4\,T\,
\,K_1(M/T)\,(1-4m_\pi^2/M^2),
\label{pipi0}
\ee
in the Boltzmann approximation.
$K_1$ is the modified Bessel function and
$\sigma(M)$ is the cross-section for pion 
annihilation given by
\be
\sigma(M) = \frac{4\,\pi\,\alpha^2}{3\,M^2}\,\sqrt{1-4m_\pi^2/M^2}\,
\sqrt{1-4m^2/M^2}\,(1+2m^2/M^2)\,|F_\pi(M)|^2,
\ee
with
\be
|F_\pi(M)|^2=\frac{m_\rho^4}
{(M^2-m_\rho^2+{\s Re}\Pi^{\rho R})^2+
M^2\Gamma_\rho(M)^2}
\ee
which is known as the pion form factor. Apart from the thermal modification of the 
$\rho$ meson properties, the rate given by Eq.~(\ref{pipi0}) is 
equivalently derivable from the kinetic theory result given by Eq.~(\ref{dilrate1}).

One should also add 
the contributions to the thermal dilepton yield coming from the decay of 
vector mesons. 
In a thermal medium the production
of an off-shell vector meson ($V$)  of four momentum 
$q$ (where $q^2=M^2$) and its subsequent
decay into a lepton pair leads to the dilepton emission rate~\cite{weldon93}
\be
dR=\frac{2M}{(2\pi)^3}\rho^V_{\mn}\,P^{\mn}\,f_{BE}(q_0)
\Gamma^{\s vac}_{V\,\ra\,l^+l^-}\,d^4q,
\ee
where $P_{\mn}=\sum_{pol}\,\epsilon_\mu\,\epsilon_\nu^\ast=-g_{\mn}+q_\mu q_\nu/q^2$
is the projection operator as applicable for an unstable vector meson $V$, and
$\Gamma^{\s vac}_{V\,\ra\,l^+l^-}$ is the partial decay width for the process
$V\,\ra\,l^+l^-$ in vacuum.
$\rho^V_{\mn}$ is the spectral function of the off-shell vector 
meson and is given by
\be
\rho^V_{\mn}=\frac{1}{\pi}\frac{{\s Im}\Pi_V^R}{(q^2-m_V^2+{\s Re}\Pi_V^R)^2+
({\s Im}\Pi_V^R)^2}\,P_{\mn}.
\ee
in the approximation $\Pi_V^{L R}=\Pi_V^{T R}=\Pi_V^R$. 
Using the relation 
$P^{\mn}\,P_{\mn}=(2J+1)$, we get the dilepton emission rate due to the
decay of an unstable vector meson of spin $J$ as 
\be
\frac{dR}{d^4q}=2\frac{(2J+1)}{(2\pi)^3}\,f_{BE}\,M\Gamma^{\s vac}_{V\,\ra\,l^+l^-}\,
\left[\frac{1}{\pi}\frac{{\s Im}\Pi_V^R}{(q^2-m_V^2+{\s Re}\Pi_V^R)^2+({\s Im}\Pi_V^R)^2}
\right].
\label{wel1}
\ee
Note that for a particle which does not decay in the collision volume {\it i.e.}
for which the total width $\Gamma_{\s tot}={\s Im}\Pi^R/M$ is small, the spectral
function in the above equation  becomes
$\delta(q^2-m_V^2)$, as it should be for a stable particle. 
As discussed before, in a  medium the width $\Gamma_{\s tot}$ 
of particle $V$ should be calculated with all the processes involving the creation
and annihilation of $V$, {\it i.e.} $\Gamma_{\s tot}=\Gamma_{V\ra\,{\s all}}-
\Gamma_{{\s all}\ra\,V}$~\cite{weldon93}.

Using similar approximations as in the case of 
pion annihilation, the invariant mass distribution of 
lepton pairs from the vector meson decays 
is obtained from Eq.~(\ref{wel1}) as,
\bea
\frac{dR}{dM}&=&\frac{2J+1}{\pi^2}\,
M^2T\,K_1(M/T)\nonumber\\
&&\times\,\frac{M\Gamma_{{\s tot}}
/\pi}{(M^2-m_V^2+{\s Re}\Pi^R_V)^2+M^2\Gamma_{{\s tot}}^2}
M\Gamma_{V\,\ra\,l^+\,l^-}^{{\s vac}}
\label{decay0},
\eea
with 
\be
\Gamma_{V\,\ra\,l^+\,l^-}^{{\s vac}}=
\frac{4\pi\alpha^2}{3g_\rho^2}\frac{m_V^4}{M^3}
\sqrt{1-4m^2/M^2}\,(1+2m^2/M^2)
\ee
where $m$ is the mass of the lepton.

We will now discuss the results of static dilepton emission rates.
The rates for BR and Nambu scaling scenarios are obtained by putting
$e^2$ times Eqs.~(\ref{parametrho}) and (\ref{parametomg}) 
in Eq.~(\ref{rcorr}) which is then inserted in Eq.~(\ref{drcor_bol})
using Eq.~(\ref{cormm}). The in-medium masses in this case is obtained from
Eq.~(\ref{anst}) with $\lambda$=1/6(1/2) for BR(Nambu) scaling. The dilepton rate in the
QHD model is obtained by adding the contributions from pion annihilation (Eq.~(\ref{pipi0})) 
and from $\omega$ decay using Eq.~(\ref{decay0}).
The effective masses of 
the $\rho$ and $\omega$ in this case are as given in Eq.~(\ref{pmasswal}).
It may be mentioned here that since we are considering dilepton production
from pion annihilation with a $\rho$ dominated form factor,
adding the contribution from $\rho$ decay to lepton pairs 
(which also contains the width due to $\rho\leftrightarrow\pi\pi$) separately
will induce a double counting.

In Fig.~(\ref{drate150}) we display the invariant mass distribution of
$e^+e^-$ pairs. The dilepton yield from $q\bar{q}$ annihilation is denoted by 
thick dots. The dotted line indicates the result obtained from the 
parametrization of the electromagnetic current correlation function 
in the `$\rho$' and `$\omega$' channels, when the medium effects are ignored.
A large shift towards the lower invariant mass region  of the $\rho$ peak
is seen in the Nambu scaling (solid line) as compared to the BR scaling
(dashed line) consistent with the relative shift in the spectral functions in the
two cases as discussed before. In the QHD model calculations 
(dot-dash line) the two peaks corresponding to
$\rho$ and $\omega$ masses are visible in the spectra~\cite{ja,pr}.
The separation between the two peaks is due to different mass shift of the $\rho$ and
$\omega$. 
\bef
\centerline{\psfig{figure=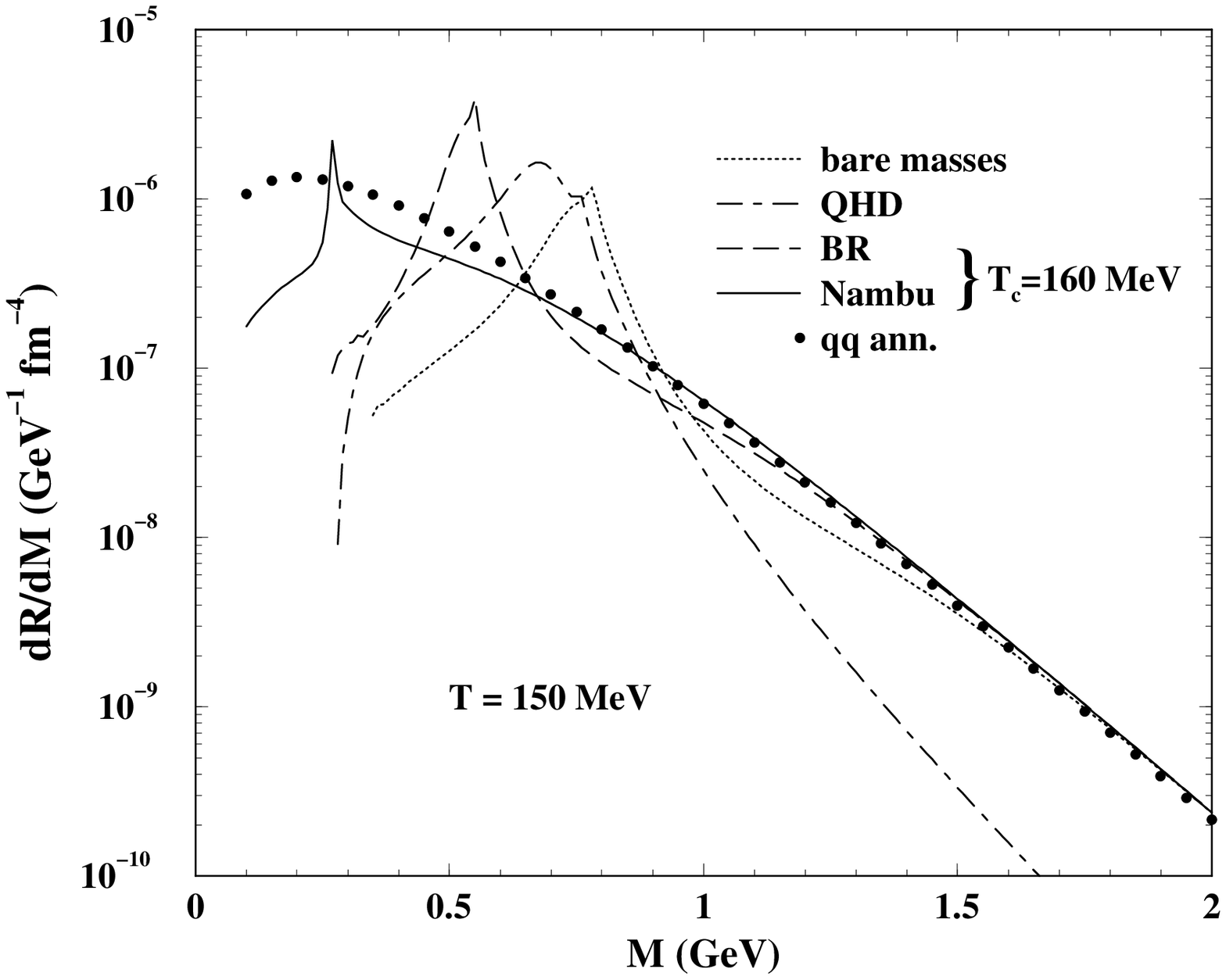,height=6cm,width=9cm}}
\caption{
Thermal dilepton spectra at $T=150$ MeV. Thick dots indicate
dilepton emission rate from QGP.
Thin dotted line represents dilepton yield 
from hot hadronic gas without medium effects. 
The result with the in-medium 
effects in the QHD model is shown by the dot-dashed line. 
Long-dashed (solid) line indicates
dilepton spectrum with BR (Nambu) scaling mass variation scenario.
}
\label{drate150}
\eef
Measurement of such separation in hadronic masses 
($\Delta m=m_{\omega}^*-m_{\rho}^*$)
would signal the in-medium effects. The validity of such results could be 
tested in URHICs by the CERES~\cite{CERES,itzhak} collaboration in future. 
Similar shift
at zero temperature but finite baryon density could be detected 
by HADES~\cite{hades} and CEBAF~\cite{book}.  
Effects of the continuum contribution in the spectral function on the dilepton
spectra is clearly visible for $M\geq 1$ GeV since the value
of the continuum threshold in vacuum is 1.3 GeV. 
Due to the continuum contribution the dilepton rates from hadronic matter
and QGP shine equally brightly in the mass range $M\geq$ 1 GeV.
The lepton pair spectra at $T=180$ MeV is shown in Fig.~(\ref{drate180}).
Since the effective mass of the $\rho$ in QHD and BR scaling scenario
is almost same in this case (see Fig.~(\ref{massall})), the corresponding rates
are very similar near the $\rho$ peak.
\bef
\centerline{\psfig{figure=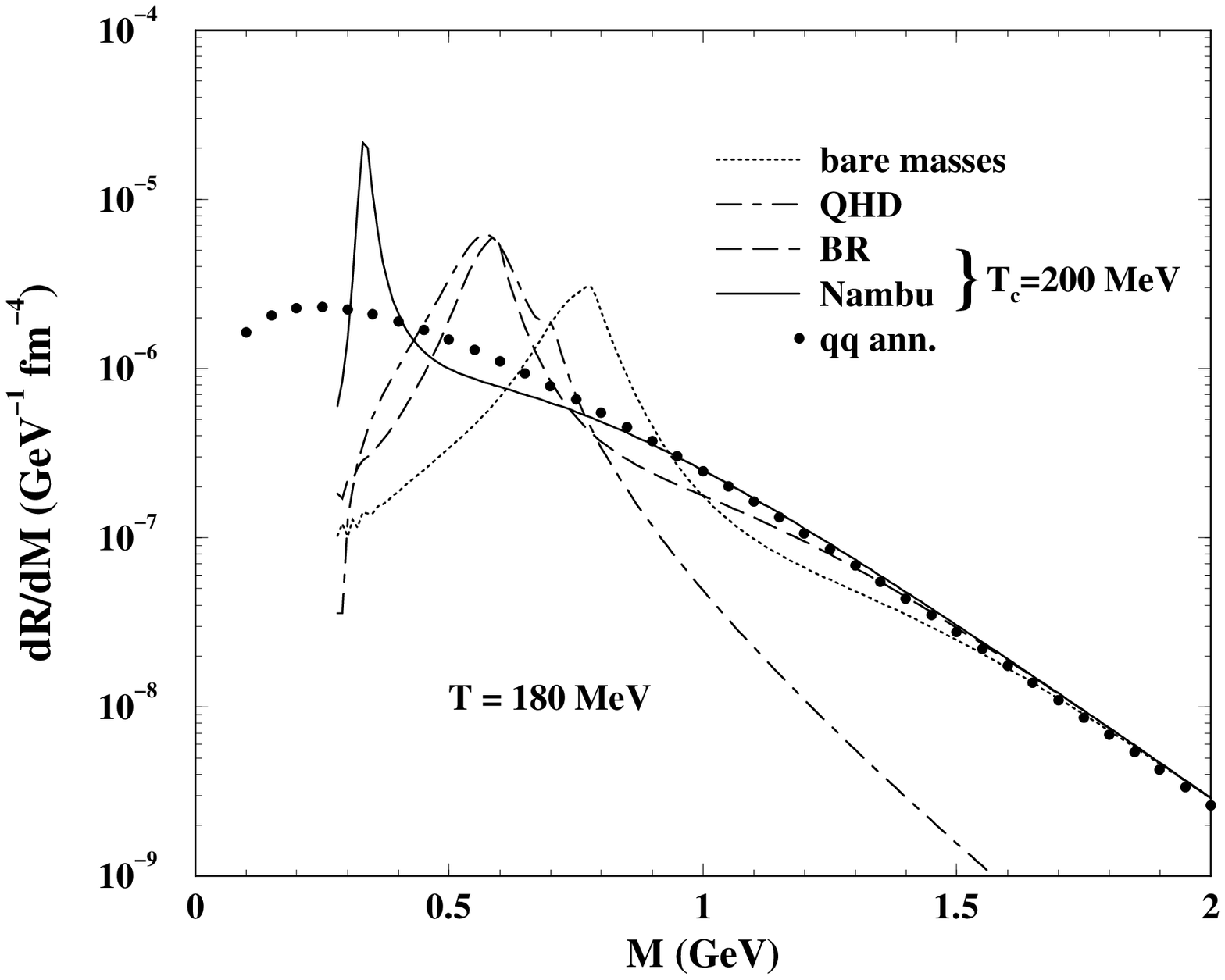,height=6cm,width=9cm}}
\caption{
Thermal dilepton spectra at $T=180$ MeV. 
}
\label{drate180}
\eef

The dilepton invariant mass distribution at $T=T_c$ is shown in 
Fig.~(\ref{dratetc}). All the peaks in the spectrum have disappeared
as expected. The rates obtained from the electromagnetic current
correlator is close to the rate from $q\bar{q}$ annihilation, 
indicating that the $q\bar{q}$ interaction in the vector channel
has become very weak {\it i.e.} signalling the onset of deconfinement 
~\cite{shuryakqm99,rappqm99}.
\bef
\centerline{\psfig{figure=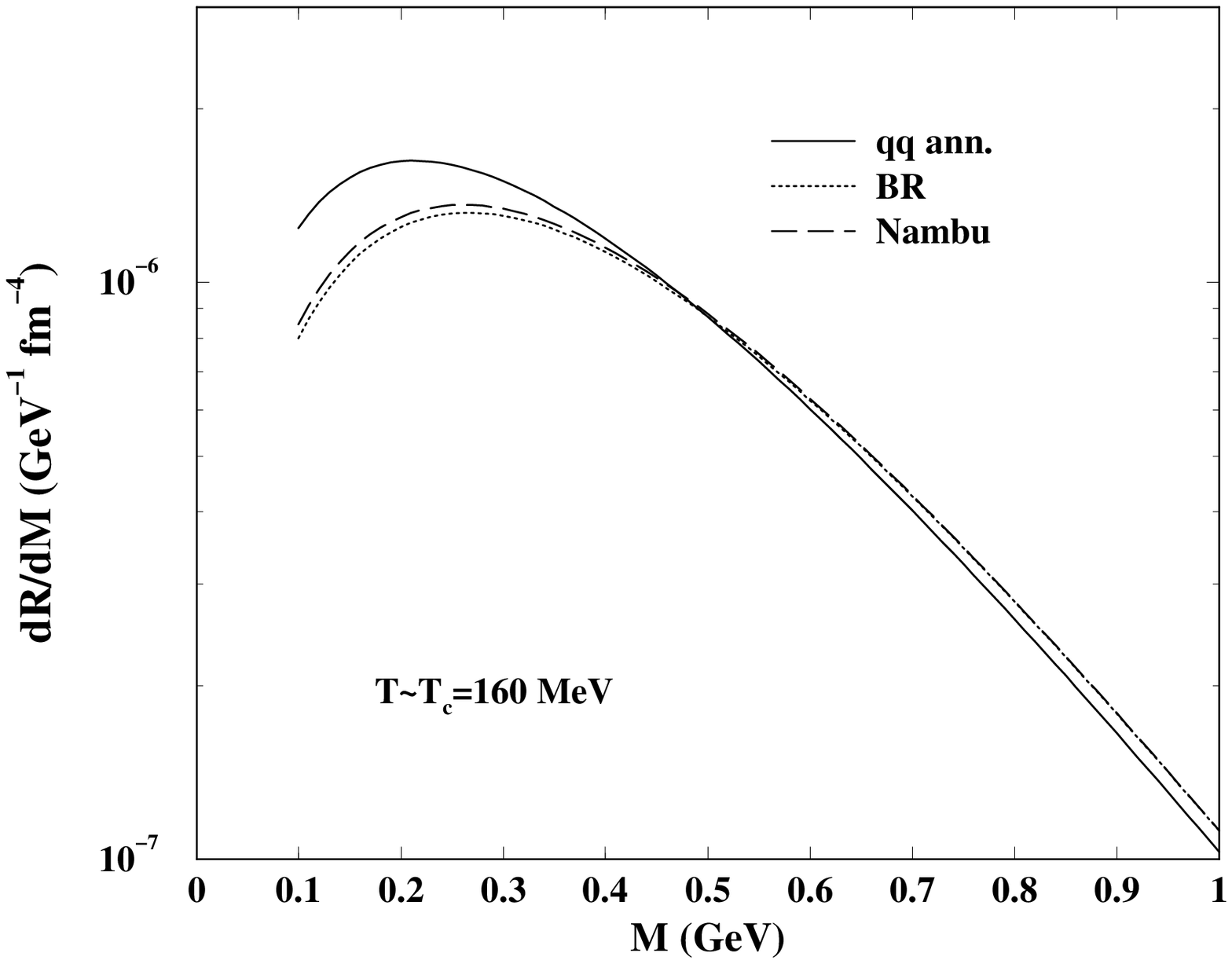,height=6cm,width=9cm}}
\caption{
Thermal dilepton spectra at $T=T_c$. 
The vector meson peaks have merged with the continuum and the rates
from QGP and hadronic matter have almost become equal. 
}
\label{dratetc}
\eef
\bef
\centerline{\psfig{figure=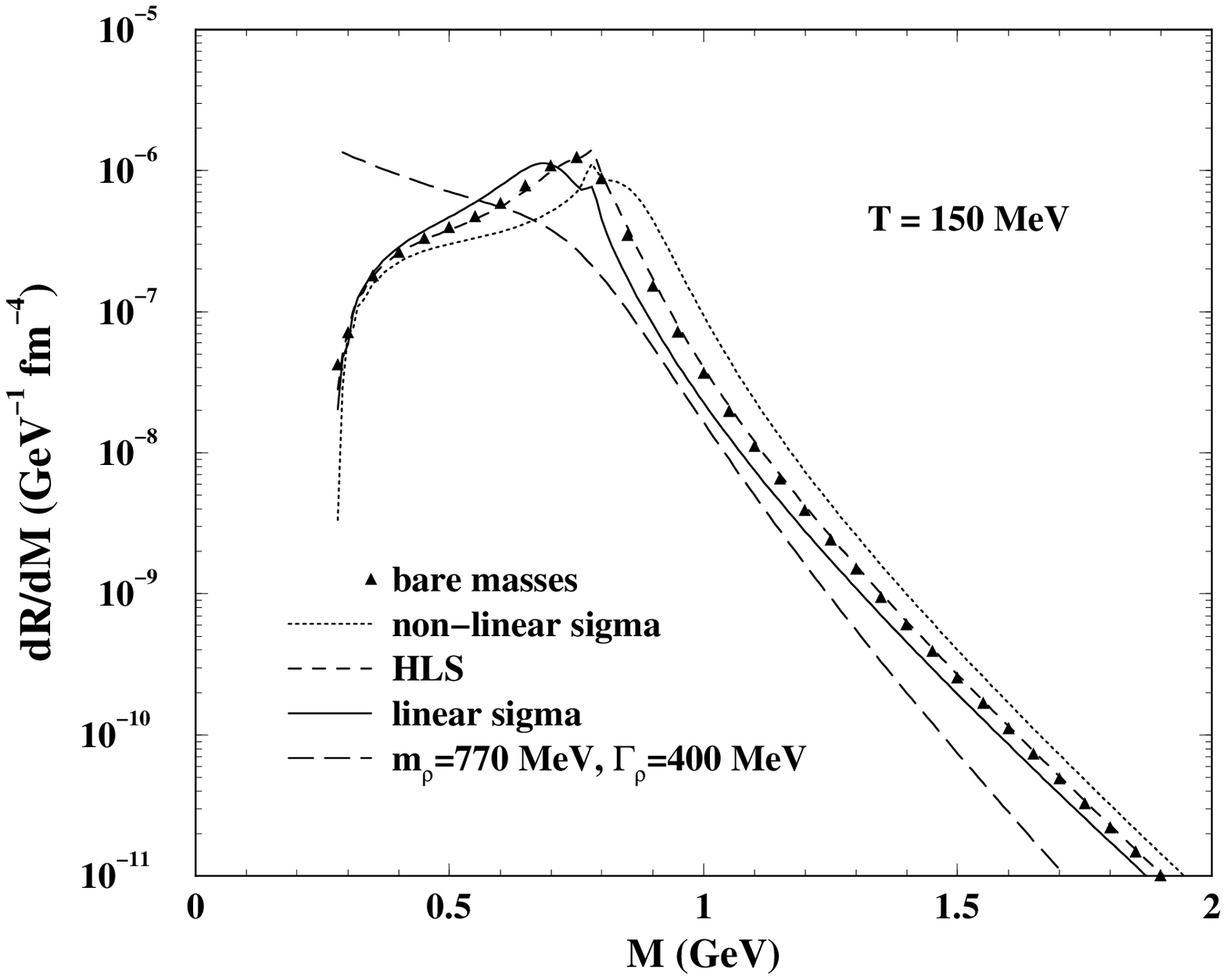,height=6cm,width=9cm}}
\caption{ The change in the invariant mass distribution 
of lepton pairs due to the finite temperature
effects on the hadronic masses in Gauged Linear, Non-Linear Sigma Model and
Hidden Local Symmetry approach at T= 150 MeV. The long-dashed line
indicates dilepton spectra for $\Gamma_\rho=400$ MeV and $m_\rho=770$
MeV (see text).
}
\label{dilrate_oth}
\eef

In Fig.~(\ref{dilrate_oth}) we compare the dilepton emission rate
at $T=150$ MeV for vacuum mass of $\rho$ with the rates in which the effective
masses are obtained in the framework of gauged LSM, NLSM and HLS approach.
The positive shift of $\rho$ mass in NLSM is reflected in the peak position  
of the spectra towards larger value of $M$ (dotted line). A very small 
change in the $\rho$ mass in HLS approach does not cause any visible
change in the invariant mass distribution of dileptons.
The long-dashed line indicates lepton pair distribution for $m_\rho=770$ MeV
and $\Gamma_\rho=400$ MeV. The large width of $\rho$ leads to the
disappearance of the $\rho$ peak from the spectra, indicating that $\rho$
ceases to exist as a quasi-particle. However, as mentioned before, it is
interesting to note that the photon spectra is insensitive to such
drastic broadening of the $\rho$ meson.

\chapter{Electromagnetic Spectra with Space-Time Evolution}

Matter produced in highly relativistic collisions of heavy ions
can be either in the form
of a hot hadronic gas or a quark gluon plasma.
So far we have discussed the rate of photon and dilepton  emission 
per unit time
from unit volume from a thermal system made up of quark matter or  hadronic matter
at a fixed temperature. However, the highly excited state of matter produced at
a high temperature will expand and cool emitting photons and dileptons in
the process. This process will continue as long as the mean free paths of the
constituents become comparable to the size of the system and freeze-out occurs.
So, in order to compare the yield with experiments we must integrate the
static rates over the space time volume of the collision from formation till
freeze-out. The total yield is hence
\be
\frac{dN}{d\Gamma}=\begin{array}{c}
{\small {freeze-out}}\\{\displaystyle\int}\\{\small {formation}}\end{array}\,
d^4x\frac{dR(E^\ast,T(x))}{d\Gamma}
\label{sptime}
\ee 
where $d\Gamma$ stands for invariant phase space elements:
$d^3q/E$ for photons and $d^4q$ for dileptons.
$T(x)$ is the local temperature, assuming that the produced
matter is in local thermal equilibrium. $x$ is the space-time coordinate
and $d^4x=d{\cal V}dt$ where ${\cal V}$ is the three-volume.
Since we are dealing with relativistic situations one must account for the fact that the thermal rates
are evaluated in the rest frame of the emitting matter and hence the momenta of the
emitted photons or dileptons are expressed in that frame.  Therefore,
the invariant rate
is a function of 
$E^\ast$, the energy of the photon or lepton pair 
in the rest frame of the emitting matter which in this case 
is the fluid element.
In a fixed frame like the laboratory or the centre of mass frame, where
the 4-momentum of the photon or lepton pair is $q_\mu=(E,\vec q)$ 
and the emitting matter element $d^3x$ moves with a velocity
$u_\mu=\gamma(1,\vec v)$, we have $E^\ast=u_\mu q^\mu$. 
As is evident from
Eq.~(\ref{sptime}) the key ingredients that we now need in order to
carry through our program are firstly, the
initial conditions {\it viz.} the initial time and the initial temperature,
and secondly, the temperature as a function of space-time or, in other
words the equation of state (EOS). These are handled by the relativistic
hydrodynamic equations which are basically the equations of conservation
of energy-momentum and particle current. 
We have seen in Chapter~3 that the hadrons redress themselves in the 
thermal medium thereby changing their vacuum masses. This feature is taken
into account in the evolution dynamics through the EOS.
As a result the cooling law in the hadronic sector will be quite different
from the QGP which is made up of weakly interacting quarks and gluons which are
essentially massless.  
We will discuss these aspects in Secs.~4.1 and 4.2. 
We then go on to evaluate the photon and dilepton
yield in typical collision scenarios at the CERN SPS and BNL RHIC energies
in Section 4.3. Our
principal contention in the whole exercise is to comment on
the initial probable state of matter, QGP or hadronic matter on the basis of
our calculations in comparison with available data. We will compare the
results of our evaluation of the thermal photon spectra with the
data from the WA80 and WA98 experiments. The dilepton data will be compared with
the results of the CERES experiment.

\section{Relativistic Hydrodynamics} 

\bef
\centerline{\psfig{figure=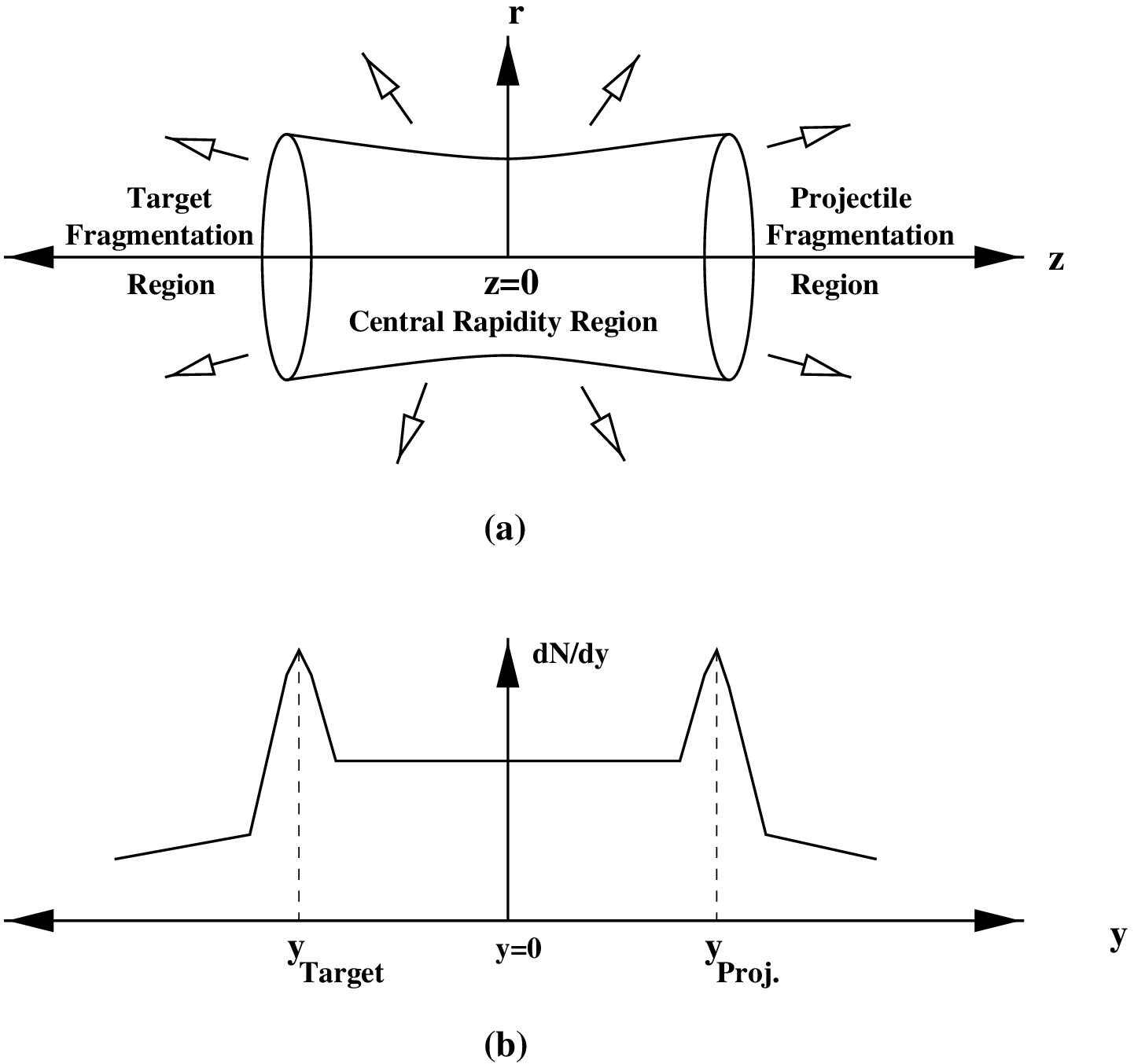,height=8cm,width=8cm}}
\caption{
Schematic diagram of (a) the collision geometry in the central collisions 
of identical nuclei and (b) the central plateau in the rapidity density.
}
\label{plateau}
\eef
A basic ingredient of the hydrodynamic description of the 
collision volume is the existence of a strong interaction 
time scale,
\be
\tau_i\sim \frac{1}{\Lambda_{QCD}}\sim {\s 1 fm/c}\,\sim\,\tau_{\s formation}
\ee
In any hadronic collision the produced fragments can only interact
after a proper time $\tau_i$ has elapsed after their collisions.
Thus, there is another time scale in the problem, 
the so called transit time , which is defined as
\be
\tau_{\s transit}\sim\frac{2R_A}{\gamma_{\s cm}}
\ee
$R_A$ is the nuclear radius, $\gamma_{\s cm}$ is the Lorentz factor.
If the value of $\gamma_{\s cm}$ (which is a function of the 
collision energy)
is such that $\tau_{\s transit}<\tau_{\s formation}$ then most of the
secondaries are formed after the nuclei pass through each other.
Consequently these secondaries
will not contribute to the energy density of the 
fluid in the central region (see Fig.~(\ref{plateau}a)). Such a scenario 
may be realized in the highly relativistic collisions of heavy ions.
This particular feature has been taken into account 
in Bjorken's hydrodynamic model~\cite{bjorken}.
 
It has been observed experimentally that the particle spectra  for
the secondaries produced in $N-N$ collisions exhibit a central plateau in the
rapidity space. This kind of behaviour
is due to the frame independence symmetry
of the hydrodynamic expansion of the system~\cite{chiu}. 
Bjorken assumed that the same kind of
plateau will also be observed in nucleus-nucleus collisions~\cite{bjorken}. 
In terms of the initial condition this means that the energy density, 
pressure etc. (all the thermodynamic quantities) will be a function of 
the initial thermalization (proper) time $\tau_i$
only and {\it will not} depend on the space time rapidity $\eta$
(defined later). This initial symmetry of the thermodynamic quantities
is preserved throughout the evolution scenario. 
If the particle rapidity density is flat
or invariant under Lorentz boosts then the entropy density ($s$)
will be independent of the rapidity. 

The evolution of the fluid is governed by the energy-momentum
conservation equation
\be
\partial_\mu\,T^{\mn}=0
\label{eom1}
\ee
where 
\be
T^{\mn}=(\epsilon+P)u^{\mu}u^{\nu}\,+\,g^{\mn}P
\ee
 is the 
energy-momentum tensor for ideal fluid; $\eps$ and $P$ are the energy density
and pressure of the fluid element. For an isentropic flow the entropy 
conservation reads
\be
\partial_\mu\,s^{\mu}=0
\label{eom2}
\ee
where $s^\mu=s\,u^\mu$ is the entropy current. Considering fluid motion along
the $z$ (beam) direction, $u^\mu=(\cosh y_f,0,0,\sinh y_f)$ where the
fluid rapidity $y_f$ is given by
\be
y_f\equiv {1\over 2}\ln\frac{1+v_z}{1-v_z},
\ee
$v_z$ being
the fluid velocity.
Let us consider the situation in a two dimensional
sub-space ($t-z$ plane).
We will make a change of variables from ($t,z$) to ($\tau,\eta$) where
\be
\tau\equiv\sqrt{t^2-z^2};\,\,\,\,\,\eta\equiv\frac{1}{2}\ln\frac{t+z}{t-z},
\ee
$\eta$ being the space-time rapidity.
As a result Eqs.~(\ref{eom1}) and (\ref{eom2}) become
\be
\frac{\partial}{\partial\tau}\left(s\tau \cosh(y_f-\eta)\right)+
\frac{\partial}{\partial\eta}\left(s\sinh(y_f-\eta)\right)=0
\label{eom3}
\ee
\be
\frac{\partial}{\partial\tau}\left(T\tau \sinh(y_f-\eta)\right)+
\frac{\partial}{\partial\eta}\left(T\cosh(y_f-\eta)\right)=0.
\label{eom4}
\ee
In order to solve these equations we need to know how the fluid velocity $v_z$
depends on the space-time coordinates, {\it i.e.} we need $y_f$ as a function of 
$\eta$. We will make use of the observation
that the amount of particle production is intimately connected with
the stopping of the colliding nuclei. At high collision energies 
the nuclei would slow down considerably, but due to
the large rapidity gap between the colliding nuclei, the produced matter
will still span a large rapidity interval as seen in Fig.~(\ref{plateau}b).
The rapidity distribution of secondaries is likely to exhibit a plateau-like
structure in the central region as conjectured by Bjorken. In this situation
it is reasonable to assume that in the central rapidity region the longitudinal
velocity of produced matter exhibits scaling behaviour, 
\be
v_z=z/t.
\ee
We then
have $y_f=\eta$ and $\tau$ becomes the proper time 
of the fluid frame which
is related to the centre of mass frame by a Lorentz transformation along the $z$-axis
with velocity $z/t$. 
Putting $y_f=\eta$ in Eqs.~(\ref{eom3}) and (\ref{eom4}) we get
\be
\frac{\partial}{\partial\tau}\left(s\tau\right)=0
\label{enscale}
\ee
and
\be
\frac{\partial T}{\partial\eta}=0.
\label{t-eta}
\ee
Equation~(\ref{t-eta}) implies that $T$ is independent of $\eta$ and so are all the
thermodynamic quantities; they depend only on $\tau$. This gives rise to
the frame independence or boost-invariant expansion. From Eq.~(\ref{enscale})
we have 
\be
s\tau=const.
\ee
which is the Bjorken's scaling solution. The resulting space-time
picture of the collision is shown in Fig.~(\ref{t-z}) of Chapter~1.
It may be noted that the above results were obtained without any specific
input from the equation of state. 
It is thus a general result that one dimensional similarity flow is 
necessarily isentropic even if there is a phase transition.
For a relativistic massless
gas with statistical degeneracy $g_k$, 
$s$ and $T$ are related through the equation of state:
\be
s=4\frac{\pi^2}{90}g_kT^3.
\label{sT}
\ee
Putting this expression for entropy density in the Bjorken scaling 
solution we get the cooling law 
\be
T^3\tau =const.
\ee
which is routinely used to evaluate the signals of QGP. 
The initial temperature of the system is determined by observing that
the variation 
of temperature from its initial value  $T_i$ to final value 
$T_f$ (freeze-out temperature) with proper time ($\tau$) is governed 
by the entropy conservation (Eq.~(\ref{enscale})) 
\be
s(T)\tau=s(T_i)\tau_i\,.
\label{entro1}
\ee
The entropy density is then expressed in terms of the observed particle (pion) multiplicity.
Using Eqs.~(\ref{sT}) and (\ref{entro1}) one gets the initial temperature as
\be
T_i^3=\frac{2\pi^4}{45\zeta(3)\pi\,R_A^2 4a_k\tau_i}\frac{dN_\pi}{dy}
\label{initemp}
\ee
where $dN_\pi/dy$ is the total pion multiplicity, $R_A$ is the radius
of the system,
$\tau_i$ is the initial thermalization time, 
and $\zeta(3)$ is the Reimann zeta function. 
$a_k=({\pi^2}/{90})\,g_k$ is the degeneracy of the produced system
and $k$ stands for either QGP or a hot hadronic gas. 
The rapidity density for the secondaries is obtained as~\cite{heinz},
\be
\left(\frac{dN}{dy}\right)_{A-A}=
A^{\alpha_r}\left(\frac{dN}{dy}\right)_{p-p}
\ee
where $\alpha_r$ is known as the rescattering parameter.
$dN/dy\mid_{p-p}$ can be parametrized
to fit the experimental data in the central region as
\be
\left(\frac{dN}{dy}\right)_{p-p}=0.8\ln(\sqrt{s}).
\ee
The assumption of a central plateau in the rapidity distribution is not
experimentally observed in nucleus-nucleus collisions at the presently available
energies. Hence the boost invariant hydrodynamics may not be a valid concept
at these energies. The concept of complete stopping in 
Landau model~\cite{landau} is not valid either at these energies. 
The physical situation
may be in-between the boost invariant model of Bjorken and the Landau
model of complete stopping, which means that there may be an overlap between
the formation zone and the collision zone. 
However, for its simplicities the  Bjorken model will be used 
to describe the space time evolution of matter formed in URHICs.
Appropriate generalization has been made to take into account the temperature 
dependent hadronic masses in the evolution.

It is worth mentioning that though significant stopping of baryons have
been observed in relativistic collisions of heavy ions in the recently
concluded experiments at the CERN SPS
we will assume the central region in
rapidity space to be approximately free of baryons. 
With strong stopping the net baryon density in the central region can
be appreciable but the matter is still not baryon rich because strong
stopping also causes large secondary particle production and baryons 
form only a small fraction of all particles. In other words, the ratio
of the baryon chemical potential to the temperature is small.
However, at RHIC and LHC the net baryon number
can certainly be neglected in the central rapidity region. 

\section{Initial Conditions and Equation of State}

The set of hydrodynamic equations is not closed by itself;
the number of unknown variables exceeds the number of equations by one.
One thus needs to postulate a functional relation between any two  
variables so that the system becomes deterministic. 
The most natural course is to look for
such a relation between the pressure $P$ and the energy density $\epsilon$. 
Under the assumption of local thermal equilibrium, this functional relation
between $P$ and $\epsilon$ is the EOS.
Obviously, different EOS's will govern the hydrodynamic
flow quite differently~\cite{pasi} and as far as the search for QGP is concerned, 
the goal is to look for
distinctions in the observables due to the different EOS's (corresponding
to the novel state of QGP vis-a-vis that for the usual hadronic matter). 
It is thus imperative to understand in what respects the two EOS's differ 
and how they affect the evolution in space and time. Recently, the sensitivity
of the photon emission rate on various evolution scenarios has been studied 
in Ref.~\cite{pp}.

A physically intuitive way of understanding the role of the EOS in governing
the hydrodynamic flow lies in the fact that the velocity of sound $c_{s}^{2}
=(\partial P/\partial\epsilon)_s$ sets an intrinsic scale in the hydrodynamic
evolution. One can thus write a simple parametric form for the EOS:
$P=c_{s}^{2}(T)\epsilon$. For an ideal gas of massless constituents $c_s^2=1/3$.
Inclusion of interactions, however, may drastically alter the value of 
$c_{s}^{2}$~\cite{ah}.
In our calculation we assume the MIT bag model equation of state
for the QGP where the energy density and pressure are given by
\be
\epsilon_Q=g_Q{\pi^2T^4\over 30}+B,
\ee
and
\be
P_Q=g_Q\frac{\pi^2}{90}T^4-B.
\ee
The effective degrees of freedom in QGP, $g_Q=37$ for two flavours.
The entropy density $s_Q$ is given by $s_Q=2g_Q(\pi^2/45)T^3$. Putting
$a_k\equiv a_Q=(\pi^2/90)g_Q$ the initial temperature for a system
produced as QGP can be determined from Eq.~(\ref{initemp}).

In the hadronic phase we have to be more careful about the presence
of heavier particles and the change in their masses due to finite temperature
effects.
The ideal limit of treating the hot hadronic matter as a gas of pions 
originated from the expectation that in the framework of local 
thermalization the system would be dominated by the lowest mass hadrons 
while the higher mass resonances would be
Boltzmann suppressed. 
Indirect justification of this assumption comes from the experimental
observation in high energy collisions that most of the secondaries are pions.
Nevertheless, the temperature of the system is 
higher than
$m_{\pi}$ during a major part of the evolution and at these temperatures the
suppression of the higher mass resonances may not be complete. It may 
therefore be more realistic to include higher mass resonances in the 
hadronic sector, their relative
abundances being governed by the condition of (assumed) thermodynamic 
equilibrium.
The hadronic phase is taken to consist of $\pi$, $\rho$, $\omega$, $\eta$, $a_1$  
mesons and nucleons. The nucleons and heavier mesons may play an important
role in the EOS in a scenario where mass of the hadrons decreases
with temperature. 

The energy density and pressure
for such a system of mesons and nucleons are given by
\be
\epsilon_H=\sum_{h={\s mesons}} \frac{g_h}{(2\pi)^3} 
\int d^3p\,E_h\,f_{BE}(E_h,T)
+\frac{g_N}{(2\pi)^3} 
\int d^3p\,E_N\,f_{FD}(E_N,T)
\ee
and
\be
P_H=\sum_{h={\s mesons}} \frac{g_h}{(2\pi)^3} 
\int d^3p\frac{p^2}{3\,E_h}f_{BE}(E_h,T)
+\frac{g_N}{(2\pi)^3} 
\int d^3p\frac{p^2}{3\,E_N}f_{FD}(E_N,T)
\ee
where the sum is over all the mesons under consideration and $N$ stands
for nucleons and $E_h=\sqrt{p^2 + m_h^2}$.          
The entropy density is then
\be
s_H=\frac{\epsilon_H+P_H}{T}\,\equiv\,4a_{\s{eff}}(T)\,T^3
= 4\frac{\pi^2}{90} g_{\s{eff}}(m^\ast(T),T)T^3
\label{entro}
\ee
where  $g_{\s eff}$ is the effective statistical degeneracy.
Thus, we can visualize the finite mass of the hadrons
having an effective degeneracy $g_{\s{eff}}(m^\ast(T),T)$. 
Because of the temperature dependence of the effective degeneracy
Eq.~(\ref{initemp}) has to be solved self-consistently in order to
calculate the initial temperature of the system initially 
produced as a hot hadronic gas. We thus solve the equation
\be
\frac{dN_\pi}{dy}=\frac{45\zeta(3)}{2\pi^4}\pi\,R_A^2 4a_{\s{eff}}(T_i)T_i^3\tau_i
\label{dndy}
\ee
where
$a_{\s{eff}}(T_i)=({\pi^2}/{90})\,g_{\s{eff}}(m^\ast(T_i),T_i)$ .
The change in the expansion dynamics
as well as the value of the initial temperature due
to medium effects enters the calculation of the
photon emission rate through the effective statistical degeneracy.

If  the  energy/entropy density in the fireball
immediately  after  the  so-called  ``formation  time" $\tau_i$ is
sufficiently high, then the matter exists in the form of  a  QGP.
As  the  hydrodynamic expansion starts, the system begins to cool
until  the  critical  temperature  $T_c$  is  reached at a time
$\tau_Q$.  At  this
instant, the phase transition  to  the  hadronic  matter  starts.
Assuming  that  the  phase transition is a first
order one, the released latent heat maintains the temperature  of
the  system  at  the  critical temperature $T_c$, even though the
system continues to expand; the cooling due to expansion is compensated
by the latent heat liberated during the process.
Together with the possible explosive
events,  we are neglecting  the  scenarios  of  supercooling or
superheating. This process continues until all the matter has
converted  to the hadronic phase at a time $\tau_H$, still at $T=T_c$; 
from then on,
the system continues to expand, governed by  the  EOS  of  the  hot
hadronic  matter  till the freeze-out temperature $T_f$ is reached
at the proper time $\tau_f$. Thus the
appearance of the so called mixed phase at $T=T_c$, when QGP  and
hadronic  matter  co-exist,  is a direct consequence of the first
order phase transition. 
The  possibility of the mixed phase affects the bulk
features of the evolution process and plays an important role
in QGP  diagnostics.

In the mixed phase, the relative proportion of QGP  and  hadronic
matter must be a function of time; initially the system consists
entirely of QGP and at the end, entirely of hot  hadronic  matter.
If we denote the fraction of the QGP
by $f_Q(\tau)$ then the hadronic fraction in the mixed phase 
is $f_H(\tau)=1-f_Q(\tau)$ so that
$f_Q(\tau_Q)=1$ and $f_H(\tau_H)=1$. 
Then the entropy density in the mixed phase  
is given by
\be
s_{\s mix}(\tau)=f_Q(\tau)\,s_Q^c+f_H(\tau)\,s_H^c
\label{enmix}
\ee
where $s_Q^c$ ($s_H^c$) denotes 
the entropy density of QGP (hadronic) phase at $T_c$.
The life-time of the mixed phase is given by
\be
\tau^{\s mixed}_{\s life}=\tau_H-\tau_Q. 
\ee
The scaling law governing 
the variation of $s$ with $\tau$ must continue
to hold also in the mixed  phase; substituting
Eq.~(\ref{enmix}) in Eq.~(\ref{enscale}) we obtain for $T_i>T_c$,
\be
f_Q(\tau)=\frac{1}{r-1}\left(r\frac{\tau_Q}{\tau}-1\right)
=\frac{1}{r-1}\left(\frac{\tau_H}{\tau}-1\right)
\label{qgpfrac}
\ee
where $r$ ($=g_Q/g_{\s eff}$)
is the ratio of the degeneracy of QGP phase and the effective 
degeneracy in the hadronic phase. 
In the above equation we have used the 
relation $\tau_H=r\tau_Q$, valid for ($1+1$) dimensional isentropic expansion.

Since the entropy density $s(\tau)$ in any phase can be expressed 
as in Eq.~(\ref{enmix}) with suitable values of $f_Q(\tau)$,
the volume fractions can also be expressed  as 
\bea
f_Q(\tau)&=&\frac{s(\tau)-s_H^c}{s_Q^c-s_H^c}\nonumber\\
f_H(\tau)&=&\frac{s_Q^c-s(\tau)}{s_Q^c-s_H^c}.
\eea

If $T_i=T_c$, {\it i.e.} if the system is formed in  the  mixed  phase
with a fraction $f_0$ of the QGP phase then
\be
f_Q(\tau)=\frac{1}{r-1}\left[(1+(r-1)f_0)\frac{\tau_i}{\tau}-1\right].
\ee
The mixed phase in this case ends at a proper time
$\tau_H^m=(1+(r-1)f_0)\tau_i$.
For $s_i<s_H^c$, the value of $f_H(\tau)$ is always unity.

Now that we have all the necessary requisites, we will evaluate
the photon and dilepton yield in heavy ion collisions at the SPS and RHIC.
We will consider two possibilities:
\begin{center}
{\bf
A\,+A\,$\ra$(QGP)$\ra$(Mixed Phase)$\ra$Hadronic Phase\\
\,or\,\\
A\,+\,A\,$\ra$Hadronic Phase.\\
}
\end{center}
 The former (latter) case where the initial
state is formed in QGP (hadronic) phase will be called the `QGP scenario'
(`no phase transition scenario'). 
For the QGP sector
 we use a simple bag model equation of state (EOS) with
two flavour degrees of freedom. The temperature in the QGP phase evolves
according to Bjorken scaling law $T^3\,\tau=T_i^3\tau_i$.
The cooling law in the hadronic sector is quite different from that of the QGP
because of the presence of massive hadrons. These hadrons redress themselves in
the medium thereby changing their vacuum masses. This is accounted for
by introducing temperature dependence in the statistical 
degeneracy which takes care of the mass varying with temperature.
In Table~1 we quote the values of the initial temperatures
obtained by assuming various mass variation scenarios. The value of initial
thermalization time has been assumed as 1 fm/c both for SPS 
and RHIC energies. The multiplicity density($dN/dy$) in the two cases are
given as 600 and 1735 respectively. For a two-flavour
QGP the effective degeneracy is 37. The freezeout time $T_f$ is taken as
130 MeV~\cite{lkb} in all the calculations.
$\vartheta$ and $\delta$ 
dictate the variation of temperature with proper time for the 
hadronic matter according to the cooling law  $T=\vartheta/\tau^{\delta}$. 
The values of $\delta$ indicates a slower cooling in the hadronic phase as 
compared to that of QGP phase where $T\propto\,1/\tau^{0.33}$.
\renewcommand{\arraystretch}{1.5}
\vskip 0.2in
\begin{center}
\begin{tabular}{|l|c|c|c|c|c|c|c|}
\hline
&
\multicolumn{4}{c|}{$dN/dy$=600 $\tau_i$=1 fm}&
\multicolumn{3}{c|}{$dN/dy$=1735 $\tau_i$=1 fm} \\
\cline{2-8}
& hadronic gas &
\multicolumn{3}{c|}{QGP + Mix + Had} &
\multicolumn{3}{c|}{QGP + Mix + Had}  \\
& initial state & \multicolumn{3}{c|}{$T_i$=185 MeV $\tau_Q$=1.6 fm} &
\multicolumn{3}{c|}{$T_i$=265 MeV $\tau_Q$=4.6 fm}  \\
\cline{2-8}
& $T_i$ (MeV) & $\tau_H$ (fm) & $\vartheta$ & $\delta$ & $\tau_H$ (fm) & $\vartheta$ & $\delta$  \\
\hline
bare masses & 270 & 10.8  & 0.267 &  0.215  &  31.9  &  0.337  &  0.215  \\
QHD &  220 & 9.4  &  0.247  &  0.194  & 27.6  &  0.305  &  0.194  \\
BR  & 195 &  8.2  &  0.236  &  0.184  &  23.9  &  0.288  &  0.185  \\
Nambu & 195 & 4.7  &  0.203  &  0.151  &  13.9  &  0.239  &  0.152  \\
\hline
\end{tabular}
\end{center}
\begin{center}
Table 1 : Values of initial temperatures and evolution parameters for
SPS and RHIC.
\end{center}

\vspace*{0.1in}

In Fig.~(\ref{degfig}) we depict the variation of effective degeneracy 
as a function of temperature with and without medium effects
on the hadronic masses for various scenarios. 
We observe that for $T>140$ MeV the effective
degeneracy becomes larger due to the reduction in temperature dependent 
masses compared to the free hadronic masses. 
\bef
\centerline{\psfig{figure=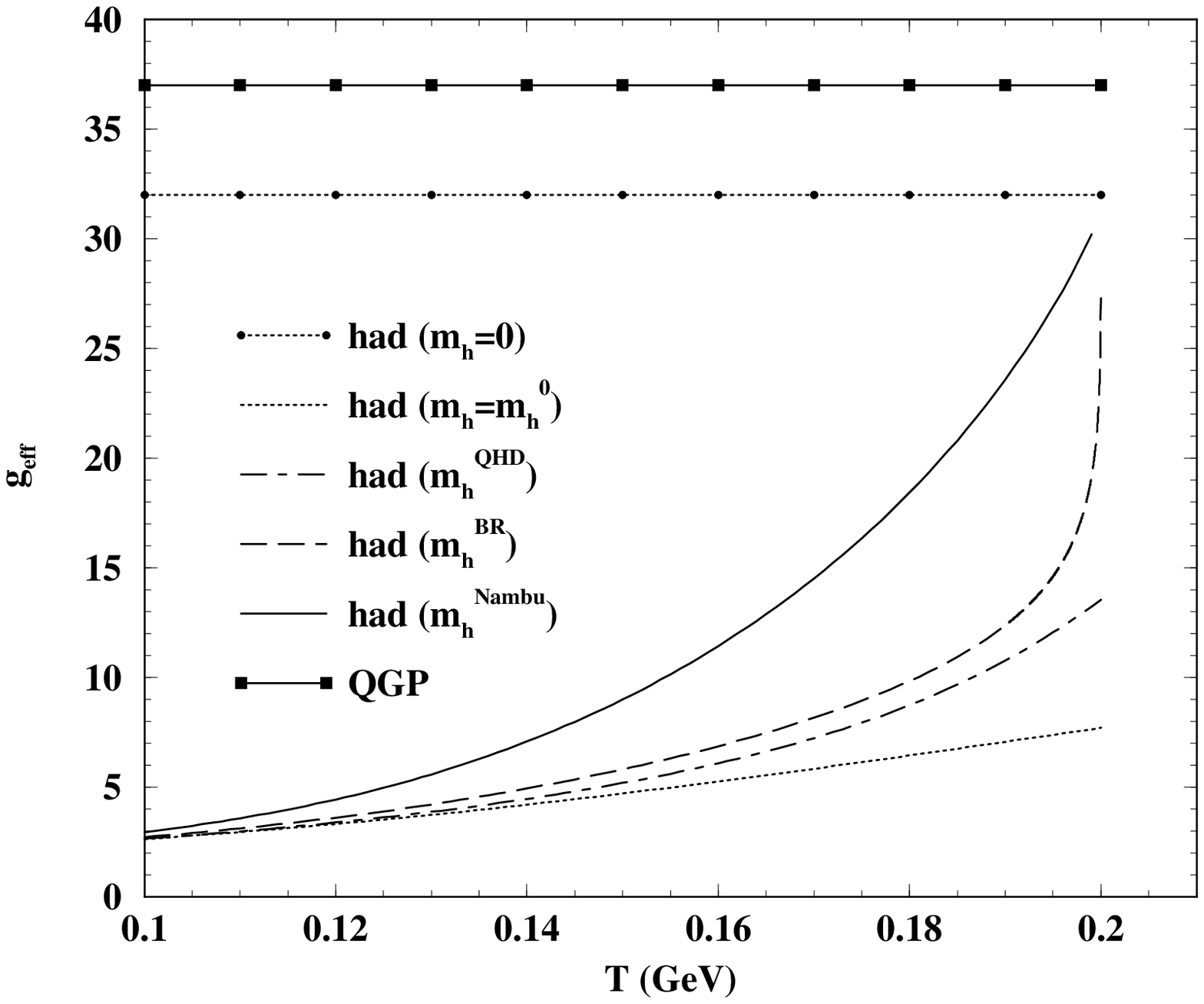,height=7cm,width=9cm}}
\caption{
Variation of effective degeneracy as a function of temperature.
}
\label{degfig}
\eef
Physically this means that the number of hadrons
in a thermal bath at a temperature $T$ is more when in-medium mass reduction
is taken into account.  Eq.~(\ref{dndy}) implies that for a given
pion multiplicity the initial temperature of the system will be 
lower (higher) when medium effects on hadronic masses are considered
(ignored). This is clearly demonstrated in Fig.~(\ref{ttausps})
where we show the variation of temperature 
with proper time for different initial conditions. 
The thick dots indicate the yield when QGP is formed initially 
at $T_i=185$ MeV and cools down according to 
Bjorken law up to a temperature $T_c$ at proper time $\tau_Q$,
at which a phase transition
takes place; it remains constant at $T_c$ up to a time
$\tau_H=9.4$ fm/c after which the temperature decreases as 
$T=0.247/\tau^{0.194}$ (when medium effects  
are taken from QHD) to a temperature $T_f$. 
If the system is considered to be formed in the hadronic phase 
then the initial temperature is obtained as $T_i=220$ MeV (270 MeV)
when in-medium effects on the hadronic masses from the QHD model is 
taken into account
(ignored). The corresponding cooling laws are
displayed in Fig.~(\ref{ttausps}). 
The above parametrizations of the cooling law in the hadronic phase
have been obtained by solving Eq.~(\ref{entro1}) self-consistently.
An initial state with the vanishing meson masses at
$T_i=195$ MeV ($\tau_i=1$ fm/c) could be realized in 
the case of BR and Nambu scaling scenarios
with pion multiplicity $dN/dy=600$. 
\bef
\centerline{\psfig{figure=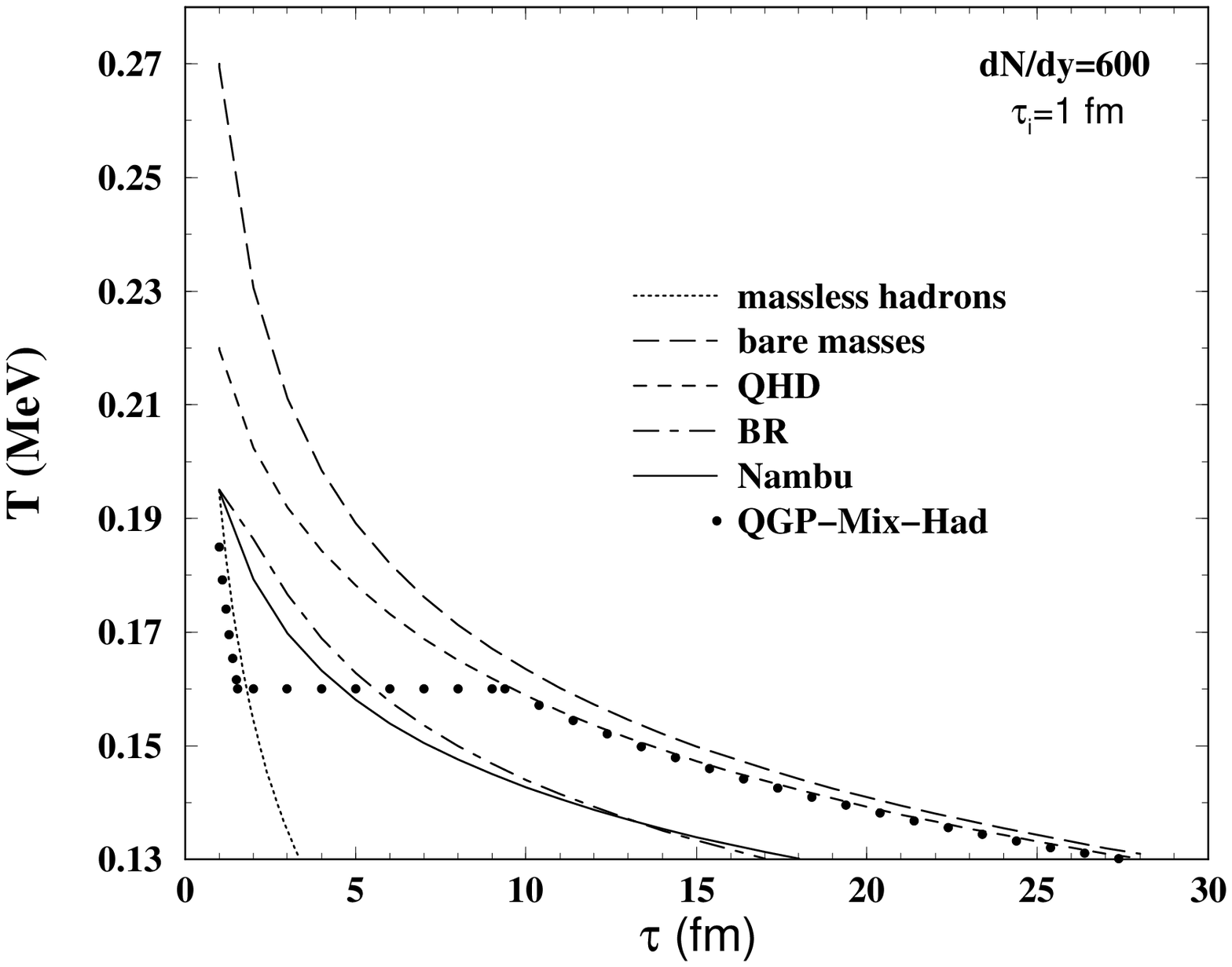,height=7cm,width=9cm}}
\caption{
Variation of temperature as a function of proper time in Pb-Pb collisions
at the SPS.
}
\label{ttausps}
\eef

\begin{figure}
\centerline{\psfig{figure=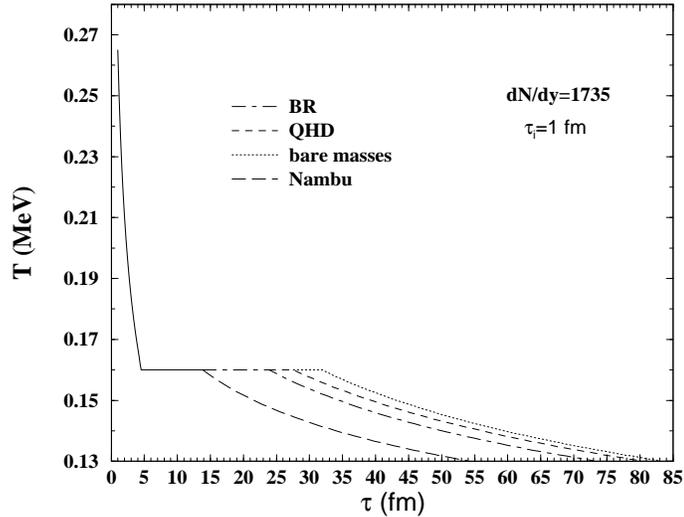,height=7cm,width=9cm}}
\caption{Variation of temperature as a function of proper time in
Pb-Pb collisions at RHIC.
The initial temperature has been determined by assuming `QGP
scenario'. The initial temperature $T_i=265$ MeV for $\tau_i=1$ fm/c 
and $dN/dy=1735$
}
\label{ttaurhc}
\end{figure}

At RHIC a scenario of a pure hot hadronic system
within the format of the model used here, appears to be
unrealistic. The initial temperature considering bare hadronic
masses turns out to be $\sim$ 340 MeV whereas for the other extreme
case of massless hadrons it is $\sim$ 290 MeV. With temperature
dependent masses the initial temperature will lie somewhere between these
two values. For such high temperatures, clearly a hot dense hadronic
system cannot be a reality, the hadrons would have melted away even for 
lower temperatures. Thus, for RHIC we have treated the case of a QGP
initial state only. The temperature profile for RHIC is depicted 
in Fig.~(\ref{ttaurhc}) where we observe that 
the length of the plateau, 
which indicates the life time of the mixed phase 
$\tau_{mix}^{life}=\tau_H-\tau_Q$,
depends on the masses of the hadrons in the hadronic phase. 
The effective degeneracy plays an important role here.
At the transition point there is a large decrease in the entropy density.
This decrease has to be compensated by the
expansion (increasing the volume) to keep the total entropy constant. 
Since we are considering
(1+1) dimensional isentropic expansion, this change in the entropy density will
be compensated by increasing $\tau$ which is a measure of the volume, so
that the total entropy remains constant. We have
seen earlier (Fig.~\ref{degfig}) that the effective degeneracy in the 
hadronic phase is the largest for the Nambu scaling and smallest for 
the bare mass scenario, resulting in smallest (largest) discontinuity in the
entropy density for the former (latter) case. Consequently the time
taken for the system to compensate the decrease of the entropy density 
in the Nambu scaling scenario is smaller compared to the case where
bare masses are considered.
Hence the life time of the mixed phase for the Nambu scaling case is 
smaller than all other cases.

\section{Electromagnetic Spectra in URHICs}

Our ultimate goal in URHICs is to distinguish between the two
possibilities described above, that of a hot hadronic gas initial state
or a QGP which undergoes a phase transition to hadronic matter via
a mixed phase of coexisting quark and hadronic matter.
In the following we will compare
the photon and dilepton spectra originating from these two scenarios.
As discussed before,
the experimentally observed photon and dilepton spectra originating from an expanding 
QGP or hadronic matter is obtained by convoluting the static
(fixed temperature) rate with expansion dynamics. 
In a first order
phase transition scenario 
the photon and dilepton spectra
from a $(1+1)$ dimensionally expanding system is obtained as
\begin{eqnarray}
\frac{dN}{d\Gamma}&=&\pi\,R_A^2\,\int\left[
\left(\frac{dR}{d\Gamma}\right)_{QGP}
\Theta(s-s_Q^c)\right.\nonumber\\
& &+\left[\left(\frac{dR}{d\Gamma}\right)_{QGP}\frac{s-s_H^c}
{s_Q^c-s_H^c}\right.\nonumber\\
& &+\left.\left(\frac{dR}{d\Gamma}\right)_{HG}\frac{s_Q^c-s}
{s_Q^c-s_H^c}\right]
\Theta(s_Q^c-s)
\Theta(s-s_H^c)\nonumber\\
& &+ 
\left.\left(\frac{dR}{d\Gamma}\right)_{HG}\Theta(s_H^c-s)\right]
\tau\,d\tau\,d\eta 
\end{eqnarray}
where 
$R_A$ is the radius of the nuclei 
and $\Theta$ functions are 
introduced to get the contribution from the QGP, mixed and hadronic gas (HG) phases.
For (1+1) dimensional expansion, 
$d^4x=\pi R_A^2\,dzdt=\pi R_A^2\, \tau d\tau d\eta$.
In the case of photons $d\Gamma$ stands for $d^3p/E (=d^2p_Tdy)$;
$p_T$ and $y$ being the transverse momentum and rapidity of the emitted
photon. For dileptons, $d\Gamma=d^4q=MdMd^2q_Tdy$ where, $M$, $q_T$ and $y$
are the invariant mass, the transverse momentum and rapidity of the
lepton pair respectively.

\subsection{Photon Spectra at SPS}

\begin{figure}
\centerline{\psfig{figure=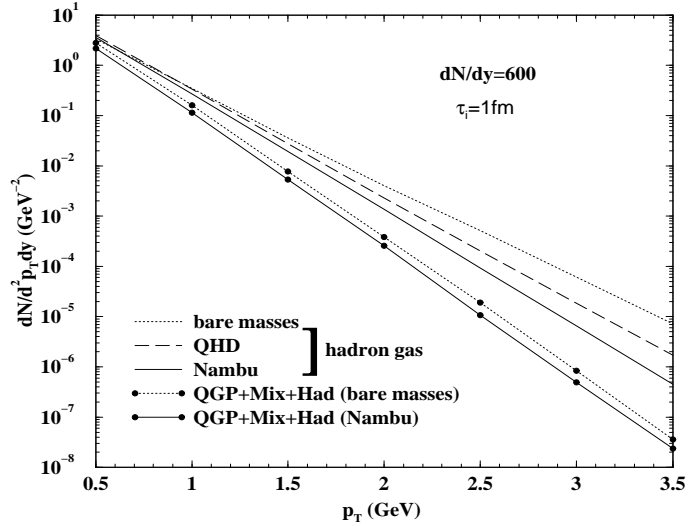,height=7cm,width=9cm}}
\caption{Total thermal photon yield corresponding to $dN/dy=600$ and
$\tau_i=1$ fm/c. 
The solid (long-dash) line indicates photon spectra  
when hadronic matter formed in the initial state 
at  $T_i=195$ MeV ($T_i=220$ MeV) 
and the medium effects are taken from Nambu scaling (QHD).
The dotted line  represents the photon spectra without 
medium effects
with $T_i=270$ MeV. The solid (dotted) line with thick dots
represent the yield for the `QGP scenario' when the hadronic
mass variations are taken from Nambu scaling (free mass).
}
\label{phosps}
\end{figure}

The thermal photon spectra at SPS energies is shown in Fig.~(\ref{phosps}).
In order to interpret the spectra corresponding to the different scenarios 
let us recall that
the effective degeneracy in the hadronic phase
$g_{\s eff}$ is obtained as a function of $T$  
by solving Eq.(\ref{entro}). A smaller (larger) value of $g_{\s eff}$ 
is obtained in the free (effective) mass scenario. 
As a result we get a larger (smaller) initial temperature
by solving Eq.(\ref{dndy}) in the free (dropping) mass scenario 
for a given multiplicity.
Naively we expect that at a given temperature if a meson mass drops
its Boltzmann factor will be enhanced and more of those mesons will
be produced leading to more photons~\cite{sourav,csong}.
However, a larger drop in the hadronic masses results in smaller initial
temperature, implying that the
space time integrated spectra crucially depends on these two
competitive factors.
Therefore, with (without) medium effects one integrates an enhanced (depleted) 
static rate over smaller (larger) temperature range for a fixed
freeze-out temperature ($T_f=130$ MeV).
In the present calculation the enhancement in the photon emission due to 
the higher initial temperature in the free mass scenario (where static rate is
smaller) overwhelms the enhancement of the rate due to negative shift
in the vector meson masses (where the initial temperature is smaller).
Accordingly, in the case of bare mass ( Nambu scaling) scenario the 
photon yield is the highest (lowest). In case of the QHD model,
the photon yield lies between the above two limits.
In the `QGP scenario' the  photon
yield with in-medium mass is lower than the case where bare masses of hadrons are
considered. However, the difference is considerably less than the
`no phase transition scenario'.   
This is because, in this case
the initial temperature is determined by the quark and gluon 
degrees of freedom and the only difference between the bare and effective mass
scenarios is due to the different lifetimes of the mixed phase.  
In Fig.~(\ref{phosps}), the photon spectra
in the `no phase transition scenario' overshines the ones from 
the `QGP scenario'. 

\begin{figure}
\centerline{\psfig{figure=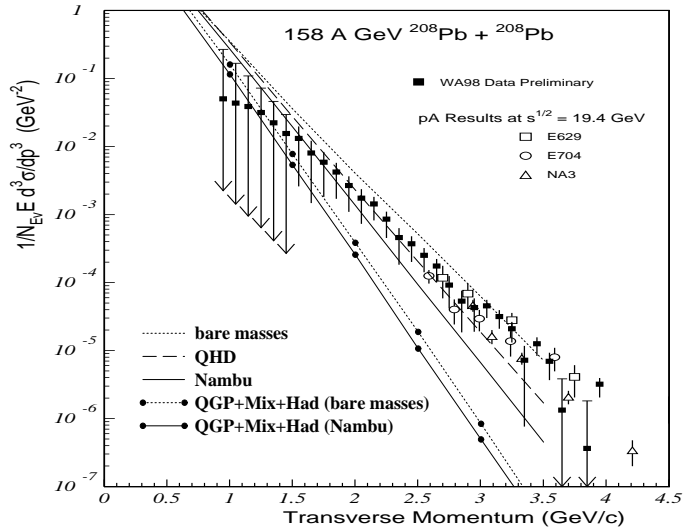,height=7cm,width=9cm}}
\caption{Same as Fig.~(\protect\ref{phosps}) with data from WA98 experiment.}
\label{phodat}
\end{figure}

We now compare the above set of curves to the
preliminary single photon data obtained by the WA98 Collaboration~\cite{WA98} in Fig.~(\ref{phodat}). 
We find that for the
entire range of $p_T$ the photon spectra in the `no phase transition
scenario' with medium effects evaluated in the QHD as well as Nambu scaling 
scenario
explain the data reasonably well. However, the yield with the QHD 
model parameters seems to overpredict in the low $p_T$ region by a small
amount. A QGP initial state as well as a hadronic gas initial state
with vacuum properties of hadrons seem to be improbable. The thermal photon
spectra for Pb-Pb collisions at SPS have also been evaluated in Ref.~\cite{pp}
for a variety of evolution scenarios without introducing medium effects.

\begin{figure}[t]
\centerline{\psfig{figure=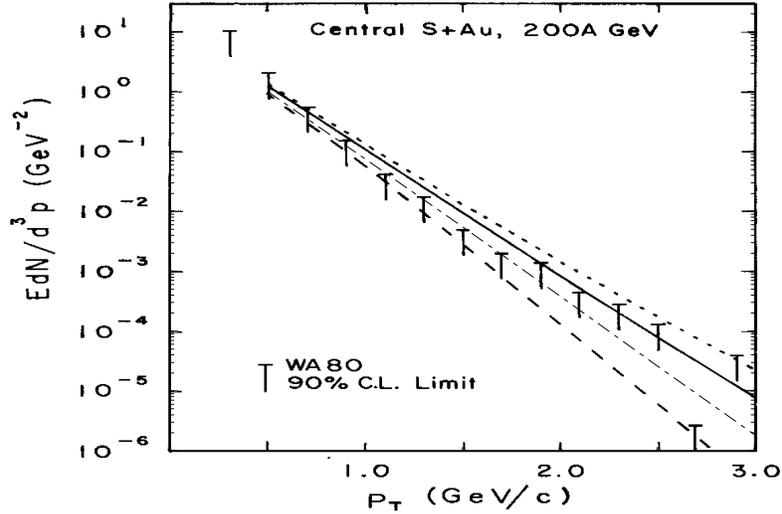,height=8cm,width=12cm}}
\caption{Total thermal photon yield in S-Au collisions at 200 A GeV
at CERN SPS. The long-dashed  line shows the results  for a QGP initial
state with $T_i$=190 MeV.
The solid (dot-dash) line indicates photon spectra  
in the  hadronic gas scenario
at  $T_i=220$ MeV ($T_i=195$ MeV) 
with medium effects calculated from QHD (Nambu scaling).
The short-dashed line represents the photon spectra from a hadronic
gas with bare masses
and $T_i=270$ MeV. The initial time $\tau_i$ in all these cases have been taken as
1.2 fm/c. The data correspond to the upper bound obtained by WA80 Collaboration.
}
\label{sau_data}
\end{figure}

We have also compared the thermal photon spectra evaluated~\cite{sspbpb} in the QGP
and hadronic gas scenarios with the data obtained by the WA80~\cite{wa80}
collaboration. This is shown in Fig.~(\ref{sau_data}). The experimental data in this case is the upper bound of
the thermal photons in S-Au collisions at 200 A GeV at the CERN SPS.
The pion multiplicity is given as $dN/dy=$225. Our conclusions are very similar
to the Pb-Pb case. The photons evaluated in the hadronic
gas scenario with bare masses (short-dashed line) can be ruled out. 
Also, in such a scenario when  medium effects
from QHD are incorporated, the yield (solid line) just crosses the upper bound.
However, thermal photons in the `no phase transition scenario' with
medium effects according to Nambu scaling (dot-dashed line) turns out to be a viable description.
Quite a few attempts have been made to explain the WA80 upper bound
using various models. Using relativistic hadron transport model
Li {\it et al}~\cite{li_wa80} have shown that the photon spectra with
either free or in-medium meson masses in dense matter
 do not exceed the upper limit.
With in-medium modifications of mesons calculated in the hidden
local symmetry approach, Halasz {\it et al}~\cite{halasz} have reached 
similar conclusions. However, using (3+1) dimensional expansion as well as
various probable equation of states Sollfrank {\it et al}~\cite{solf}
does not claim evidence for a phase transition. The WA80 data have also
been explained by Hui {\it et al}~\cite{hui} in a hadron and string cascade
model. These results are in sharp contrast to the calculations
performed by Srivastava and Sinha~\cite{sinha94}, Shuryak {\it et al}~\cite{shyak1,shyak2},
Dumitru {\it et al}~\cite{dumitru}, Arbex {\it et al}~\cite{arbex},
Neumann {\it et al}~\cite{neumann} and
Song~\cite{csong}
who have concluded that the WA80 upper limits are satisfied 
only if a QGP is assumed to have been formed in the initial stages.

\subsection{Dilepton Spectra at SPS}

\begin{figure}
\centerline{\psfig{figure=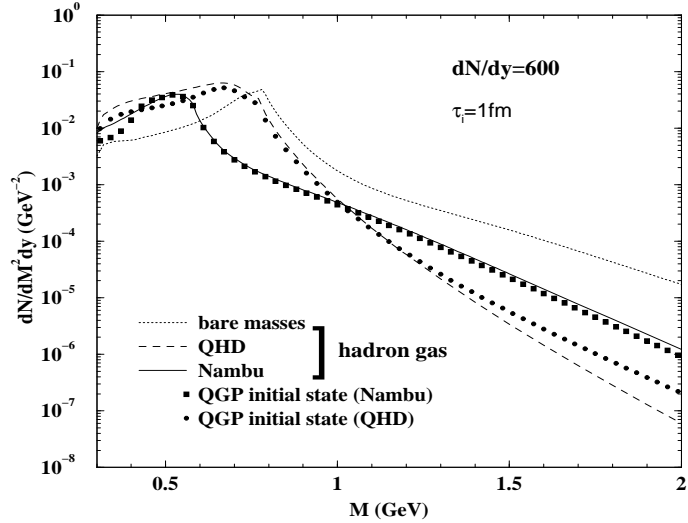,height=7cm,width=9cm}}
\caption{Total thermal dilepton yield corresponding to $dN/dy=600$ and
$\tau_i=1$ fm/c. 
The solid (long-dash) line indicates dilepton spectra  
when hadronic matter formed in the initial state 
at  $T_i=195$ MeV ($T_i=220$ MeV) 
and the medium effects are taken from Nambu scaling (QHD).
The dotted line  represents the yield from hadronic gas scenario with bare masses and
$T_i=270$ MeV.
The thick dots
represent the yield for the `QGP scenario' with mass modifications
included.}
\label{dilsps}
\end{figure}

The space time integrated dilepton spectra  for
the `QGP scenario' and `no phase transition scenario' 
with different models of mass variation are shown in Fig.~(\ref{dilsps}).
The shifts in the invariant mass distribution of the spectra
due to the reduction in the hadronic masses, particularly
the $\rho$ meson mass, according
to the different models are distinctly visible. Their is no appreciable
 difference between the QGP and `no phase transition' scenarios
which is expected because one is looking at low values of
the invariant mass where the dominant contribution in either case
will be from hadronic matter. Dileptons originating from the 
high temperature QGP phase will show up at higher values of $M$.

\begin{figure}
\centerline{\psfig{figure=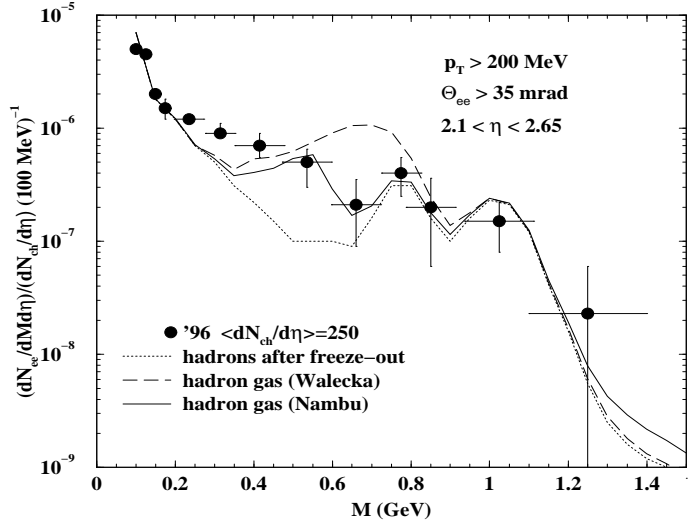,height=7cm,width=9cm}}
\caption{Our calculations compared with the inclusive $e^+e^-$ spectra measured by CERES in 158 A GeV
Pb-Au collisions in the '96 run.
}
\label{dildat}
\end{figure}

Let us now make a comparison of our predictions with the dielectron
spectrum in Pb-Au collisions at 158 A GeV as measured by the 
CERES Experiment~\cite{ceres_data}.
In this case one has to incorporate~\cite{solf}
the kinematical cuts in the momenta of the electron and
the positron relevant for the CERES detector. In Fig.~(\ref{dildat})
we have plotted the dilepton yield normalized to $dN_{ch}/dy$ within the
rapidity interval 2.1 to 2.65 in the `no phase transition scenario' with
the spectral functions in the QHD and Nambu scaling scenarios.
We observe that dileptons from an evolving hot hadronic gas with
medium modifications calculated in the Nambu scaling scenario
clearly fits the data beyond $M \sim$ 500 MeV. The QHD model calculations
under similar conditions overshoots the data in the region of the 
shifted $\rho$ peak between 500 and 700 MeV.

\vskip 0.7in
Though not a direct signal of QGP,
the observed enhancement of low-mass dileptons compared to the yield from
hadronic decays at freeze-out (shown by the dotted curve in Fig.~(\ref{dildat}))
has triggered a host of theoretical activity. Though the pion annihilation
channel $\pi^+\pi^- \ra l^+l^-$ via the $\rho$ meson accounts for a large fraction of
this enhancement, it turns out that a quantitative explanation of the data 
requires the incorporation of medium modifications of the vector mesons.
Li, Ko and Brown~\cite{lkb} were the first to achieve an excellent agreement with
the CERES data using a decreased $\rho$ mass in the dense fireball. 
Rapp {\it et al}~\cite{rcw}
have used a large broadening of the $\rho$ meson line shape and hence a shorter
life time due to scattering
off baryons in order to explain the data. However, both approaches rely
on a high baryon density for the dropping mass or the enlarged width of the $\rho$ meson
but the role of baryons is still a debatable issue. It is worth mentioning that
we have assumed zero baryon density in our calculations. Koch~\cite{koch} also
finds very little effect due to baryons and are able to obtain a reasonable
explanation of the data. It is worth emphasizing here that as yet it
has not been possible to explain~\cite{solf,dksdil} the observed low-mass enhancements
 of dileptons measured in the Pb-Au collisions as well as in the S-Au
collisions at the CERN SPS in a scenario which does not incorporate 
medium induced dropping of the vector meson mass, the $\rho$ meson mass
in particular~\cite{shyak2}. 

\subsection{Photon and Dilepton Spectra at RHIC} 

\begin{figure}
\centerline{\psfig{figure=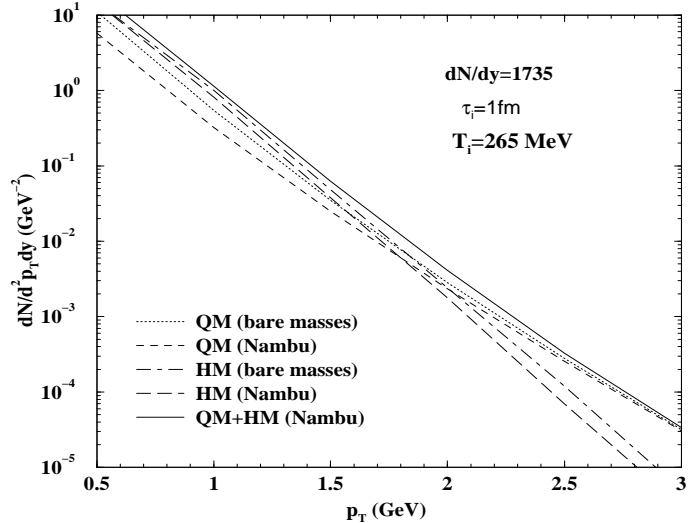,height=7cm,width=9cm}}
\caption{Thermal photon spectra at RHIC energies. A first order
phase transition scenario has been considered.
}
\label{phorhc}
\end{figure}
Finally we study the electromagnetic probes for RHIC energies.
As discussed earlier, in this case a scenario of a pure hot hadronic system 
appears to be unrealistic. 
So we have treated the case of a QGP initial state only.
The baryonic chemical potential has been taken to be zero, which in this case is 
certainly a valid assumption as far as the central rapidity region is concerned.
The thermal photon yield for RHIC is displayed in Fig.~(\ref{phorhc}).
The solid line represents the total thermal photon yield 
originating from initial QGP state, mixed phase and the pure hadronic
phase. The short dash line indicates photons from quark 
matter (QM) (= pure QGP phase + QGP part of the mixed phase)
and the long dash line represents photons from hadronic matter (HM)
(= hadronic part of the mixed phase + pure hadronic phase).
In all these cases the effective masses of the hadrons have been taken 
from Nambu scaling. For $p_T>2$ GeV photons from QM overshines
those from HM since most of these high $p_T$ photons originate
from the high temperature QGP phase.
The dotted and the dot-dash lines indicate photon yields from QM
and HM respectively with bare masses in the hadronic sector. 
The HM contribution for the bare mass is larger than the
effective mass (Nambu) scenario because of the larger value of the life
time of the mixed phase in the earlier case (see Table~1).
It is important to note that for $p_T>2$ GeV, the difference in the QM 
and HM contribution in the effective mass scenario is more that the
bare mass scenario.
\begin{figure}
\centerline{\psfig{figure=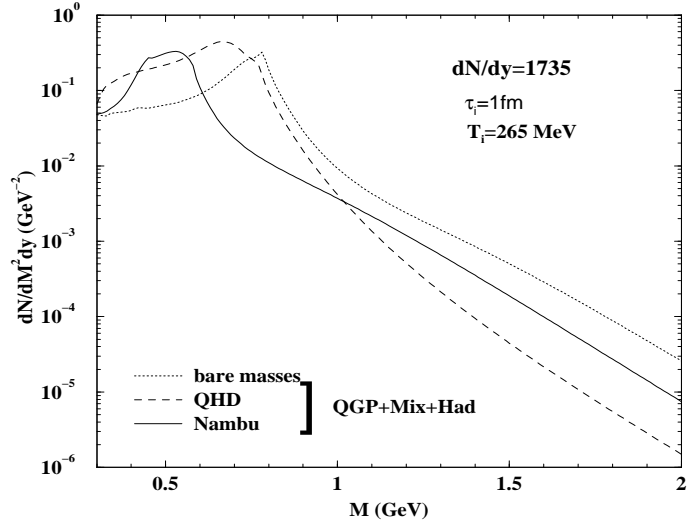,height=7cm,width=9cm}}
\caption{Thermal dilepton spectra at RHIC energies in a phase
transition scenario.
}
\label{dilrhc}
\end{figure}

Thermal dilepton yield at RHIC energies for QGP initial state 
for different mass variation scenarios are shown in 
Fig.~(\ref{dilrhc}). The shape of the peak in the dilepton
spectra in case of QHD is slightly different (broader) from the other
cases because of the larger mass separation between $\rho$ and $\omega$
mesons in this case (see Fig.~(\ref{massall})). The dilepton yield beyond the vector meson
peak is larger in the bare mass scenario because of larger initial temperature
as seen in Table~1.

\chapter{Summary and Discussions}

As emphasized in Chapter~1, electromagnetic signals are the only direct probes
of evolving strongly interacting systems and hence the study of their
spectra is of tremendous importance as far as the detection of QGP in URHICs is
concerned. In this thesis we have studied photon and dilepton production
from a thermal source composed of strongly interacting
matter, QGP or a hot hadronic gas, likely to be produced in the ultra-relativistic collisions
of heavy ions.

In Chapter~2 we have reviewed the formulation of photon and lepton pair production 
from a thermal medium in the framework of Thermal Field Theory.
We have seen that the production rate per unit time per unit volume
is proportional to the electromagnetic current correlation function
or, more generally, to the spectral function of the photon in the medium.
Neglecting the possible reinteractions of the virtual photon on its way
out of the medium, the emission rate is found to be proportional to the 
retarded self energy of the photon. In Chapter~2 we have also reviewed
how Hard Thermal Loop (HTL) resummation technique has been used to 
cancel the infra-red divergences so as to obtain finite rates of 
photon emission due to quark and gluon interactions in the QGP.

As mentioned in the introduction, our main emphasis has been to study
the effect of the medium modifications of hadronic properties, particularly
of the $\rho$ and $\omega$ mesons, on the spectra of photons and dileptons
obtained in URHICs. In Chapter~3 we have made a detailed study of 
the static (fixed temperature) rates of photon and dilepton emission from hadronic matter
incorporating medium modifications of the masses and decay widths
of vector mesons evaluated within various well-known models.
In the Quantum Hadrodynamic (QHD) model the $\rho$ and $\omega$ masses
are found to decrease differently. The disentanglement of the
$\rho$ and $\omega$ peaks in the dilepton spectrum resulting
from URHICs would be an excellent evidence of in-medium mass
shift of vector mesons~\cite{CERES} and/or the validity of 
such model calculations for the situation under consideration.
We have studied the modification of the spectral functions of the $\rho$
and $\omega$ mesons using QCD sum rules at finite temperature.
Here, the hadronic spectral
function in vacuum have been
parametrized using the experimental data on $e^+e^-\rightarrow hadrons$
in terms of a resonance and a continuum.
Due to the lack of our
understanding of the critical behaviour of scalar and tensor condensates 
the vector meson masses and continuum threshold
as a function of temperature were taken to vary according to Brown-Rho 
and Nambu scaling. The interesting possibility of the disappearance
of the $\rho$ and $\omega$ peaks in the dilepton spectra due to temperature induced
lowering of the continuum threshold was investigated.
We have also made a brief study of the $\rho$ mass modifications within the
gauged linear and non-linear sigma models and the hidden local symmetry
approach. The changes in these models are observed to be small.
The static rates of photon and dilepton production from interacting 
hadronic matter have been evaluated.
It is observed that the in-medium effects on the dilepton 
and the photon spectra are prominently visible for QHD model
and Brown-Rho and Nambu scaling scenarios. 
It is interesting to note that the dilepton
spectra spectra is affected by both the changes in the decay width as
well as in the mass of the vector mesons
whereas the photon spectra is affected
only by the change in the mass of the vector mesons and is rather 
insensitive to the change in width. The effects
of the continuum on the dilepton spectra are seen to be substantial.

In Chapter~4 the photon and dilepton yield from a longitudinally expanding
system likely to be produced in URHICs at the SPS
was contrasted between a QGP initial state and a hot hadronic gas
initial state scenarios. It was observed that the incorporation of
medium modifications resulted in  substantial changes in the evaluation
of the initial temperature as well as the cooling profile. 
When compared with the preliminary Pb-Pb data obtained by the WA98 Collaboration
we find that the photon yield in the hot hadronic gas scenario with
medium effects calculated from QHD as well as Nambu scaling scenarios
explain the data reasonably well. Similar conclusions were reached when
we compared the thermal photon yield in S-Au collisions at 200 AGeV to
the upper limits obtained by the WA80 Collaboration. The thermal photon
spectra from a hadronic gas with spectral modifications evaluated in the
Nambu scaling scenario falls within the experimental upper limits. The
corresponding QHD model calculations does not seem to qualify as a viable
possibility in this case.

We have also compared the results
of the dielectron yield to the  Pb-Au data obtained by the CERES Collaboration.
Here again the hot hadronic gas scenario with medium effects in 
the Nambu scaling scenario seem to fit the data fairly well. This is 
certainly due to the increased production of low mass electron pairs originating
mainly from the decay of $\rho$ mesons with decreased mass and enhanced width in the medium.
This is however not a direct signal of QGP because
the contribution from the QGP phase is 
very small in the low invariant mass region. 

It is interesting to note that
so far the observed low mass enhancement in the dilepton data could
not be explained without the consideration of $\rho$ mesons with a depleted mass.
In this connection it is worth mentioning that CERES is planning to substantially 
improve the mass resolution to values of the order of the natural
line width of the $\omega$ meson (8.5 MeV). It should then be possible to directly
measure the yield of the vector mesons $\rho$, $\omega$ and even $\phi$ 
($\Gamma_\phi$=4.5 MeV) including any possible changes in their mass
or decay width~\cite{itzhak}.
Also, the mass shifts of vector mesons
at zero temperature but finite baryon density could be detected 
by HADES~\cite{hades} and CEBAF~\cite{book}.

In the following we will discuss some general aspects regarding
photon and dilepton emission from evolving matter with reference
to the calculations presented in this thesis:

The photon production from QGP has been evaluated using 
HTL resummation based on the assumption $g_s<<1$, which is
impossible to meet in URHICs even at the highest energy to
be available at the CERN LHC in future. 
The strong coupling constant is likely to attain a value $g_s\sim$ 2 at
RHIC/LHC. Evaluation of the photon emission rates at such high values 
of the strong coupling is a formidable task.
In this respect the development of methods suitable for
addressing non-perturbative effects near and 
above the QCD phase
transition point is of paramount importance. 
Extension of the 
self-consistent resummation
scheme developed in $\phi^4$ theory~\cite{chiku} to 
non-abelian gauge theory~\cite{abs,bir} would be a 
very important step
towards the understanding of the phenomena near the QCD phase 
transition. 

The exact value of the critical temperature ($T_c$) for 
deconfinement phase transition is still uncertain. 
However, recent lattice simulation~\cite{ukawa}
for two flavour QCD indicates a value of $T_c$ for chiral transition 
$\sim 130-160$ MeV.  We have taken $T_c=160$ MeV, although 
it is not known whether the values of $T_c$ for the chiral and deconfinement 
transition are the same or not. 
The value of the initial thermalization time $\tau_i$ is unfortunately also
an unknown quantity. 
We take $\tau_i=1$ fm/c as a canonical value following Bjorken~\cite{bjorken}.  
A similar value of $\tau_i$ has been 
considered in the literature~\cite{janepr,csong,solf}.
The freeze-out temperature $T_f$ is  another parameter
which is not precisely known. This is important because
the yield from the hadronic phase depends substantially on the
value of $T_f$. We have also considered a baryon-free
system based on the assumption that the ratio of the baryon chemical
potential to the temperature is reasonably small. This might not be
totally justified at SPS energies, but at RHIC a baryon-free
central region is certainly a possibility.

We have assumed thermodynamic equilibrium in our calculations.
Such a situation may not be realized 
practically~\cite{ijmpa,pkr,geiger,xnwang,janeprl,rqmd}. Also,
emissions from the pre-equilibrium era may pollute the kinematical domains
where one looks for signals of QGP~\cite{pkr}.
Unfortunately, although considerable progress has been made~\cite{an,mt,mb},
the general techniques for solving non-equilibrium quantum field theoretical problems is
still in the early stages of development~\cite{kajantie}.

We have assumed that the produced system behaves as an ideal, non-dissipative
fluid. Incorporation of viscosity breaks the time-reversality of the
evolution. The ensuing generation of entropy during the temporal
evolution invalidates the role of $dN/dy$ as a handy constant of motion
and consequently affects the estimation of the initial temperature~\cite{ssvisc}.

In this thesis we have neglected transverse expansion of the evolving matter.
Incorporation of medium modifications in a 3+1 dimensionally expanding system
is a non-trivial task. We intend to take up this project in the near future.

\vskip 0.3in

We conclude with a few comments based on the experimental results 
from URHICs at the CERN SPS. From the measured hadron yields and momentum
spectra including transverse and directed flow effects there is a strong
evidence of production of a collectively behaving strongly interacting matter with a 
large volume and finite lifetime. The thermal photon spectra as well as
the pattern of low-mass enhancement of $e^+e^-$
pairs observed in S-Au, Pb-Pb and Pb-Au collisions also lend support to
the picture of a collective thermal system of hadrons. On the other hand,
the observation of a significant enhancement in multistrange baryon production
as well as the anomalous suppression of $J/\Psi$ in Pb-Pb collisions
cannot be explained without invoking partonic degrees of freedom 
and deconfined matter. At this critical and interesting juncture we
look forward to the experiments at RHIC where larger and hotter systems
are likely to be produced. 
Along with quantitative gains in all the signals there is a distinct 
possibility of direct detection of electromagnetic
radiation from the QGP phase which will pave the way for an unambiguous conclusion
regarding the formation of this novel form of matter. 

\newpage
\addcontentsline{toc}{chapter}{Appendix: Invariant Amplitudes for Photon Production}
\chapter*{Appendix}
\section*{Invariant Amplitudes for Photon Production}
\setcounter{equation}{0}
\def\theequation{A.\arabic{equation}}
\def\thefigure{A.\arabic{figure}}

We list the invariant amplitudes for photon production from
hadronic matter consisting of $\pi$, $\rho$, $\omega$, $\eta$ and $a_1$
mesons which have been considered in the evaluation of the photon
yield from hot hadronic matter. The medium modifications of the
$\rho$ and $\omega$ mesons have been taken into account in the
respective propagators.

\vskip .2in
\centerline{\underline{{(1) $\pi^+(p_1)+\pi^-(p_2)\,\ra\,\rho^0(p_3)+\gamma(p_4)$}
}}
\vskip .2in
\begin{figure}[h]
\centerline{\psfig{figure=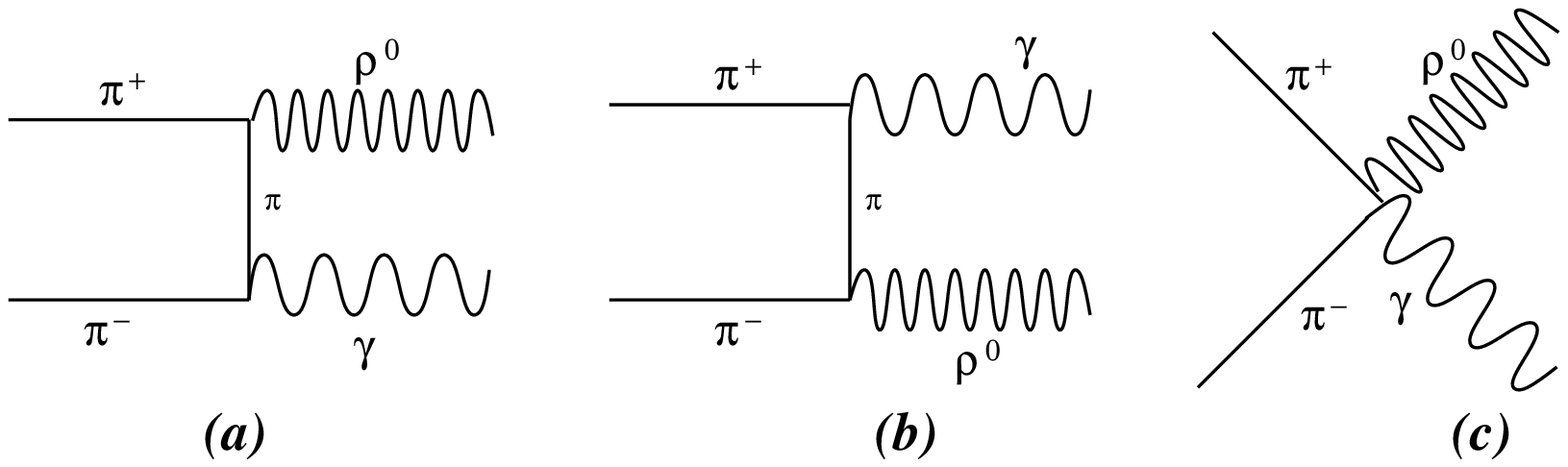,height=3cm,width=10cm}}
\caption{
Feynman diagrams for  $\pi^+\pi^-\,\ra\,\rho^0\gamma$
}
\label{reac1}
\end{figure}

\begin{eqnarray} 
\ov {\vert {\cal M}_a \vert^2}&=&\frac{16e^2g_{\rho \pi \pi}^2}
{(t-m_{\pi}^2)^2}\,
m_{\pi}^2\left[m_{\pi}^2-\frac{(m_{\pi}^2+m_{\rho}^2-t)^2}{4m_{\rho}^2}
\right]\nonumber\\
\ov {\vert {\cal M}_b \vert^2}&=&\frac{16e^2g_{\rho \pi \pi}^2}
{(u-m_{\pi}^2)^2}\,
m_{\pi}^2\left[m_{\pi}^2-\frac{(m_{\pi}^2+m_{\rho}^2-u)^2}{4m_{\rho}^2}
\right]\nonumber\\
\ov {|{\cal M}_c|^2}&=&12e^2g_{\rho \pi \pi}^2
\end{eqnarray} 
\begin{eqnarray} 
2{\mathrm {Re}}\ov {[{\cal M}_a^{\ast}{\cal M}_b]}
&=&\frac{8e^2g_{\rho \pi \pi}^2(s-2m_{\pi}^2)}
{(t-m_{\pi}^2)(u-m_{\pi}^2)}
\left[(s-2m_{\pi}^2)-\frac{(m_{\pi}^2+m_{\rho}^2-t)(m_{\pi}^2
+m_{\rho}^2-u)}{2m_{\rho}^2}\right]\nonumber\\
2{\mathrm {Re}}\ov {[{\cal M}_a^{\ast}{\cal M}_c]}
&=&\frac{8e^2g_{\rho \pi \pi}^2}{(t-m_{\pi}^2)}
\left[(s-2m_{\pi}^2)-\frac{(m_{\pi}^2+m_{\rho}^2-u)(m_{\pi}^2+m_{\rho}^2-t)}
{2m_{\rho}^2}\right]\nonumber\\
2{\mathrm {Re}}\ov {[{\cal M}_b^{\ast}{\cal M}_c]}
&=&\frac{8e^2g_{\rho \pi \pi}^2}{(u-m_{\pi}^2)}
\left[(s-2m_{\pi}^2)-\frac{(m_{\pi}^2+m_{\rho}^2-u)(m_{\pi}^2+m_{\rho}^2-t)}
{2m_{\rho}^2}\right]
\end{eqnarray} 
\vskip .5cm
\vskip .2in
\centerline{\underline{{(2) $\pi^0(p_1)+\pi^{\pm}(p_2)\,\ra\,\rho^{\pm}
(p_3)+\gamma(p_4)$}
}}
\vskip .2in
\begin{figure}[h]
\centerline{\psfig{figure=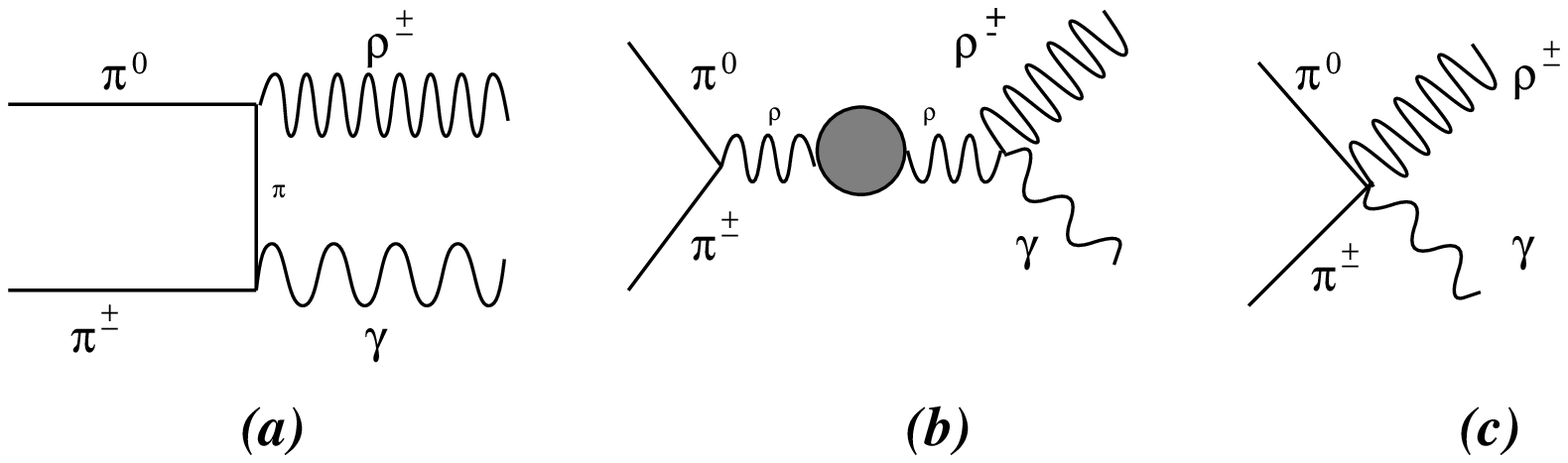,height=3cm,width=10cm}}
\caption{
Feynman diagrams for  $\pi^0\pi^{\pm}\,\ra\,\rho^{\pm}\,\gamma$
}
\label{reac2}
\end{figure}

\begin{eqnarray} 
\ov {\vert {\cal M}_a \vert^2}&=&\frac{16e^2g_{\rho \pi \pi}^2}{(t-m_{\pi}^2)^2}\,
m_{\pi}^2\left[m_{\pi}^2-\frac{(m_{\pi}^2+m_{\rho}^2-t)^2}{4m_{\rho}^2}
\right]\nonumber\\
\ov {|{\cal M}_b|^2}&=&\frac{e^2g_{\rho \pi \pi}^2}
{\left[(s-m_{\rho}^2)^2+m_{\rho}^2\Gamma_{\rho}^2\right]}
\left[2(t-u)^2+(4m_{\pi}^2-s)\left\{4m_{\rho}^2-\frac{(s-m_{\rho}^2)^2}
{m_{\rho}^2}\right\}\right]\nonumber\\
\ov {|{\cal M}_c|^2}&=&3e^2g_{\rho \pi \pi}^2
\end{eqnarray} 
\begin{eqnarray} 
2{\mathrm {Re}}\ov {[{\cal M}_a^{\ast}{\cal M}_b]}
&=&\frac{4e^2g_{\rho \pi \pi}^2(s-m_{\rho}^2)}
{(t-m_{\pi}^2)[(s-m_{\rho}^2)^2+m_{\rho}^2\Gamma_{\rho}^2]}
\left[2m_{\pi}^2(t-u)-s(s-4m_{\pi}^2)\right.\nonumber\\
&&+\left.\frac{(s-4m_{\pi}^2)(s-m_{\rho}^2)(m_{\pi}^2+m_{\rho}^2-t)}
{2m_{\rho}^2}\right]\nonumber\\
2{\mathrm {Re}}\ov {[{\cal M}_a^{\ast}{\cal M}_c]}
&=&\frac{4e^2g_{\rho \pi \pi}^2}{(t-m_{\pi}^2)}
\left[(s-2m_{\pi}^2)-\frac{(m_{\pi}^2+m_{\rho}^2-u)(m_{\pi}^2+m_{\rho}^2-t)}
{2m_{\rho}^2}\right]\nonumber\\
2{\mathrm {Re}}\ov {[{\cal M}_b^{\ast}{\cal M}_c]}
&=&\frac{e^2g_{\rho \pi\pi}^2}{m_{\rho}^2}
\frac{(t-u)(5m_{\rho}^2-s)(s-m_{\rho}^2)}{[(s-m_{\rho}^2)^2+m_{\rho}^2
\Gamma_{\rho}^2]}
\end{eqnarray} 

\vskip .2in
\centerline{\underline{{(3) $\pi^{\pm}(p_1)+\rho^0(p_2)\,\ra\,\pi^{\pm}
(p_3)+\gamma(p_4)$}
}}
\vskip .2in
\begin{figure}[htb]
\centerline{\psfig{figure=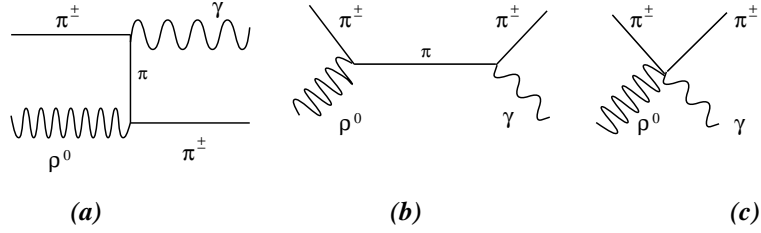,height=3cm,width=10cm}}
\caption{
Feynman diagrams for  $\pi^{\pm}\rho^0\,\ra\,\pi^{\pm}\,\gamma$
}
\label{reac3}
\end{figure}

\begin{eqnarray} 
\ov {\vert {\cal M}_a \vert^2}&=&\frac{16e^2g_{\rho \pi \pi}^2}
{3(u-m_{\pi}^2)^2}\,
m_{\pi}^2\left[m_{\pi}^2-\frac{(m_{\pi}^2+m_{\rho}^2-u)^2}{4m_{\rho}^2}
\right]\nonumber\\
\ov {\vert {\cal M}_b \vert^2}&=&\frac{16e^2g_{\rho \pi \pi}^2}
{3(s-m_{\pi}^2)^2}\,
m_{\pi}^2\left[m_{\pi}^2-\frac{(m_{\pi}^2+m_{\rho}^2-s)^2}{4m_{\rho}^2}
\right]\nonumber\\
\ov {|{\cal M}_c|^2}&=&4e^2g_{\rho \pi \pi}^2
\end{eqnarray} 
\begin{eqnarray} 
2{\mathrm {Re}}\ov {[{\cal M}_a^{\ast}{\cal M}_b]}
&=&\frac{4e^2g_{\rho \pi \pi}^2(t-2m_{\pi}^2)}
{3(u-m_{\pi}^2)(s-m_{\pi}^2)}
\left[(t-2m_{\pi}^2)-\frac{(m_{\pi}^2+m_{\rho}^2-s)(m_{\pi}^2
+m_{\rho}^2-u)}{2m_{\rho}^2}\right]\nonumber\\
2{\mathrm {Re}}\ov {[{\cal M}_a^{\ast}{\cal M}_c]}
&=&\frac{4e^2g_{\rho \pi \pi}^2}{3(u-m_{\pi}^2)}
\left[(t-2m_{\pi}^2)-\frac{(m_{\pi}^2+m_{\rho}^2-u)(m_{\pi}^2+m_{\rho}^2-s)}
{2m_{\rho}^2}\right]\nonumber\\
2{\mathrm {Re}}\ov {[{\cal M}_b^{\ast}{\cal M}_c]}
&=&\frac{4e^2g_{\rho \pi \pi}^2}{3(s-m_{\pi}^2)}
\left[(t-2m_{\pi}^2)-\frac{(m_{\pi}^2+m_{\rho}^2-s)(m_{\pi}^2+m_{\rho}^2-u)}
{2m_{\rho}^2}\right]
\end{eqnarray} 

\vskip .2in
\centerline{\underline{(4) $\pi^{\pm}(p_1)+\rho^{\mp}(p_2)\,\ra\,\pi^0(p_3)
+\gamma(p_4)$}
}
\vskip .2in
\begin{figure}[htb]
\centerline{\psfig{figure=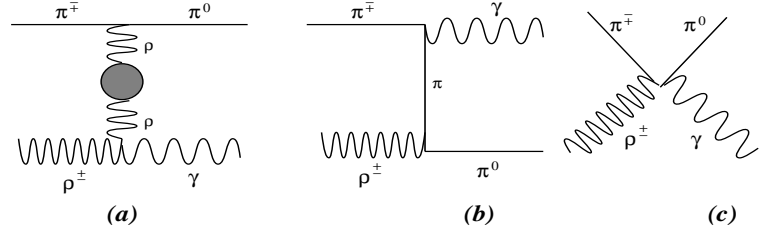,height=3cm,width=10cm}}
\caption{
Feynman diagrams for  $\pi^{\pm}\rho^{\mp}\,\ra\,\pi^0\,\gamma$
}
\label{reac4}
\end{figure}
\begin{eqnarray} 
\ov {|{\cal M}_a|^2}&=&\frac{e^2g_{\rho \pi \pi}^2}
{3\left[(t-m_{\rho}^2)^2+m_{\rho}^2\Gamma_{\rho}^2\right]}
\left[2(s-u)^2+(4m_{\pi}^2-t)\left\{4m_{\rho}^2-\frac{(t-m_{\rho}^2)^2}
{m_{\rho}^2}\right\}\right]\nonumber\\
\ov {|{\cal M}_b|^2}&=&\frac{16e^2g_{\rho \pi \pi}^2}{3(u-m_{\pi}^2)^2}\,
m_{\pi}^2\left[m_{\pi}^2-\frac{(m_{\pi}^2+m_{\rho}^2-u)^2}{4m_{\rho}^2}
\right]\nonumber\\
\ov {|{\cal M}_c|^2}&=&e^2g_{\rho \pi \pi}^2
\end{eqnarray} 
\begin{eqnarray} 
2{\mathrm {Re}}\ov {[{\cal M}_a^{\ast}{\cal M}_b]}
&=&\frac{4e^2g_{\rho \pi \pi}^2(t-m_{\rho}^2)}
{3(u-m_{\pi}^2)[(t-m_{\rho}^2)^2+m_{\rho}^2\Gamma_{\rho}^2]}
\left[2m_{\pi}^2(u-s)-t(t-4m_{\pi}^2)\right.\nonumber\\
&&+\left.\frac{(t-4m_{\pi}^2)(t-m_{\rho}^2)(m_{\pi}^2+m_{\rho}^2-u)}
{2m_{\rho}^2}\right]\nonumber\\
2{\mathrm {Re}}\ov {[{\cal M}_a^{\ast}{\cal M}_c]}
&=&\frac{e^2g_{\rho \pi\pi}^2}{3m_{\rho}^2}
\frac{(u-s)(5m_{\rho}^2-t)(t-m_{\rho}^2)}{[(t-m_{\rho}^2)^2+m_{\rho}^2
\Gamma_{\rho}^2]}\nonumber\\
2{\mathrm {Re}}\ov {[{\cal M}_b^{\ast}{\cal M}_c]}
&=&\frac{4e^2g_{\rho \pi \pi}^2}{3(u-m_{\pi}^2)}
\left[(t-2m_{\pi}^2)-\frac{(m_{\pi}^2+m_{\rho}^2-s)(m_{\pi}^2+m_{\rho}^2-u)}
{2m_{\rho}^2}\right]
\end{eqnarray} 

\vskip .5cm
\vskip .2in
\centerline{\underline{(5) $\pi^0(p_1)+\rho^{\pm}(p_2)\,\ra\,\pi^{\pm}
(p_3)+\gamma(p_4)$}
}
\vskip .2in
\begin{figure}[htb]
\centerline{\psfig{figure=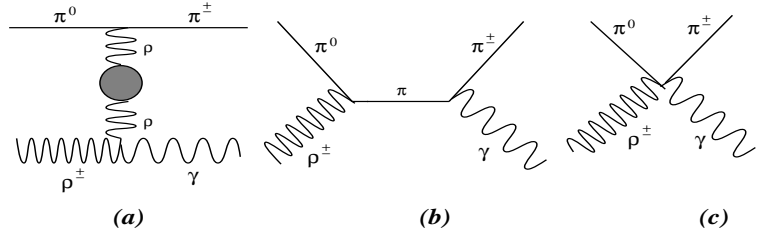,height=3cm,width=10cm}}
\caption{
Feynman diagrams for  $\pi^0\rho^{\pm}\,\ra\,\pi^{\pm}\,\gamma$
}
\label{reac5}
\end{figure}
\begin{eqnarray} 
\ov {|{\cal M}_a|^2}&=&\frac{e^2g_{\rho \pi \pi}^2}
{3\left[(t-m_{\rho}^2)^2+m_{\rho}^2\Gamma_{\rho}^2\right]}
\left[2(s-u)^2+(4m_{\pi}^2-t)\left\{4m_{\rho}^2-\frac{(t-m_{\rho}^2)^2}
{m_{\rho}^2}\right\}\right]\nonumber\\
\ov {|{\cal M}_b|^2}&=&\frac{16e^2g_{\rho \pi \pi}^2}{3(s-m_{\pi}^2)^2}\,
m_{\pi}^2\left[m_{\pi}^2-\frac{(m_{\pi}^2+m_{\rho}^2-s)^2}{4m_{\rho}^2}
\right]\nonumber\\
\ov {|{\cal M}_c|^2}&=&e^2g_{\rho \pi \pi}^2
\end{eqnarray} 
\begin{eqnarray} 
2{\mathrm {Re}}\ov {[{\cal M}_a^{\ast}{\cal M}_b]}
&=&\frac{4e^2g_{\rho \pi \pi}^2(t-m_{\rho}^2)}
{3(s-m_{\pi}^2)[(t-m_{\rho}^2)^2+m_{\rho}^2\Gamma_{\rho}^2]}
\left[2m_{\pi}^2(s-u)-t(t-4m_{\pi}^2)\right.\nonumber\\
&&+\left.\frac{(t-4m_{\pi}^2)(t-m_{\rho}^2)(m_{\pi}^2+m_{\rho}^2-s)}
{2m_{\rho}^2}\right]\nonumber\\
2{\mathrm {Re}}\ov {[{\cal M}_a^{\ast}{\cal M}_c]}
&=&\frac{e^2g_{\rho \pi\pi}^2}{3m_{\rho}^2}
\frac{(s-u)(5m_{\rho}^2-t)(t-m_{\rho}^2)}{[(t-m_{\rho}^2)^2+m_{\rho}^2
\Gamma_{\rho}^2]}\nonumber\\
2{\mathrm {Re}}\ov {[{\cal M}_b^{\ast}{\cal M}_c]}
&=&\frac{4e^2g_{\rho \pi \pi}^2}{3(s-m_{\pi}^2)}
\left[(t-2m_{\pi}^2)-\frac{(m_{\pi}^2+m_{\rho}^2-u)(m_{\pi}^2+m_{\rho}^2-s)}
{2m_{\rho}^2}\right]
\end{eqnarray} 
\vskip .5cm
\vskip .2in
\centerline{\underline{{(6) $\pi^+(p_1)+\pi^-(p_2)\,\ra\,\eta(p_3)+\gamma(p_4)$}
}}
\vskip .2in
\bef[h]
\centerline{\psfig{figure=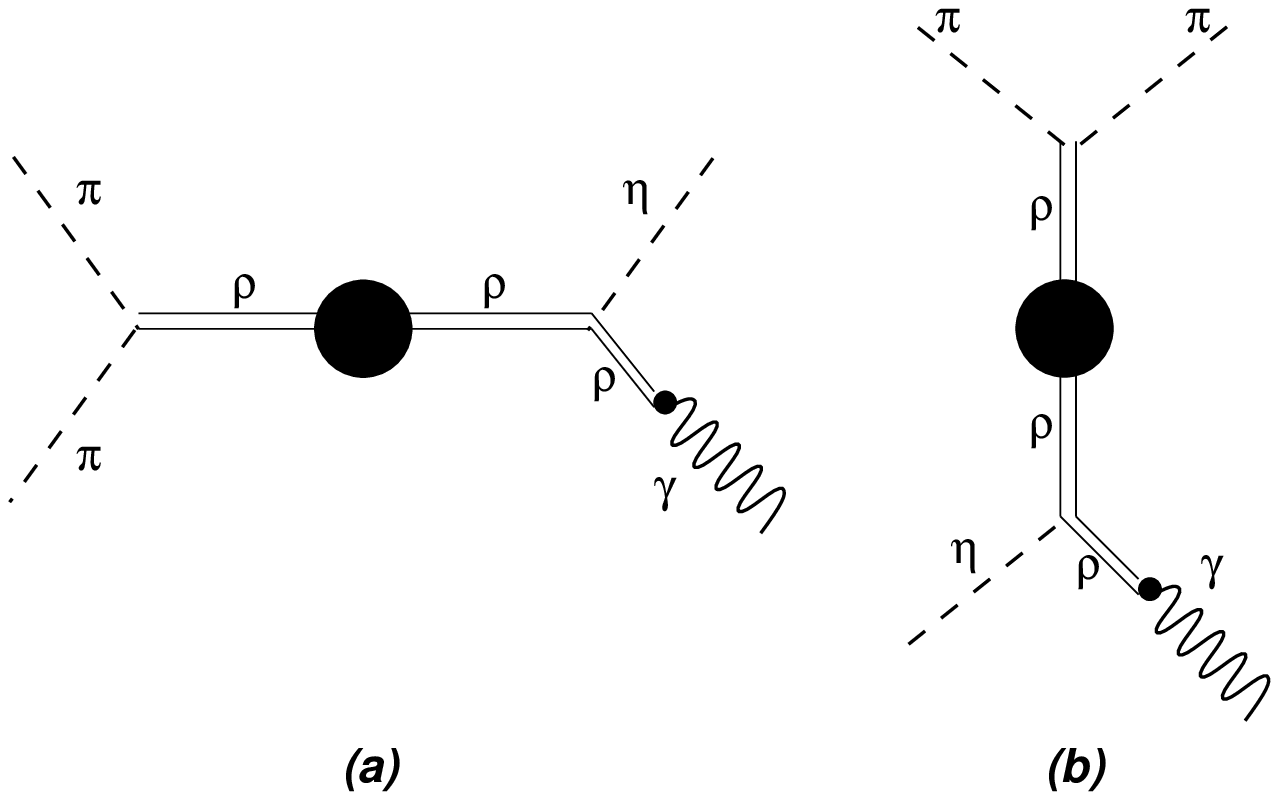,height=6cm,width=8cm}}
\caption{
Feynman diagrams for $\pi\,\pi\,\ra\,\eta\,\gamma$ and $\pi\,\eta\,\ra\,
\pi\,\gamma$.
}
\label{reac67}
\eef
\bea
|{\cal M}|^2=\frac{4\pi\,\alpha\,g_{\rho \rho \eta}^2}
{m_{\eta}^2[(s-m_{\rho}^2)^2+m_{\rho}^2\Gamma_{\rho}^2]}\,
\left[\frac{}{}s(u-m_{\pi}^2)(t-m_{\pi}^2)-m_{\pi}^2(s-m_{\eta}^2)^2\right]
\eea
\vskip .5cm
\vskip .2in
\centerline{\underline{{(7) $\pi^{\pm}(p_1)+\eta(p_2)\, \ra\,\pi^{\pm}(p_3)+\gamma(p_4)$}
}}
\vskip .2in
\bea
|{\cal M}|^2=\frac{4\pi\,\alpha\,g_{\rho \rho \eta}^2}
{m_{\eta}^2[(t-m_{\rho}^2)^2+m_{\rho}^2\Gamma_{\rho}^2]}\,
\left[\frac{}{}t(u-m_{\pi}^2)(s-m_{\pi}^2)-m_{\pi}^2(t-m_{\eta}^2)^2\right]
\eea
The coupling constant $g_{\rho \rho \eta}$ is evaluated from the following 
relations:
\bea
\Gamma(\rho\,\ra\,\eta\,\gamma)&=&\frac{(m_{\rho}^2-m_{\eta}^2)^3}
{96\pi\,m_{\eta}^2\,m_{\rho}^3}\,g_{\eta \gamma \rho}^2\nonumber\\
g_{\eta \gamma \rho}&=&\frac{e\,g_{\rho \rho \eta}}{g_{\rho \pi \pi}}
\nonumber\\
\frac{g_{\rho \pi \pi}^2}{4\pi}&=&2.9\nonumber
\eea
and $\Gamma(\rho\,\ra\,\eta\,\gamma) = (57 {\pm} 10.5)$ keV.

\bef[h]
\centerline{\psfig{figure=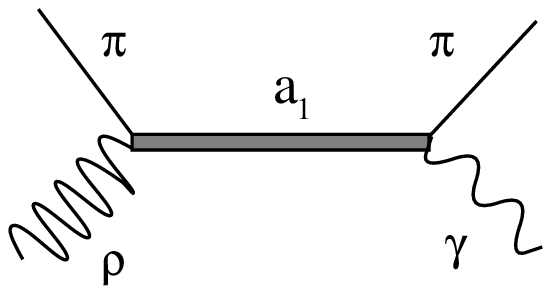,height=2cm,width=4cm}}
\caption{
Feynman diagram for $\pi\,\rho\,\ra\,a_1\,\ra\,\pi\,\gamma$.
}
\eef
\vskip .2in
\centerline{\underline{{(8) $\pi + \rho \ra\, a_1 \,\ra\pi + \gamma$}
}}
\vskip .2in
\be
|{\cal M}|^2=\frac{4\pi\alpha g_{a_1\rho\pi}^2}{g^2}
\frac{1}{(s-m_{a_1}^2)^2+m_{a_1}^2\Gamma_{a_1}^2}
\left[X_1\,+\,X_2\,-\,X_3\right]
\ee
where
\begin{eqnarray*}
X_1&=&\frac{f_{a_1\rho\pi}^2}{4}
\left(x^2+\frac{z}{m_\rho^2}xy+2sy-\frac{s}{m_\rho^2}y^2\right)\nonumber\\
X_2&=&\frac{g_{a_1\rho\pi}^2}{4}
\left(sm_\rho^2x^2-\frac{1}{4}x^2z^2-sxyz+\frac{1}{4m_\rho^2}z^3yx+s^2y^2
-\frac{s}{4m_\rho^2}z^2y^2\right)\nonumber\\
X_3&=&\frac{1}{2}g_{a_1\rho\pi}f_{a_1\rho\pi}\left(\frac{x^2z}{2}-
\frac{z^2xy}{2m_\rho^2}+\frac{s}{2m_\rho^2}zy^2\right)\nonumber\\
\end{eqnarray*}
and
\begin{eqnarray*}
x&=&s-m_\pi^2\,\,,y=m_\rho^2-t\,\,, z=x+m_\rho^2\nonumber\\
f_{a_1\rho\pi}&=&\frac{g^2f_\pi}{Z_\pi}\left(2c+Z_\pi + 
\frac{s+m_\rho^2-m_\pi^2}{2m_{a_1}^2}-Z_\pi\kappa_6\frac{s-m_\rho^2-m_\pi^2}
{2m_{\rho}^2}\right)\nonumber\\
g_{a_1\rho\pi}&=&\frac{g^2f_\pi}{Z_\pi}\left(\frac{Z_\pi\kappa_6}{m_\rho^2} 
-\frac{1}{m_{a_1}^2}\right)\nonumber\\
\end{eqnarray*}

$g$=5.04 (from $\rho\ra e^+e^-$ decay)

$c$=-0.12, $Z_\pi=0.17$, $m_{a_1}=1.26 $ GeV,  $m_\sigma=0.7$ GeV,
$\kappa_6=1.25$ (from $\rho\ra \pi\pi$ decay)
\vskip 0.2in
\vskip .2in
\centerline{\underline{(9) $\rho(p_1)\,\ra\,\pi(p_2)\,\pi(p_3)\gamma(p_4)$,
~~$\omega(p_1)\,\ra\,\pi(p_2)\,\gamma(p_3)$}}
\vskip .2in
\begin{figure}[h]
\centerline{\psfig{figure=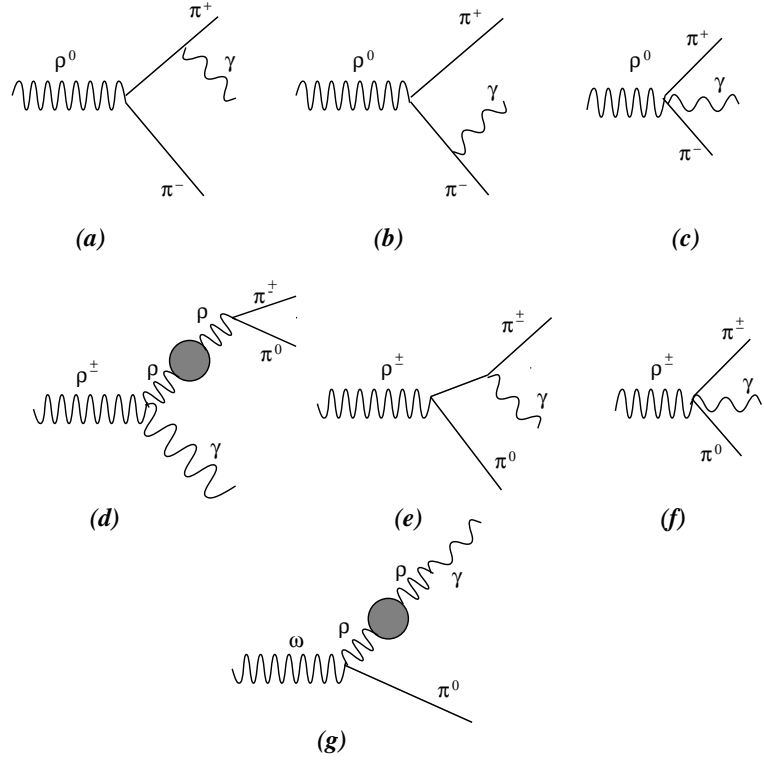,height=10cm,width=10cm}}
\caption{
Feynman diagrams for vector meson decays.
}
\label{vecdec}
\end{figure}
\begin{eqnarray} 
\ov {\vert {\cal M}_a \vert^2}&=&\frac{16e^2g_{\rho \pi \pi}^2}
{3(t-m_{\pi}^2)^2}\,
m_{\pi}^2\left[m_{\pi}^2-\frac{(m_{\pi}^2+m_{\rho}^2-t)^2}{4m_{\rho}^2}
\right]\nonumber\\
\ov {\vert {\cal M}_b \vert^2}&=&\frac{16e^2g_{\rho \pi \pi}^2}
{3(s-m_{\pi}^2)^2}\,
m_{\pi}^2\left[m_{\pi}^2-\frac{(m_{\pi}^2+m_{\rho}^2-s)^2}{4m_{\rho}^2}
\right]\nonumber\\
\ov {|{\cal M}_c|^2}&=&4e^2g_{\rho \pi \pi}^2
\end{eqnarray} 
\begin{eqnarray} 
2{\mathrm {Re}}\ov {[{\cal M}_a^{\ast}{\cal M}_b]}
&=&\frac{8e^2g_{\rho \pi \pi}^2(u-2m_{\pi}^2)}
{3(t-m_{\pi}^2)(s-m_{\pi}^2)}
\left[(u-2m_{\pi}^2)-\frac{(m_{\pi}^2+m_{\rho}^2-t)(m_{\pi}^2
+m_{\rho}^2-s)}{2m_{\rho}^2}\right]\nonumber\\
2{\mathrm {Re}}\ov {[{\cal M}_a^{\ast}{\cal M}_c]}
&=&\frac{8e^2g_{\rho \pi \pi}^2}{3(t-m_{\pi}^2)}
\left[(u-2m_{\pi}^2)-\frac{(m_{\pi}^2+m_{\rho}^2-s)(m_{\pi}^2+m_{\rho}^2-t)}
{2m_{\rho}^2}\right]\nonumber\\
2{\mathrm {Re}}\ov {[{\cal M}_b^{\ast}{\cal M}_c]}
&=&\frac{8e^2g_{\rho \pi \pi}^2}{3(s-m_{\pi}^2)}
\left[(u-2m_{\pi}^2)-\frac{(m_{\pi}^2+m_{\rho}^2-s)(m_{\pi}^2+m_{\rho}^2-t)}
{2m_{\rho}^2}\right]
\end{eqnarray} 

\begin{eqnarray} 
\ov {|{\cal M}_d|^2}&=&\frac{e^2g_{\rho \pi \pi}^2}
{3\left[(u-m_{\rho}^2)^2+m_{\rho}^2\Gamma_{\rho}^2\right]}
\left[2(t-s)^2+(4m_{\pi}^2-u)\left\{4m_{\rho}^2-\frac{(u-m_{\rho}^2)^2}
{m_{\rho}^2}\right\}\right]\nonumber\\
\ov {|{\cal M}_e|^2}&=&\frac{16e^2g_{\rho \pi \pi}^2}{3(t-m_{\pi}^2)^2}\,
m_{\pi}^2\left[m_{\pi}^2-\frac{(m_{\pi}^2+m_{\rho}^2-t)^2}{4m_{\rho}^2}
\right]\nonumber\\
\ov {|{\cal M}_f|^2}&=&e^2g_{\rho \pi \pi}^2
\end{eqnarray} 
\begin{eqnarray} 
2{\mathrm {Re}}\ov {[{\cal M}_d^{\ast}{\cal M}_e]}
&=&\frac{4e^2g_{\rho \pi \pi}^2(u-m_{\rho}^2)}
{3(t-m_{\pi}^2)[(u-m_{\rho}^2)^2+m_{\rho}^2\Gamma_{\rho}^2]}
\left[2m_{\pi}^2(t-s)-u(u-4m_{\pi}^2)\right.\nonumber\\
&&+\left.\frac{(u-4m_{\pi}^2)(u-m_{\rho}^2)(m_{\pi}^2+m_{\rho}^2-t)}
{2m_{\rho}^2}\right]\nonumber\\
2{\mathrm {Re}}\ov {[{\cal M}_d^{\ast}{\cal M}_f]}
&=&\frac{e^2g_{\rho \pi\pi}^2}{3m_{\rho}^2}
\frac{(t-s)(5m_{\rho}^2-u)(u-m_{\rho}^2)}{[(u-m_{\rho}^2)^2+m_{\rho}^2
\Gamma_{\rho}^2]}\nonumber\\
2{\mathrm {Re}}\ov {[{\cal M}_e^{\ast}{\cal M}_f]}
&=&\frac{4e^2g_{\rho \pi \pi}^2}{3(t-m_{\pi}^2)}
\left[(u-2m_{\pi}^2)-\frac{(m_{\pi}^2+m_{\rho}^2-t)(m_{\pi}^2+m_{\rho}^2-s)}
{2m_{\rho}^2}\right]
\end{eqnarray} 

\be
\ov {|{\cal M}_g|^2}=\frac{2\pi\alpha}{3}\,\left(\frac{g_{\opr}}{g_{\rpp}}\right)^2
\,\frac{m_{\rho}^4}{m_{\pi}^2}\,\frac{(m_{\omega}^2-m_{\pi}^2)^2}
{[(t-m_{\rho}^2)^2+m_{\rho}^2 \Gamma_{\rho}^2]}
\ee


\end{document}